\documentclass[referee]{raa}
\usepackage{graphicx,times}
\usepackage{longtable,lscape}
\begin{document}

   \title{Flux and Spectral Variation Characteristics of 3C 454.3 at the GeV Band}

   \volnopage{Vol.0 (200x) No.0, 000--000}      %%preserved for Editor. DOn't remove!
   \setcounter{page}{1}          %%starting page, preserved for Editor. DOn't remove!

   \author{Hai-Ming Zhang
      \inst{1}
   \and Jin Zhang
      \inst{2\dag}
   \and Rui-Jing Lu
      \inst{1}
   \and Ting-Feng Yi
      \inst{3}
   \and Xiao-Li Huang
      \inst{1}
   \and En-Wei Liang
	 \inst{1}
   }
%% Here is an example of three authors come from different institutes.
%% For single author or all the authors from an institute, use "\inst{}" only

   \institute{Guangxi Key Laboratory for Relativistic Astrophysics, Department of Physics, Guangxi University,
		  Nanning 530004, China;\\
%% Please give the E-mail address of the author, to whom future correspondence and
%% offprint requests will be sent.
        \and
             Key Laboratory of Space Astronomy and Technology, National Astronomical Observatories, Chinese Academy of Sciences, Beijing 100012, China; {\it jinzhang@bao.ac.cn}\\
        \and
             Department of Physics, Yunnan Normal University, Kunming 650500, China.\\
   }

   \date{Received~~201X month day; accepted~~201X~~month day}

\abstract{We analyze the long-term lightcurve of 3C 454.3 observed with \emph{Fermi}/LAT and investigate its relation to the flux in the radio, optical, and X-ray bands. By fitting
the 1-day binned GeV lightcurve with multiple Gaussian function (MGF), we propose that the typical variability timescale in the GeV band is 1--10 days. The GeV flux variation is accompanied by the spectral variation characterized as flux-tracking, i.e., ``harder when brighter". The GeV flux is correlated with the optical and X-ray fluxes, and a weak correlation between $\gamma$-ray flux and radio flux is also observed. The $\gamma$-ray flux is not correlated with the optical linear polarization degree for the global lightcurves, but they show a correlation for the lightcurves before MJD 56000. The power density spectrum of the global lightcurve shows an obvious turnover at $\sim7.7$ days, which may indicate a typical variability timescale of 3C 454.3 in the $\gamma$-ray band. This is also consistent with the derived timescales by fitting the global lightcurve with MGF. The spectral evolution and an increase of the optical linear polarization degree along with the increase of the $\gamma$-ray flux may indicate that the radiation particles are accelerated and the magnetic field is ordered by the shock processes during the outbursts. In addition, the nature of 3C 454.3 may be consistent with the self-organized criticality system, similar to Sagittarius A$^{*}$, and thus the outbursts are from plasmoid ejections driven by magnetic reconnection. This may further support the idea that the jet radiation regions are magnetized.
}

   \authorrunning{H. M. Zhang \& J. Zhang et al. }            %author_head in even pages
   \titlerunning{Characteristics of 3C 454.3 at the GeV Band}  % title_head in odd pages
   \maketitle
%% The author head (on even pages) and the title head (on odd pages) will be
%% automatically extracted from \author{} and \title{}. Whenever the title is too long,
%% you will be asked to supply a shorter one by inserting either \authorrunning{} or
%% \titlerunning{} before \maketitle. Anyway, you can specify your own heads.
%%
%%
%% Note: In the following text body of your manuscript, please note several differences from
%%       other major journals:
%% (1) \subsection{Please Capitalize the First Letter of Each Notional Word in Subsection Title}
%% (2) Please Capitalize the First Letter of Each Notional Word in all tables' captions

%
%________________________________________________ sections below
%

%%%%%%%%%%%%%%%%%%%%%%%%%%%%%%%%%%%%%%%%%%%%%%%%%%%%%%%%%%%%%%%%
\section{Introduction}           %% first-level sections will be auto-capitalized
\label{sect:intro}

Variabilities in multi-wavelength with timescales from one year to several hours, even as short as a few minutes in the GeV-TeV band in some extreme cases, are observed in blazars (e.g., Fossati et al. 2008; Abdo et al. 2010; Aleksi\'{c} et al. 2011; Arlen et al. 2013; Liao et al. 2016; Hong et al. 2017). The physical origin of the variability is still debated. The variability may be related with the activities of central engine, e.g., newly-emerging component (Arlen et al. 2013; Jorstad et al. 2013), or the change of jet physical property, e.g., the variations of Lorenz factors and particle acceleration mechanisms (Villata et al. 2007; Raiteri et al 2011; Nalewajko 2013; Zhu et al. 2016), or the geometric mechanism, e.g., the fractal helical structure jets (Larionov et al. 2010), or an emission blob smaller than the jet cross section, e.g., `jet-in-jet' or spine-layer model (e.g., Ghisellini \& Tavecchio 2008; Chen 2017). The quasi-periodic oscillation (QPO), such as the 12-year QPO in OJ 287 was proposed due to a binary system of supermassive black holes in the center (Sillanpaa et al. 1988, 1996; Lehto \& Valtonen 1996).

The flux variation is always accompanied with the spectral variation. This may shed light on the particle acceleration and the properties of the radiation regions. The observed relations between spectral index and flux in blazars are diverse. Cui (2004) reported that the spectral evolution behaviors of Mrk 421 in the X-ray band are very complex and different during different flares. An increase of the synchrotron peak flux along with the increase of its peak position is observed in Mrk 421 and the other four TeV BL Lacs (PKS 0548--322, 1H 1426+418, Mrk 501 and 1ES 1959+650, Massaro et al. 2008; Tramacere et al. 2009), and the similar feature has also been reported in the multiwavelength radiation mechanism research of blazars (Zhang et al. 2012, 2013). However, Nalewajko (2013) did not find a uniform feature of the spectral variation in the GeV band with a sample of 40 brightest $\gamma$-ray flares observed by \emph{Fermi}/LAT.

3C 454.3 is one of the most active blazars in multi-wavelengthes from the radio to the $\gamma$-ray bands (Ackermann et al. 2010; Jorstad et al. 2010, 2013; Raiteri et al. 2011; Wehrle et al. 2012; Britto et al. 2016). An extraordinary outburst at the $\gamma$-ray band in 2009 December was observed with \emph{Fermi}/LAT, which makes 3C 454.3 to be the brightest $\gamma$-ray source in the sky for over one week, and since then several brighter outbursts in the $\gamma$-ray band were detected by \emph{Fermi}/LAT (Ackermann et al. 2010; Pacciani et al. 2014; Britto et al. 2016). By observing the parsec-scale (pc) jet in 3C 454.3 with the Very Long Baseline Array (VLBA) during its pronounced flaring in 2005-2008, Jorstad et al. (2010) suggested that a superluminal knot emerges from the core, which yields a series of optical and high-energy outbursts. The similar phenomenon was observed again in 3C 454.3 during its $\gamma$-ray outbursts in 2009 and 2010 (Jorstad et al. 2013). These factors may imply that the variability of 3C 454.3 is connected with its central engine activities.

In this paper, we dedicate to present a comprehensive analysis on the GeV emission of 3C 454.3 with the \emph{Fermi}/LAT observation data and investigate its relation to the emission in X-ray, optical and radio bands. The reduction of the \emph{Fermi}/LAT data is given in Section 2. Variability behaviors in the GeV band are described in Section 3. Using the well-sampled observation data of \emph{Fermi}/LAT, we investigate the spectral evolution in the GeV band in Section 4. The cross-correlation analysis of variability among multiple
wavelength lightcurves is presented in Section 5. A discussion and summary are reported in Section 6 and Section 7, respectively.

\section{\emph{Fermi}/LAT Data Reduction}
The data observed with \emph{Fermi}/LAT (Pass 8 data) from 2008 August 6 (Modified Julian Day, MJD 54684) to 2016 February 16 (MJD 57434) for 3C 454.3 are used for our analysis. The temporal coverage of the data is 7.5 years. Our data analysis is performed with the standard analysis tool \emph{gtlike/pyLikelihood}, which is part of the \emph{Fermi} Science Tools software package\footnote{https://fermi.gsfc.nasa.gov/ssc/data/analysis/software/} (ver. v10r0p5). The P8R2-SOURCE-V6 set of instrument response functions (IRFs) was used.

Photons with energies from 0.1 to 200 GeV were taken into account for our analysis. The significance of the $\gamma$-ray signal from the source is evaluated with the maximum-likelihood test statistic (TS), where the events with TS $\geq$ 21 were taken. They are selected from the region of interest (ROI) with radius of 10$^{\circ}$, centered at the position of 3C 454.3. The Galactic longitude and latitude are $86.1^{\circ}$ and $-38.2^{\circ}$, respectively. The isotropic background, including the sum of residual instrumental background and extragalactic diffuse $\gamma$-ray background, was fitted with a model derived from the isotropic background at high Galactic latitude, i.e., ``iso-P8R2-SOURCE-V6-v06.txt", and the Galactic diffuse GeV emission was modeled with ``gll-iem-v06.fits"\footnote{Taken from http://fermi.gsfc.nasa.gov/ssc/data/access/lat/BackgroundModels.html}. In order to eliminate the contamination from the $\gamma$-ray-bright Earth limb, the events with zenith angle 100$^{\circ}$ were excluded.

A power-law spectral model is used to fit the observed spectrum in  each time bin with an unbinned maximum likelihood method, i.e.,
\begin{equation}
\frac{dN}{dE}=\frac{N(\Gamma+1)E^{\Gamma}}{E_{\rm max}^{\Gamma+1}-E_{\rm min}^{\Gamma+1}},
\end{equation}
where $\Gamma$ is the photon spectral index for the events in the energy band between $E_{\rm min}$ (0.1 GeV) and $E_{\rm max}$ (200 GeV).

\section{Variability in the GeV Band}

Figure \ref{LAT}(a) shows the derived LAT lightcurve of 3C 454.3 in a time bin of 1 day. Note that no observational data with TS $\geq$ 21 are available in the time intervals from  MJD 54797 to MJD 54939 and from MJD 55758 to MJD 56228. One can observe that 3C 454.3 experienced several violent outbursts in the $\gamma$-ray band during the past $\sim$8 years. An extremely bright flare is observed with the highest luminosity of $1.60\pm0.06\times10^{50}$ erg s$^{-1}$ on 20 November 2010 (MJD 55520), which sets up at $\sim$ MJD 55478 and returns to the quiet state at $\sim$ MJD 55594, lasting more than three months with other three obvious flare peaks.

Giant outbursts are composed of many flares. We divide the global GeV lightcurve into 14 outburst episodes in order to demonstrate these flares. We do not adopt a rigid criterion to select each episode. It usually starts at a time when the flux goes up from a baseline and ends at a time when the flux goes down to the baseline. Each episode is composed of several pulses. We fit the lightcurve of each episode with multiple Gaussian functions. Our results are shown in Figure \ref{Gaussian}. The global lightcurve during the 7.5 years in the $\gamma$-ray band is fitted with 236 Gaussian components, corresponding 236 flares.
We calculate the full width at half maximum (FWHM), the peak luminosity ($L_{\rm peak}$), the total radiation energy ($E_{\gamma}$) of each flare, where $E_{\gamma}$ is derived by the integrating emission in the duration of each Gaussian component in the 0.1--200 GeV energy band. The results are listed in Table 1.

On the basis of the flare-finding with the Gaussian-fitting method, we also find that the GeV lightcurves in some episodes are composed of two components, spiked flares with short-timescale and broad flares with long-timescale, as shown in Figure \ref{Gaussian}. Most flares are the spike flares. The FWHM values of the broad components are usually tens days and they are dimmer than the spiked flares. It is possible that the two components are from the different radiation regions.

The distributions of FWHM, $L_{\rm peak}$, and $E_{\gamma}$ are shown in Figure \ref{Distributions}. It is found that the FWHM values range from $\sim$1 day to $\sim$25 days, clustered at 2--5 days. Although the time-bin selection effect (1-day in this analysis) may lead a bias on the intrinsic FWHM distribution, it still indicates that the timescales of the GeV flares should be several days. The $L_{\rm peak}$ values of these flares range from 10$^{47.6}$ erg s$^{-1}$ to 10$^{50.2}$ erg s$^{-1}$ and narrowly cluster within 10$^{48.3}$--10$^{49.2}$ erg s$^{-1}$. $E_{\gamma}$ ranges from $10^{53.0}$ to $10^{55.6}$ erg, clustering at $10^{53.6}$ to $10^{54.6}$ erg.

To further investigate the variability properties of 3C 454.3 in the GeV band, we calculate the power density spectrum (PDS) of the global lightcurve in the GeV band with the Lomb-Scargle Periodogram (LSP) algorithm (Lomb 1976; Scarle 1982). The PDS curve is shown in Figure \ref{episode_PDS}. We use a broken power-law function to fit the PDS curve of 3C 454.3, and the posterior probability density of the parameters of the model is derived by a Bayesian Markov chain Monte Carlo technique (Vaughan 2010). The derived turnover is at $\sim7.7$ days, which may correspond to a typical variability timescale of 3C 454.3 in the GeV band. This is roughly consistent with the FWHM distribution.

\section{Correlations Between Spectral Index and Flux Variations}

\begin{figure*}
\includegraphics[angle=0,scale=0.41]{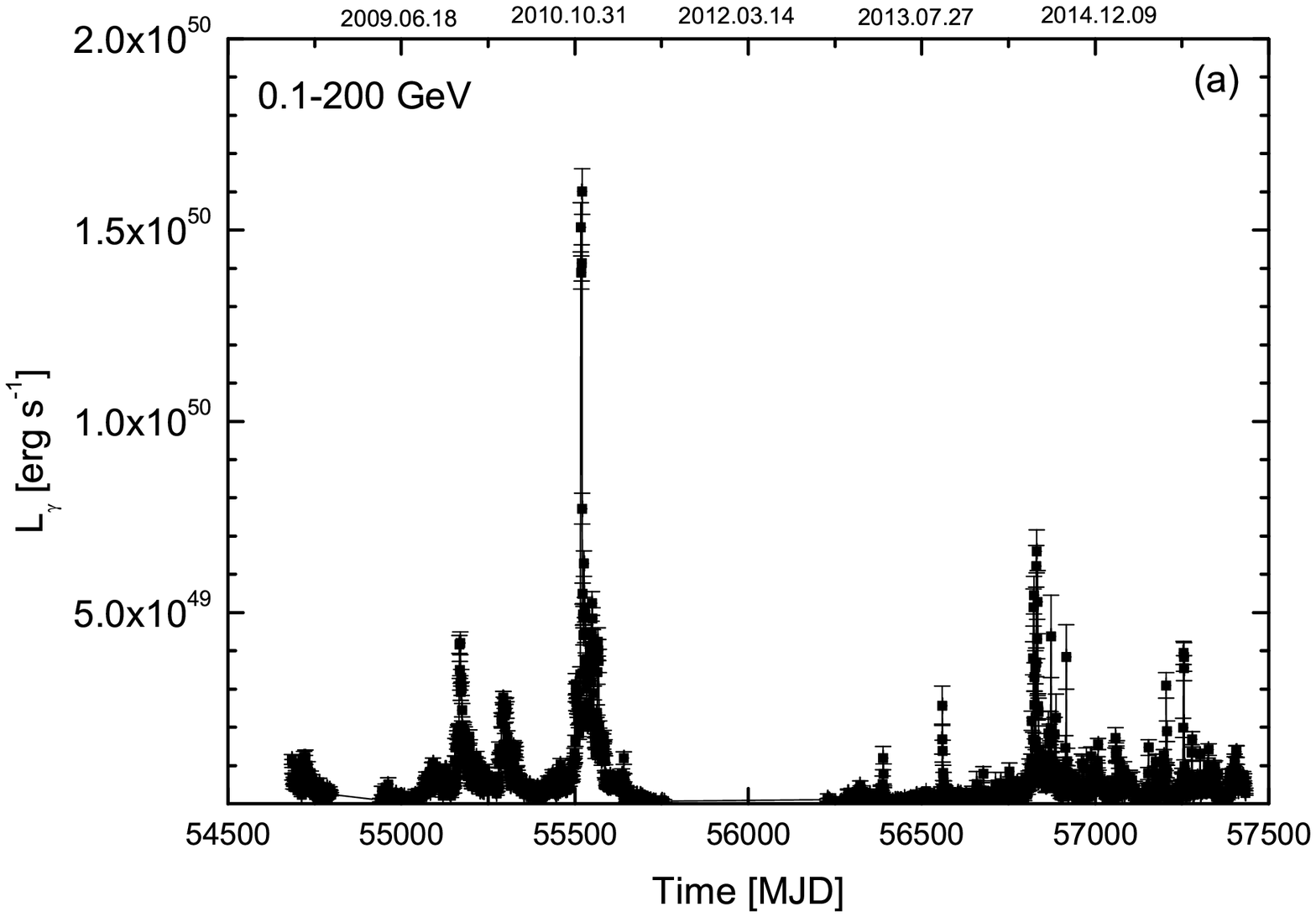}
\includegraphics[angle=0,scale=0.33]{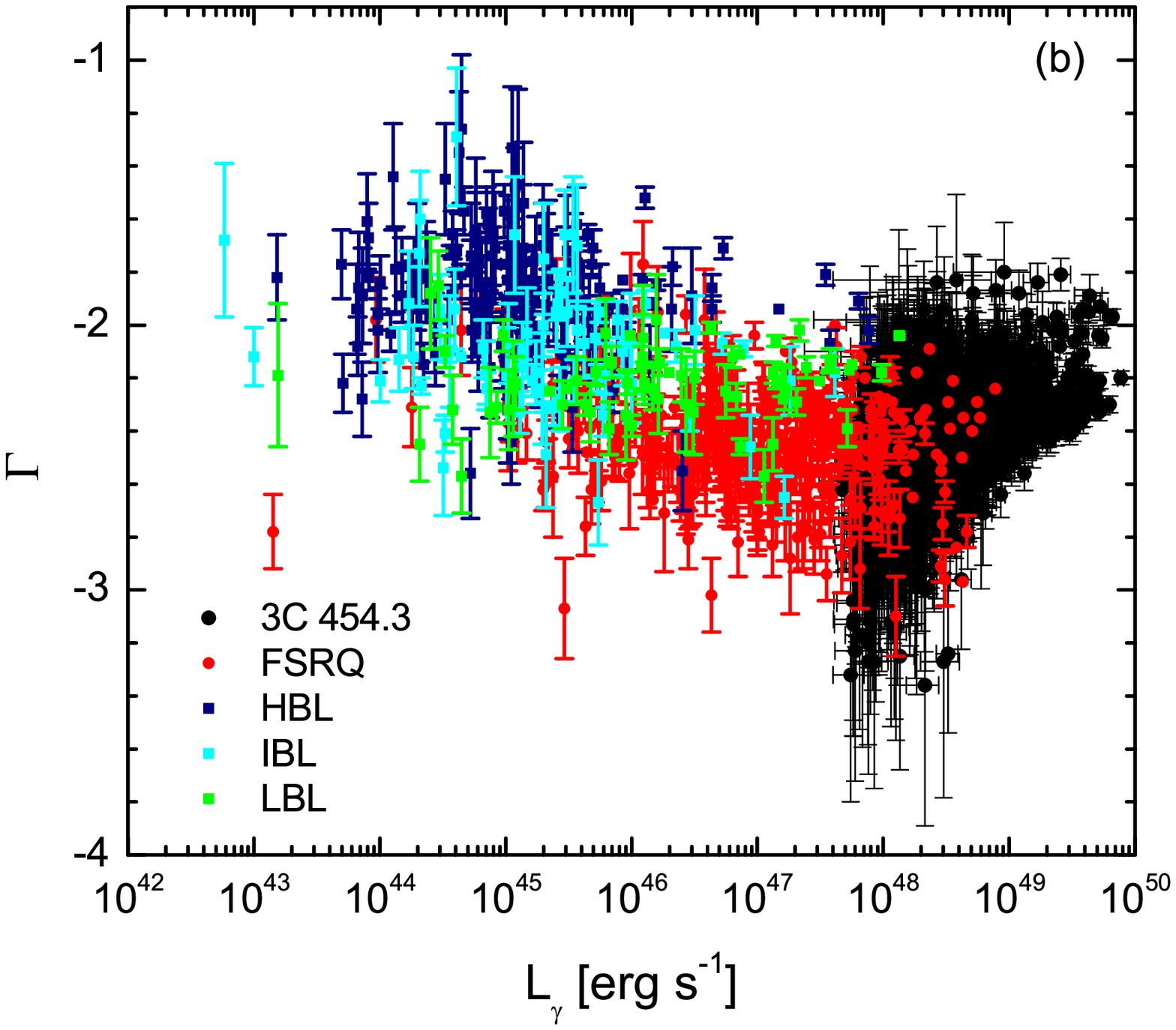}
\caption{\emph{Panel (a)}---Global lightcurve of 3C 454.3 in the GeV band, where the observed luminosity ($L_{\gamma}$) is the integral luminosity of \emph{Fermi}/LAT energy band (from 100 MeV to 200 GeV). \emph{Panel (b)}---$\Gamma$ as a function of $L_{\gamma}$ for 3C 454.3. The data of blazars are taken from Ackermann et al. (2015), where HBL, IBL, and LBL indicate high-frequency-peaked BL Lac, intermediate-frequency peaked BL Lac, and low-frequency-peaked BL Lac, respectively.}\label{LAT}
\end{figure*}

Spectral index and flux variation may shed light on the particle acceleration and radiation physics. Figure \ref{Gaussian} also shows the temporal evolution of photon spectral index ($\Gamma$) in the $\gamma$-ray band for each episode. One can find that the flux variation is accompanied by the $\Gamma$ variation. Figure \ref{LAT}(b) illustrates the variation of 3C 454.3 in the $L_{\gamma}$--$\Gamma$ plane, where the \emph{Fermi} blazars taken from Ackermann et al. (2015, see also their Figure 14) are also presented in order to make comparison. The data of 3C 454.3 distribute in the high luminosity end in the $L_{\gamma}$--$\Gamma$ plane. Its $\Gamma$ values range from $-1.80\pm0.19$ (in MJD [56974, 56975]) to $-3.36\pm0.53$ (in MJD [56922, 56923]), and are larger than $-2$ in some bright flares, such as that observed in the episodes of MJD [56545, 56575], MJD [56790, 56852], MJD [56860, 56890], and MJD [56960, 57040].

Using the Pearson correlation analysis method, we analyze the correlation between $L_{\gamma}$ and $\Gamma$ for 3C 454.3 in different flux stages. A tentative correlation with a correlation coefficient of $r=0.46$ and a chance probability of $p\sim0$ is found. To further study this issue, we investigate the $\Gamma-L_\gamma$ correlation in 34 bright outbursts, among them 6 outbursts having $L_{\rm max}/L_{\rm min}\ge10$, 13 outbursts having $5\leq L_{\rm max}/L_{\rm min}<10$, and 15 outbursts having $3\le L_{\rm max}/L_{\rm min}<5$, where $L_{\rm max}$ and $L_{\rm min}$ are the maximum and minimum luminosities in the selected outbursts. Note that we do not have a critical criterion to separate these outbursts. We just only artificially select the time intervals that have bright flares. The maximum and minimum durations among the 34 outbursts are 45 days and 3 days, respectively. The values of $L_{\rm max}$, $L_{\rm min}$, $L_{\rm max}/L_{\rm min}$ as well as the duration of each outburst are reported in Table 2.

Figure \ref{episode} shows the temporal variations of $L_{\gamma}$ and $\Gamma$ as well as $\Gamma$ as a function of $L_{\gamma}$ for the 34 selected outbursts. Most of outburst episodes exhibit a strong correlation between $\Gamma$ and $L_{\gamma}$. The correlation coefficient and chance probability of the Pearson correlation analysis with slope of linear fits for each outburst episode are reported in Table 2, where the errors of both $\Gamma$ and $L_{\gamma}$ are considered during the linear fitting. These results indicate that the variations of luminosity are accompanied with the spectral evolution in most of outburst cases for 3C 454.3, showing the behavior of flux tracking as ``harder when brighter".

We also study the correlation and possible lag behavior between $L_{\gamma}$ and $\Gamma$ in each outburst episode using the discrete cross-correlation function (DCF,  Edelson \& Krolik 1988). The results of the DCF analysis are presented in the right panels of Figure \ref{episode}. The correlation between $L_{\gamma}$ and $\Gamma$ for 28 outbursts is significant over the 95\% confidence level, and except for the outburst episodes 5 (10 days), 6 (-1 days), 8 (5 days), no lag behavior is found in other outburst episodes.

Note that the above analysis is on the basis of the one-day binning lightcurves. The fast variabilities on timescale of hours (even shorter) had been reported for 3C 454.3 (Abdo et al. 2011; Britto et al. 2016). Therefore, we also re-analyze the two outbursts that have a peak luminosity $L_{\gamma} > 5\times 10^{49}$ erg s$^{-1}$ in Figure \ref{LAT}(a) by using a time-bin of 3-hour. As illustrated in Figure \ref{flare}, the statistical correlations between $L_{\gamma}$ and $\Gamma$ are still presented during the fast flares (see also Abdo et al. 2011; Britto et al. 2016). The Pearson correlation analysis yields $r=0.59$ in a chance probability of $p=3.1\times10^{-5}$ and $r=0.76$ with $p=6.0\times10^{-9}$ for the two flares, respectively. And no lag behavior is found in the DCF analysis. These facts may imply that the behavior of ``harder when brighter" is an intrinsic property of this source, which is independent of the time-bin size of analysis.

\section{Correlations of Emission between $\gamma$-ray and Other Energy Bands}

The long-term simultaneously observed lightcurves of 3C 454.3 in X-ray, optical, and radio 43 GHz, as well as the polarization data are shown in Figure \ref{lightcurve}, which are taken from the web ``\emph{http://www.bu.edu/blazars/VLBAproject.html}"; the X-ray data were from the \emph{Swift} satellite. the optical photometric and polarization data were obtained with the 1.8 m Perkins Telescope of Lowell Observatory, which are not corrected for the galactic extinction. the 43 GHz data were taken from the VLBA observations. The radio data at 15 GHz that were obtained by the Owens Valley Radio Observatory (OVRO, \emph{http://www.astro.caltech.edu/ovroblazars/}) 40 m radio telescope (Richards et al. 2011) are also given in Figure \ref{lightcurve}. The OVRO supports an ongoing blazar monitoring program of \emph{Fermi} satellite.

As illustrated in Figure \ref{lightcurve}, 3C 454.3 shows the significant flux variation in the multiple wavelength. It was reported that its gamma-ray outbursts are usually accompanied by the flux variations in low-energy bands (e.g., Jorstad et al. 2013; Wehrle et al. 2012). In addition, its flux variation in the gamma-ray band is also observed to be correlated with the variation of the optical polarization (Jorstad et al. 2010, 2013). We make the DCF analysis of variability between $\gamma$-ray and other energy bands using the long-term lightcurves in multiple wavelength, as shown in Figure \ref{DCF}. The DCF results are calculated using the logarithm of the flux, which is less dominated by the high flux values (see also Ackermann et al. 2010). Note that no observational data are available from MJD 55758 to MJD 56228 at the gamma-ray band and there is a very large outburst at $\sim$ MJD 55520, hence we divide the global lightcurves into two segments with MJD 56000, and then we also calculate the DCF results for the two segments among multiwavelength, respectively. The results are also given in Figure \ref{DCF}. It is found that the gamma-ray flux is correlated with the fluxes in the R-band and X-ray band, and is weakly correlated with the radio flux. No clear correlation between the gamma-ray flux and the evolution of the optical linear polarization degree is found for the global and second segment lightcurves, but they are correlated during the first segment lightcurves, which is consistent with the reported results in Jorstad et al. (2010, 2013).

\section{Discussion}
Several models have been proposed to explain the origin of the erratic outbursts observed in blazars. 3C 454.3 with the significant spectral and flux variations should be the best candidate to study this issue. 3C 454.3 is a typical FSRQ and its jet orientation should point to the line of sight. It was suggested that the outbursts with different brightness may be due to a helical jet , in which blobs move at different angles to the line of sight leading to different Doppler boosting effect on the observed photons (e.g., Villata et al. 2007; Jorstad et al. 2013). As we show here that the variation of gamma-ray flux is accompanied by the variations of spectral index and optical polarization degree. These facts cannot be simply explained with the Doppler effect.

Several observations have revealed that the optical and gamma-ray outbursts of 3C 454.3 show a connection with a superluminal knot through the core (Jorstad et al. 2010, 2013). It is possible that the central engine of 3C 454.3 intermittently ejects the sequential blobs with different velocities, which induce multiple collisions and a series of outbursts, i.e., the complex lightcurves that are produced by superpositions of many flares with different timescales. The correlation between gamma-ray flux and optical linear polarization degree indicates that the gamma-ray emission is produced in a region with ordering magnetic field. It may imply that the radiation regime of this FSRQ is highly magnetized (e.g., Zhang et al. 2014). Zhang \& Yan (2011) proposed an internal-collision-induced magnetic reconnection and turbulence (ICMART) process to explain the prompt emission of gamma-ray bursts. This model suggests that the Poynting flux energy is converted to the energies of electrons and protons efficiently and the prompt emission is due to the synchrotron radiation of these electrons. This scenario may be consistent with the observed correlations between gamma-ray flux and optical linear polarization degree as well as the spectral index.

In addition, the typical variability timescale in the gamma-ray band derived in our analysis is several days.  The timescales of the order of 1 day or longer should be interpreted as the typical timescales of successive flare events, which may be due to the collisions of blobs in the internal shock, as discussed by Kataoka et al. (2001) for X-ray data of three BL Lacs. Nakagawa \& Mori (2013) also reported that the four-year lightcurve at the GeV band of 3C 454.3 shows a specific timescale of $6.8\times10^{5}$ s, and this value suggests a black hole mass of $10^8$--$10^{10} M_{\odot}$ within the framework of the internal shock models. Therefore, the magnetized shells with different velocities ejected by the central engine may collide and induce an ICMART process to accelerate the radiation particles in 3C 454.3.

Cross comparisons of the statistical properties for 3C 454.3 with other sources may give clues to the nature of these flare. Neilsen et al. (2013, 2015) analyzed the X-ray flares of Sagittarius A$^{*}$ and found that the cumulative distributions of the duration, luminosity, and energy of the X-ray flares can be described by a power-law function, with indices of $\alpha_{\rm T}=0.9\pm0.2$, $\alpha_{\rm L}=1.9^{+0.4}_{-0.3}$, and $\alpha_{\rm E}=1.5\pm0.2$, respectively. Li et al. (2015) got the similar results via a simulation analysis and suggested that these results are consistent with the theoretical prediction of the self-organized criticality (SOC) system with the spatial dimension S=3 (Aschwanden 2011, 2012, 2014). They further reported that the X-ray flares represent the plasmoid ejections driven by magnetic reconnection (similar to the solar flares) in the accretion flow onto the black hole. We made the similar analysis to the $\gamma$-ray flares of 3C 454.3, as shown in Figure \ref{SOC}. We get $\alpha_{\rm E}=1.46\pm0.02$, $\alpha_{\rm L}=1.54\pm0.08$, and $\alpha_{\rm T}=2.28\pm0.05$ (duration time defined as four times of FWHM), respectively. The derived $\alpha_{\rm E}$ and $\alpha_{\rm L}$ are roughly consistent with that of the X-ray flares in Sagittarius A$^{*}$, but $\alpha_{\rm T}$ is much larger than that of the X-ray flares in Sagittarius A$^{*}$. Based on the Gaussian flare profile assumption as we adopt in this paper, Li et al. (2015) obtained $\alpha_{\rm T}> 2.1$ for the X-ray flares in Sagittarius A$^{*}$. Wang et al. (2015) analyzed the detected sample of Neilsen et al. (2013) with a different selection of the fitting data ranges and found $\alpha_{\rm T} =1.9\pm 0.5$. These results are consistent with ours, and thus the nature of 3C 454.3 may be also consistent with the SOC system, similar to Sagittarius A$^{*}$. Therefore, the GeV flares of 3C 454.3 may also from the plasmoid ejections driven by magnetic reconnection. This further supports the idea that the jet radiation regions of FSRQs may be highly magnetized (e.g., Zhang 2014, 2015).

\section{Summary}

We have analyzed the $\gamma$-ray long-term lightcurve observed with \emph{Fermi}/LAT, and collected the simultaneous lightcurves in the radio, optical and X-ray bands of 3C 454.3 from literature. Our results are summarized below.

\begin{itemize}

\item The flux variation of 3C 454.3 in the $\gamma$-ray band is correlated with the flux variation in the R-band and X-ray band, and a weak correlation between gamma-ray flux and radio flux is also observed, indicating that the radiations at these energy bands are co-spatial.

\item The PDS of the global lightcurve in the $\gamma$-ray band is fitted with a broken power-law and yields an obvious turnover at $\sim7.7$ days, which may indicate a typical variability timescale of 3C 454.3 in the $\gamma$-ray band.

\item  The spectral evolutions accompanying the flux variations in the $\gamma$-ray band are observed, showing the behavior of ``harder when brighter".

\item The spectral evolution and an increase of the optical linear polarization degree along with the increase of the $\gamma$-ray flux may imply that the radiation particles are accelerated and the magnetic field is ordered by the shock processes. The nature of 3C 454.3 may be consistent with the SOC system, and thus the $\gamma$-ray outbursts are from the plasmoid ejections driven by magnetic reconnection.

\end{itemize}

\begin{acknowledgements}
We thank the helpful discussion with Wei Cui, Shuang-Nan Zhang, Yuan Liu, and Da-Bin Lin. This study makes use of radio, optical, and X-ray data from the VLBA-BU Blazar Monitoring Program (VLBA-BU-BLAZAR; http://www.bu.edu/blazars/VLBAproject.html), funded by NASA through the Fermi and Swift Guest Investigator Programs. This research has made use of data from the OVRO 40-m monitoring program (Richards, J. L. et al. 2011, ApJS, 194, 29) which is supported in part by NASA grants NNX08AW31G, NNX11A043G, and NNX14AQ89G and NSF grants AST-0808050 and AST-1109911. This work is supported by the National Natural Science Foundation of China (grants 11573034, 11533003, 11363002, 11373036, 11463001, and U1731239), the National Basic Research Program (973 Programme) of China (grant 2014CB845800), and En-Wei Liang acknowledges support from the special funding from the Guangxi Science Foundation for Guangxi distinguished professors (Bagui Yingcai \& Bagui Xuezhe).
\end{acknowledgements}

\begin{figure*}
\includegraphics[angle=0,scale=0.28]{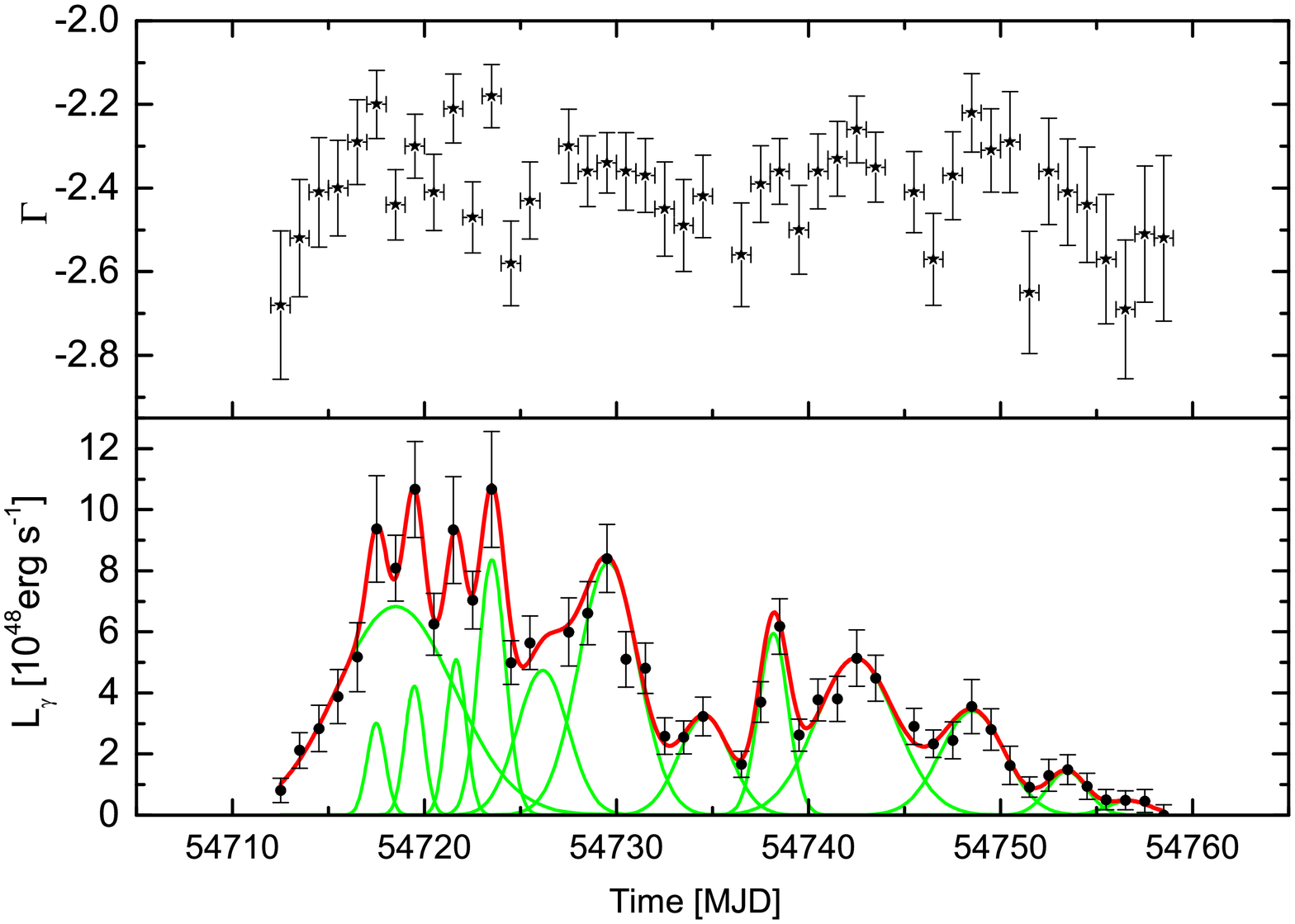}
\includegraphics[angle=0,scale=0.28]{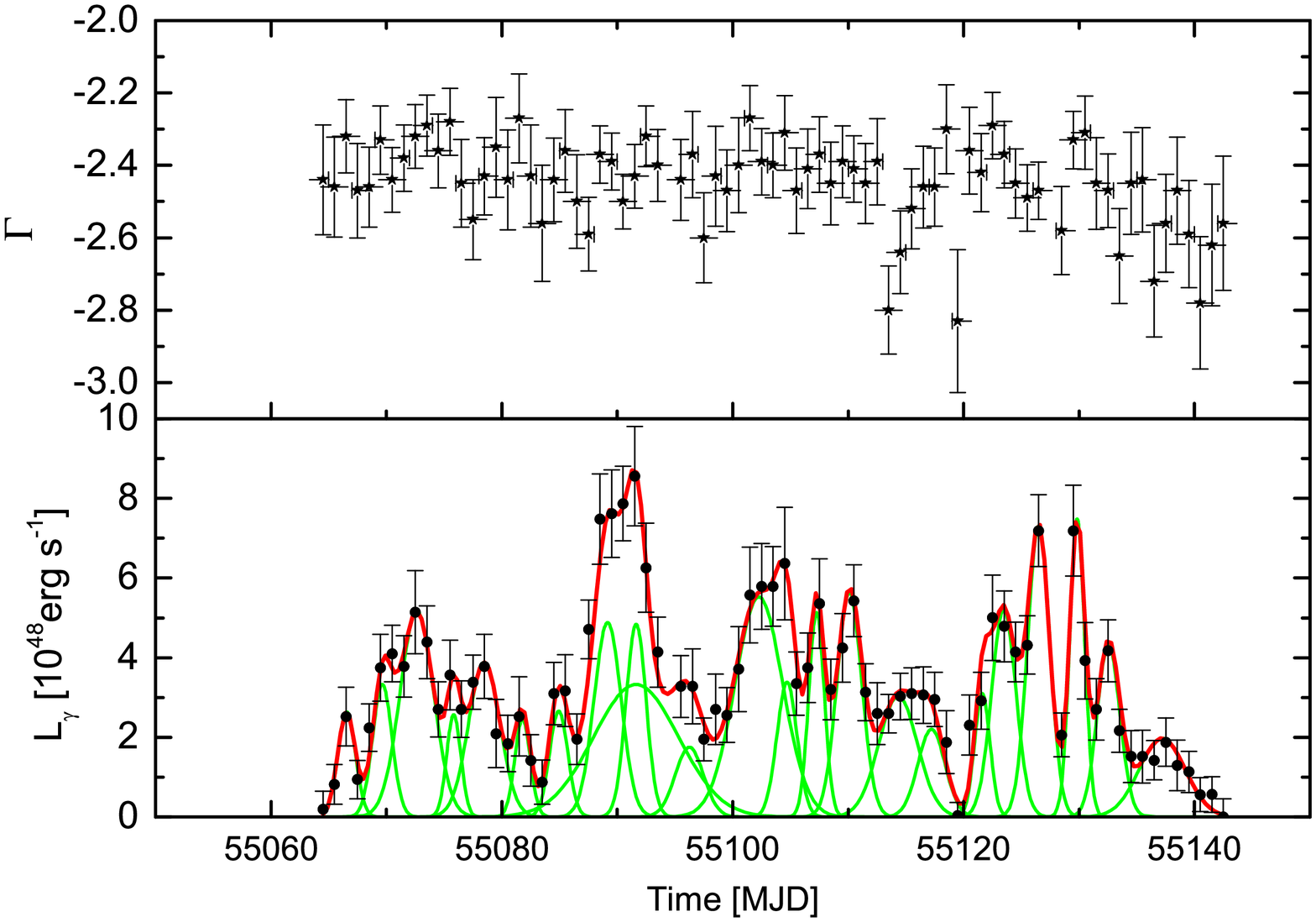}\\
\includegraphics[angle=0,scale=0.28]{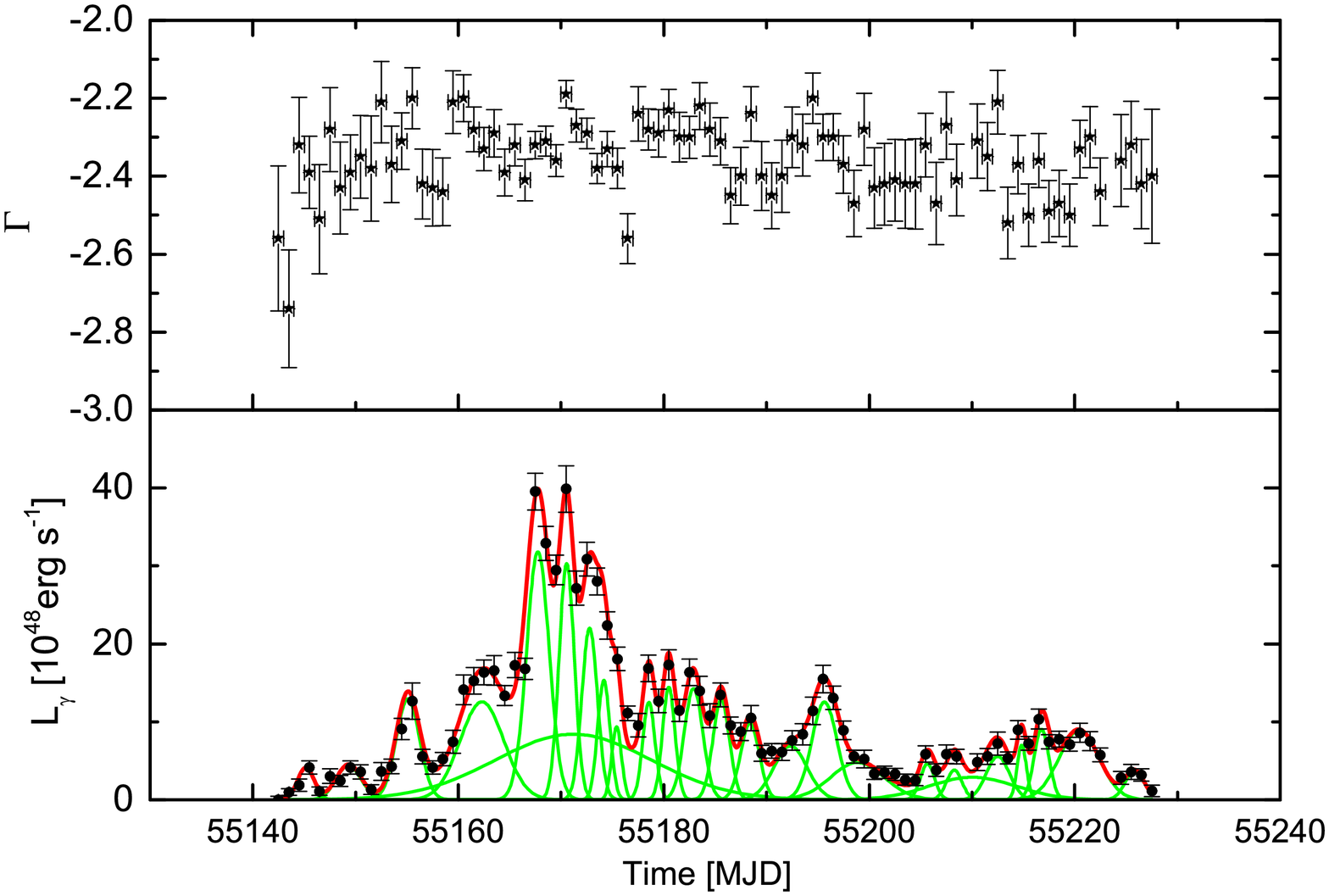}
\includegraphics[angle=0,scale=0.28]{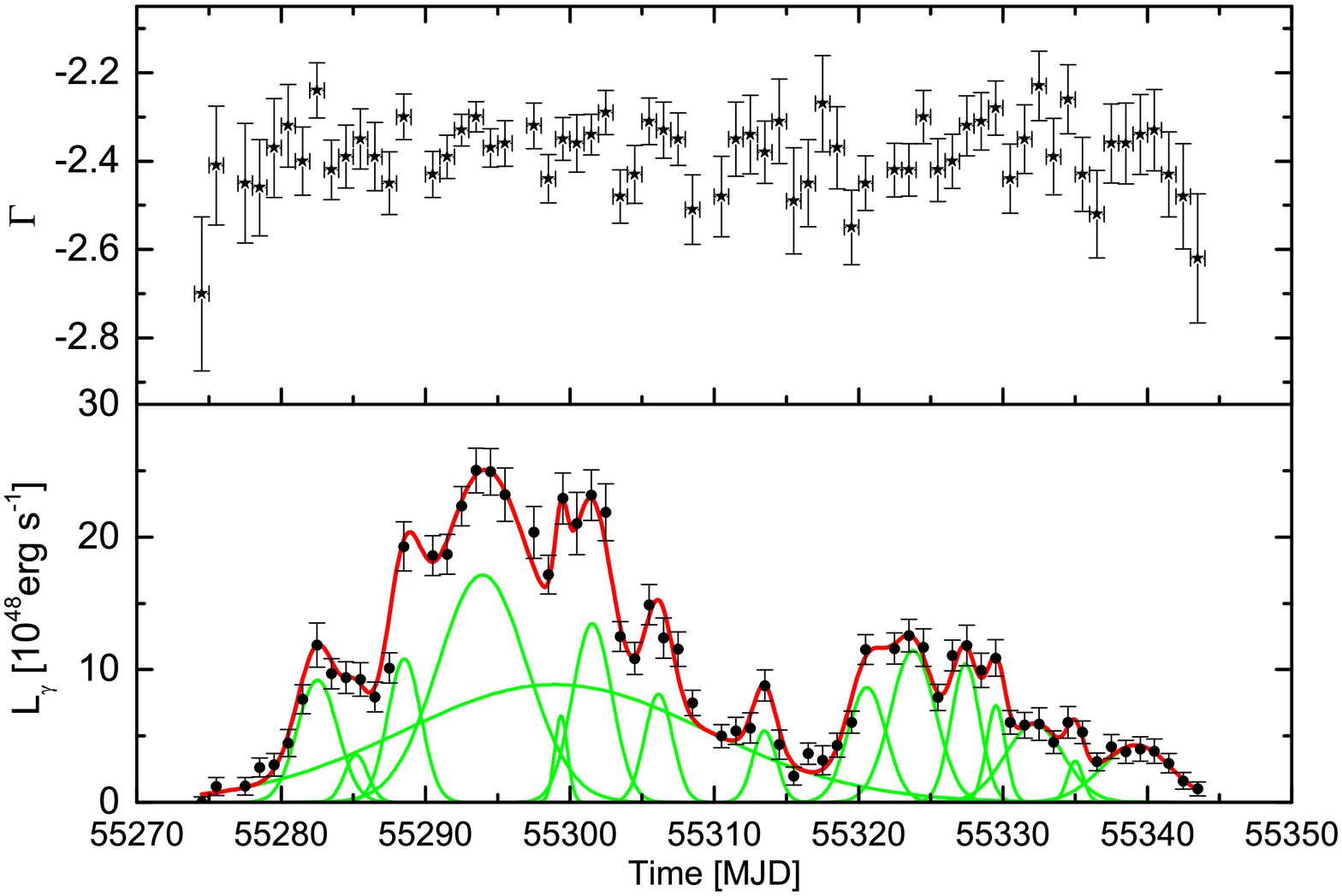}\\
\includegraphics[angle=0,scale=0.28]{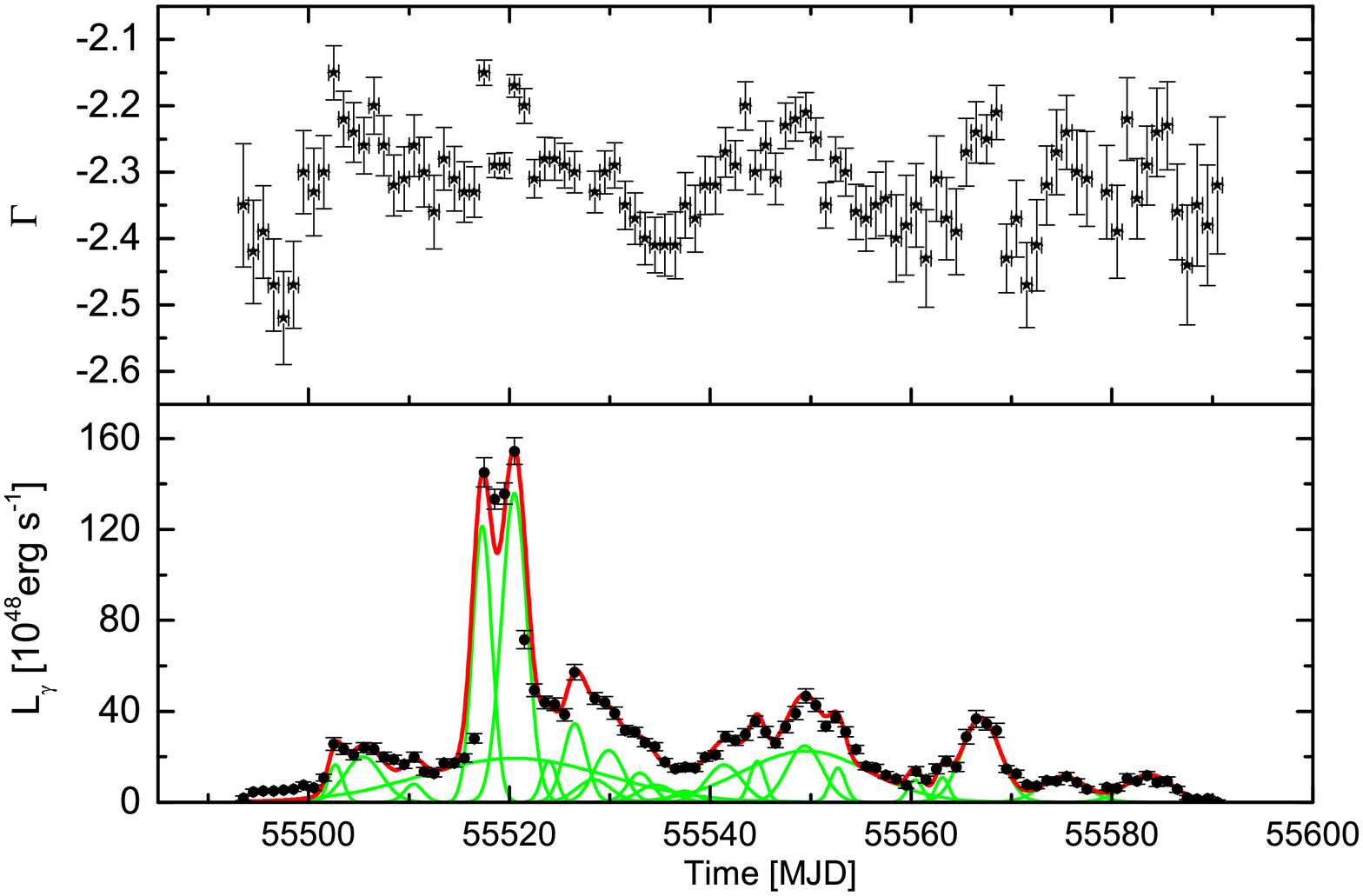}
\includegraphics[angle=0,scale=0.28]{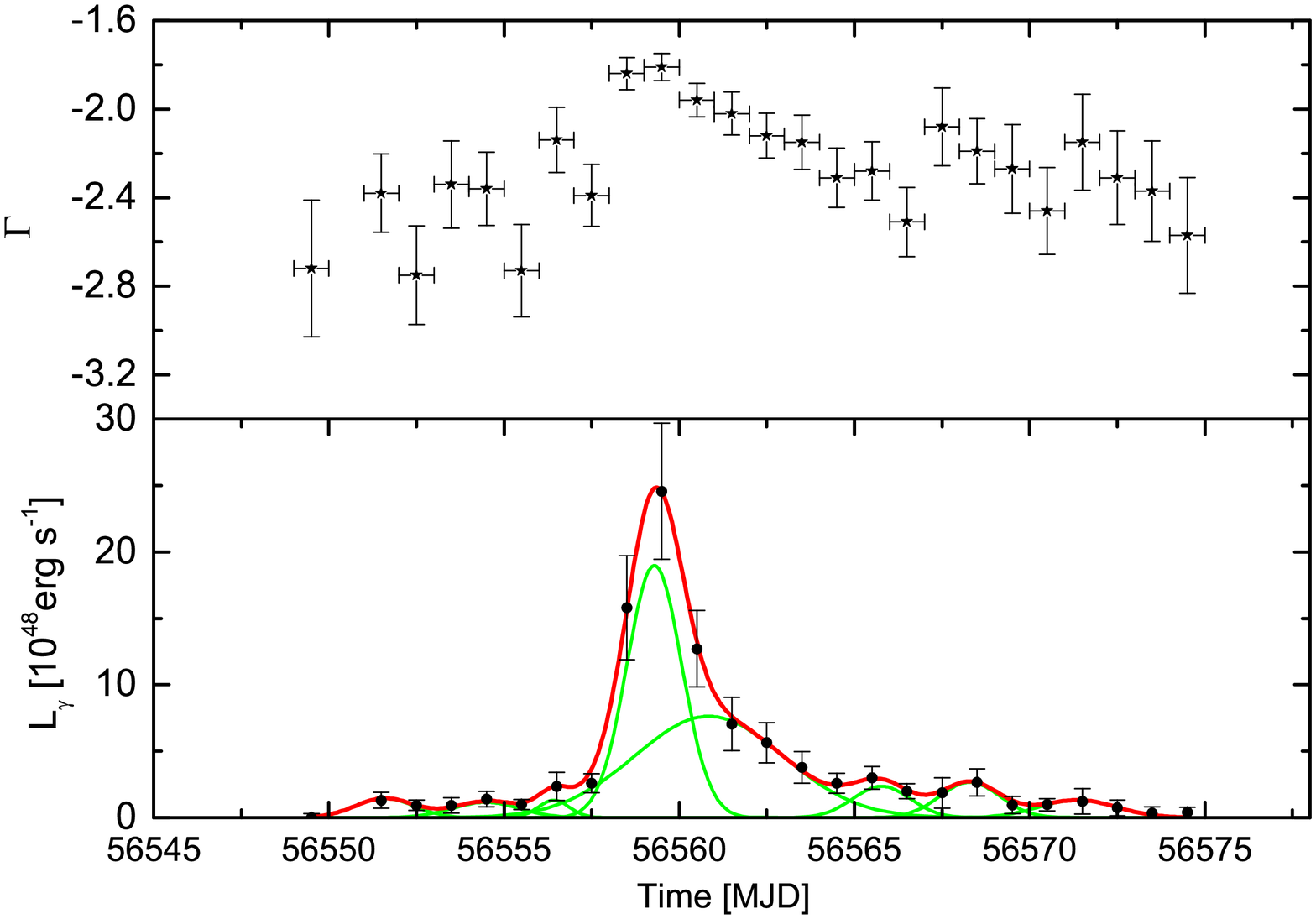}\\
\includegraphics[angle=0,scale=0.28]{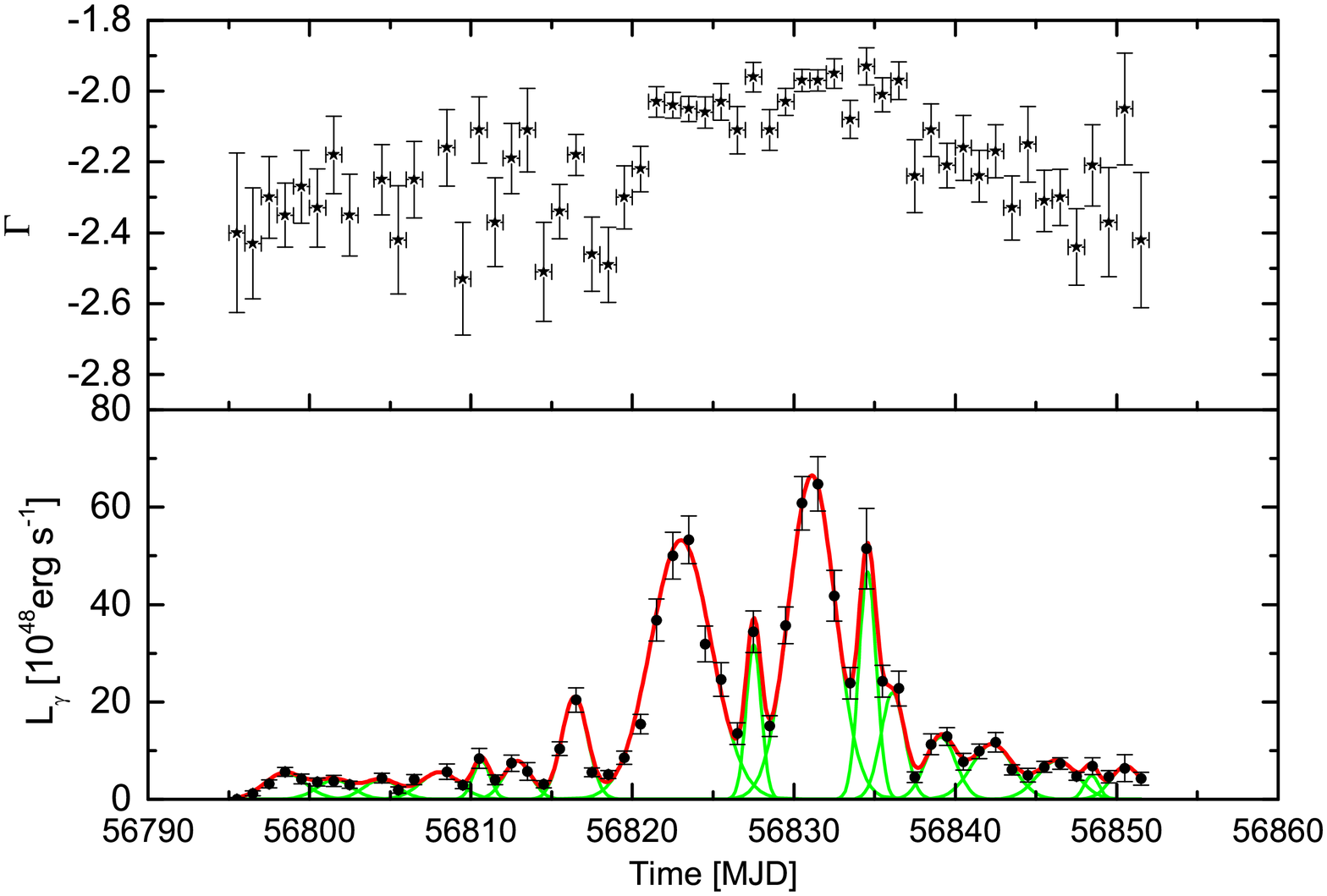}
\includegraphics[angle=0,scale=0.28]{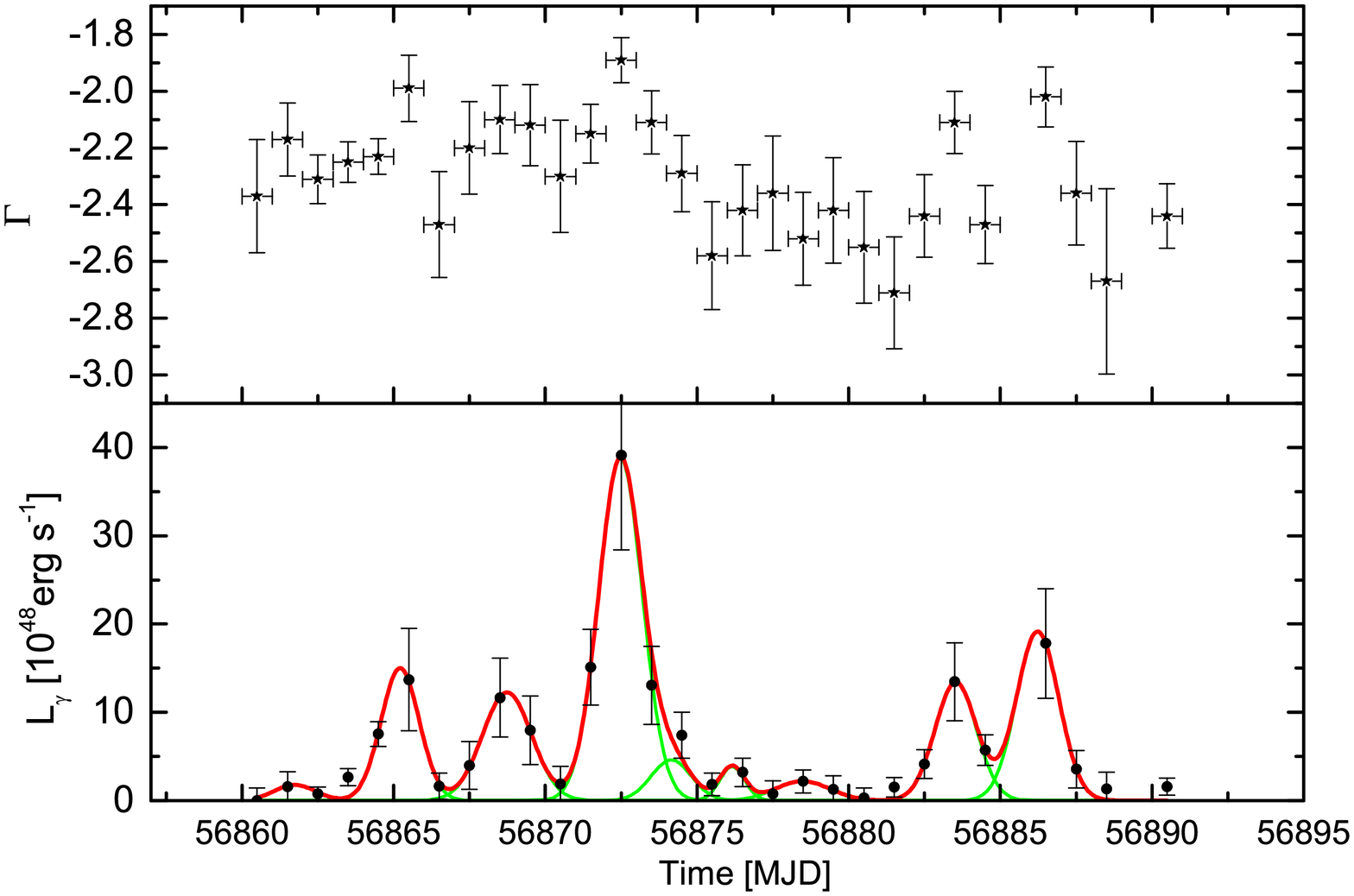}\\
\end{figure*}
\begin{figure*}
\includegraphics[angle=0,scale=0.28]{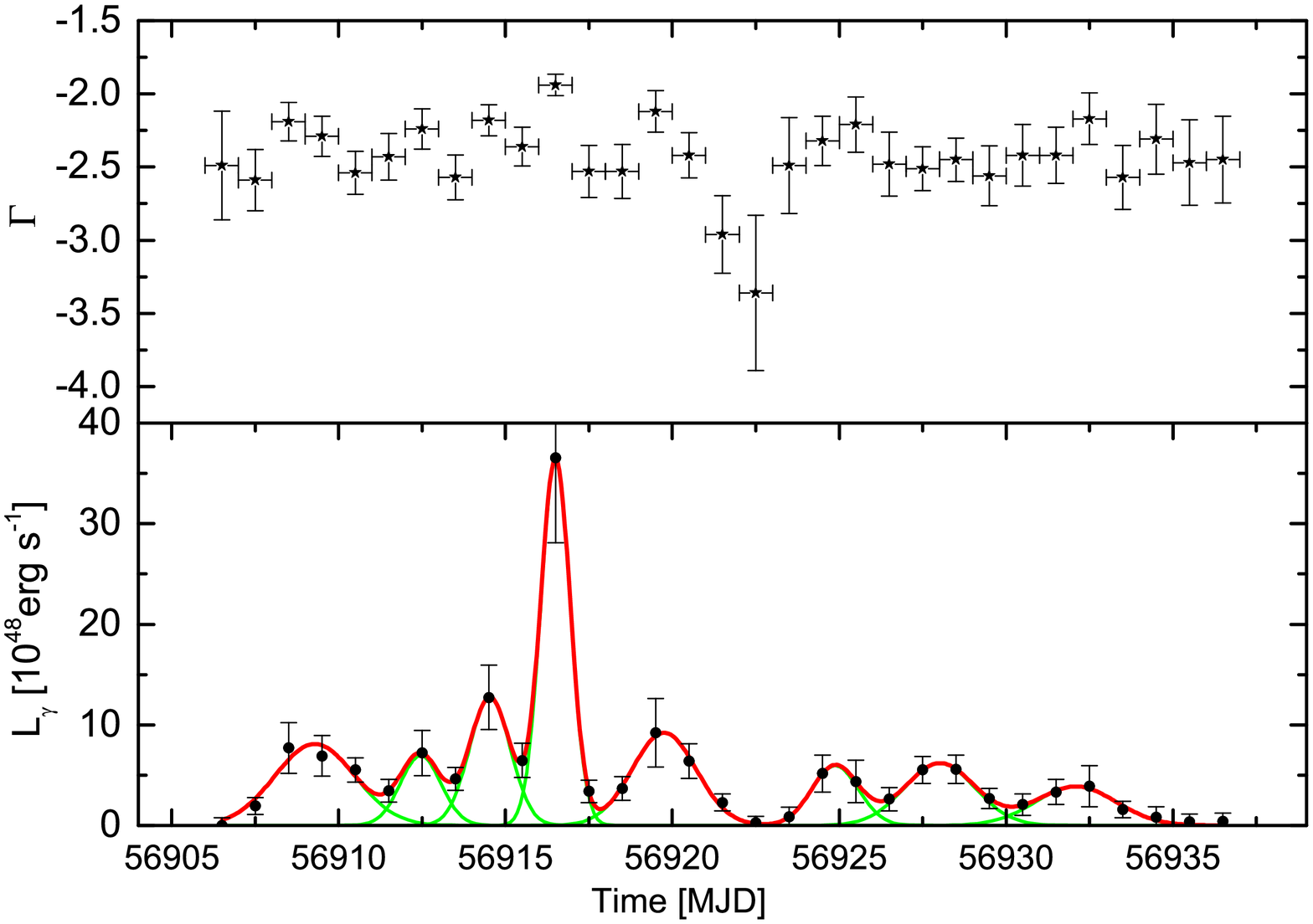}
\includegraphics[angle=0,scale=0.28]{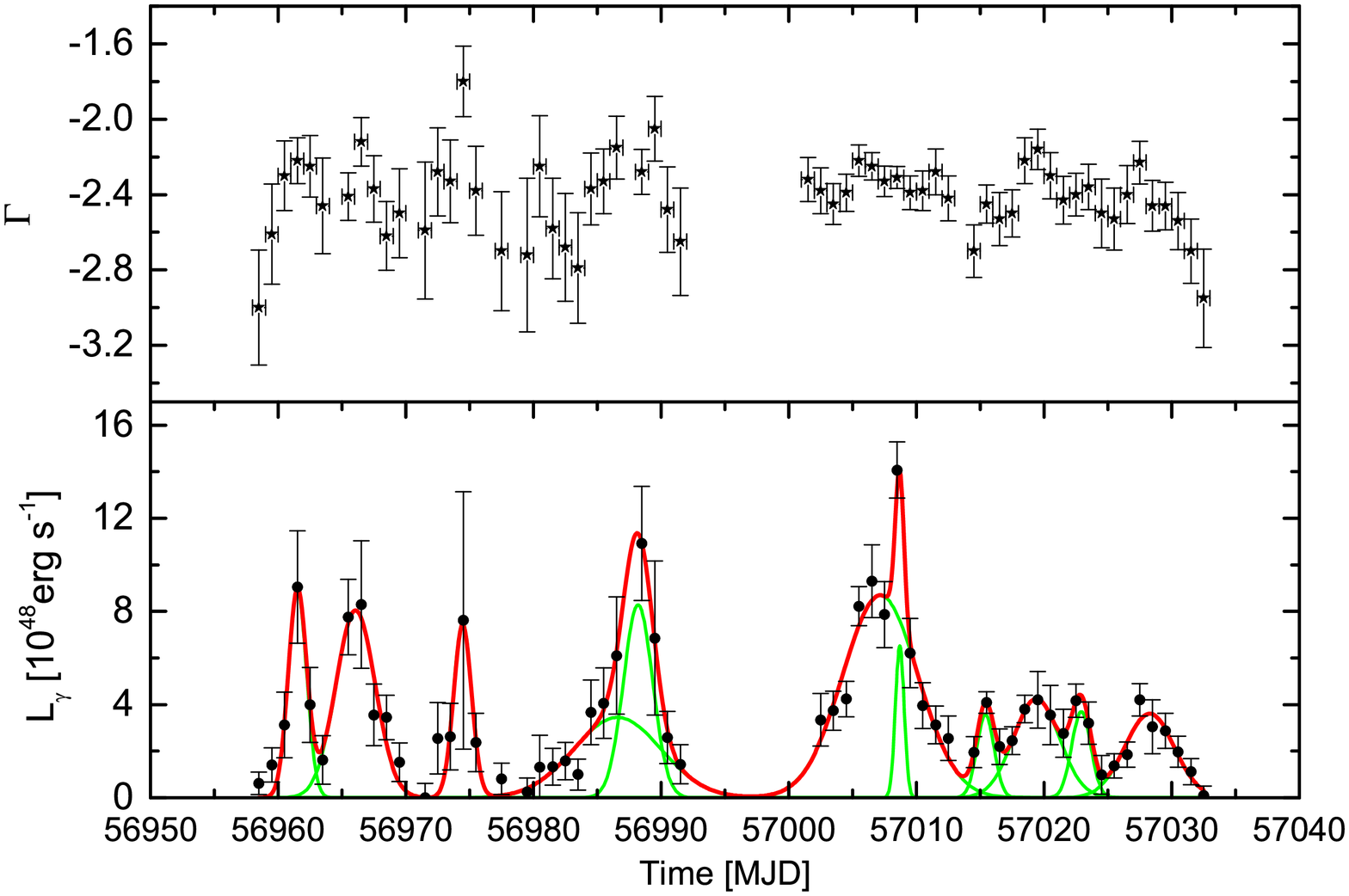}\\
\includegraphics[angle=0,scale=0.28]{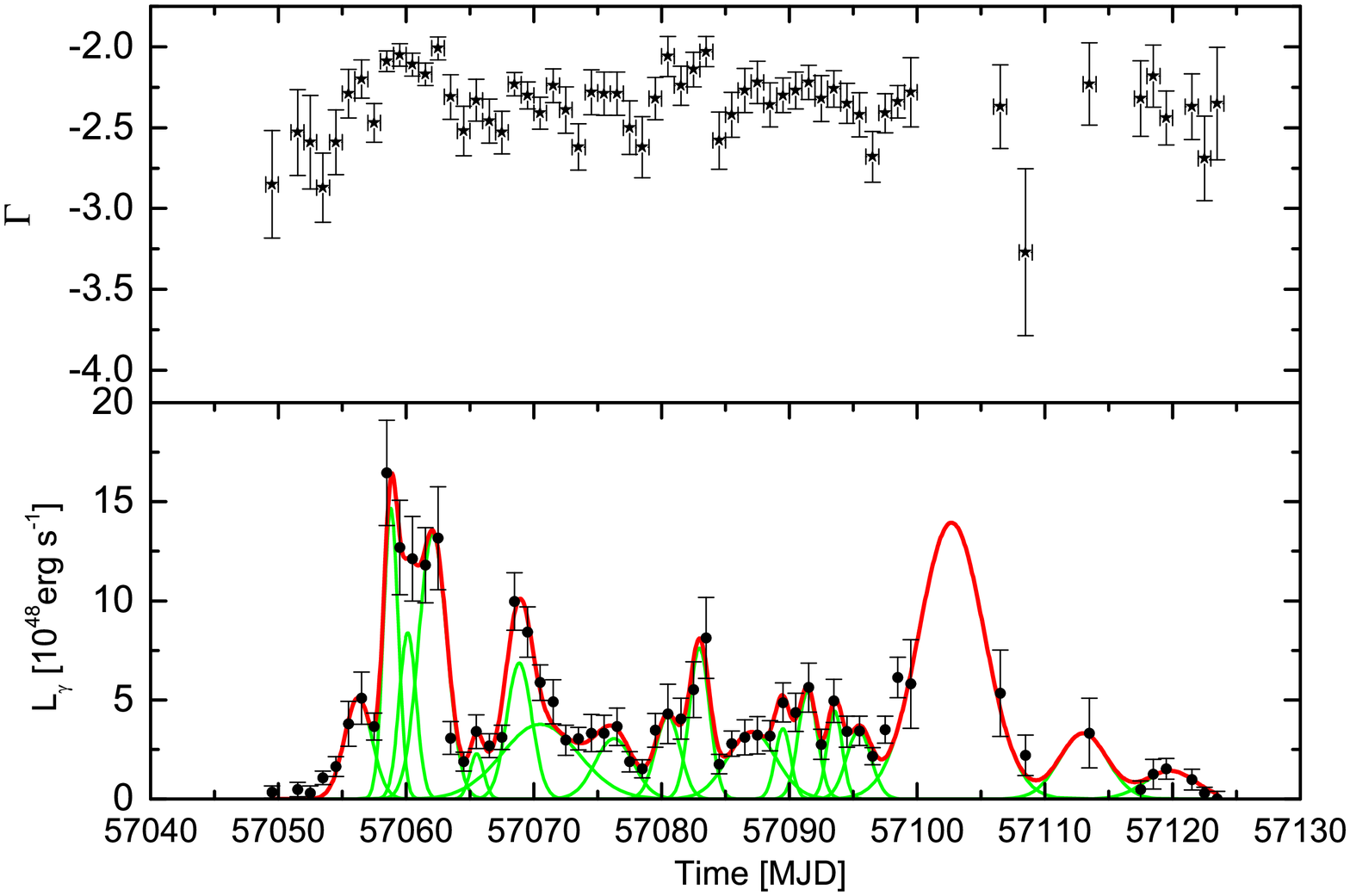}
\includegraphics[angle=0,scale=0.28]{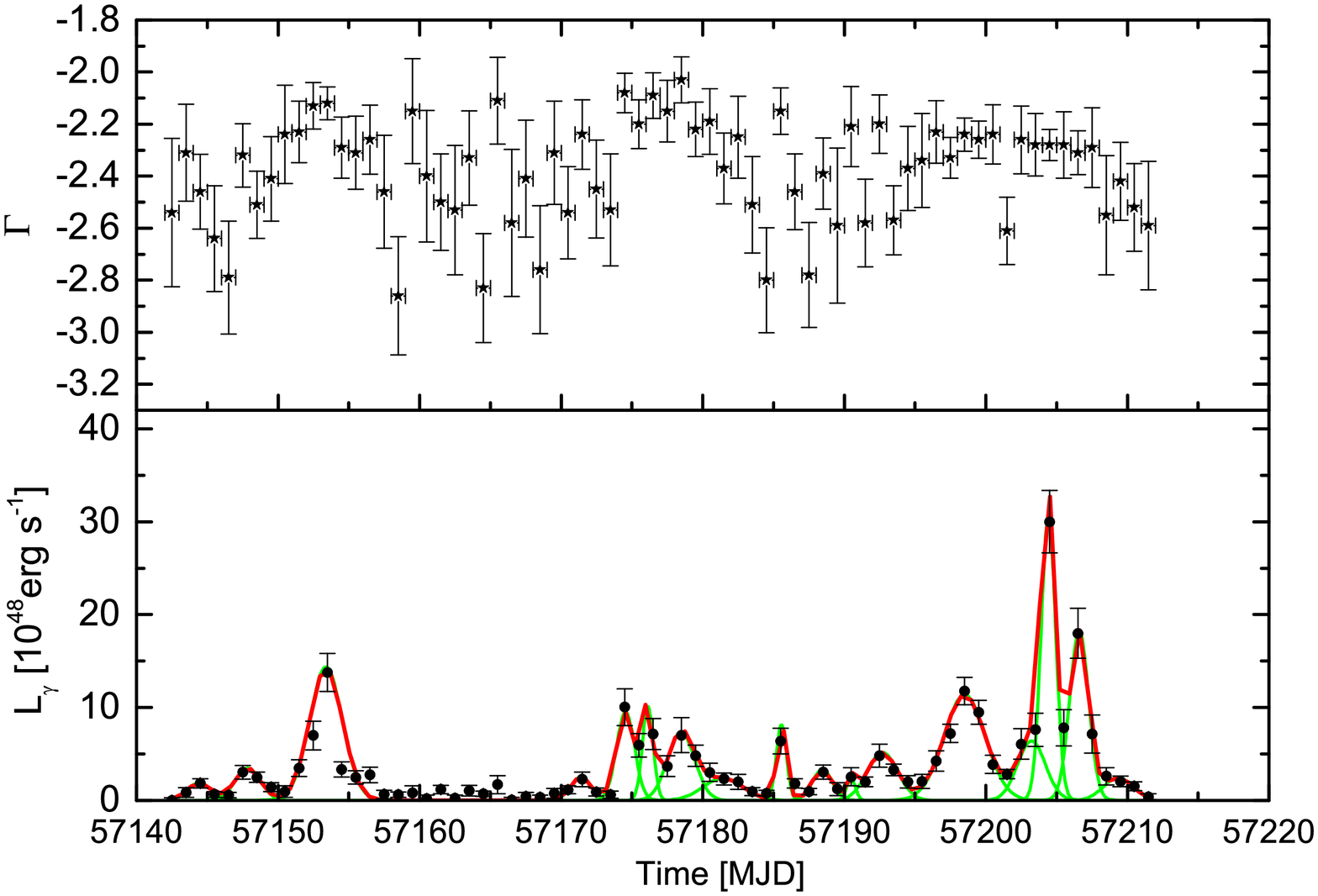}\\
\includegraphics[angle=0,scale=0.28]{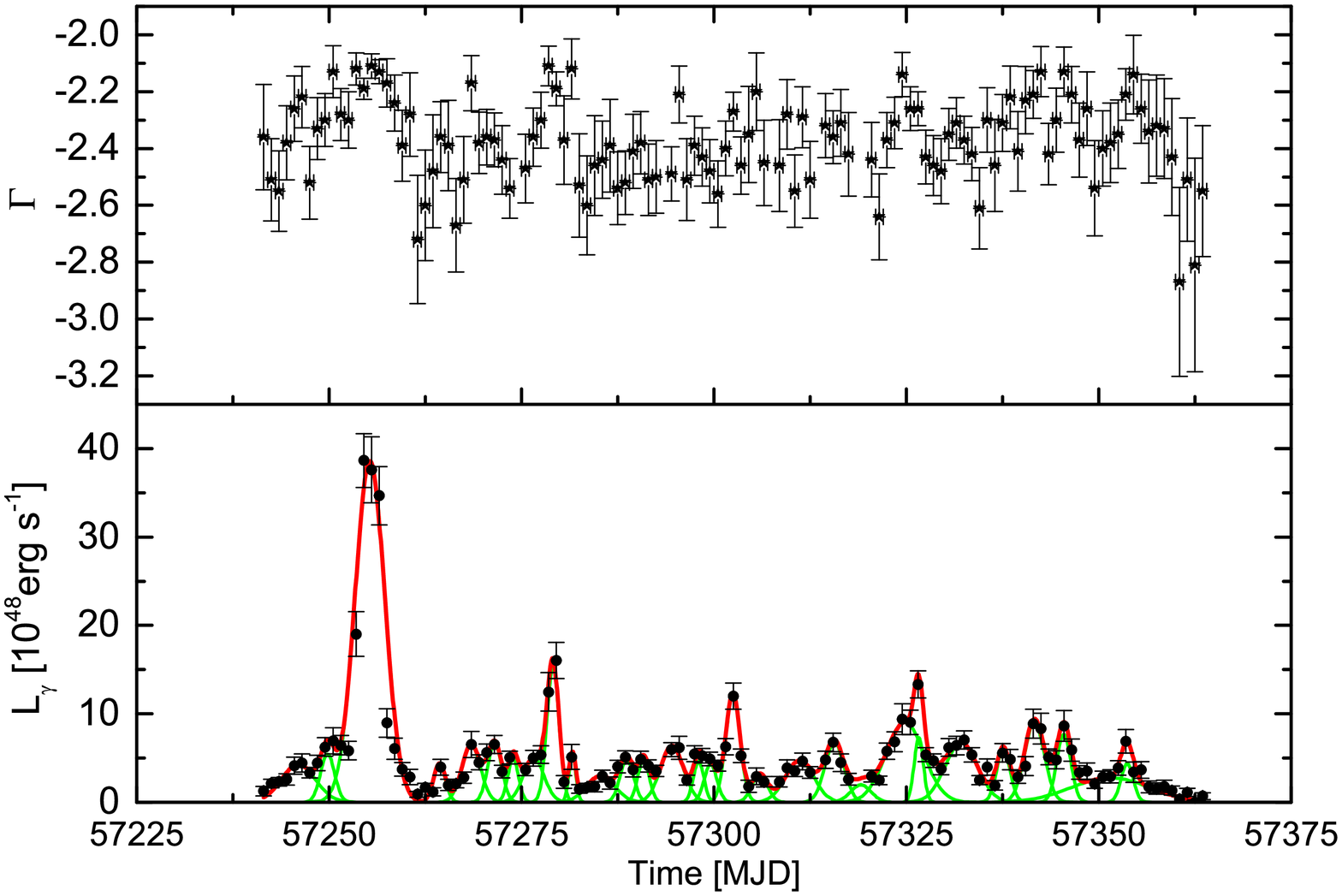}
\includegraphics[angle=0,scale=0.28]{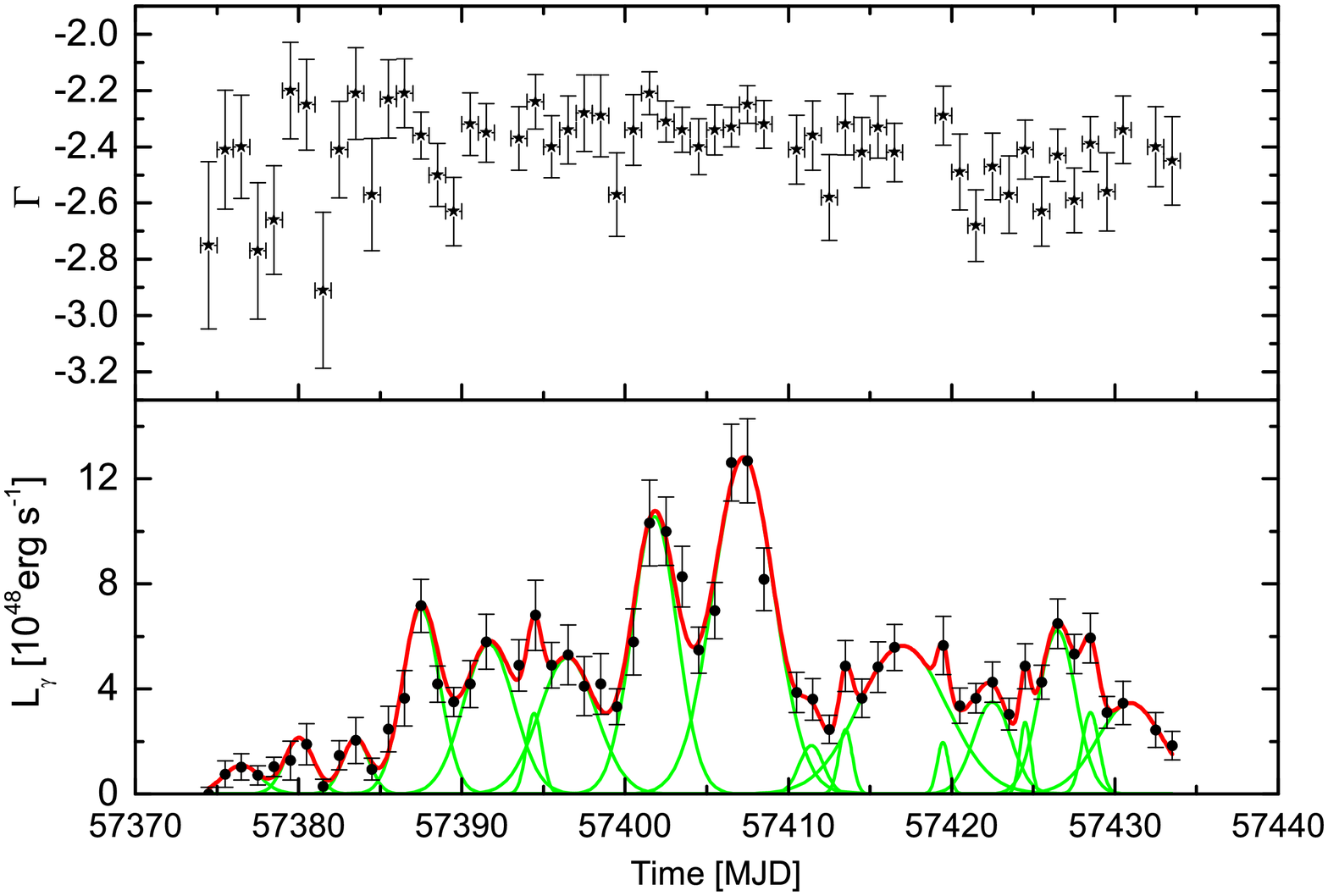}\\
\caption{The daily lightcurves of luminosity and photon spectral index for 3C 454.3. The scattered points are observation data of \emph{Fermi}/LAT. The thin green lines are given by multiple Gaussian function, and the thick red lines are the sum of all the Gaussian components. The values of FWHM, $L_{\rm peak}$, $E_{\gamma}$ for each Gaussian component are listed in Table 1. }\label{Gaussian}
\end{figure*}

\begin{figure*}
\includegraphics[angle=0,scale=0.22]{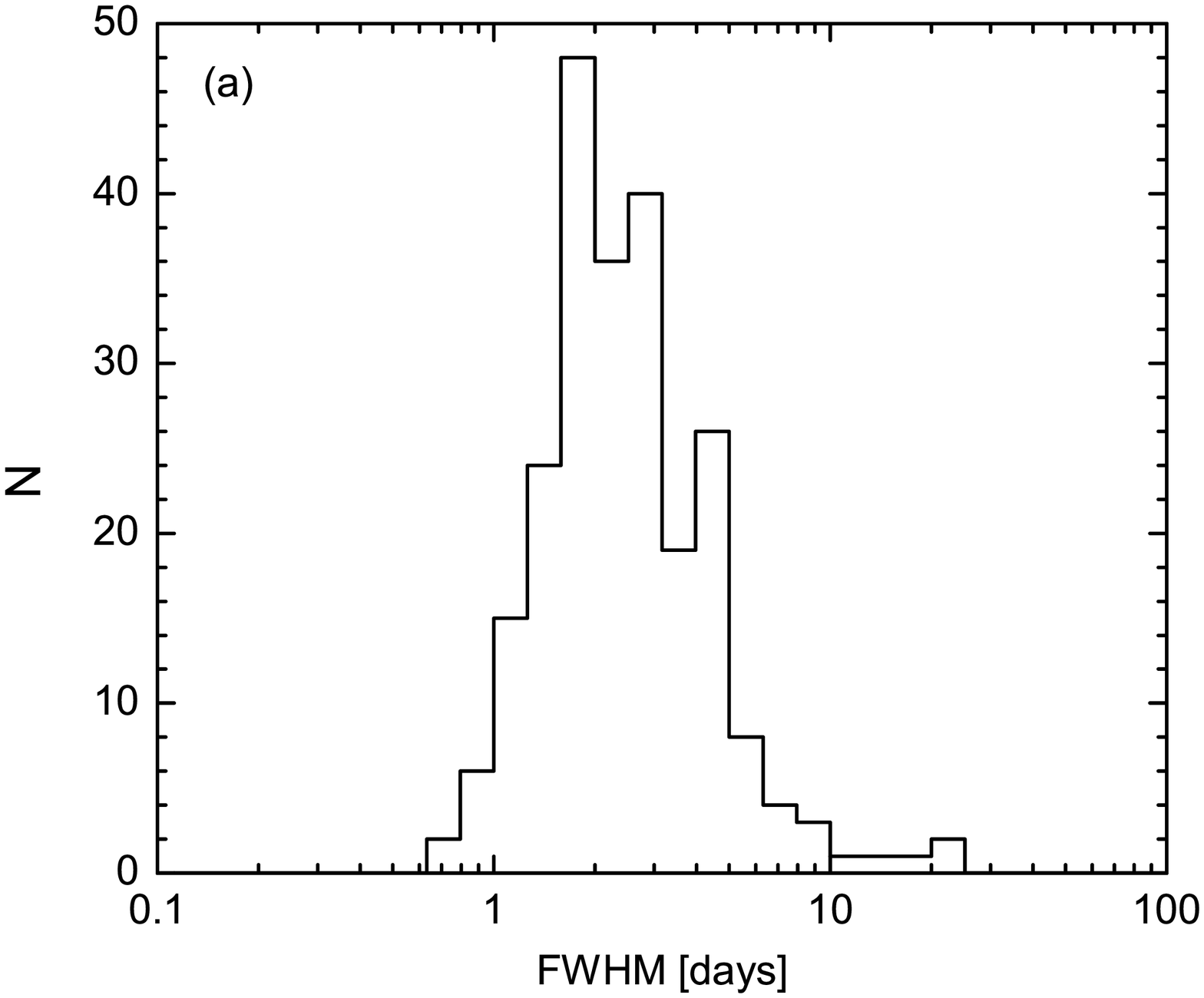}
\includegraphics[angle=0,scale=0.22]{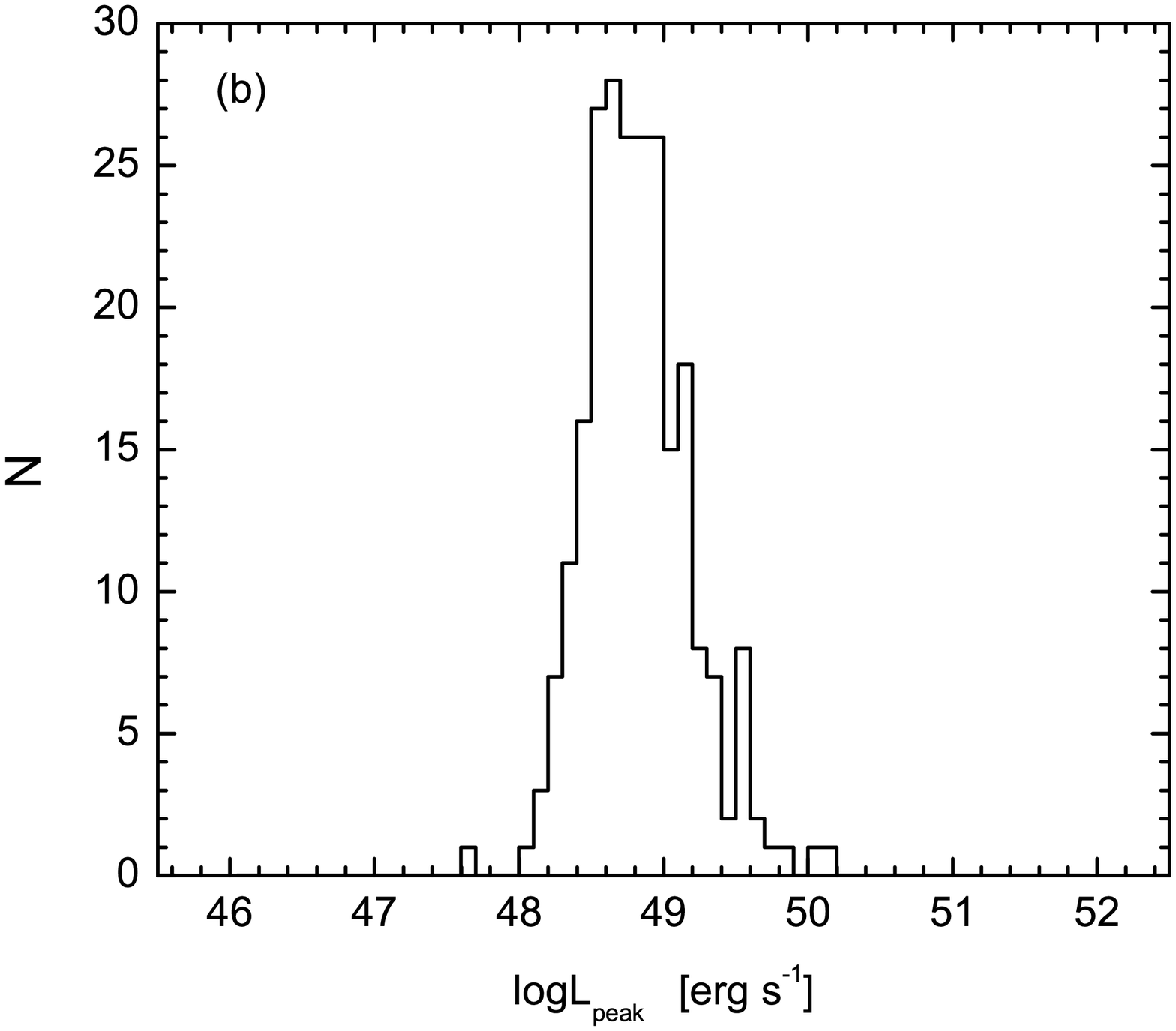}
\includegraphics[angle=0,scale=0.22]{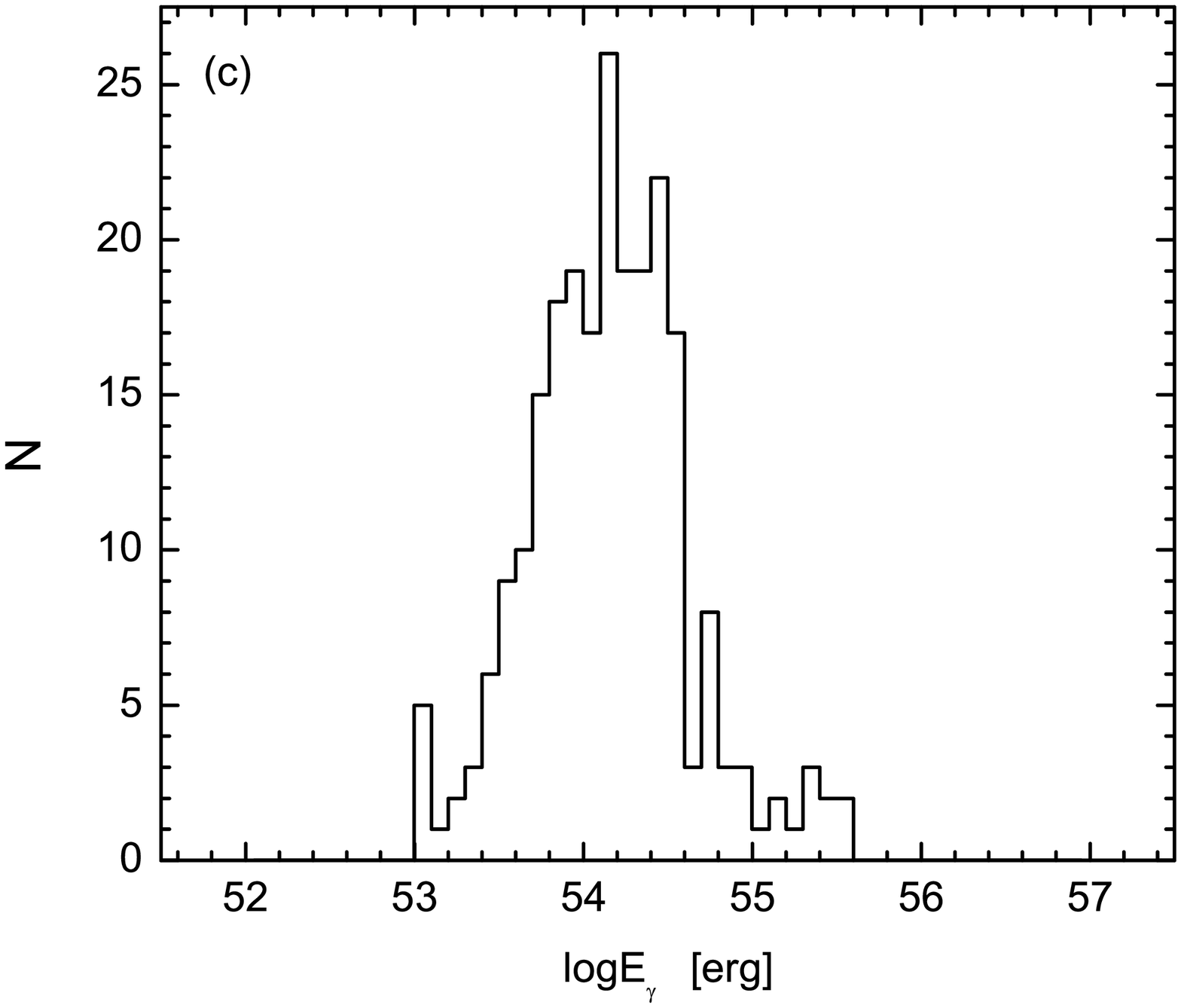}
\caption{Distributions of FWHM, $L_{\rm peak}$, and $E_{\gamma}$ of these Gaussian components in Figure \ref{Gaussian}.}\label{Distributions}
\end{figure*}

\begin{figure*}
\includegraphics[angle=0,scale=0.45]{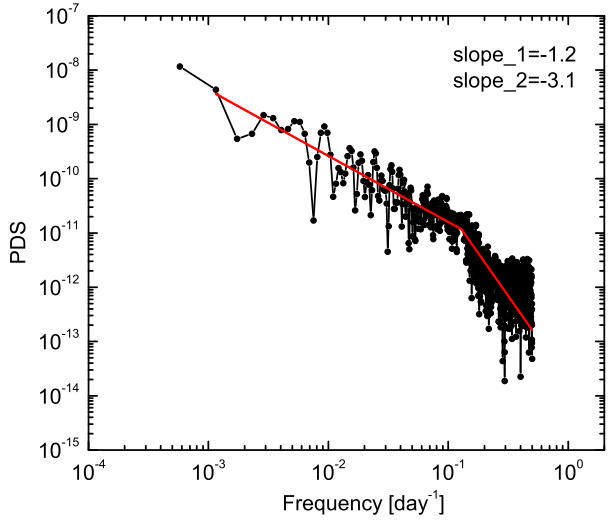}
\caption{PDS of the global lightcurve in the $\gamma$-ray band. The red line shows the fitting result with the broken power-law function.}\label{episode_PDS}
\end{figure*}

\begin{figure*}
\includegraphics[angle=0,scale=0.18]{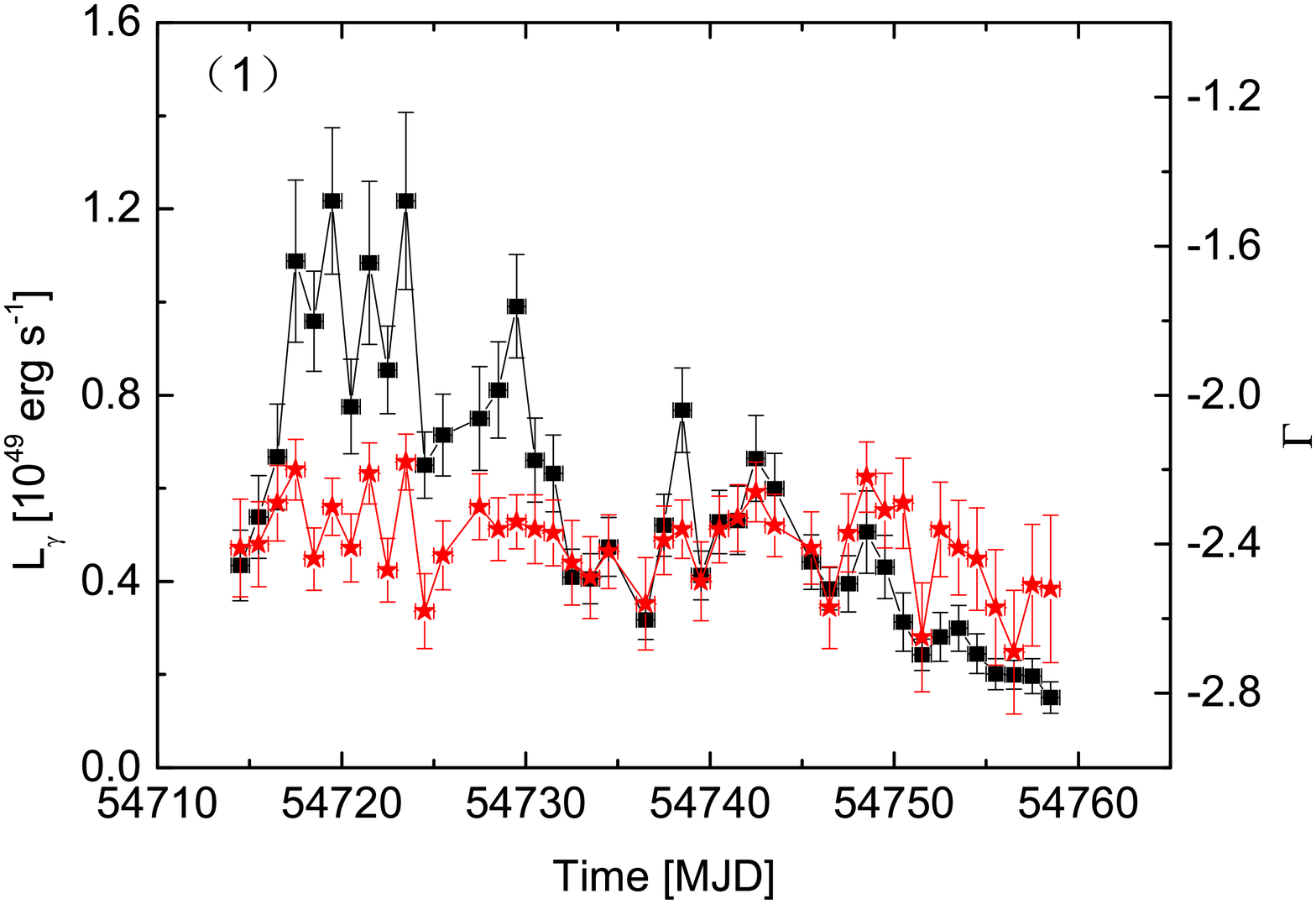}
\includegraphics[angle=0,scale=0.18]{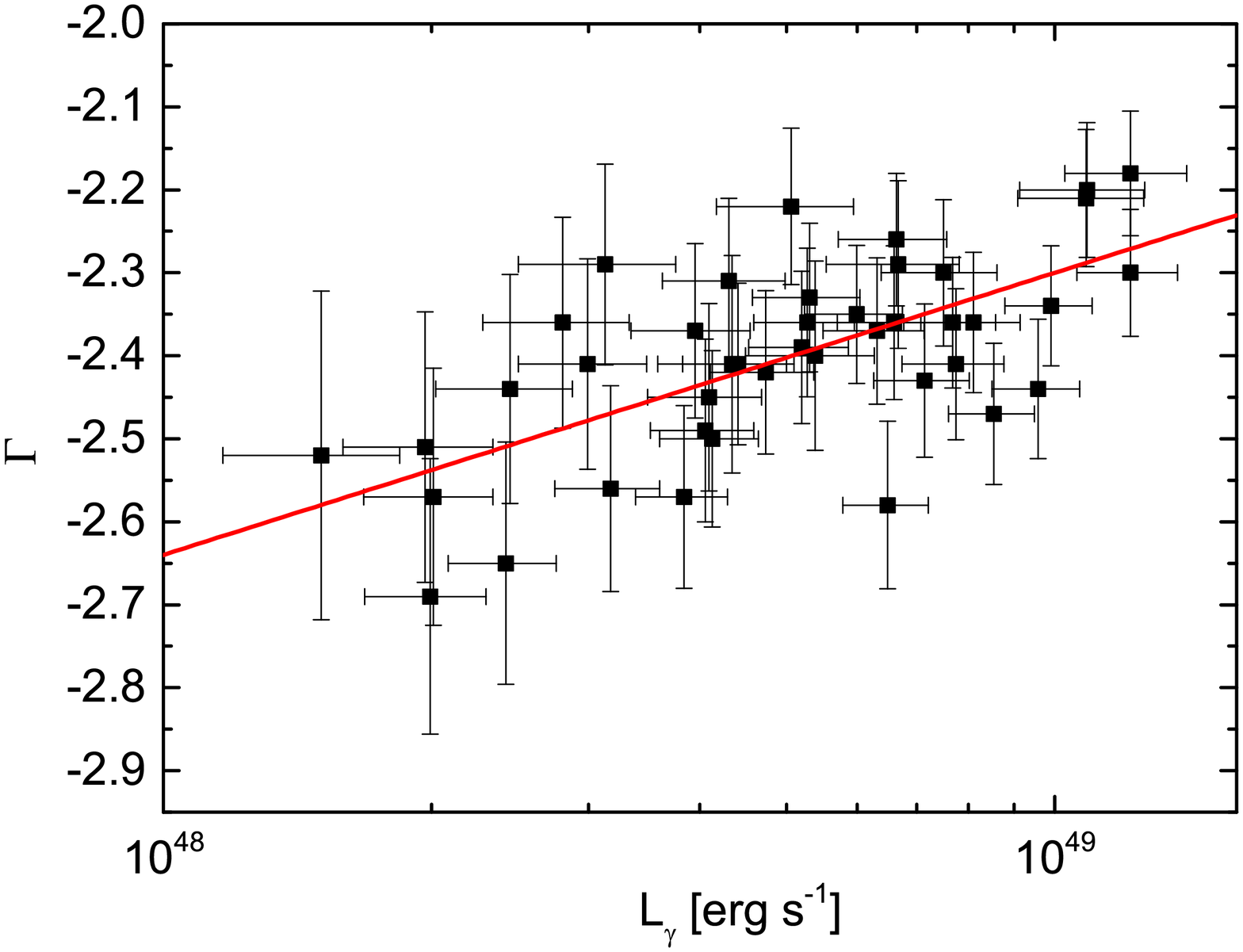}
\includegraphics[angle=0,scale=0.23]{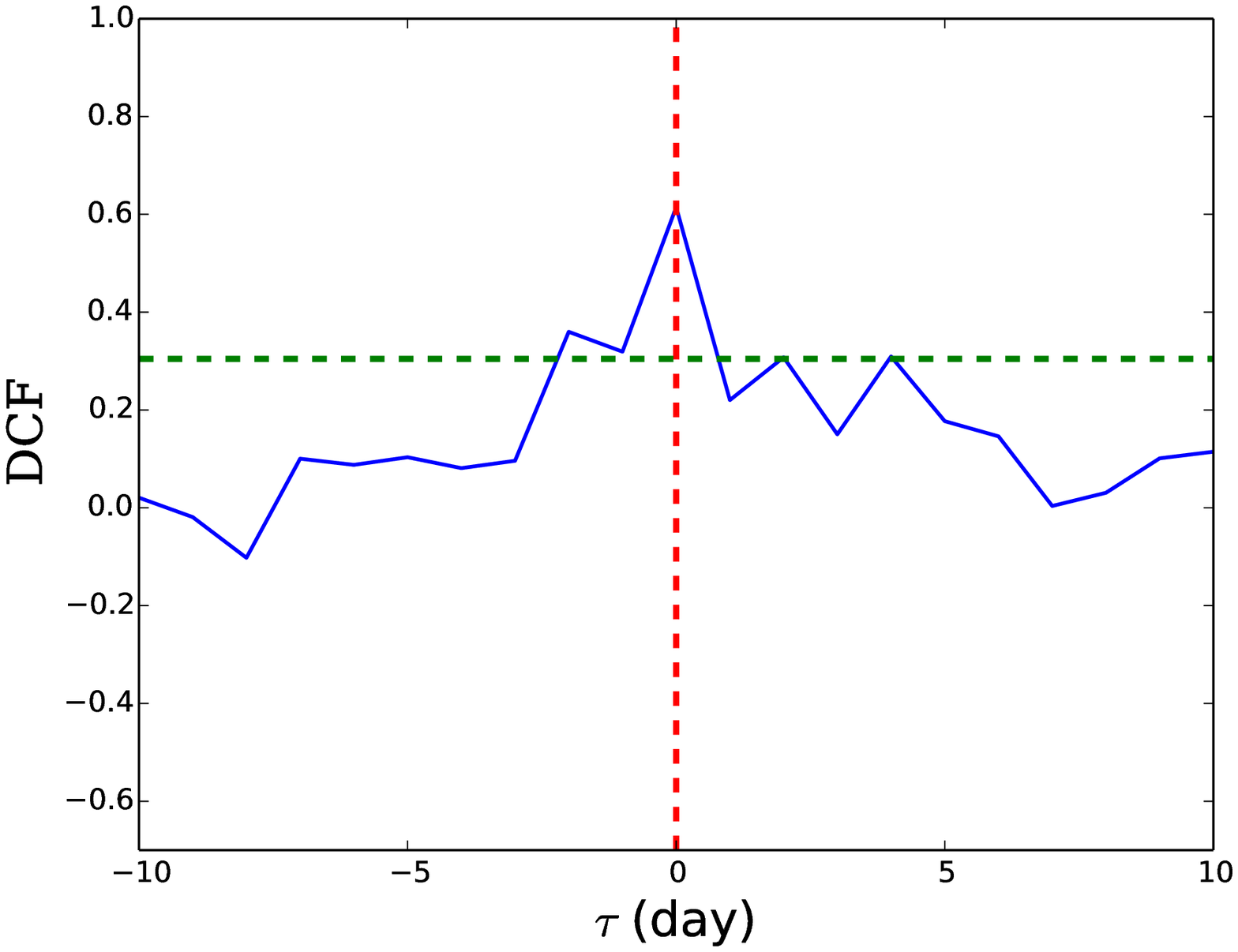}\\
\includegraphics[angle=0,scale=0.18]{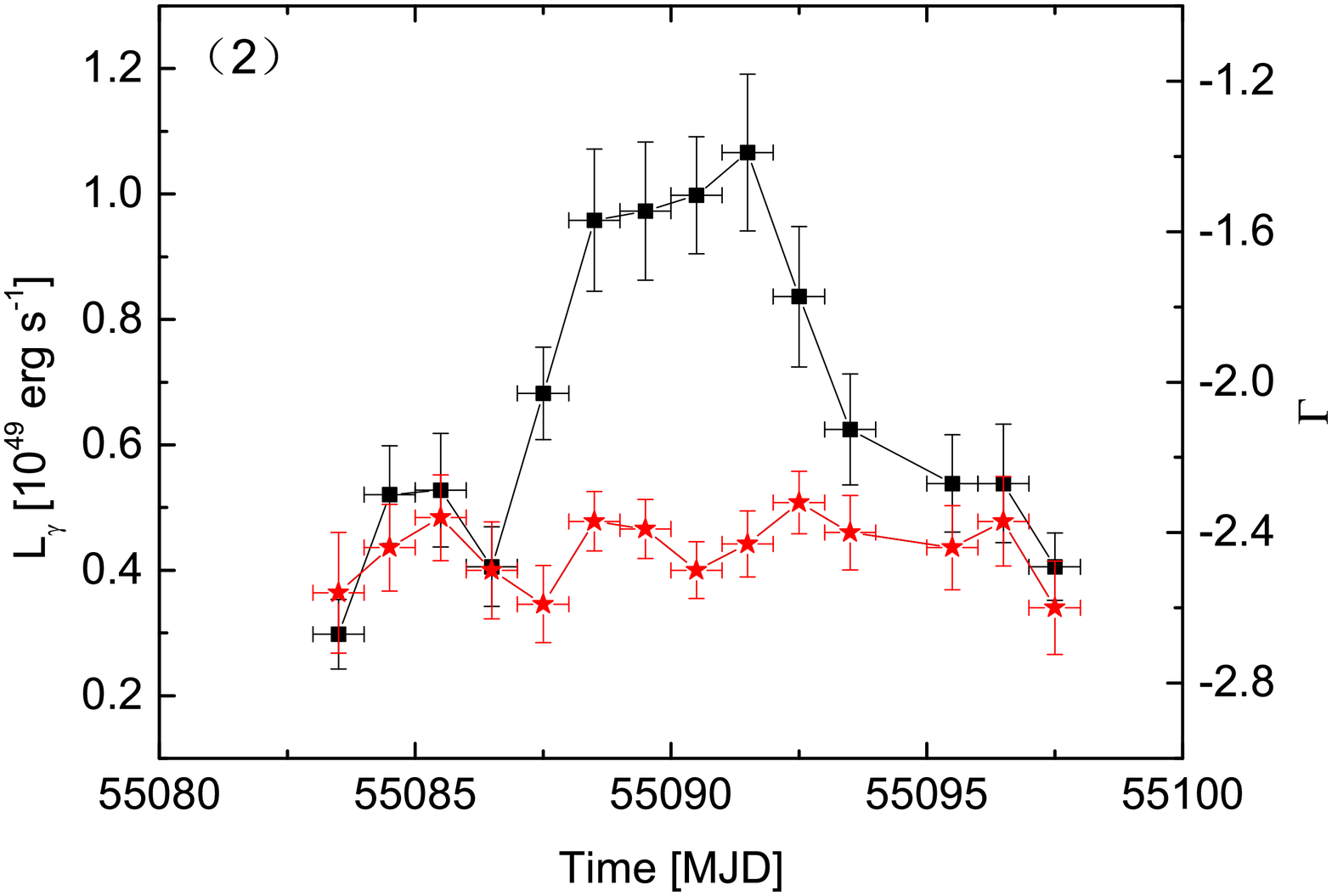}
\includegraphics[angle=0,scale=0.18]{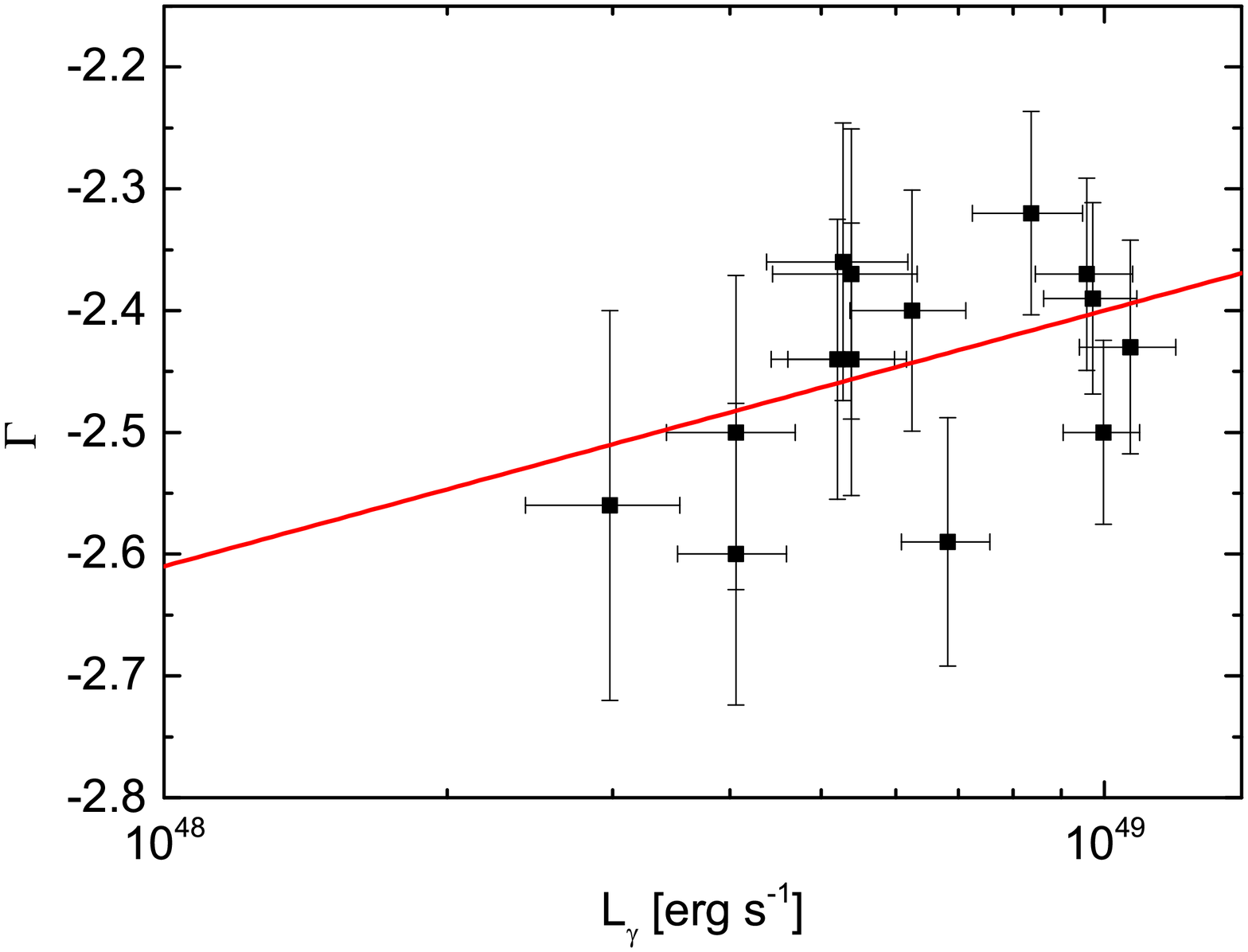}
\includegraphics[angle=0,scale=0.23]{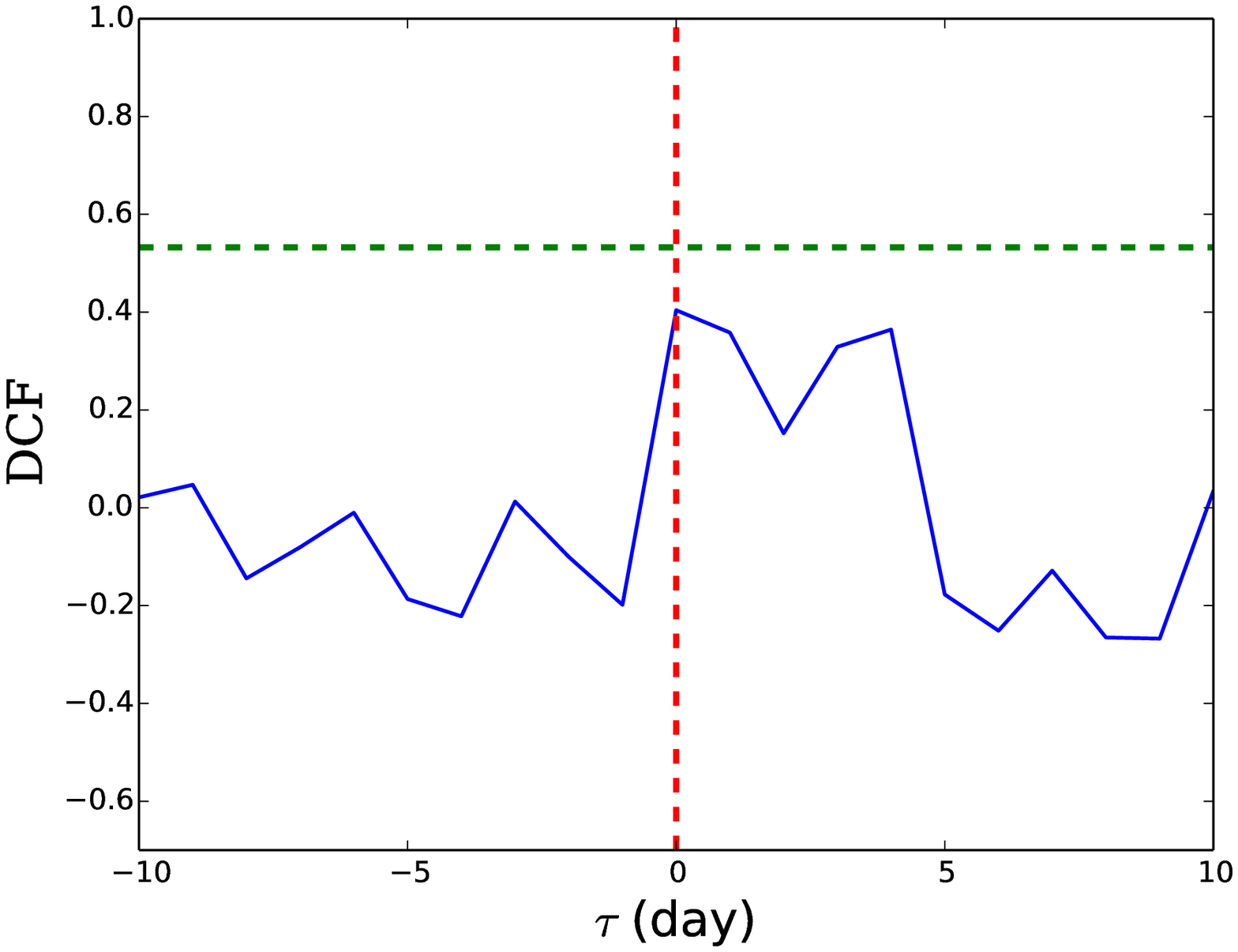}\\
\includegraphics[angle=0,scale=0.18]{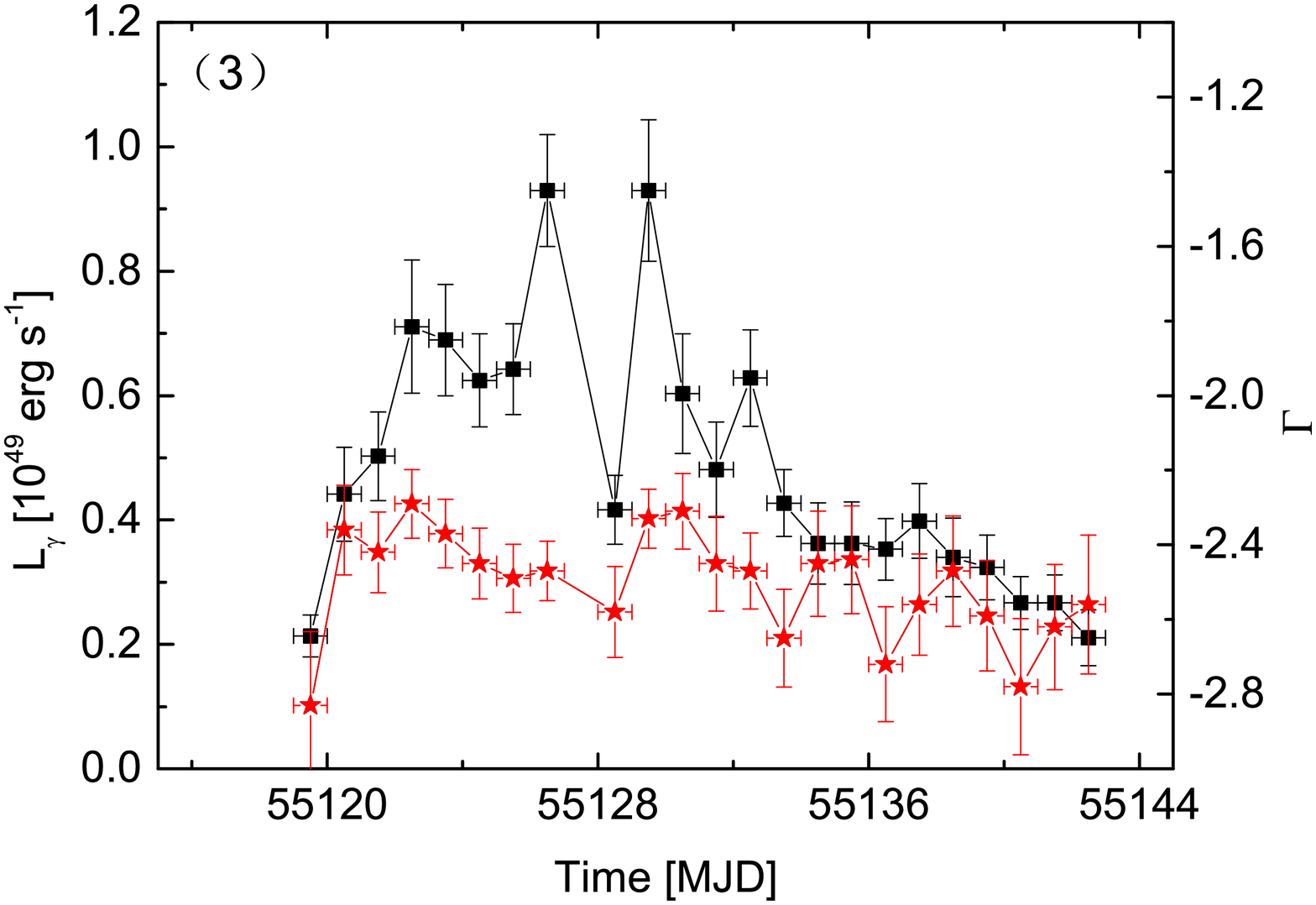}
\includegraphics[angle=0,scale=0.18]{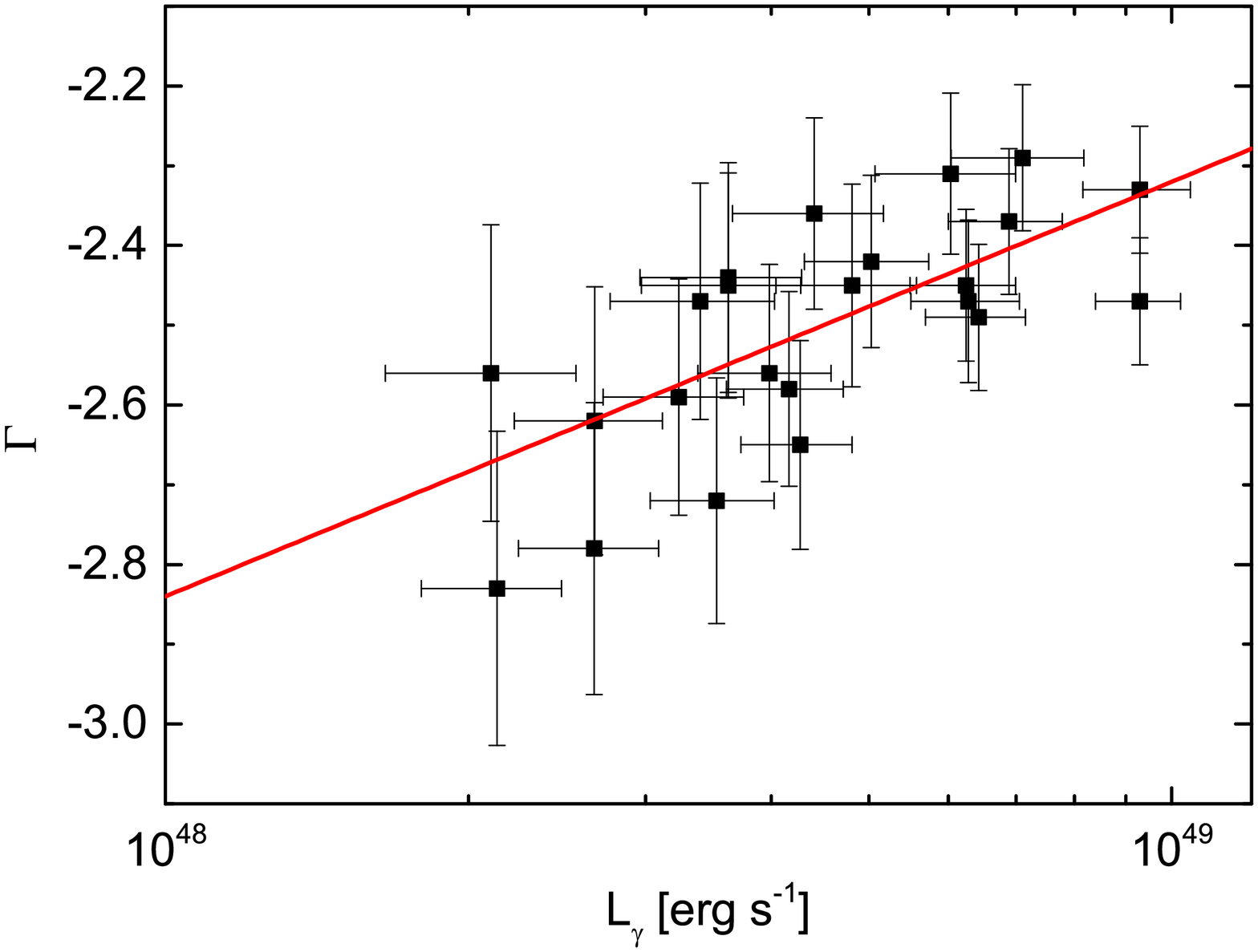}
\includegraphics[angle=0,scale=0.23]{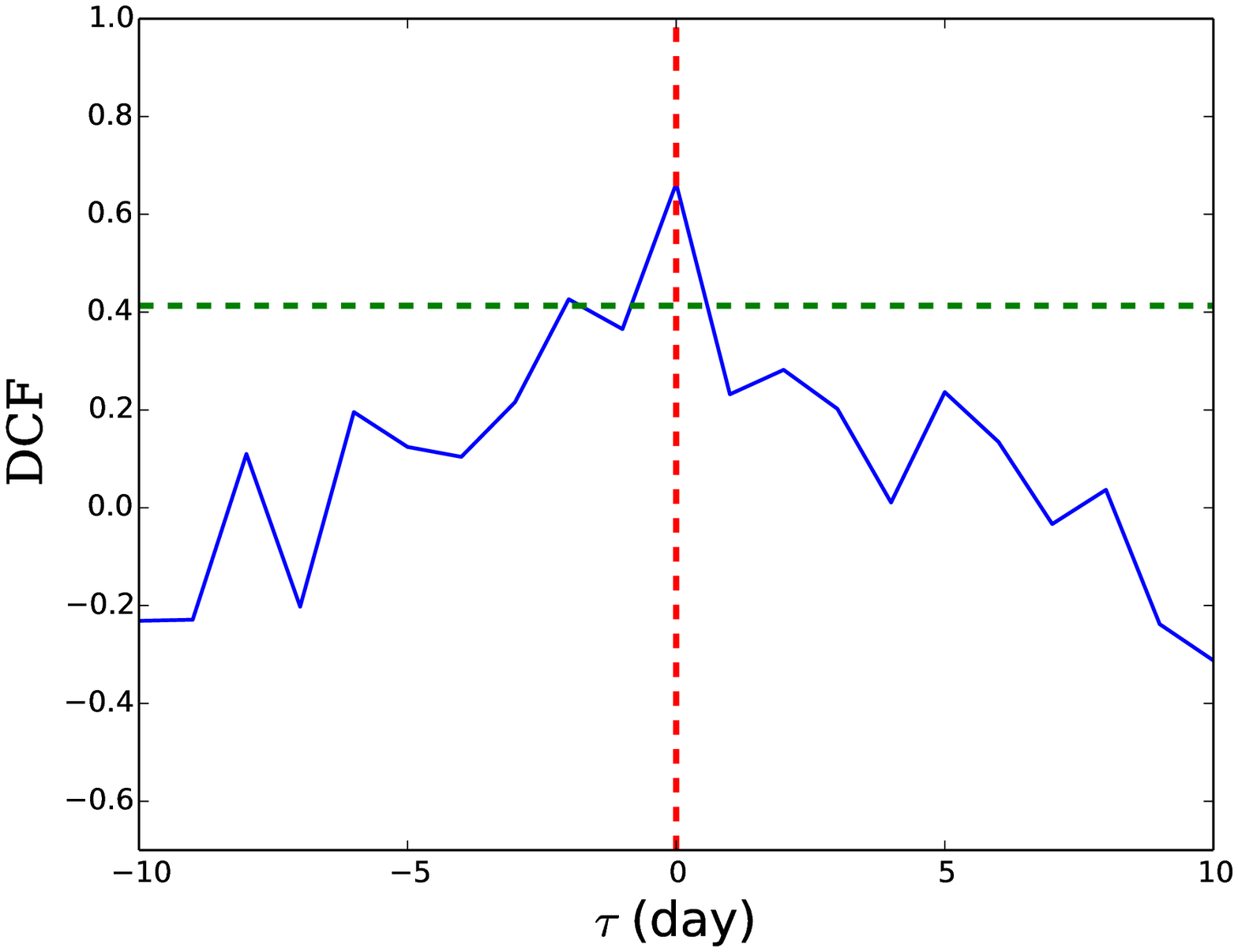}

\caption{\emph{Left Panels}---The temporal evolution of luminosity ($L_{\gamma}$, \emph{black squares}) and photon spectral index ($\Gamma$, \emph{red stars}) observed by \emph{Fermi}/LAT for 3C 454.3. \emph{Middle Panels}---$\Gamma$ vs. $L_{\gamma}$ for each outburst episode in the left panels. The \emph{red lines} are the linear fitting lines by considering the errors of both $\Gamma$ and $L_{\gamma}$, for which the slopes are listed in Table 2. \emph{Right Panels}---The DCF results between $\Gamma$ and $L_{\gamma}$, where $\Gamma$ and $L_{\gamma}$ have been normalized with the Min-Max Normalization method before the DCF analysis. The \emph{green horizontal lines} are the 95\% confidence level lines.}\label{episode}
\end{figure*}

\begin{figure*}
\includegraphics[angle=0,scale=0.18]{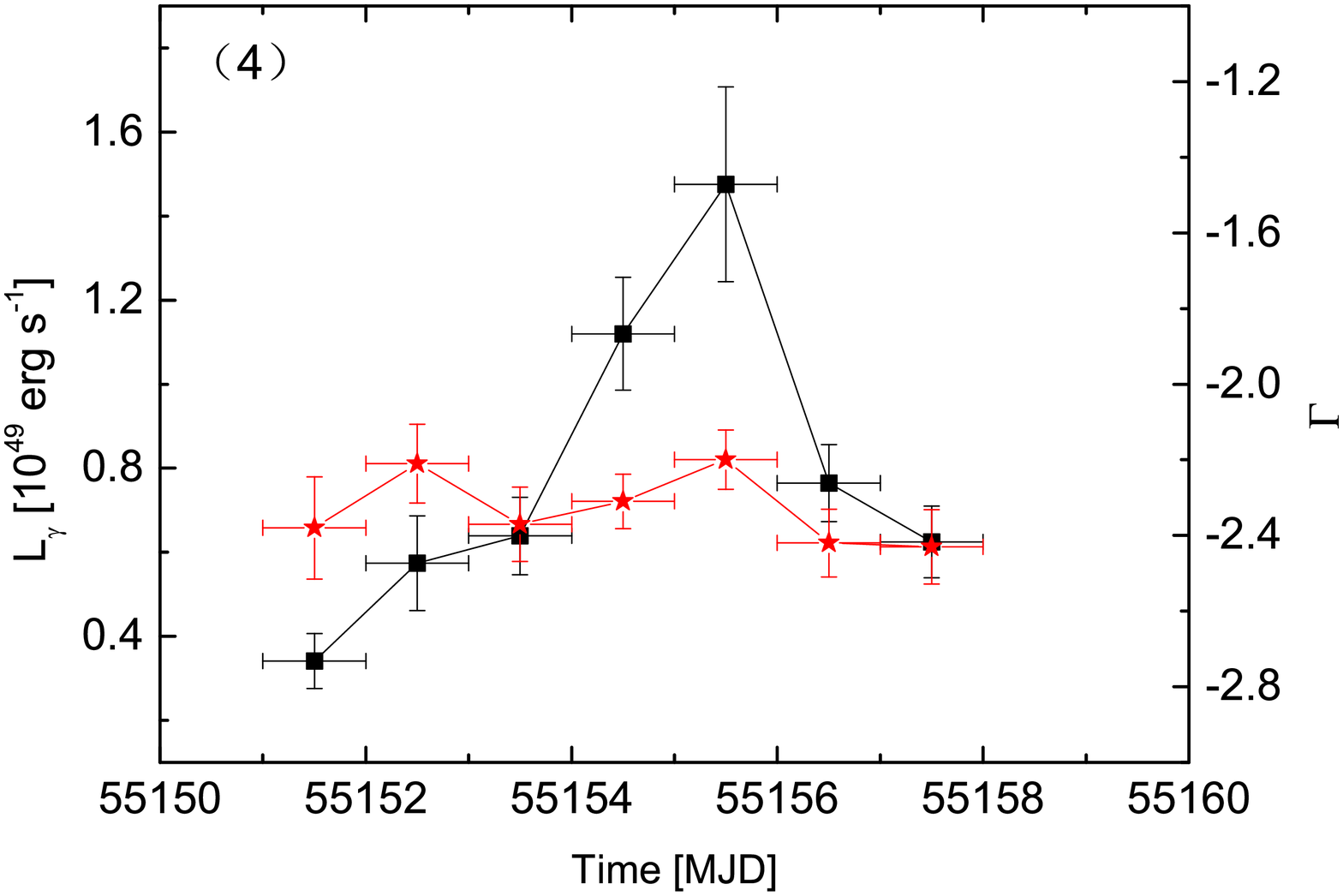}
\includegraphics[angle=0,scale=0.18]{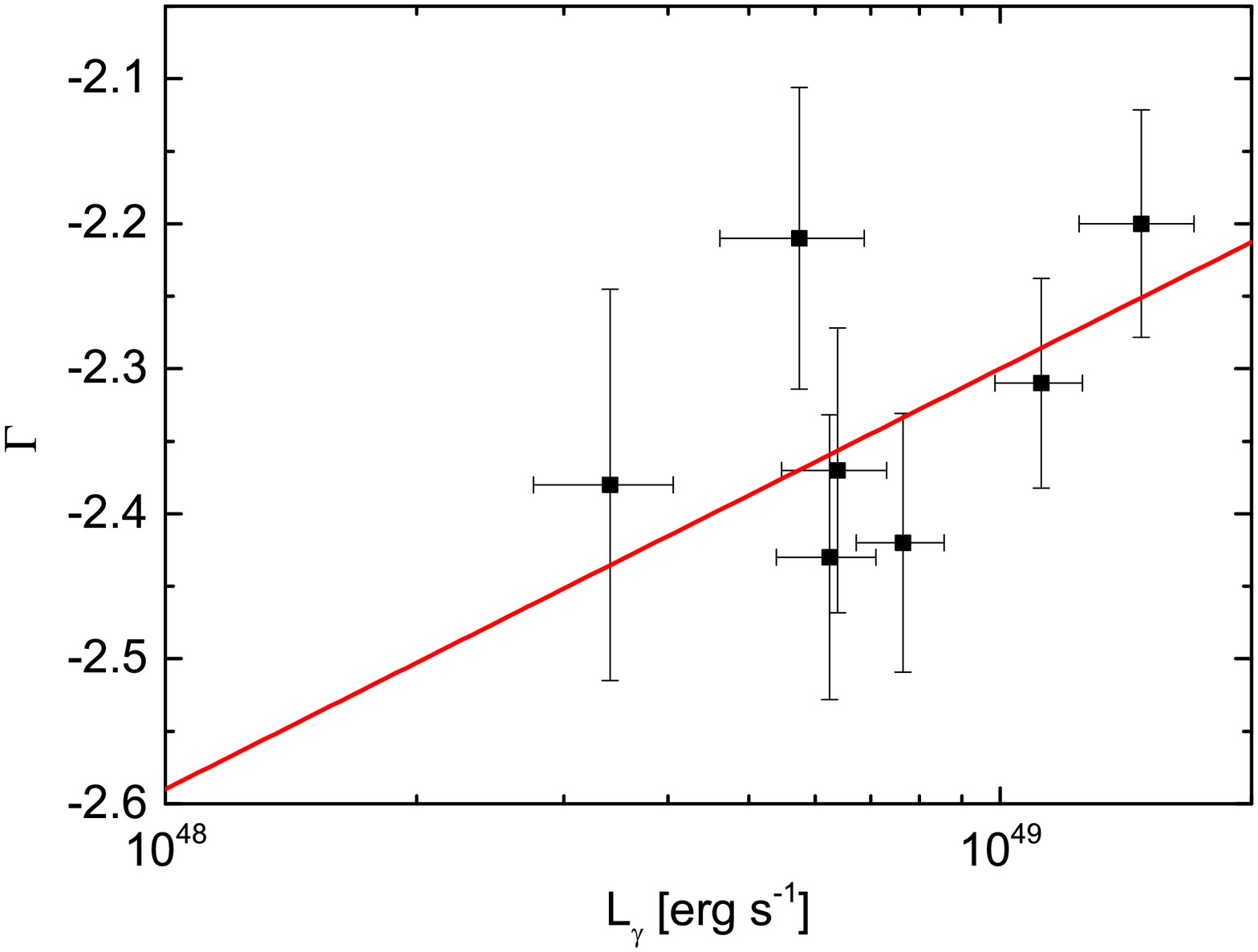}
\includegraphics[angle=0,scale=0.23]{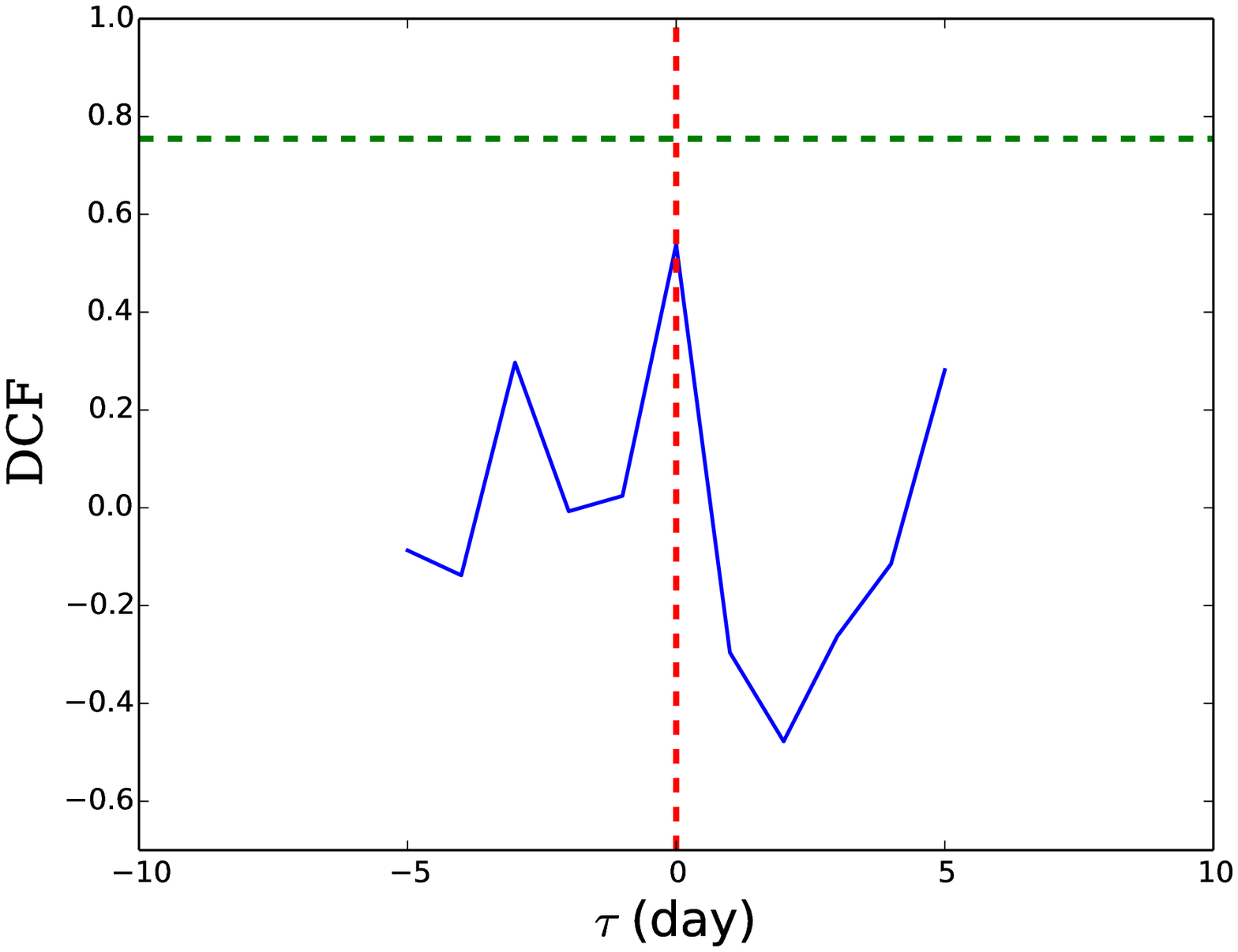}\\
\includegraphics[angle=0,scale=0.18]{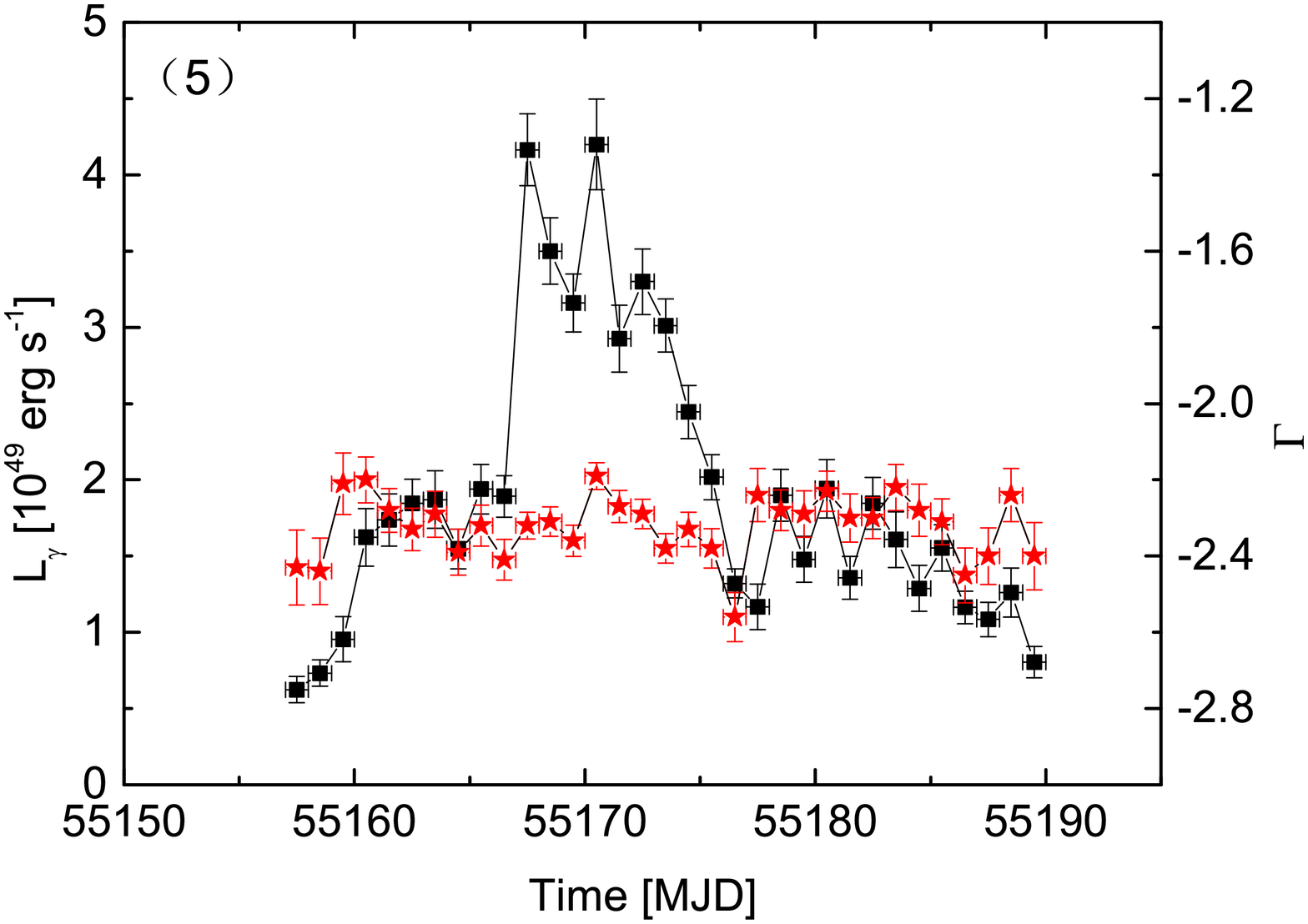}
\includegraphics[angle=0,scale=0.18]{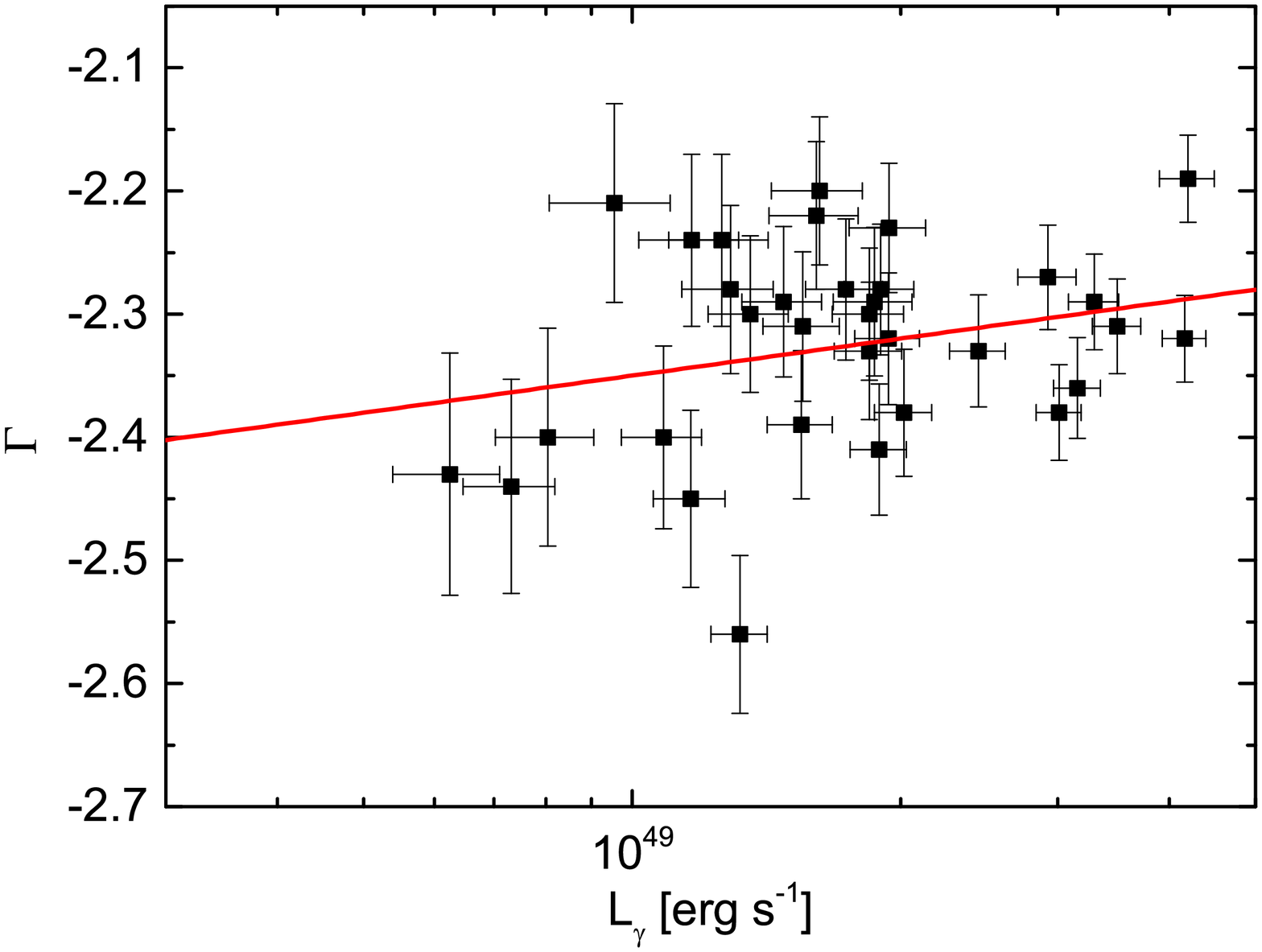}
\includegraphics[angle=0,scale=0.23]{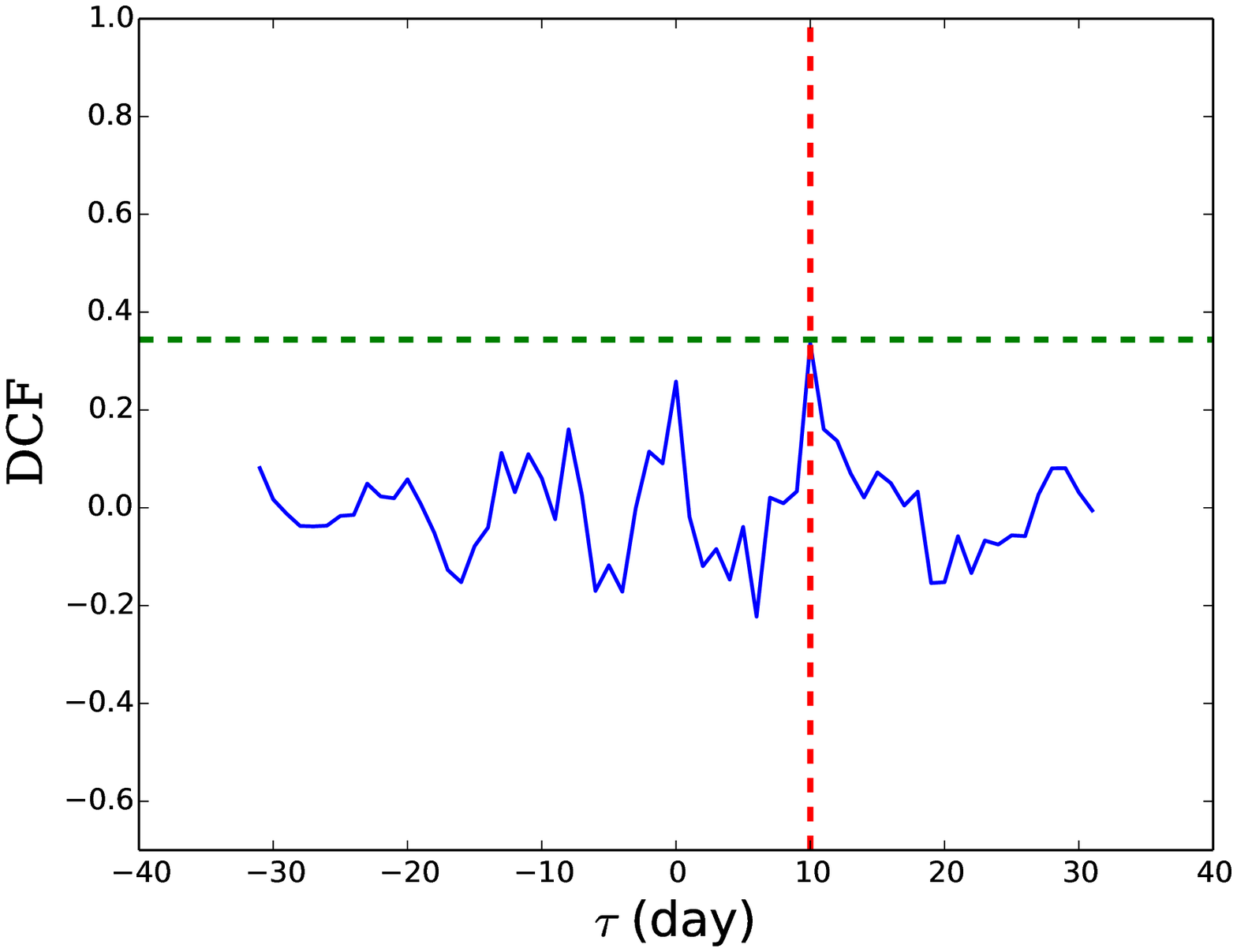}\\
\includegraphics[angle=0,scale=0.18]{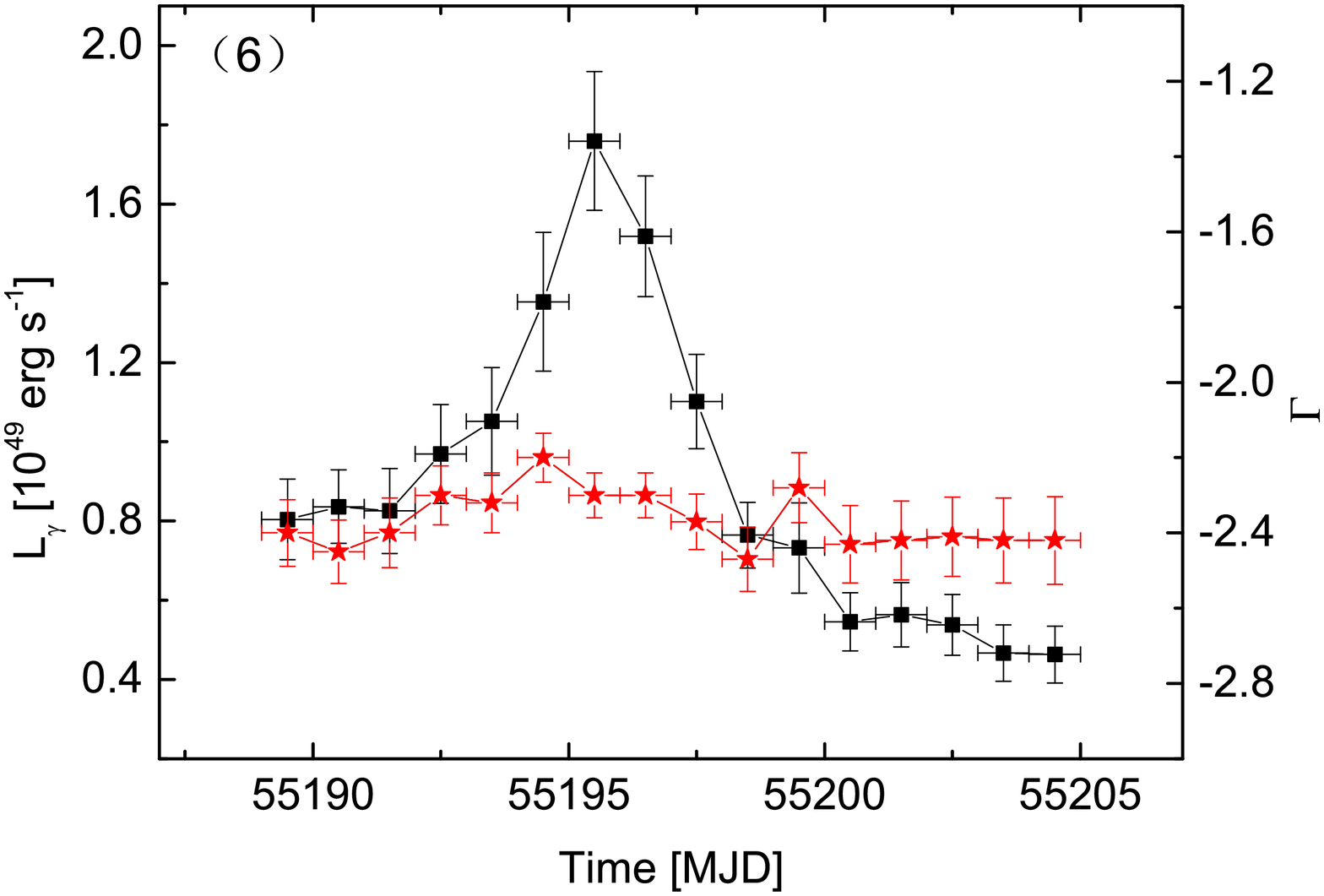}
\includegraphics[angle=0,scale=0.18]{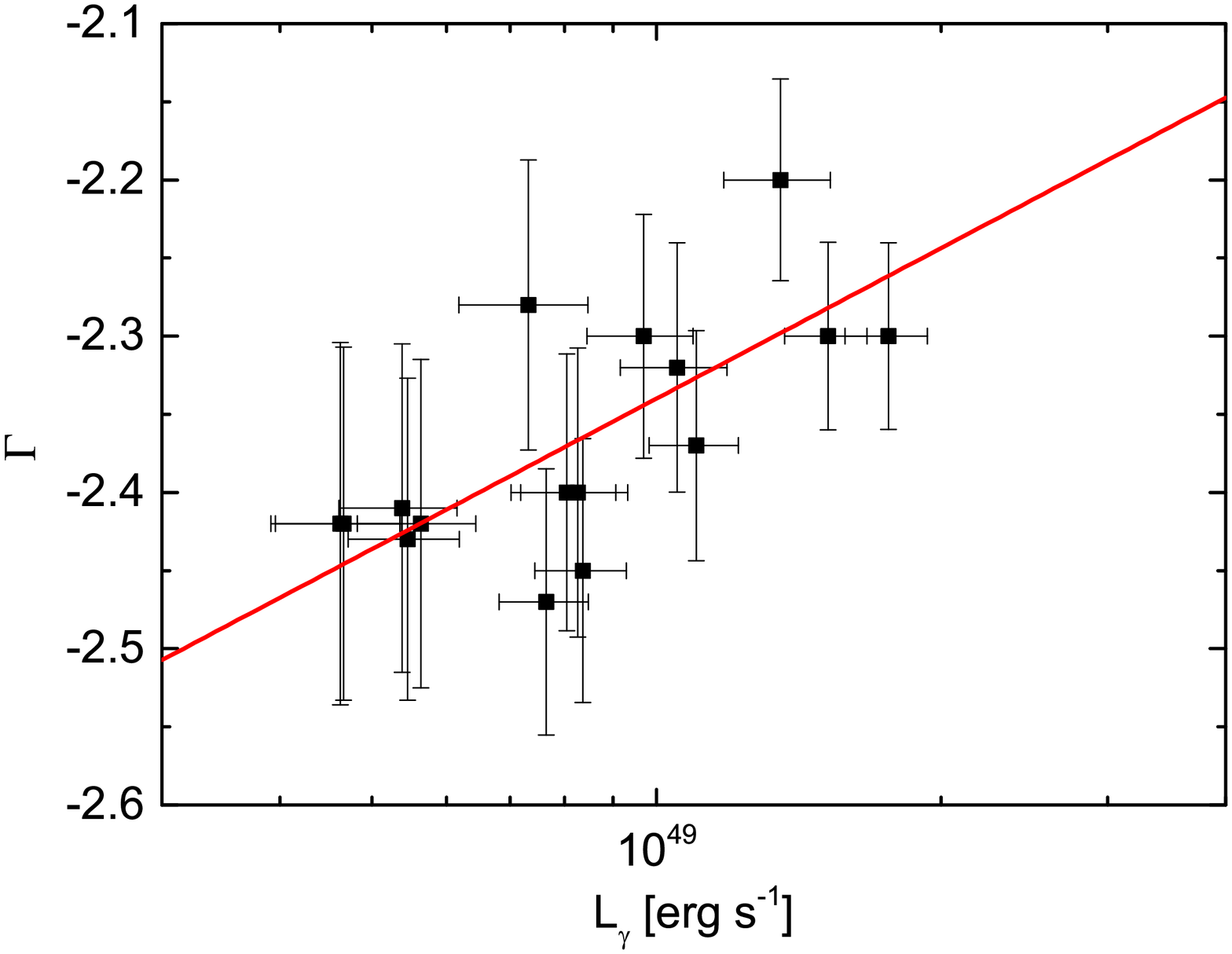}
\includegraphics[angle=0,scale=0.23]{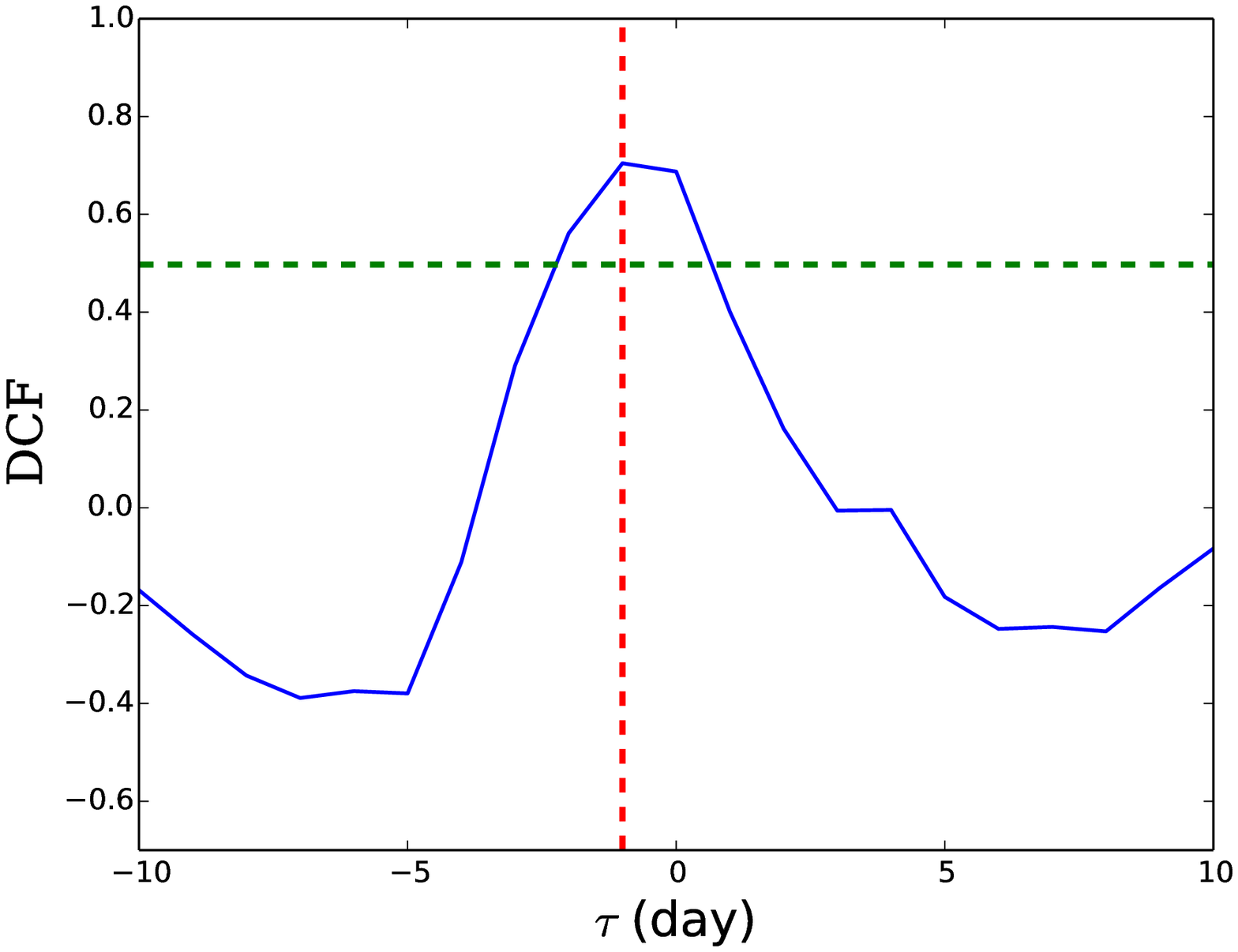}\\
\includegraphics[angle=0,scale=0.18]{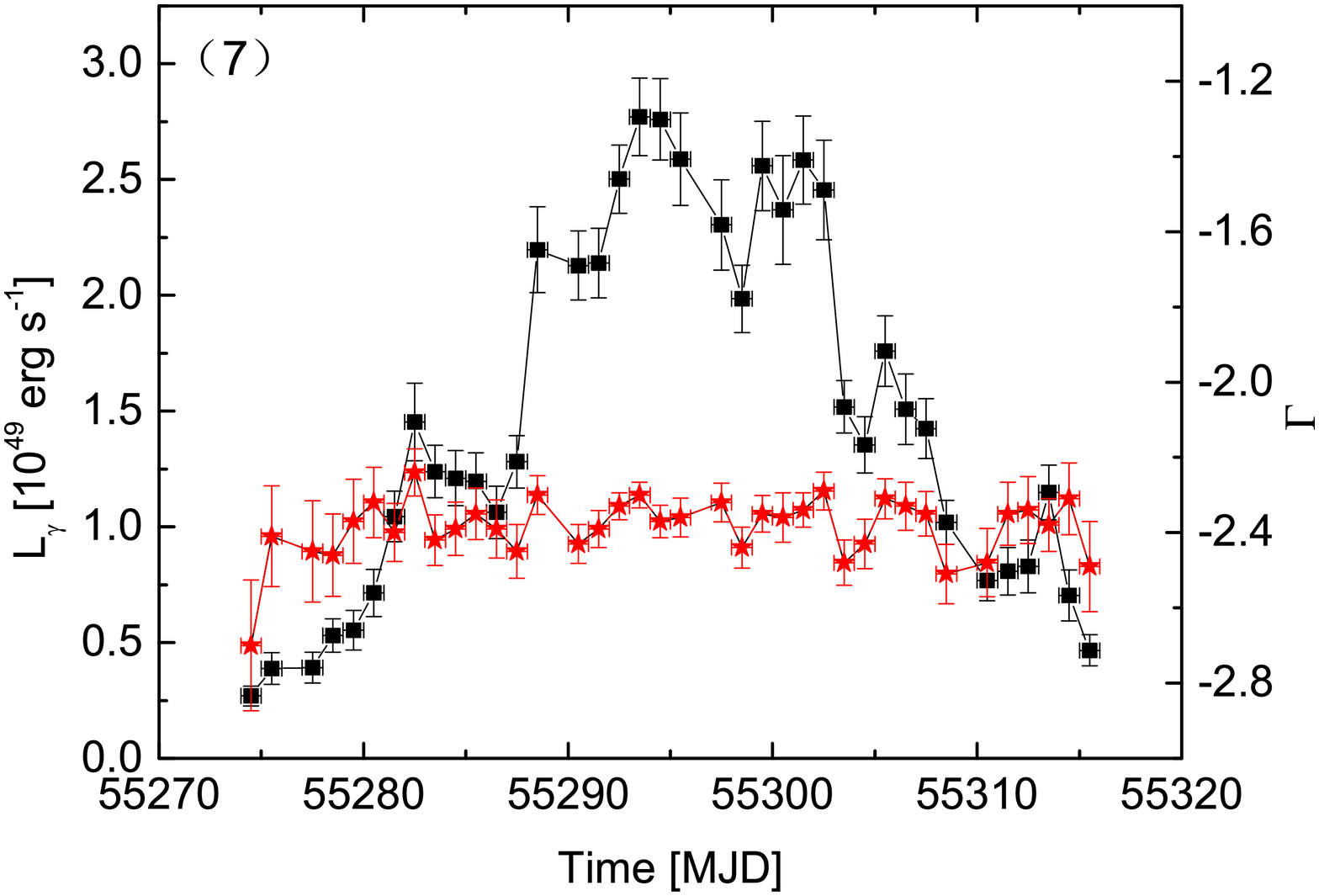}
\includegraphics[angle=0,scale=0.18]{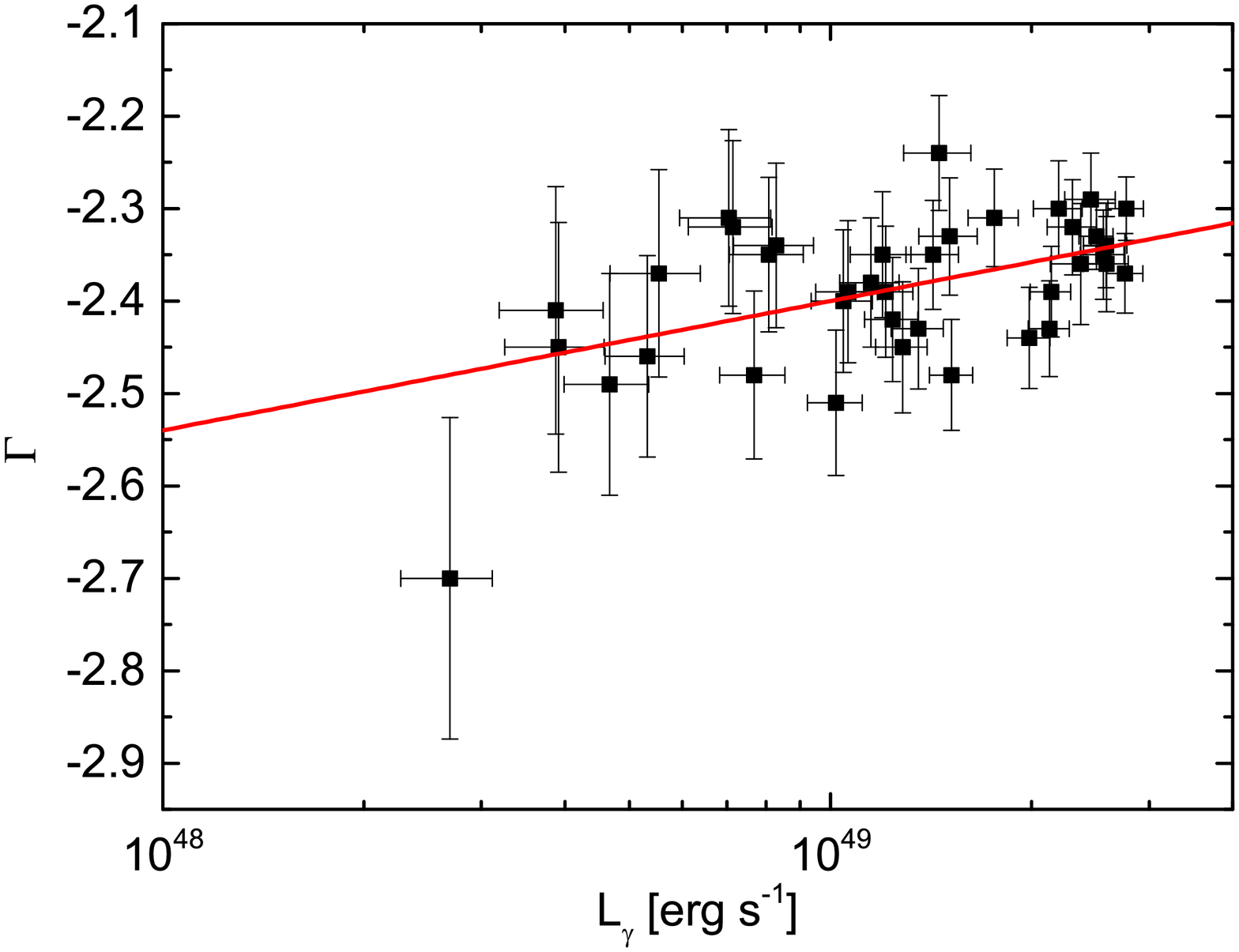}
\includegraphics[angle=0,scale=0.23]{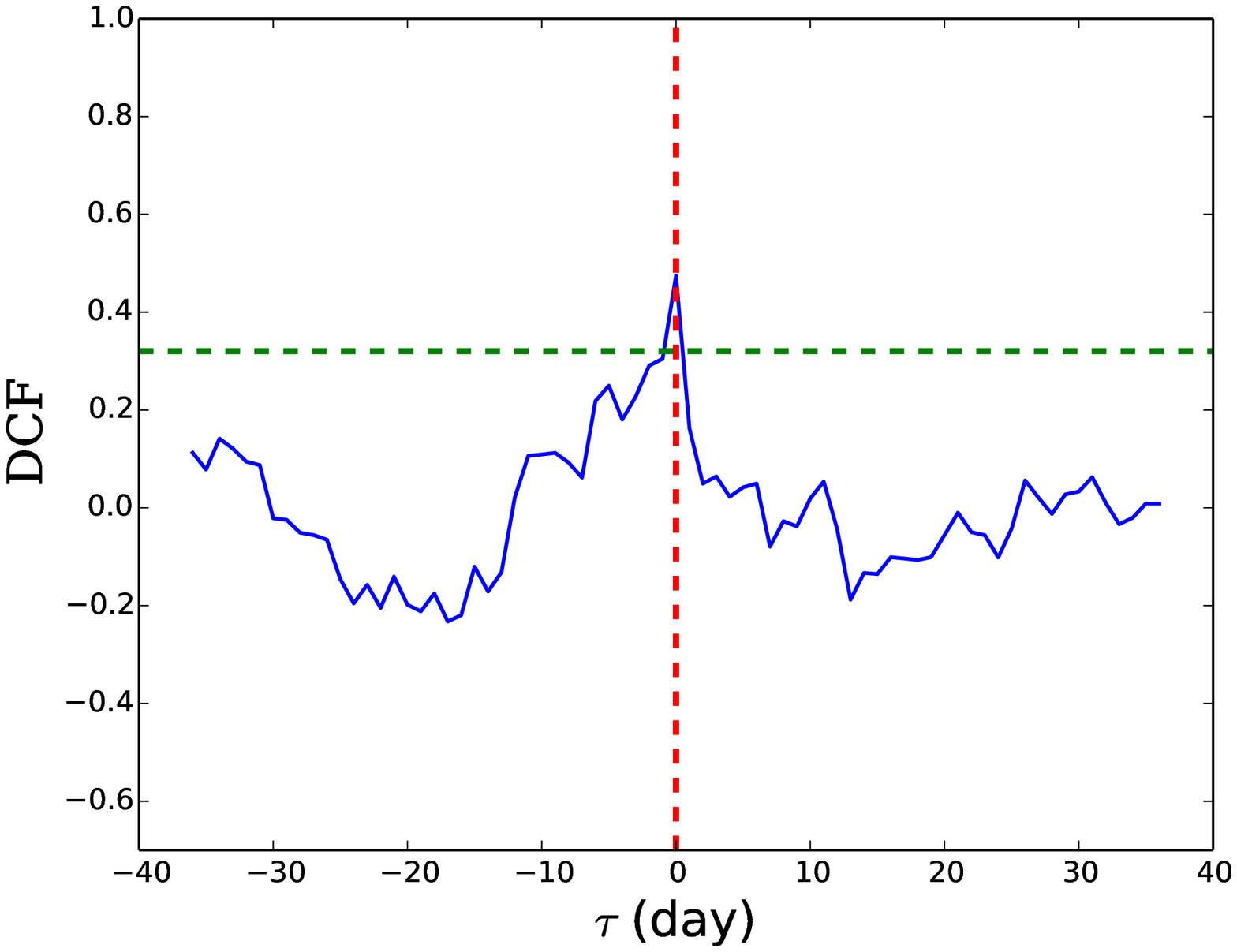}\\
\includegraphics[angle=0,scale=0.18]{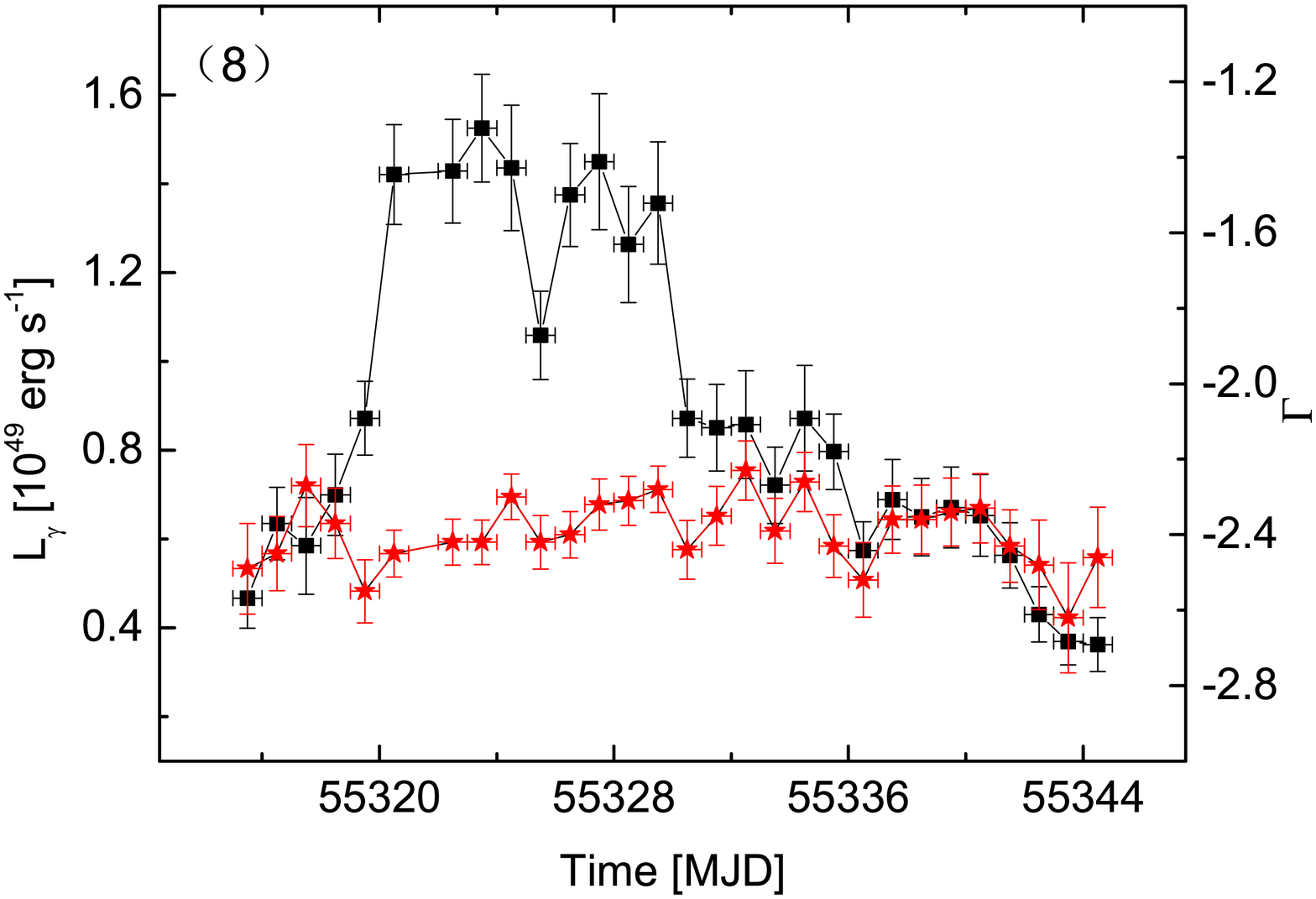}
\includegraphics[angle=0,scale=0.18]{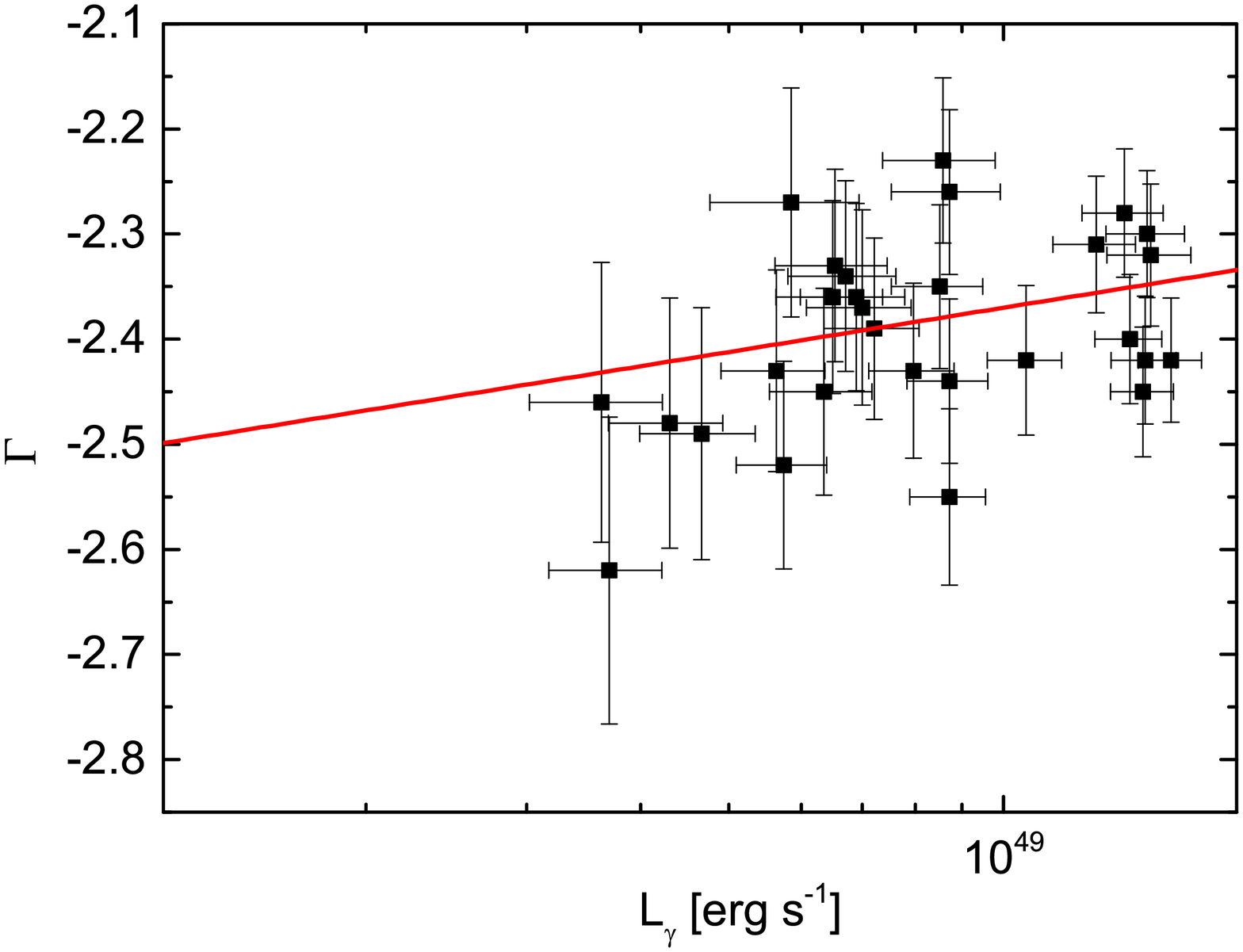}
\includegraphics[angle=0,scale=0.23]{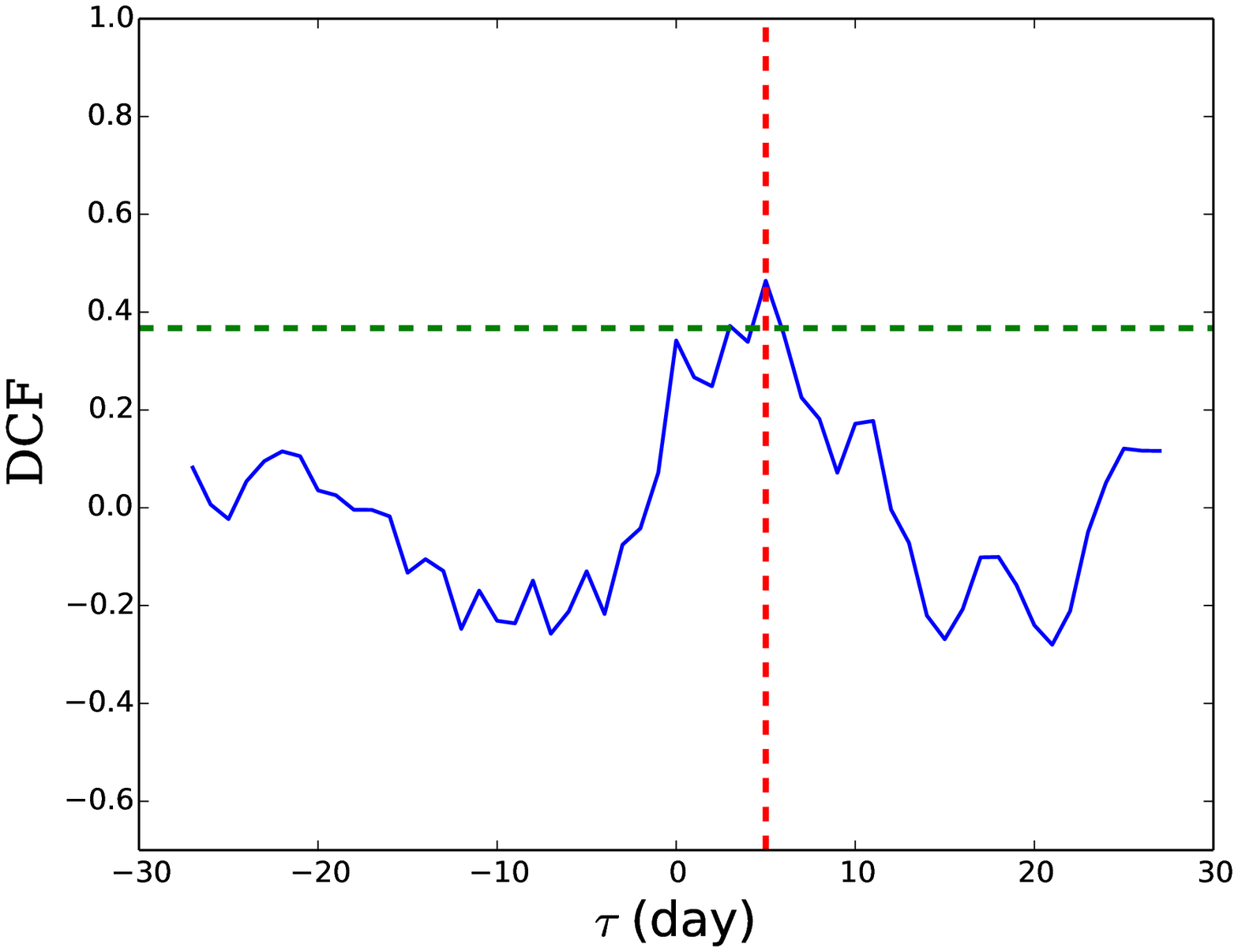}\\
\includegraphics[angle=0,scale=0.18]{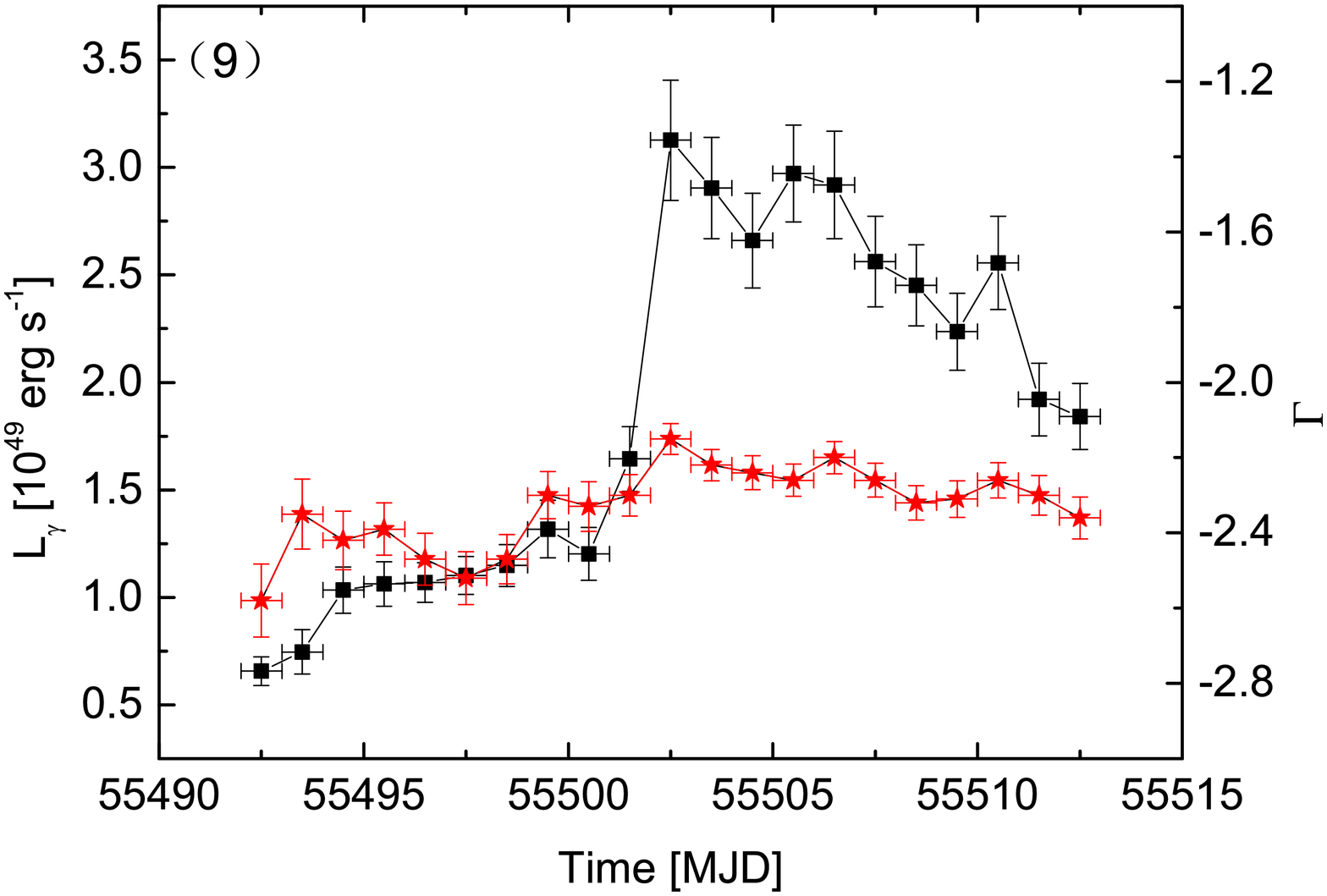}
\includegraphics[angle=0,scale=0.18]{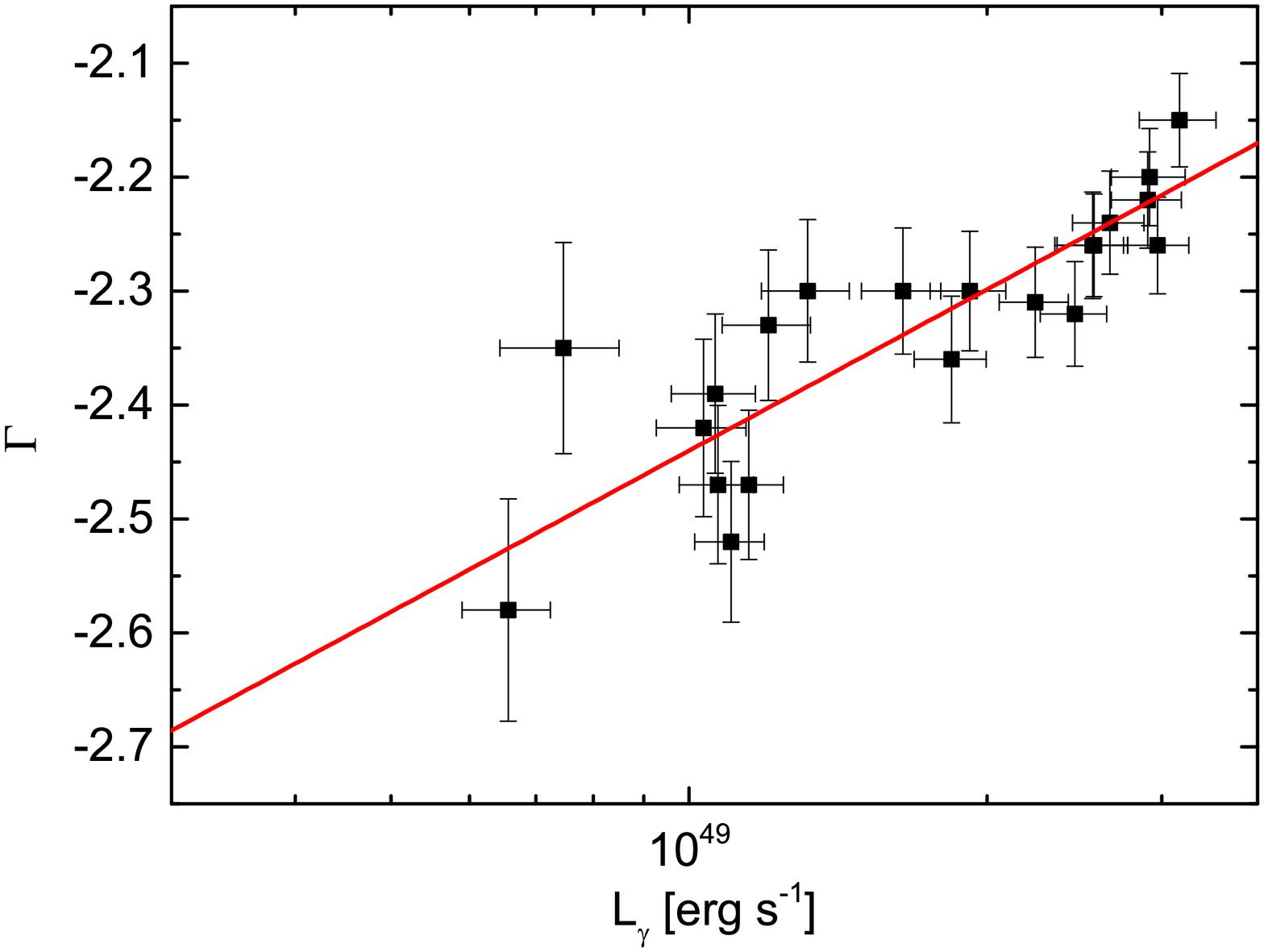}
\includegraphics[angle=0,scale=0.23]{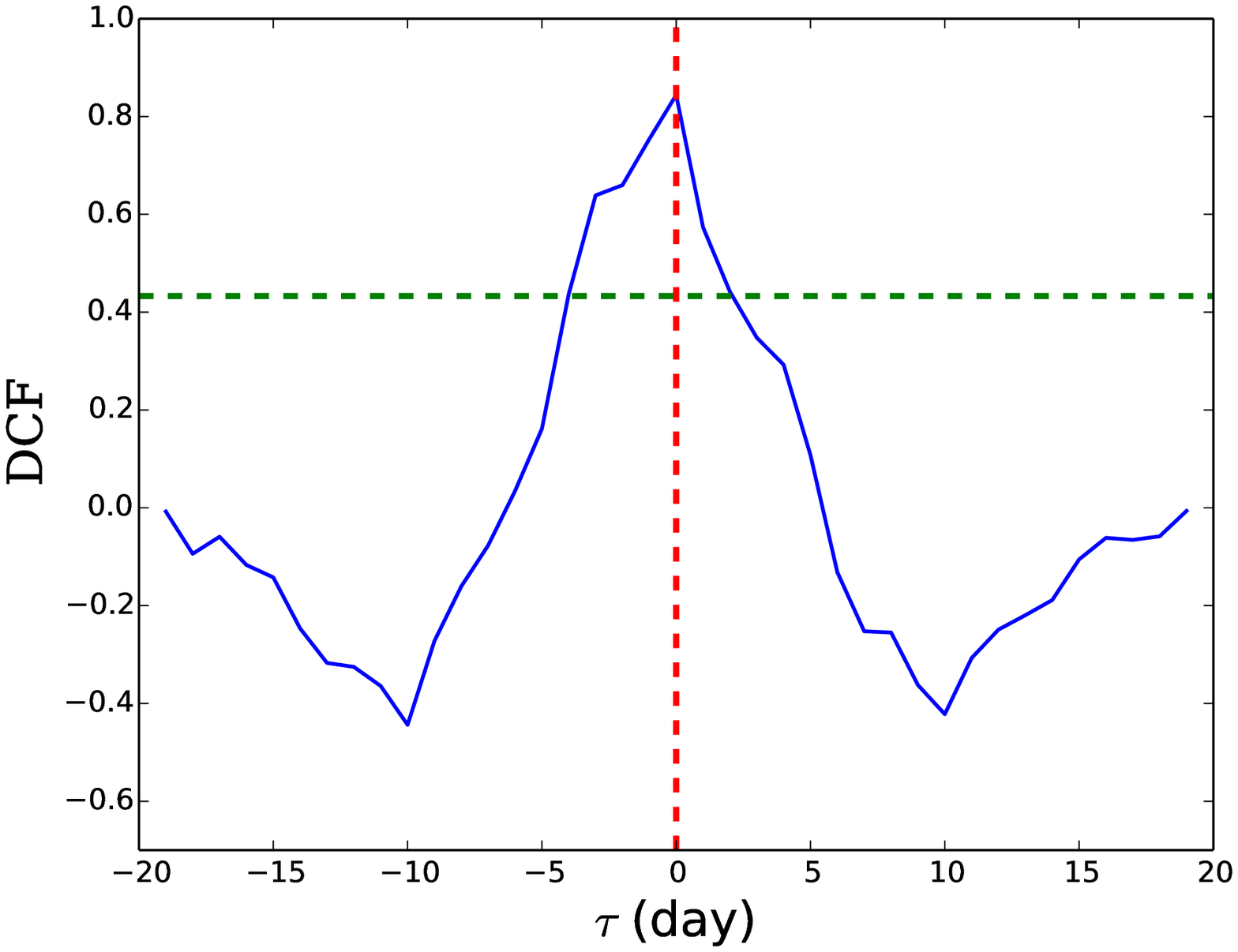}

\hfill\center{Fig.5---  continued}
\end{figure*}

\begin{figure*}
\includegraphics[angle=0,scale=0.18]{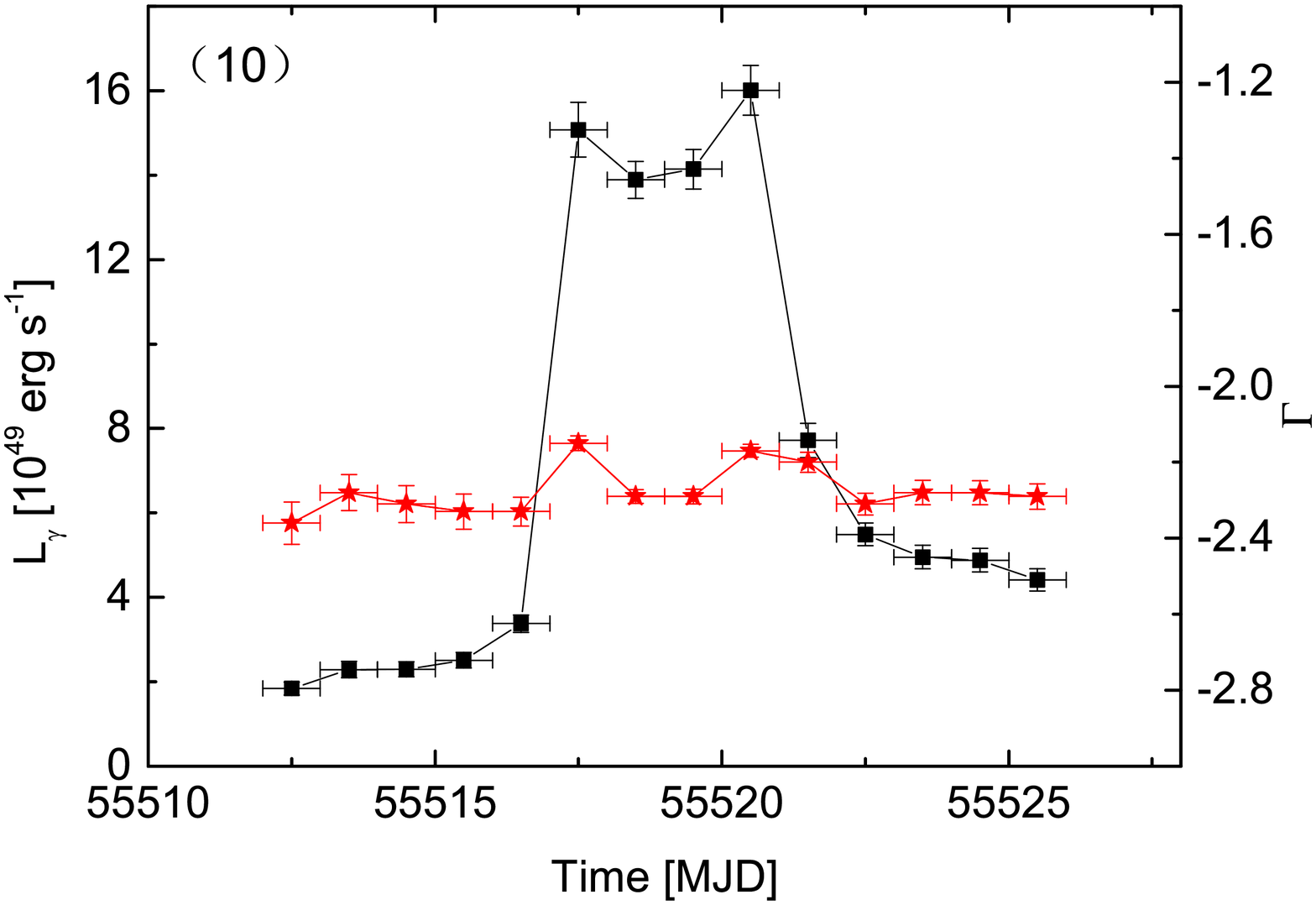}
\includegraphics[angle=0,scale=0.18]{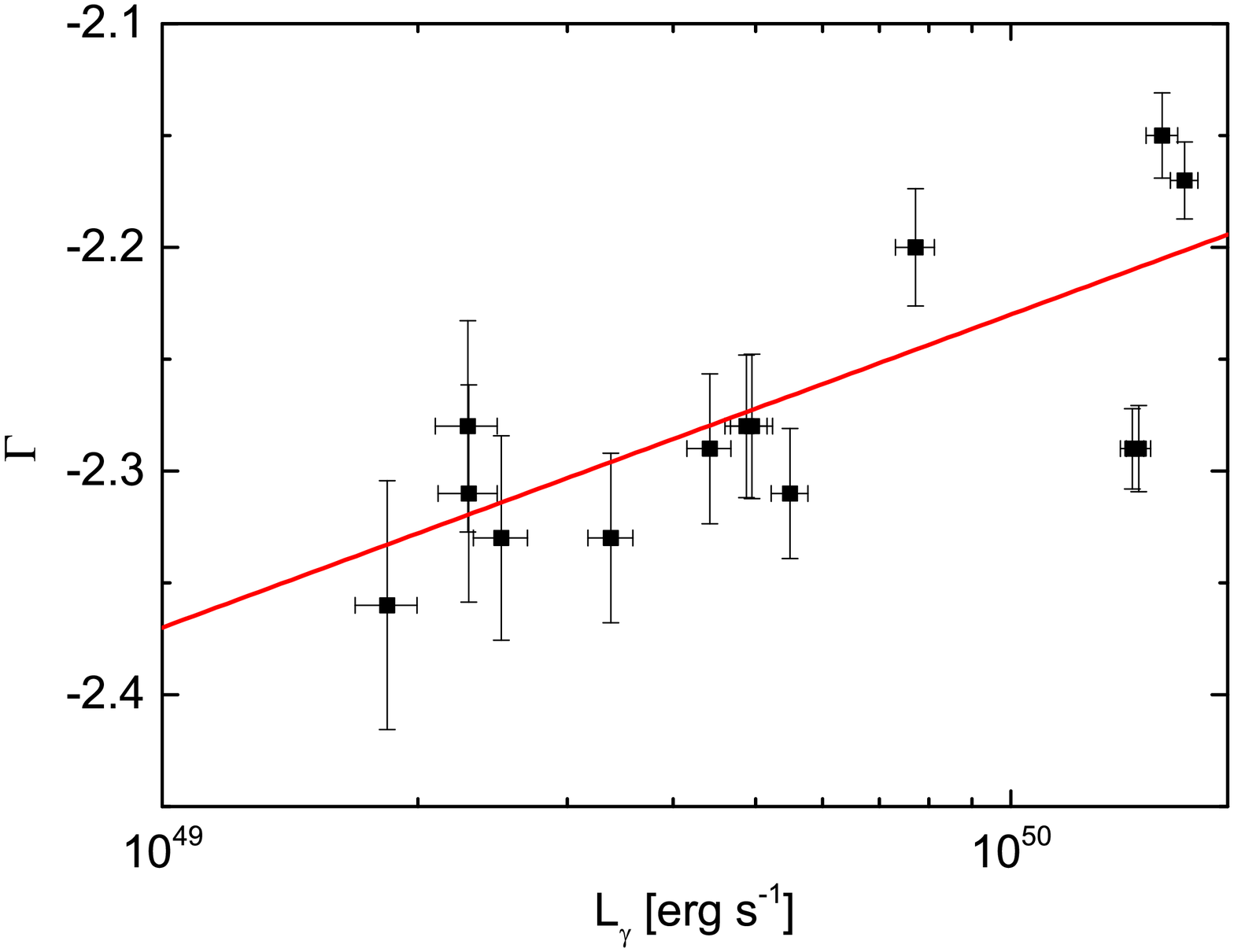}
\includegraphics[angle=0,scale=0.23]{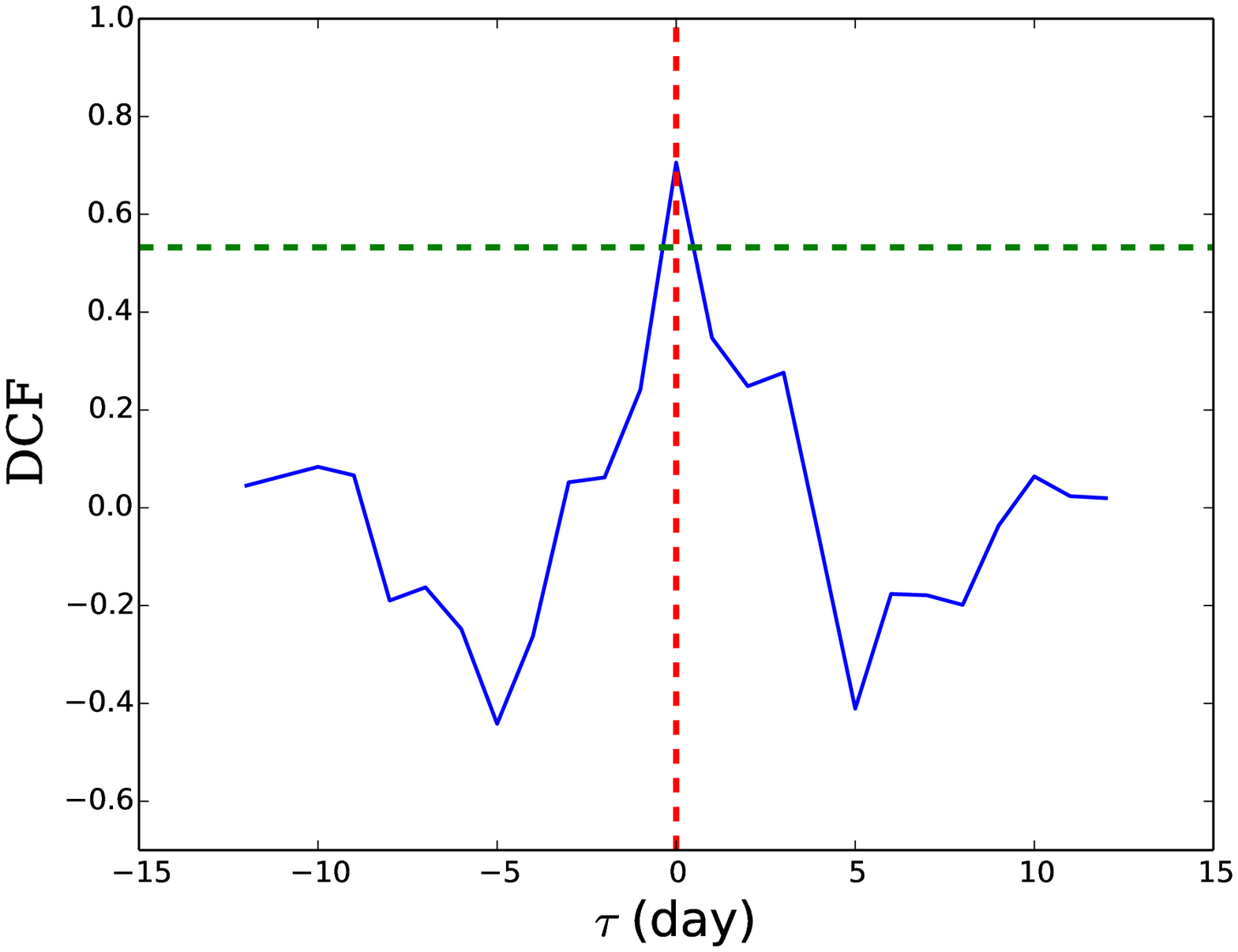}\\
\includegraphics[angle=0,scale=0.18]{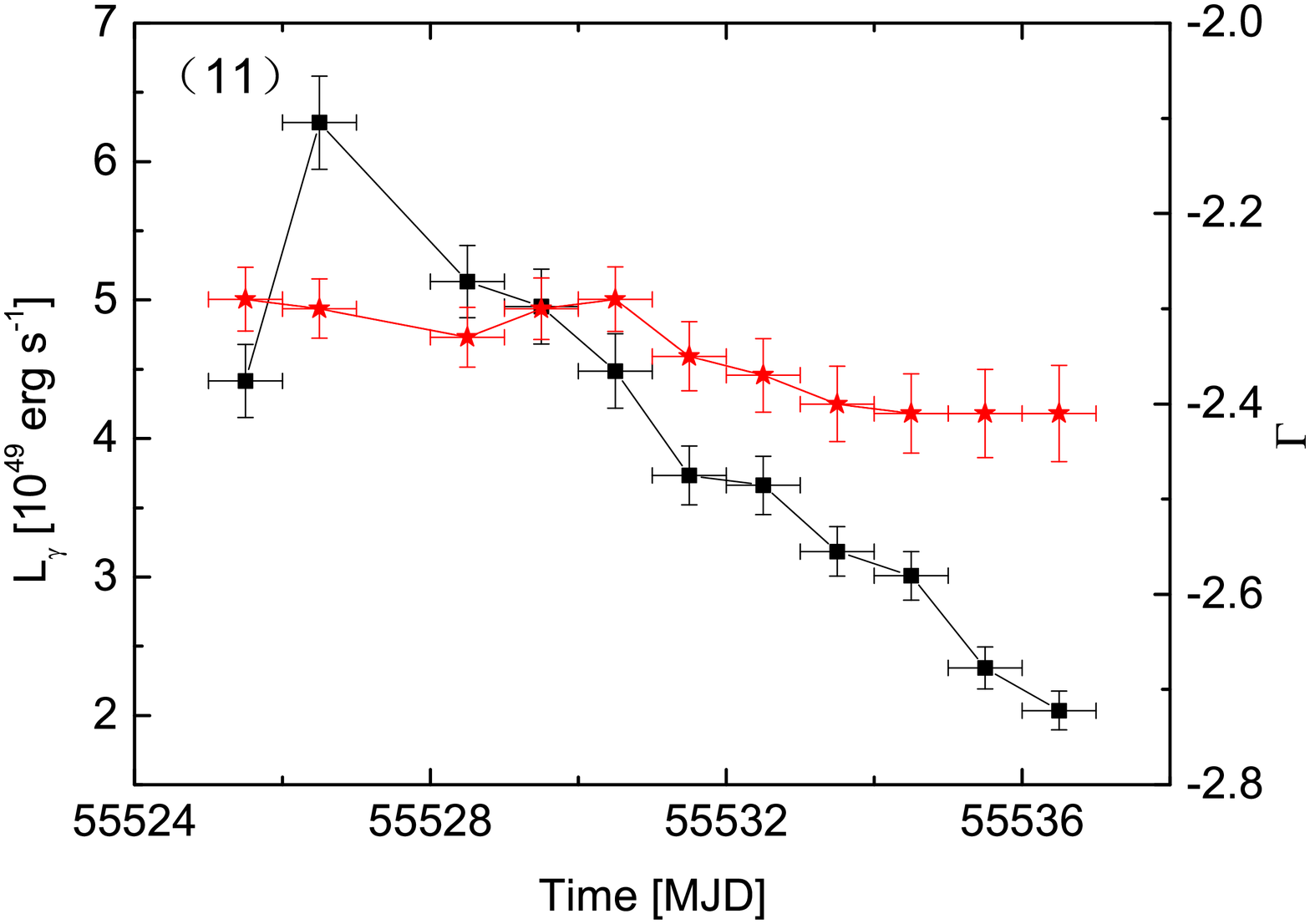}
\includegraphics[angle=0,scale=0.18]{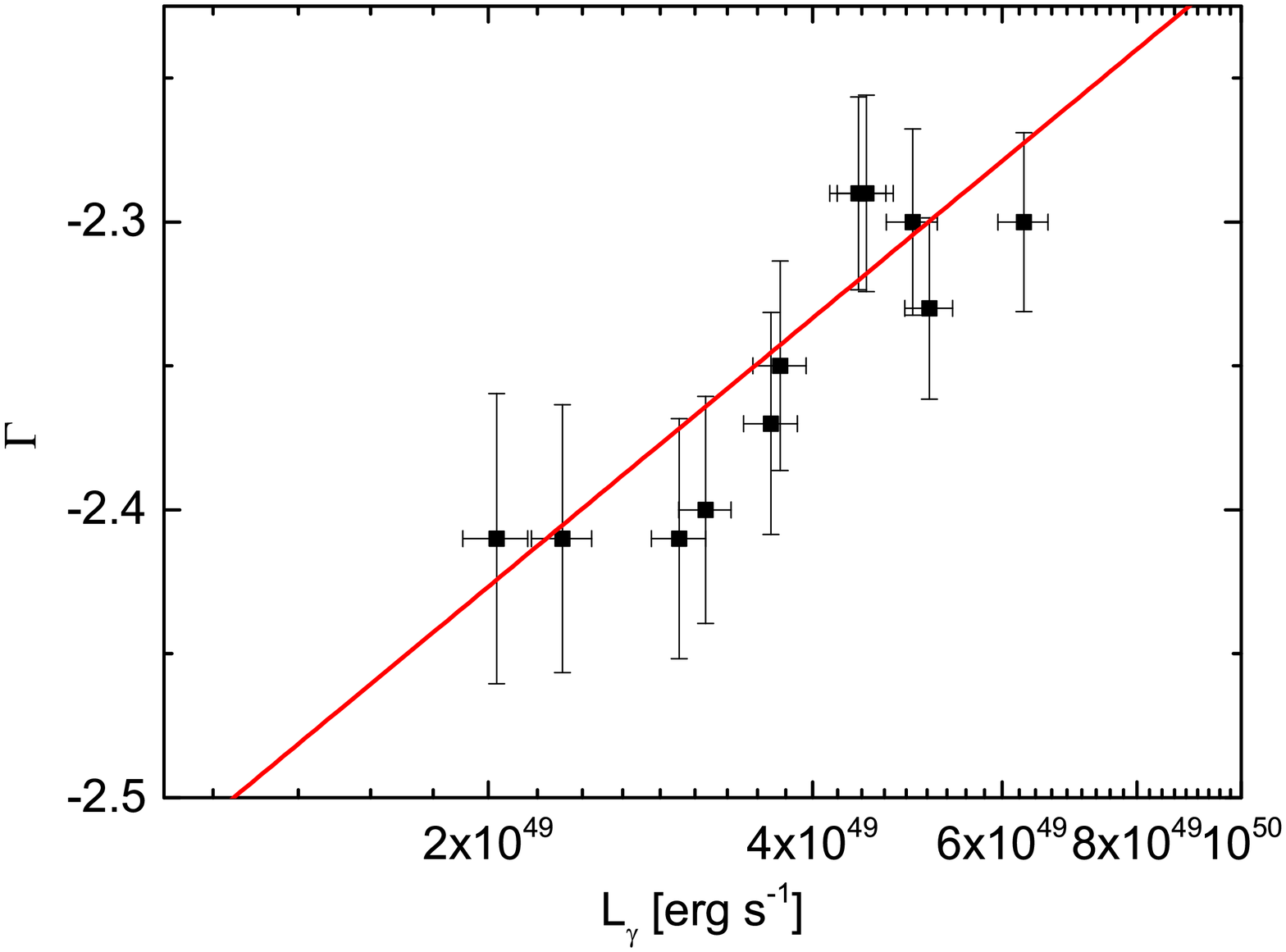}
\includegraphics[angle=0,scale=0.23]{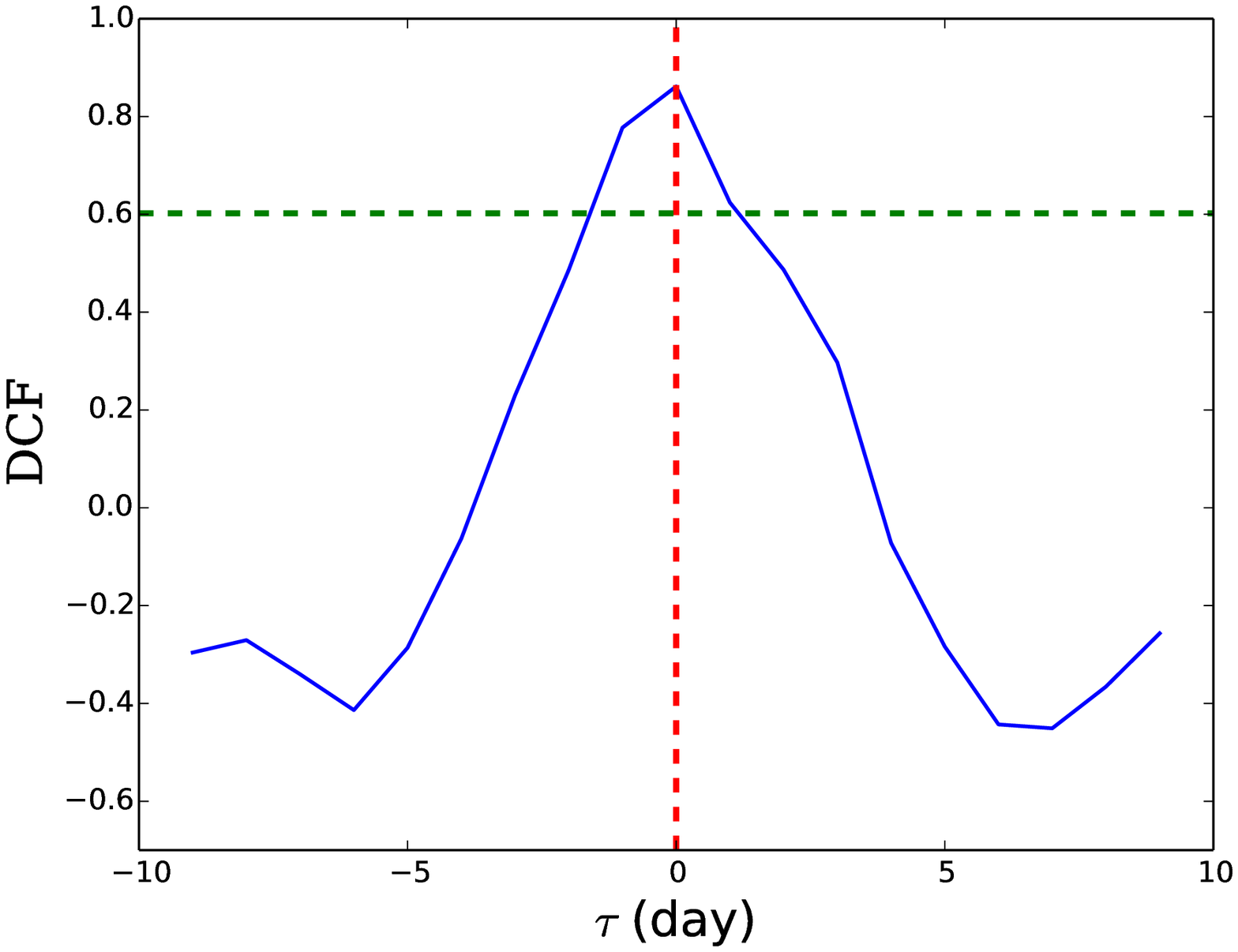}\\
\includegraphics[angle=0,scale=0.18]{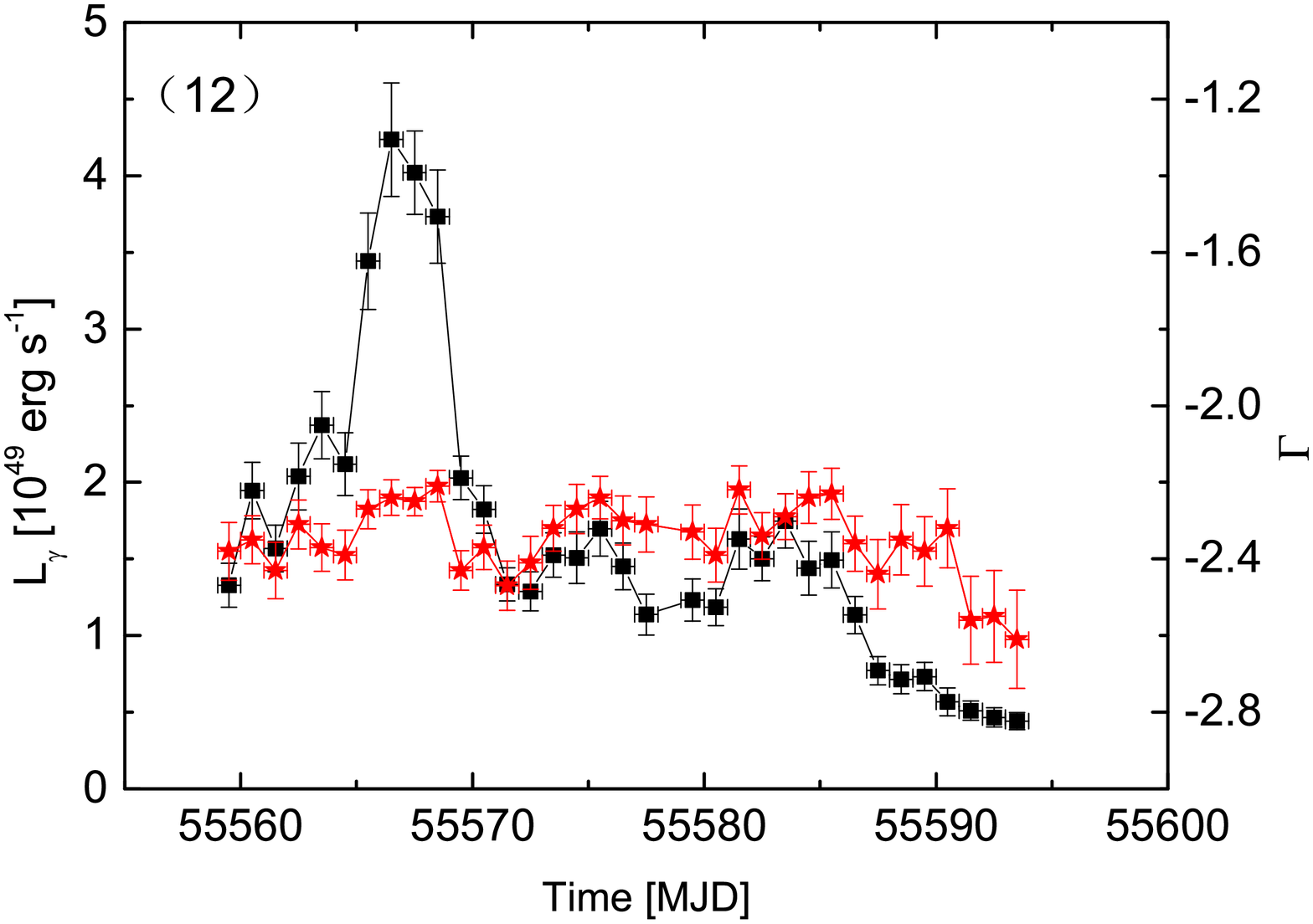}
\includegraphics[angle=0,scale=0.18]{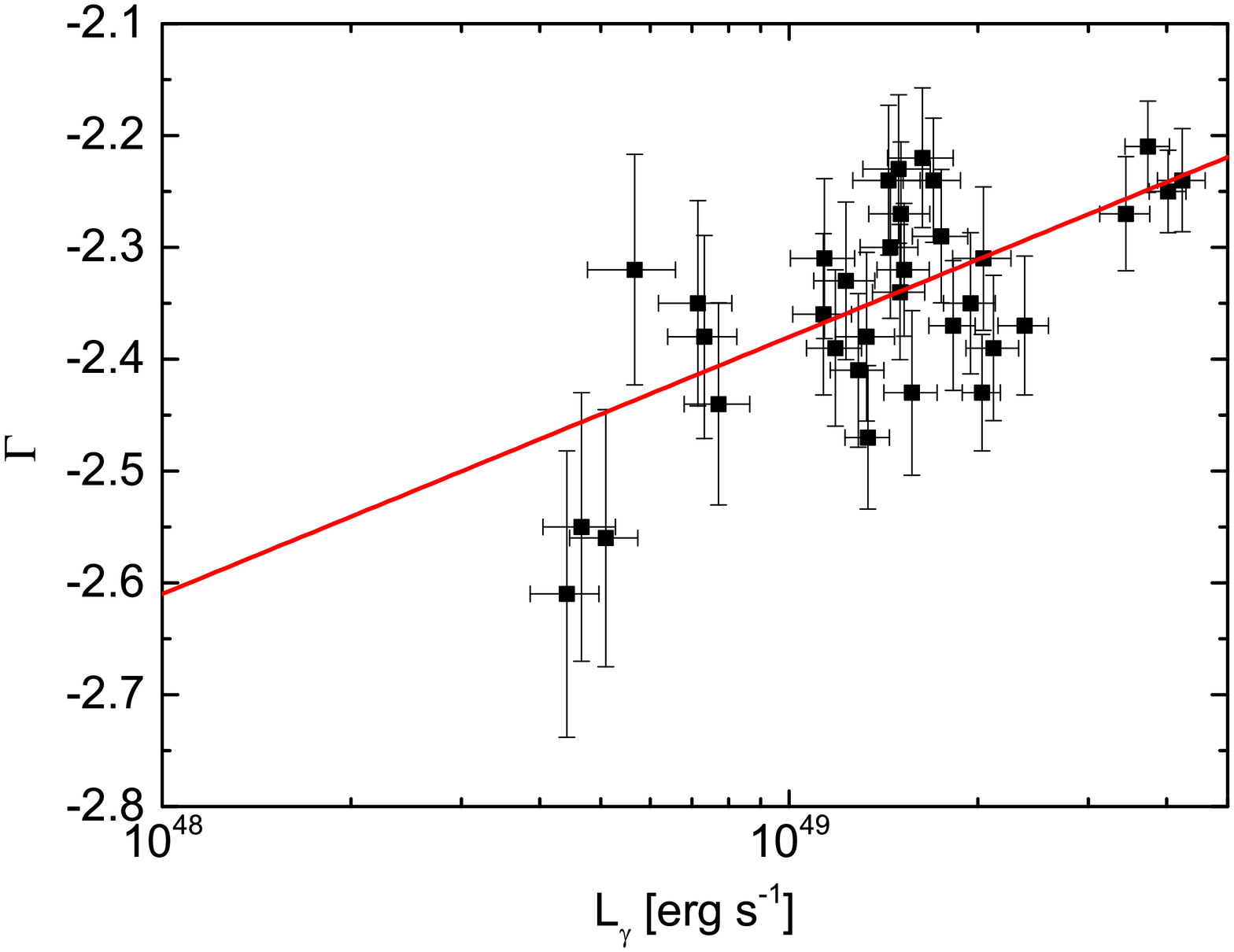}
\includegraphics[angle=0,scale=0.23]{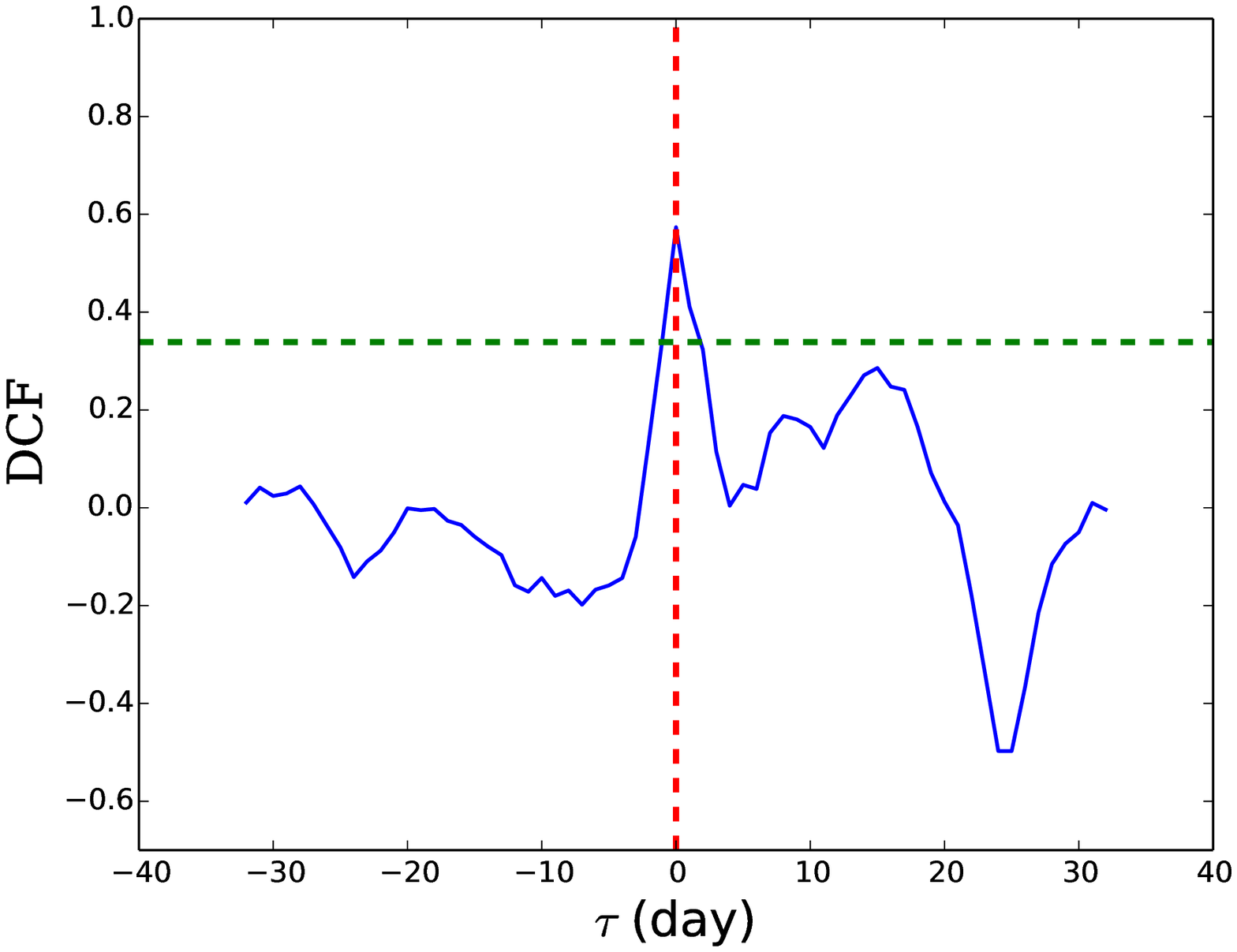}\\
\includegraphics[angle=0,scale=0.18]{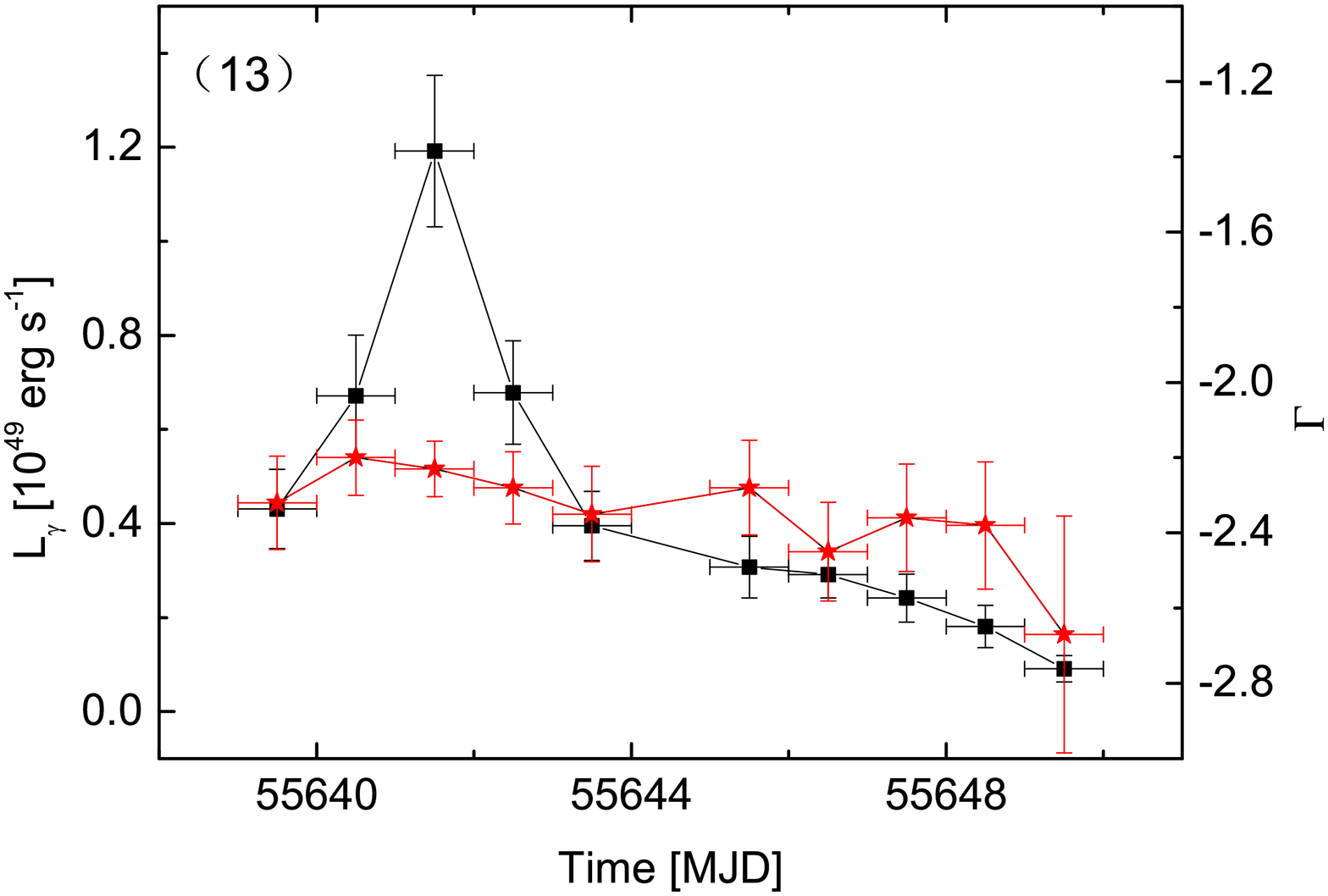}
\includegraphics[angle=0,scale=0.18]{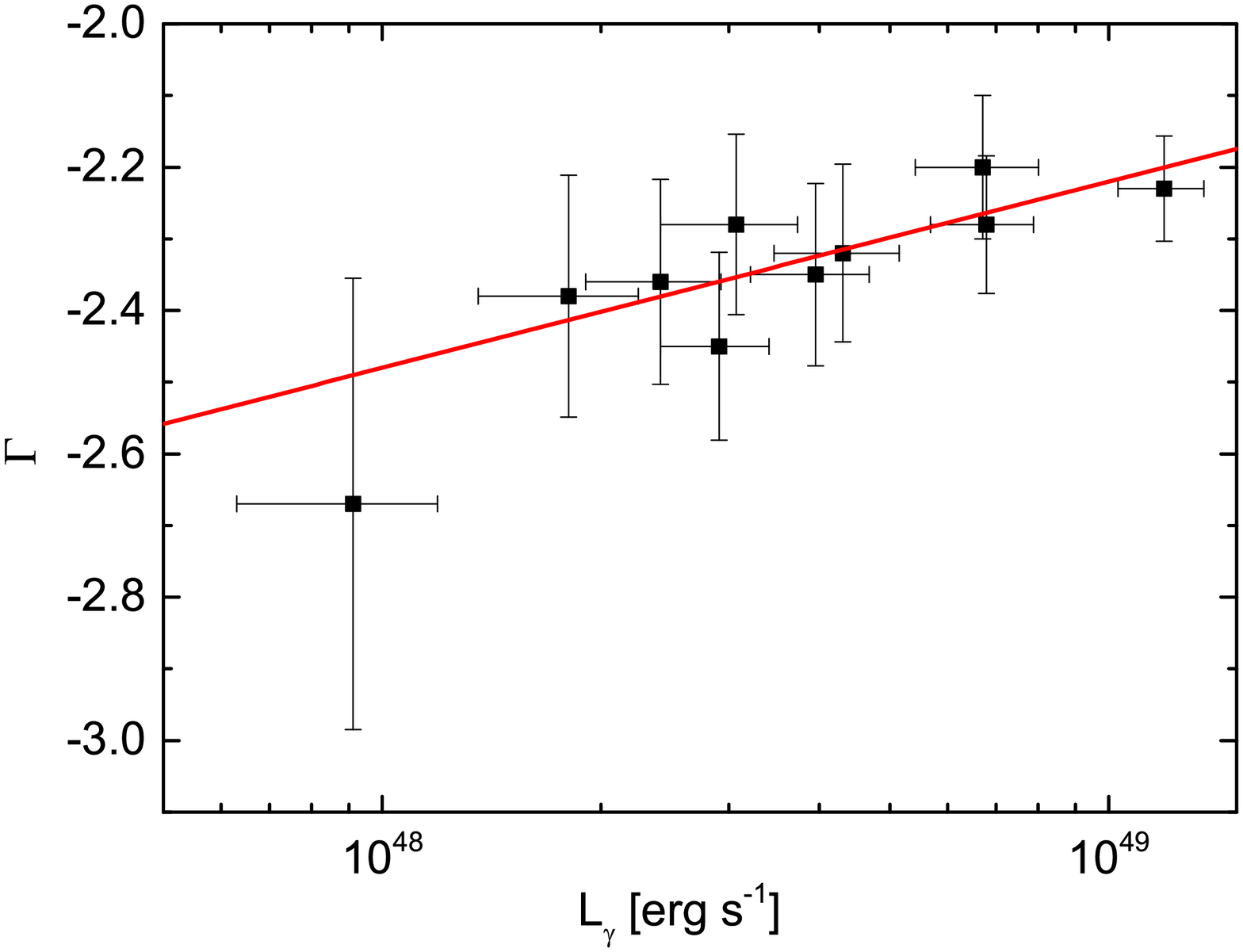}
\includegraphics[angle=0,scale=0.23]{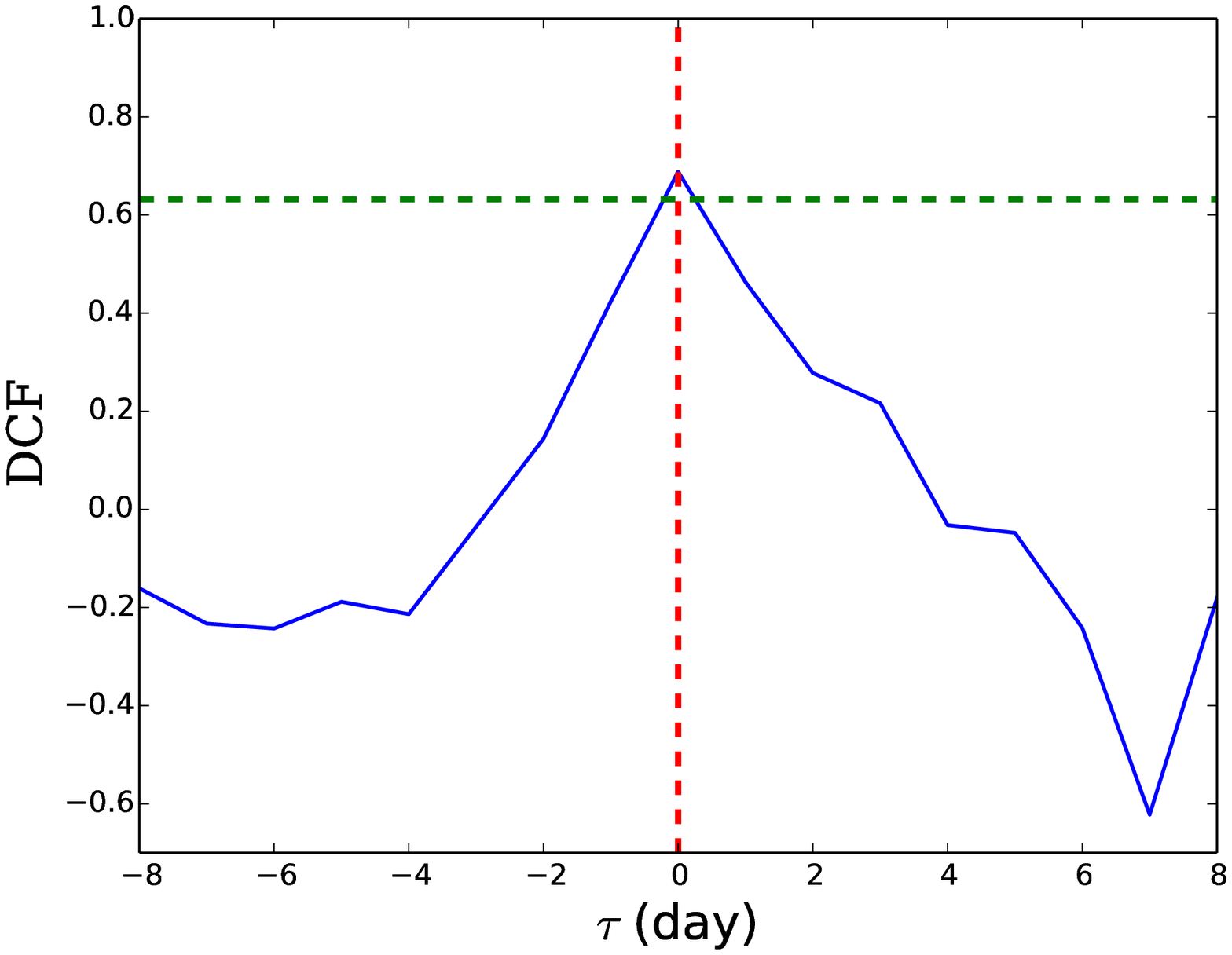}\\
\includegraphics[angle=0,scale=0.18]{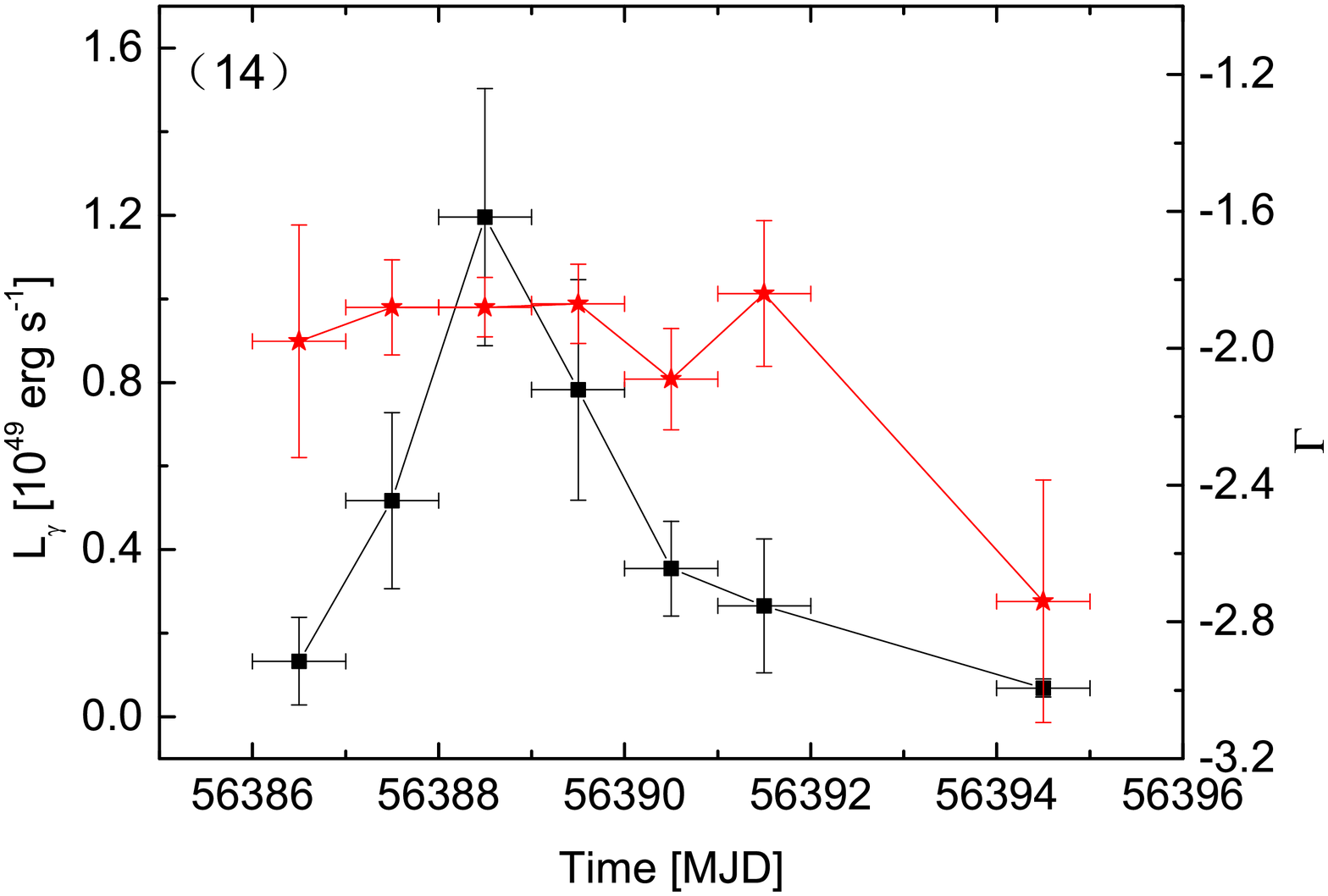}
\includegraphics[angle=0,scale=0.18]{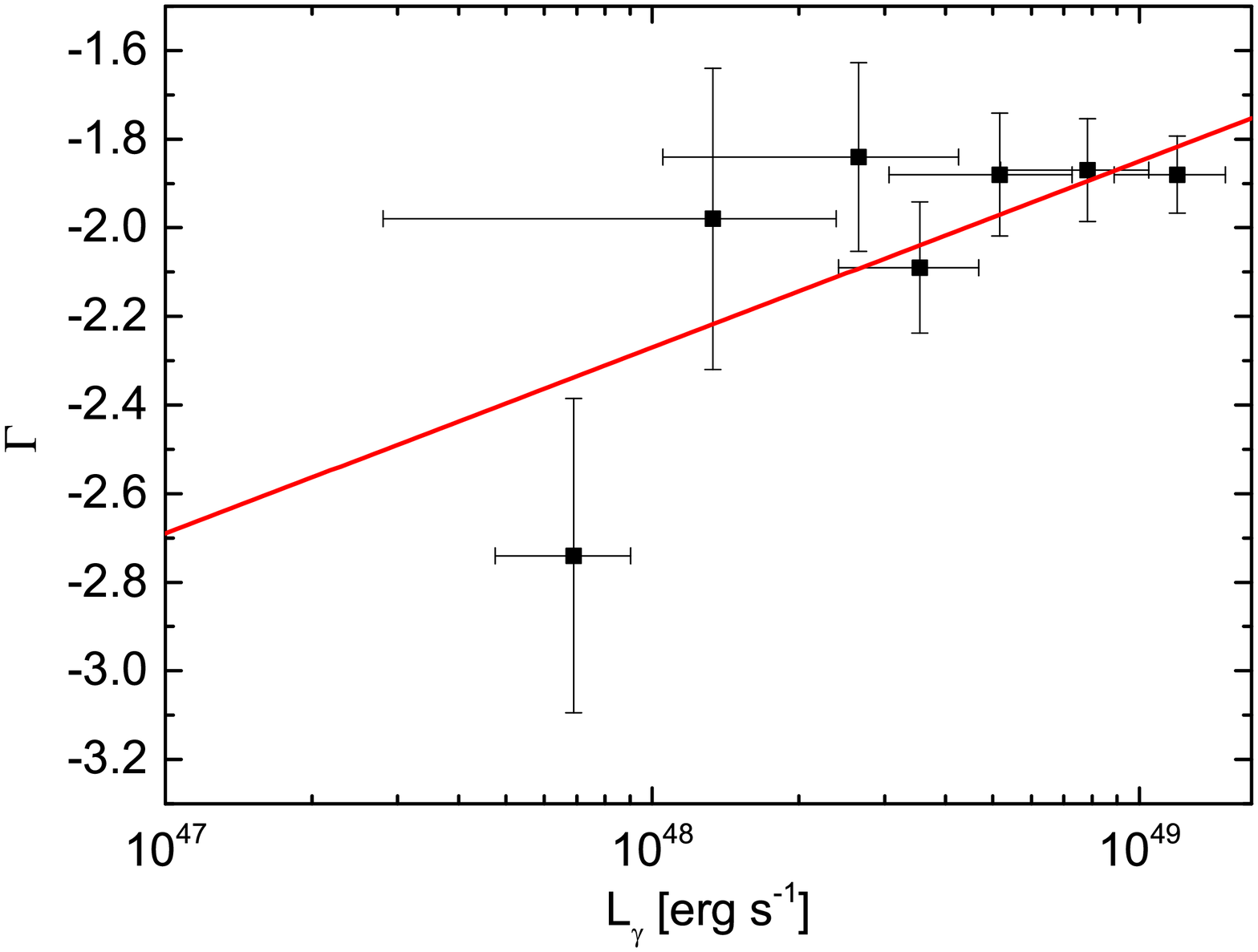}
\includegraphics[angle=0,scale=0.23]{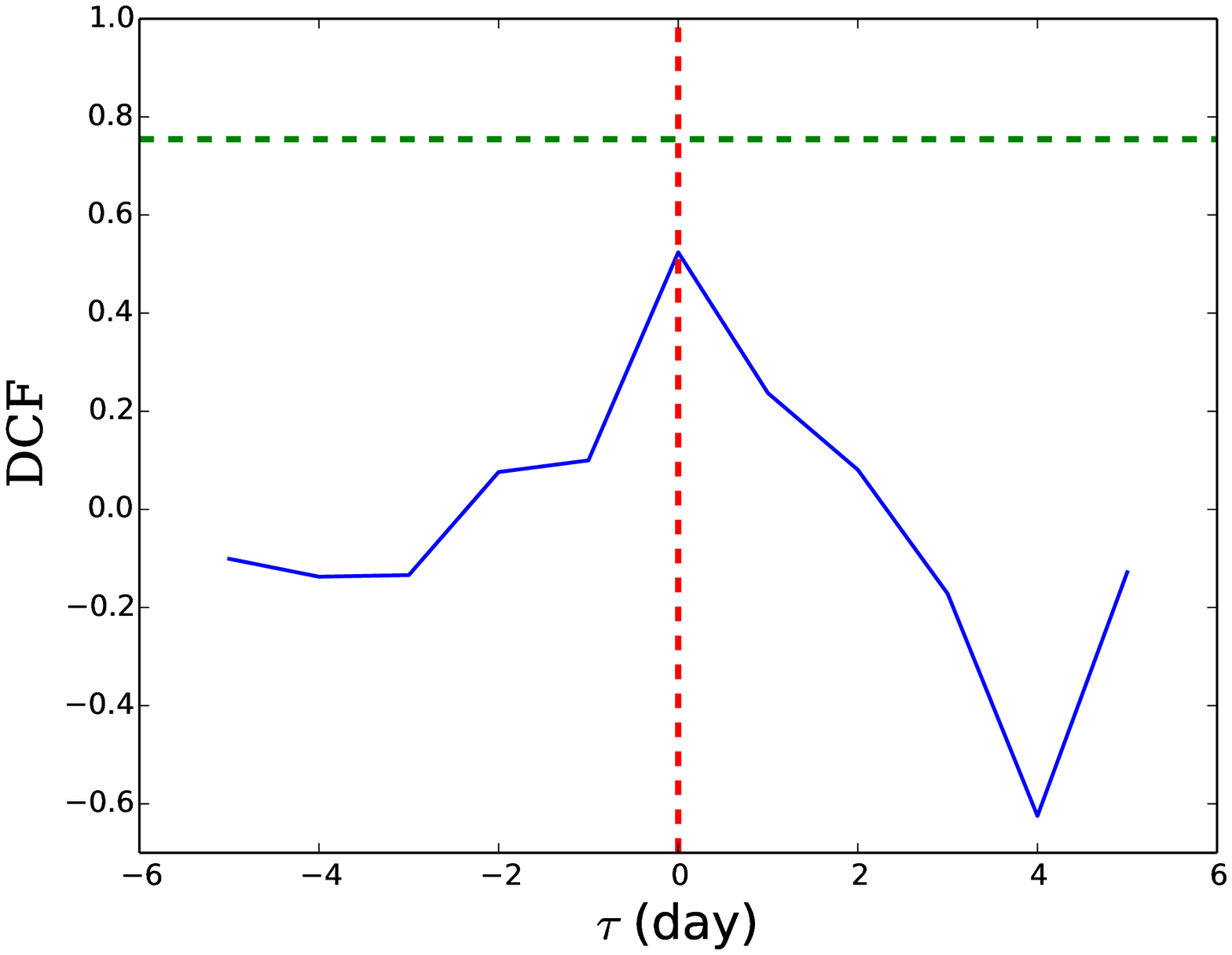}\\
\includegraphics[angle=0,scale=0.18]{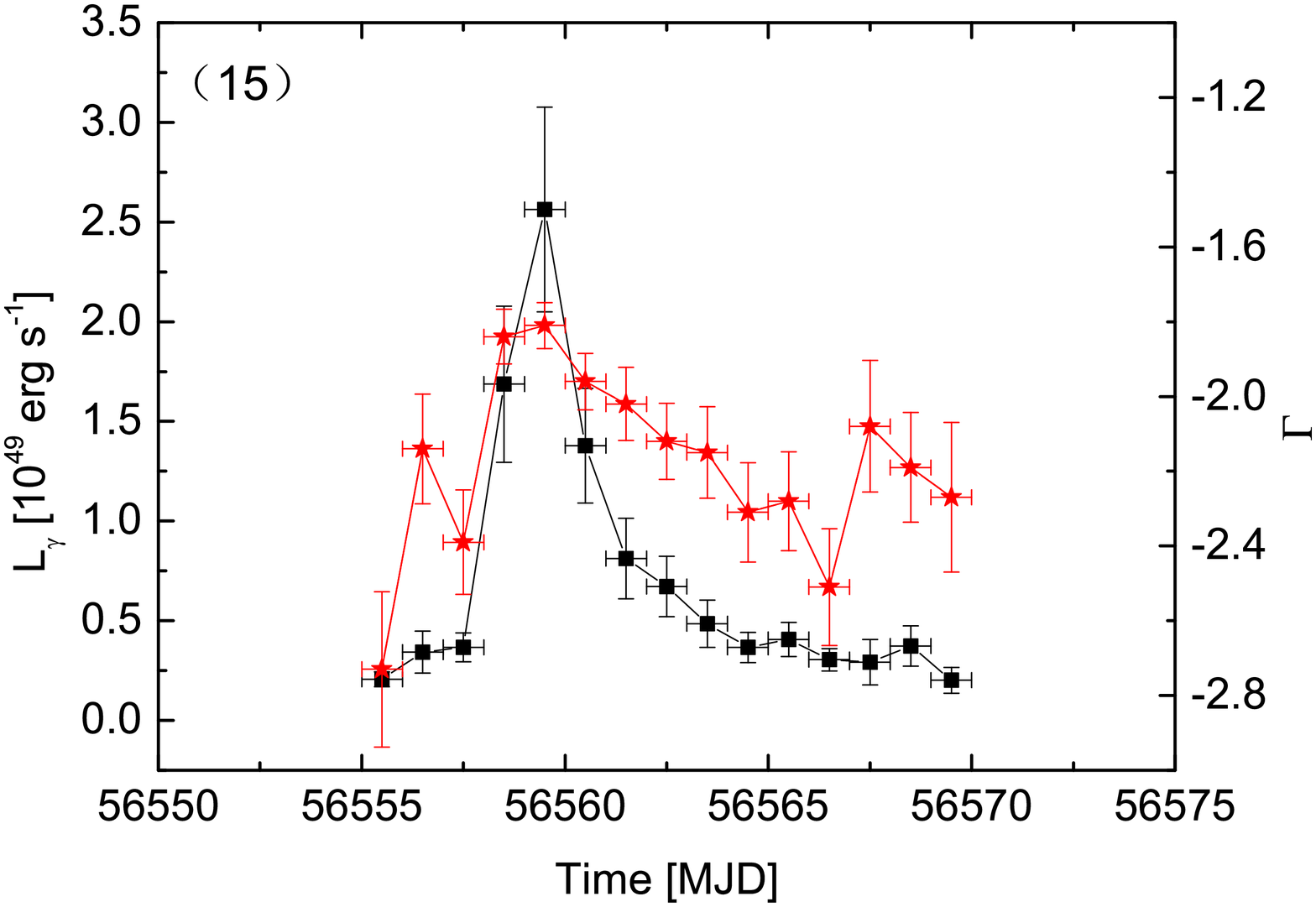}
\includegraphics[angle=0,scale=0.18]{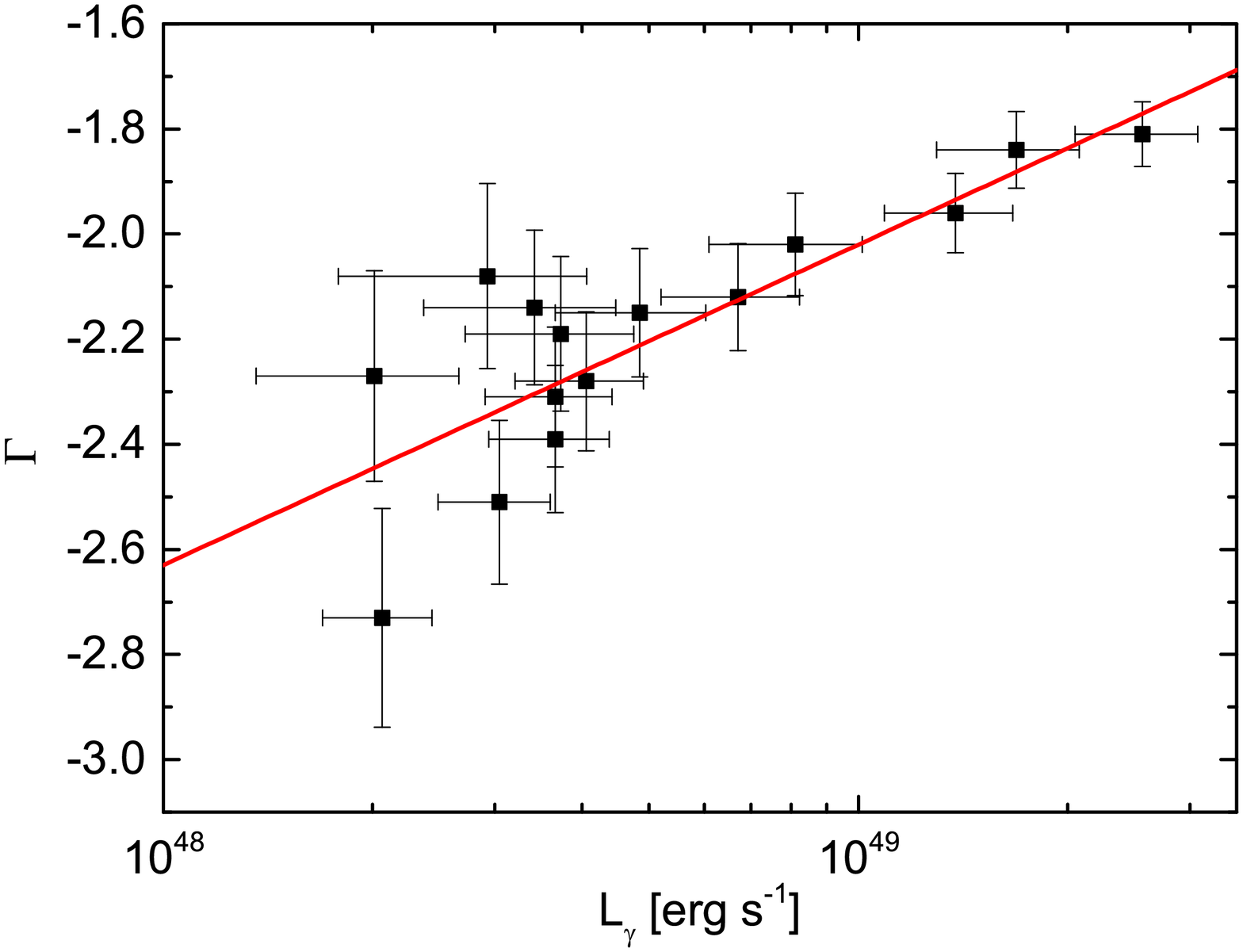}
\includegraphics[angle=0,scale=0.23]{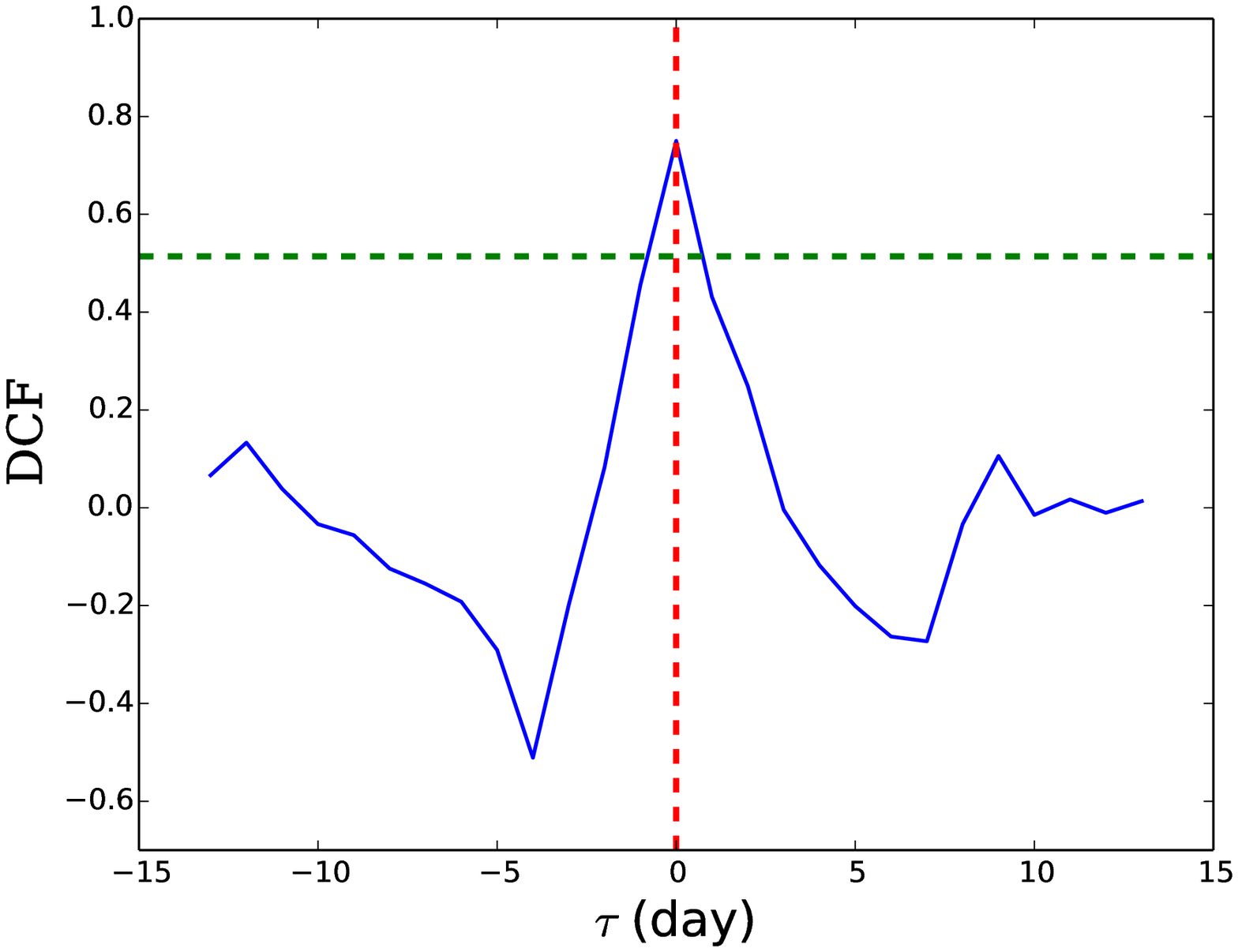}\\

\hfill\center{Fig.5---  continued}
\end{figure*}

\begin{figure*}
\includegraphics[angle=0,scale=0.18]{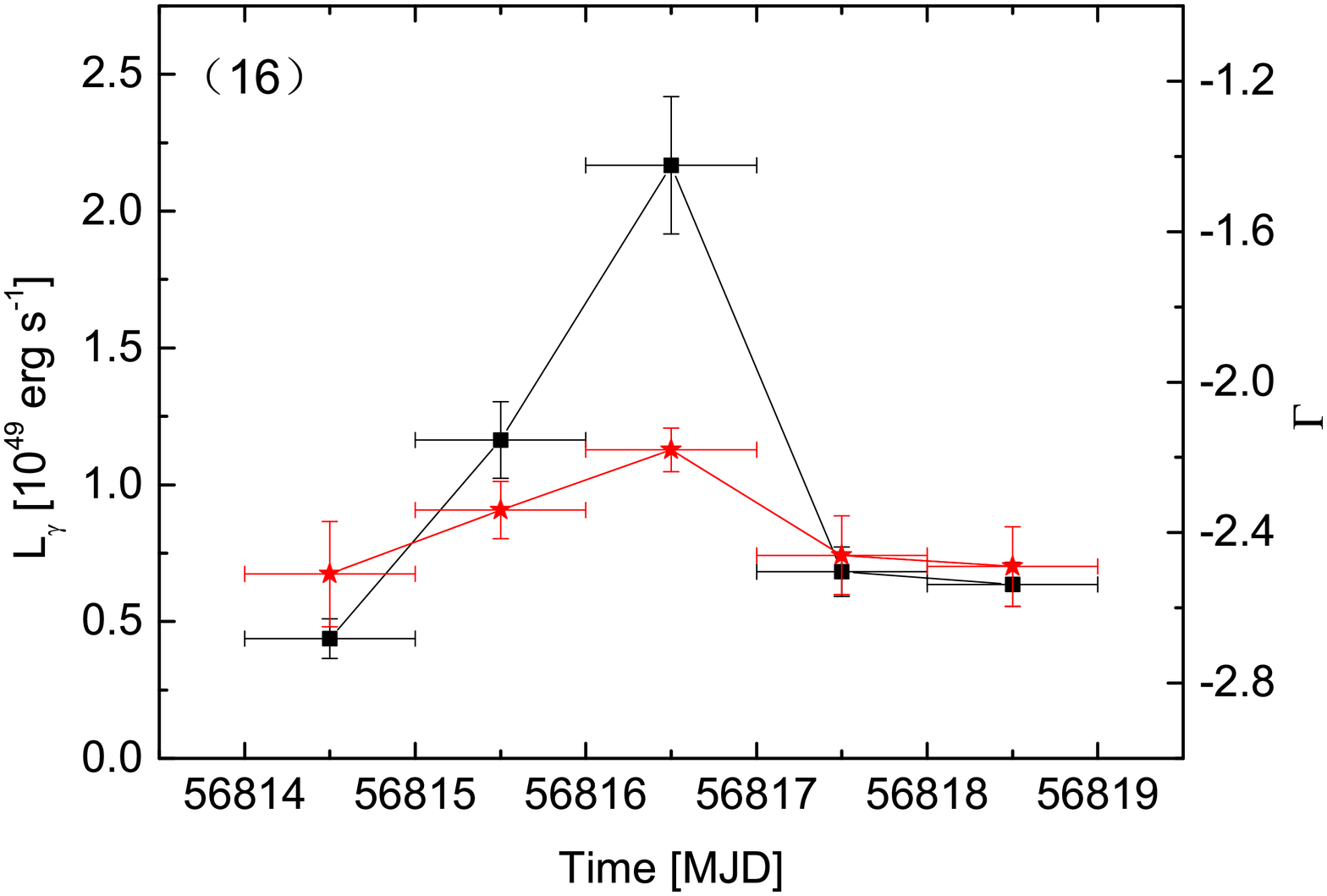}
\includegraphics[angle=0,scale=0.18]{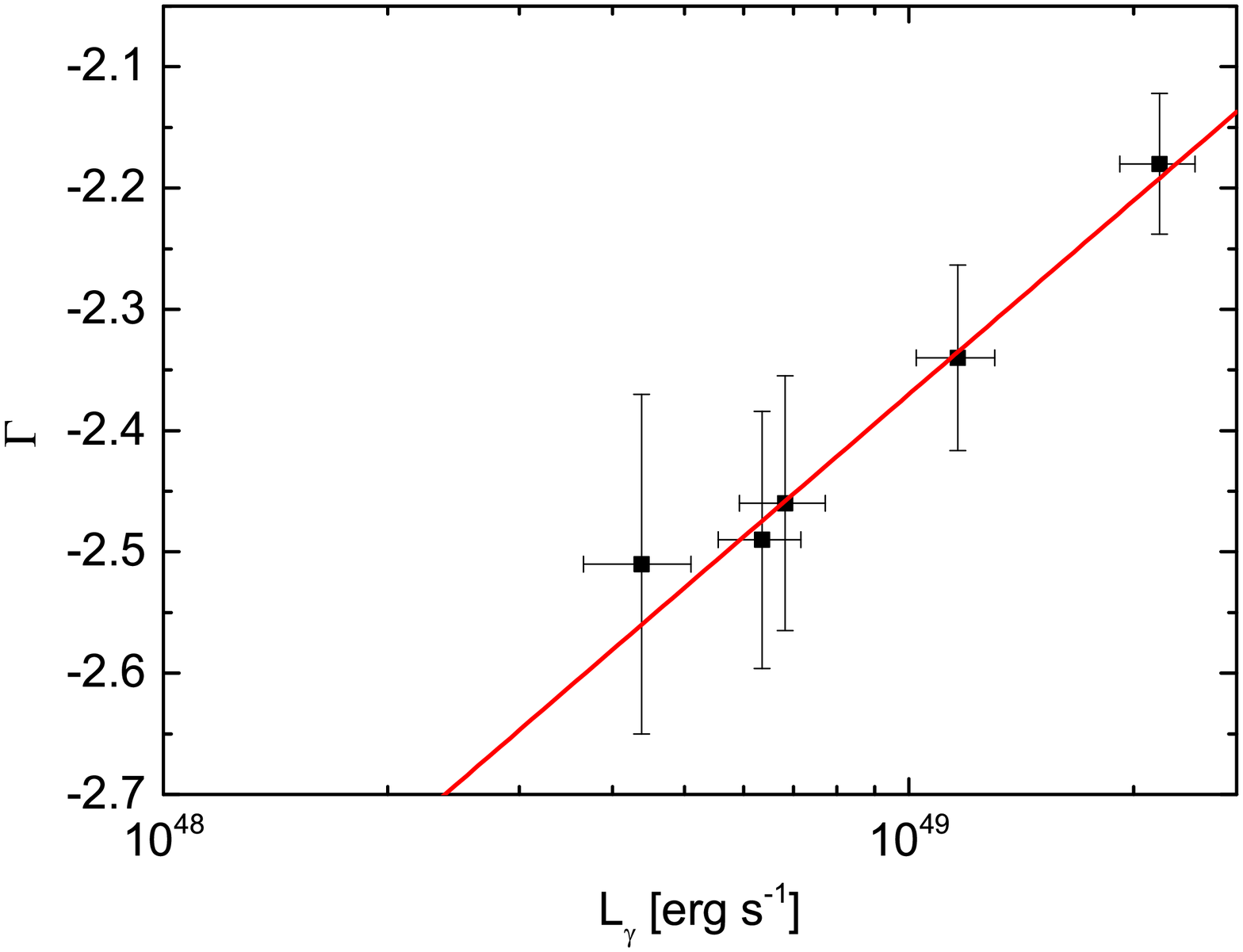}
\includegraphics[angle=0,scale=0.23]{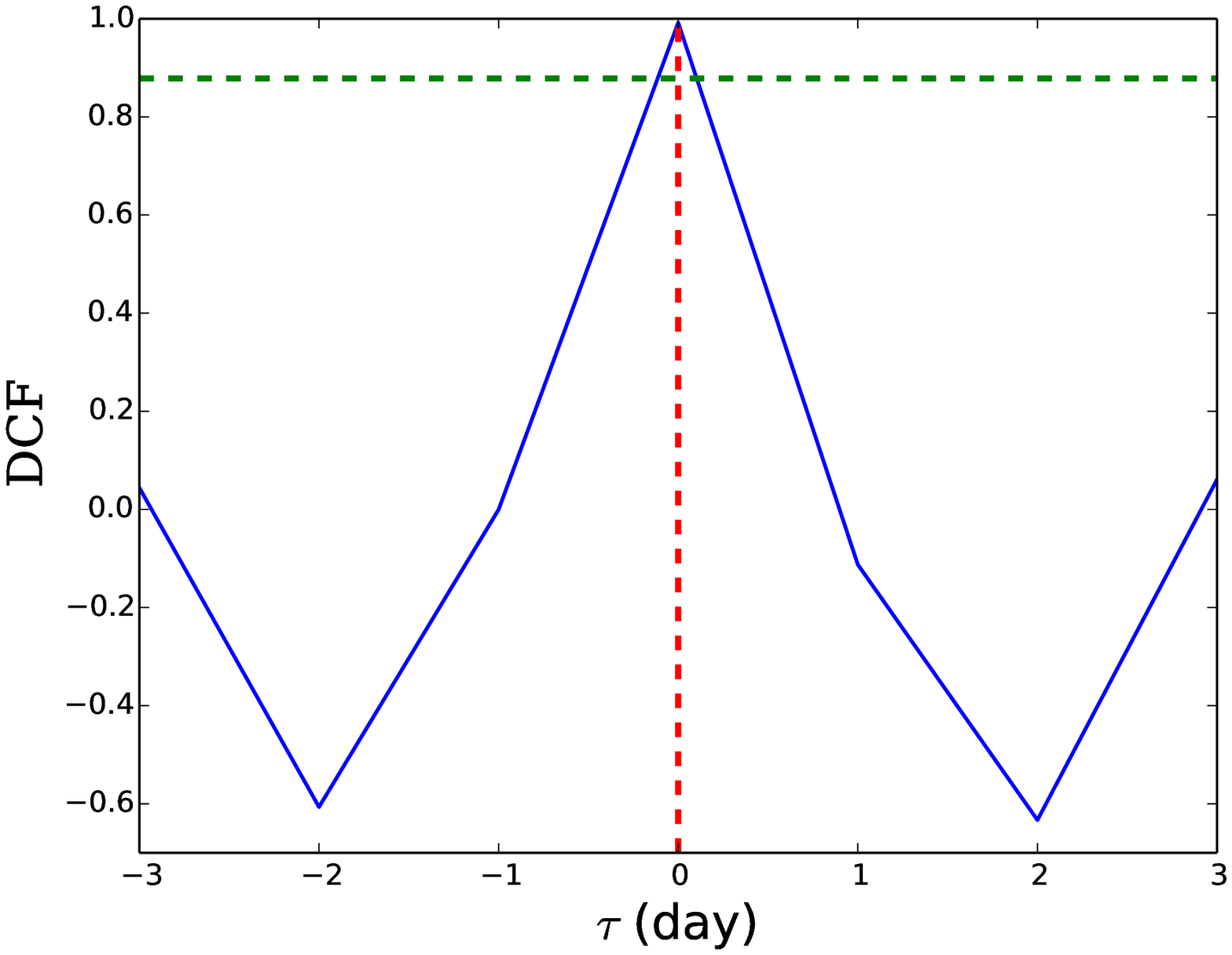}\\
\includegraphics[angle=0,scale=0.18]{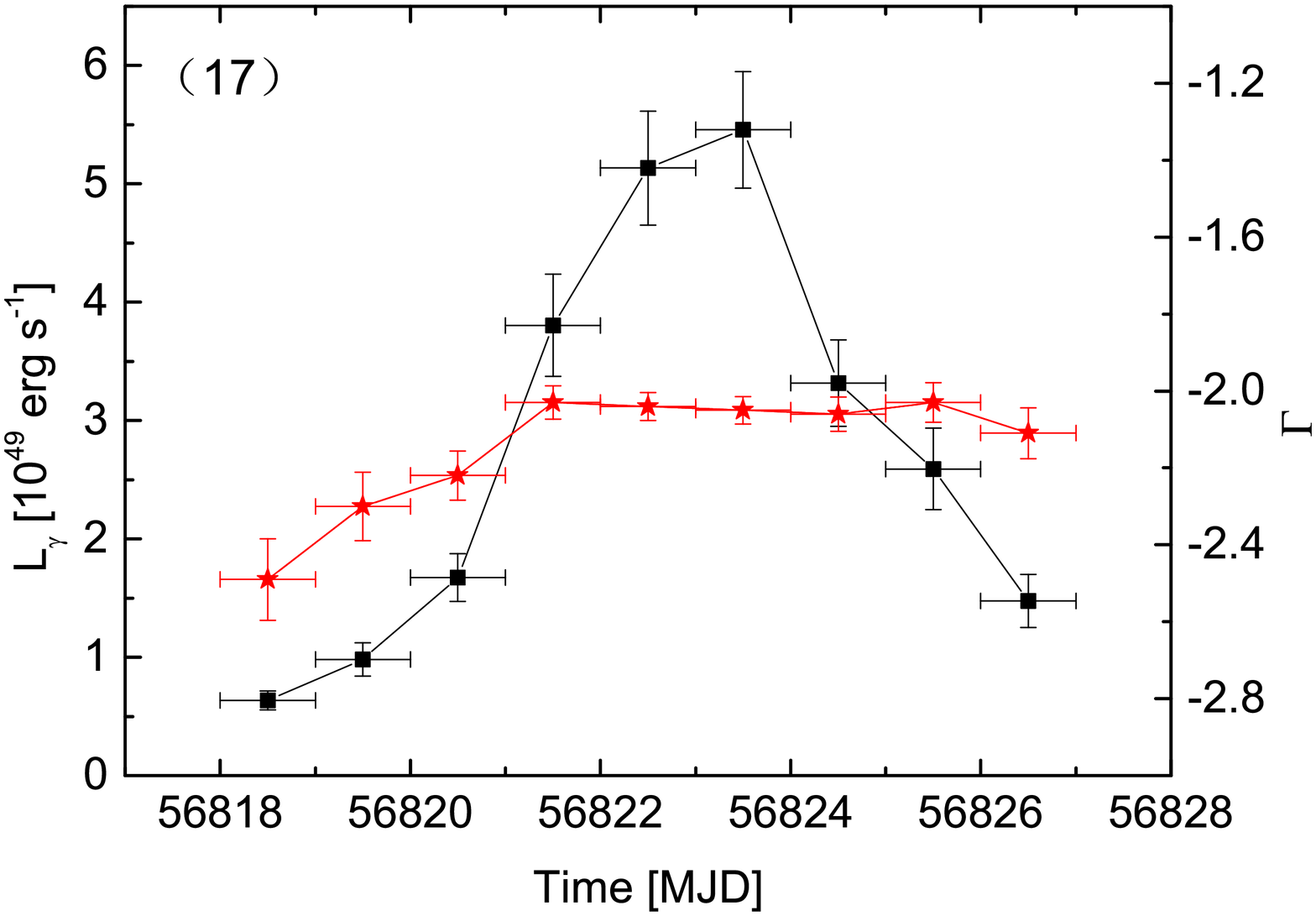}
\includegraphics[angle=0,scale=0.18]{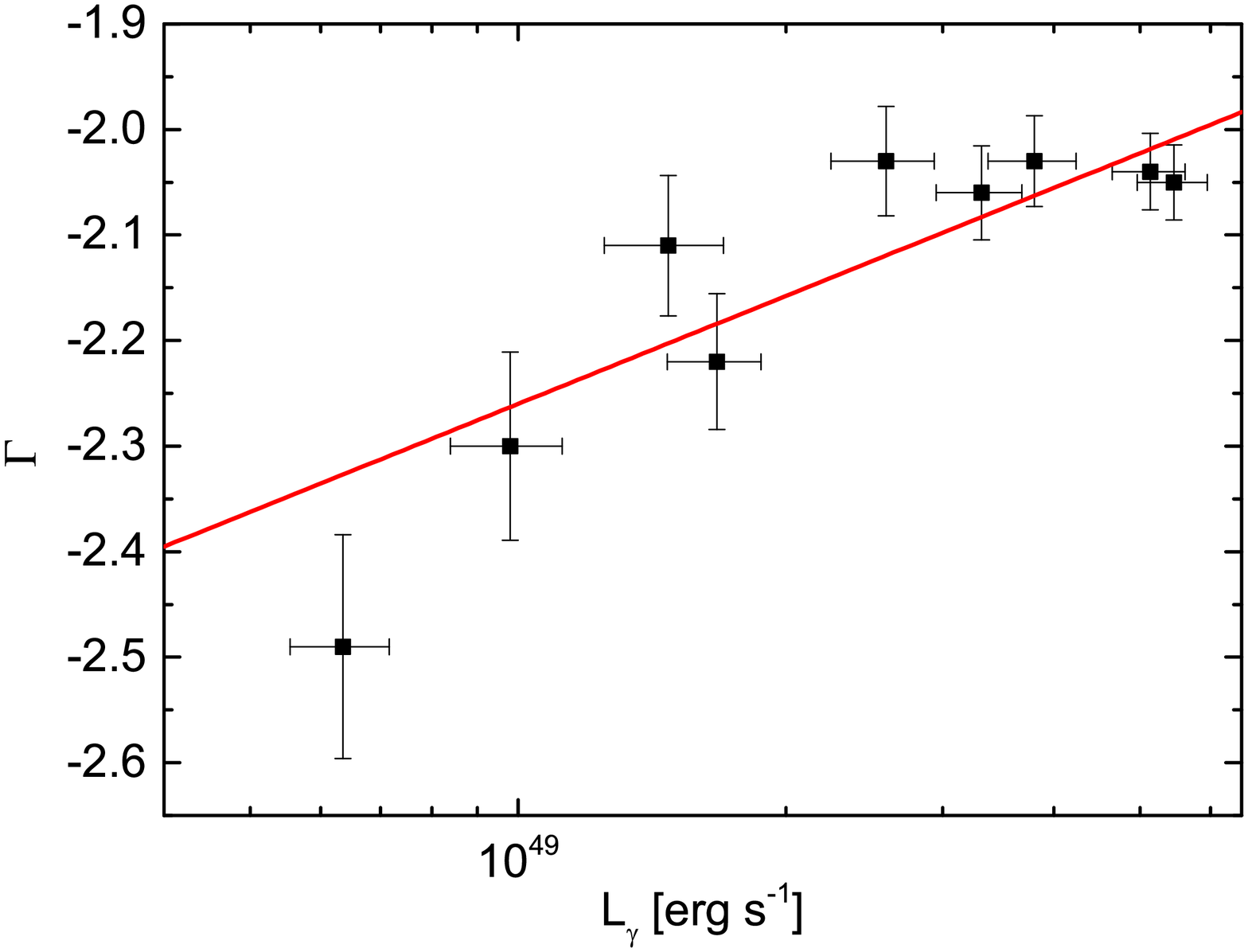}
\includegraphics[angle=0,scale=0.23]{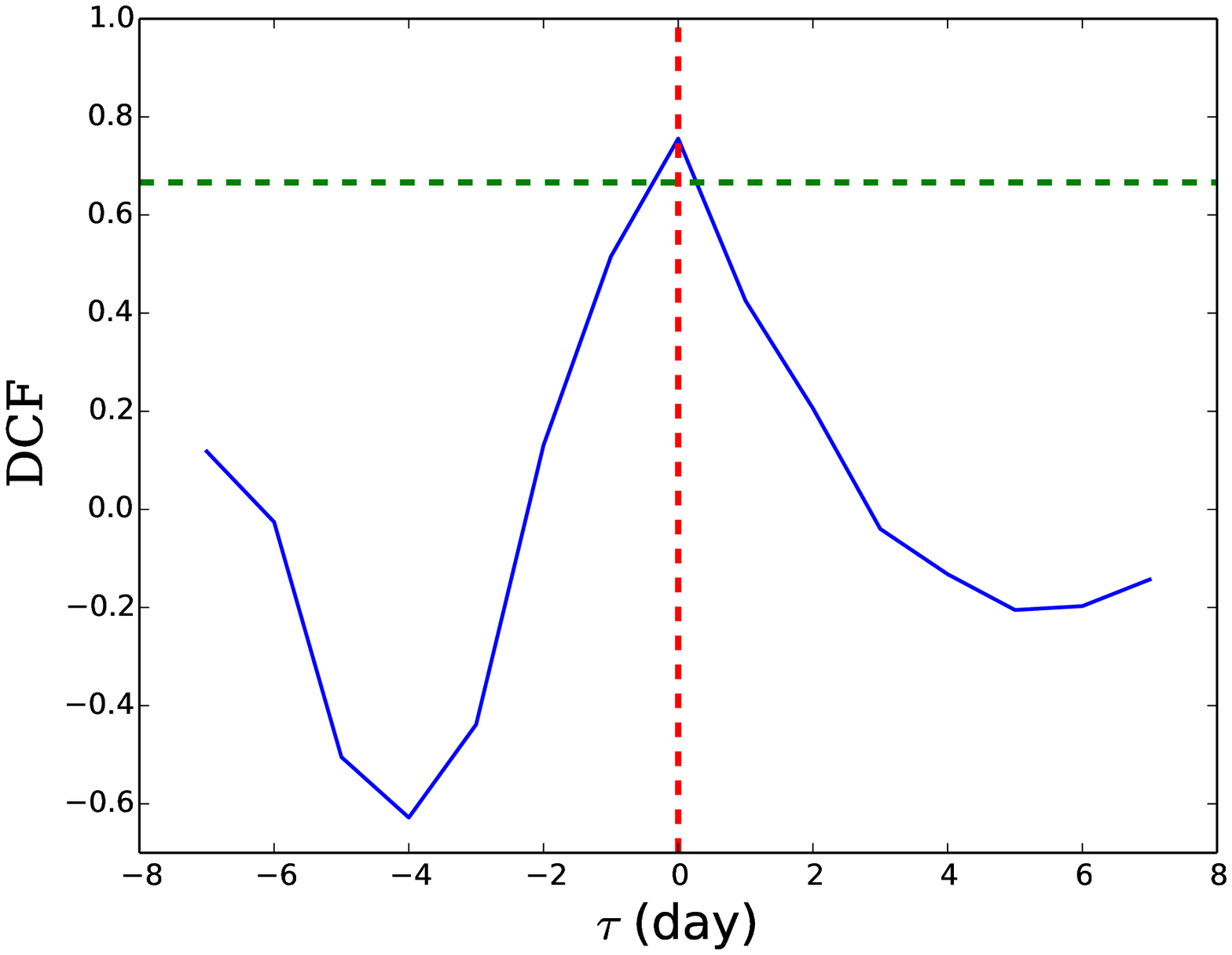}\\
\includegraphics[angle=0,scale=0.18]{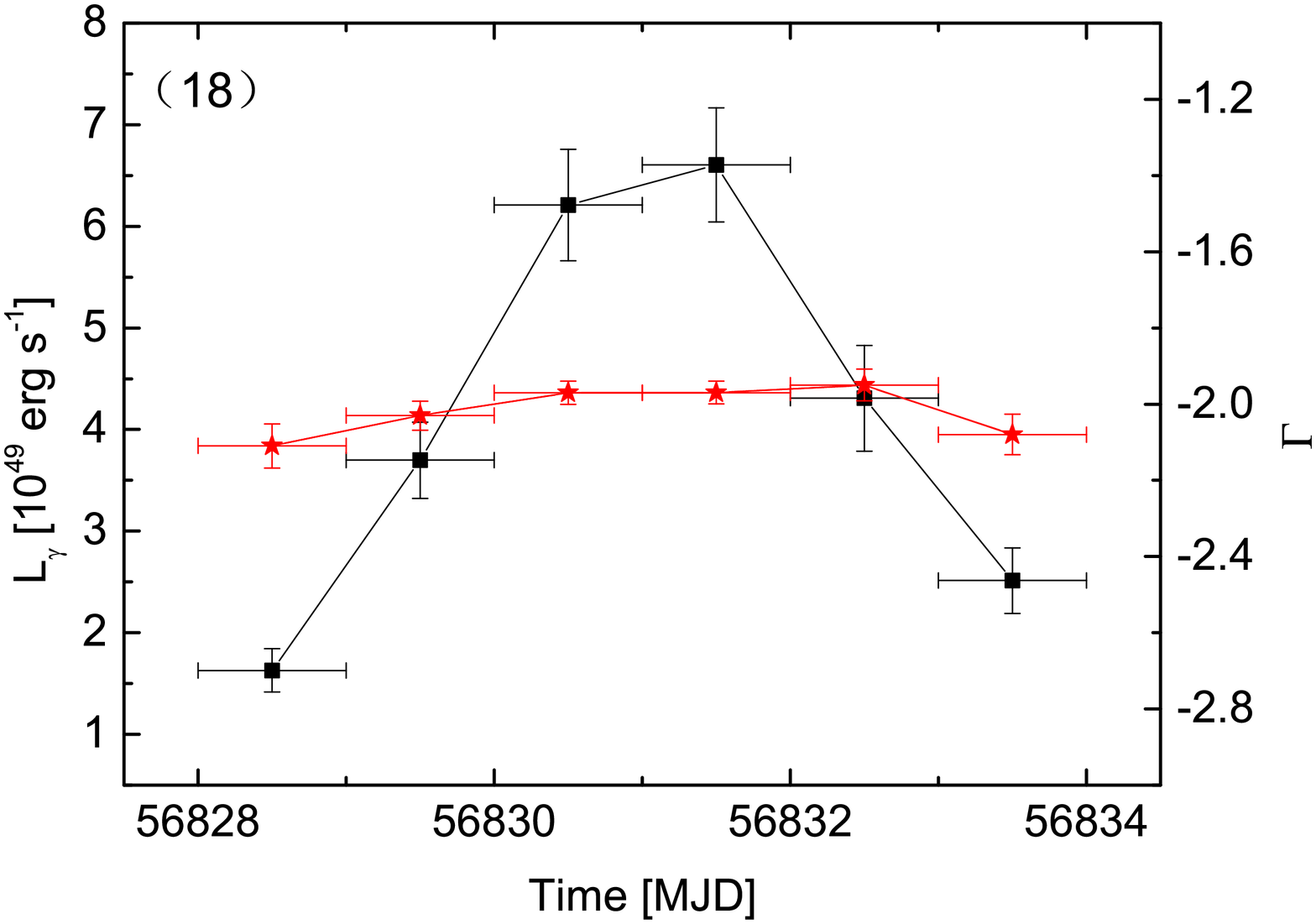}
\includegraphics[angle=0,scale=0.18]{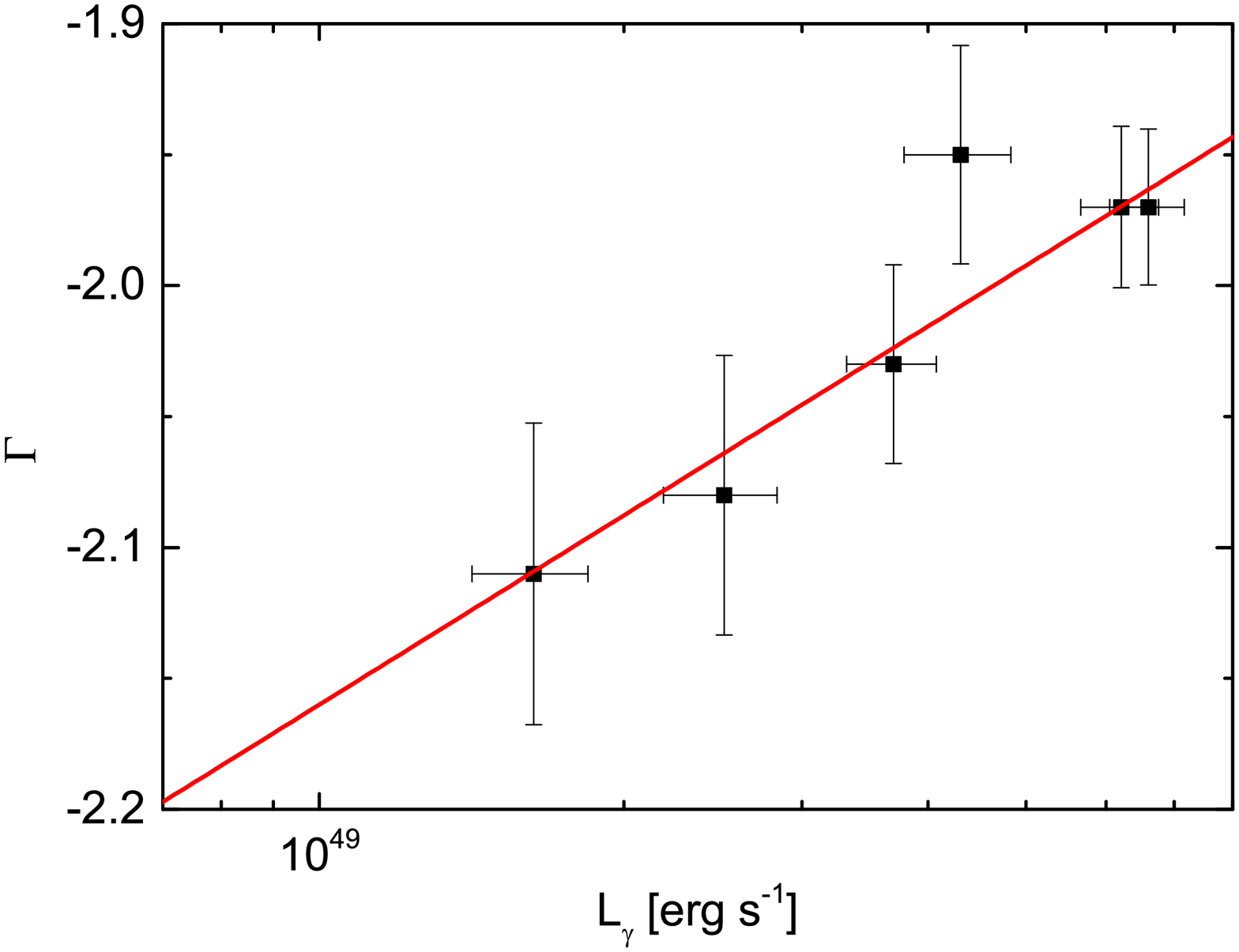}
\includegraphics[angle=0,scale=0.23]{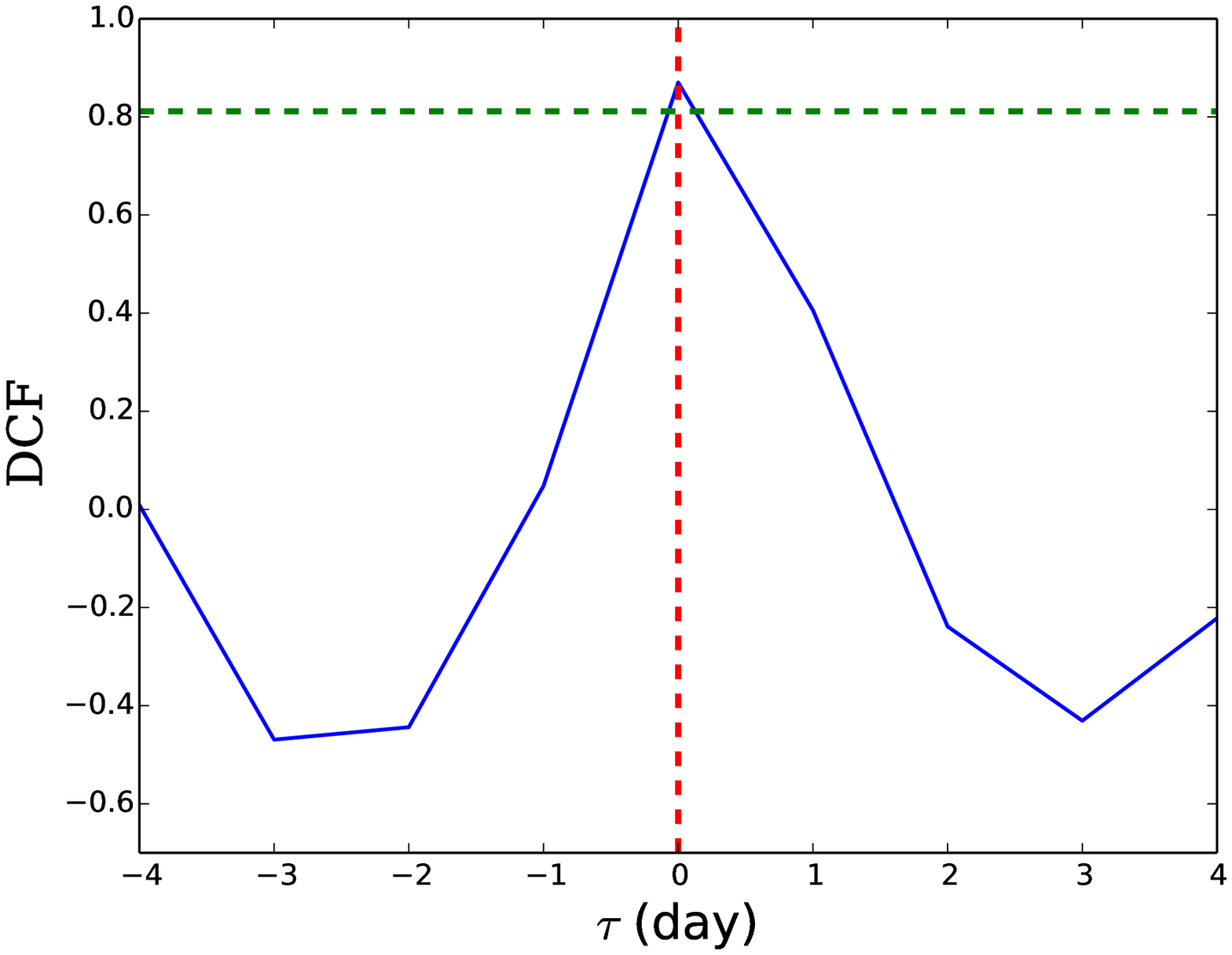}\\
\includegraphics[angle=0,scale=0.18]{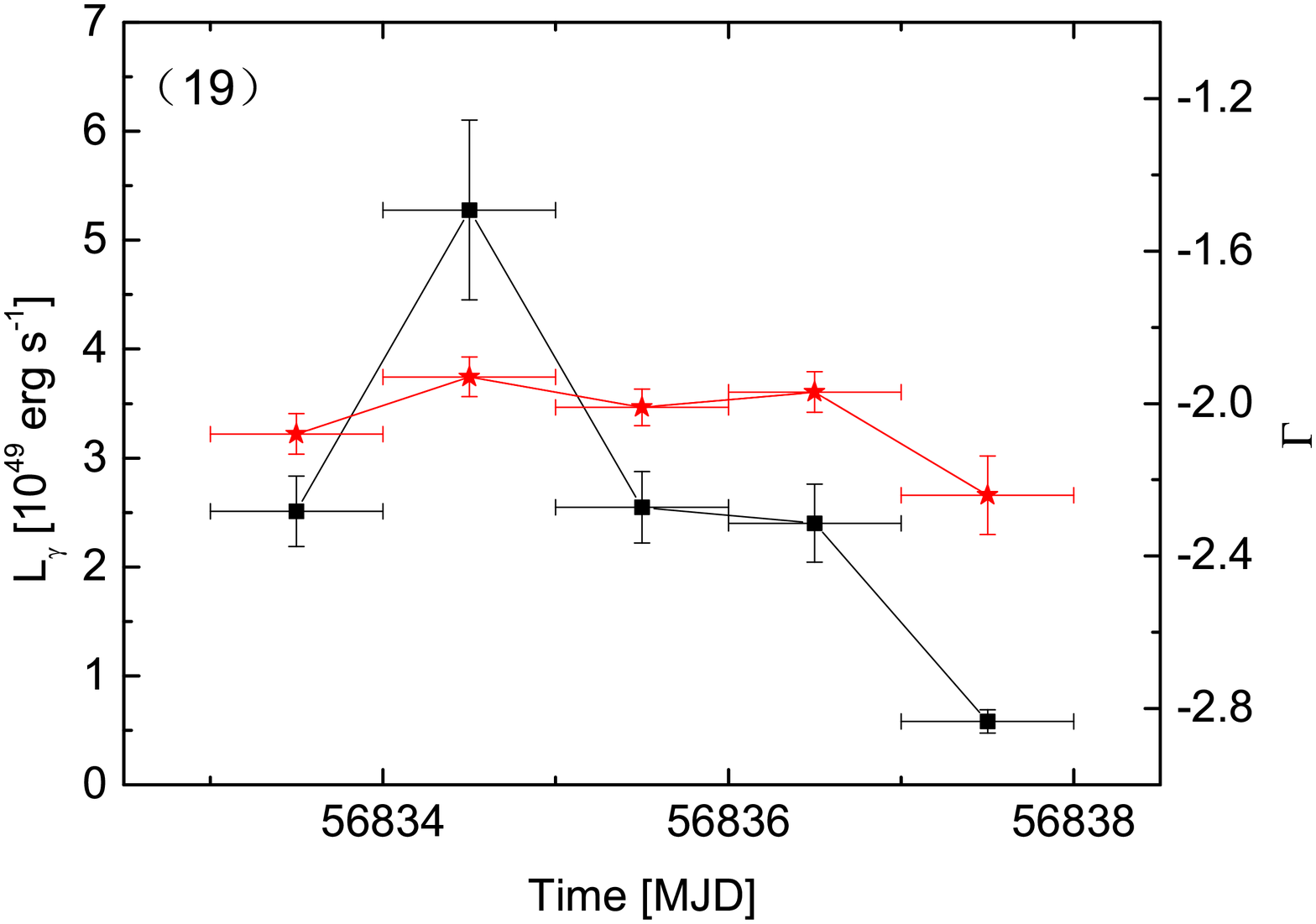}
\includegraphics[angle=0,scale=0.18]{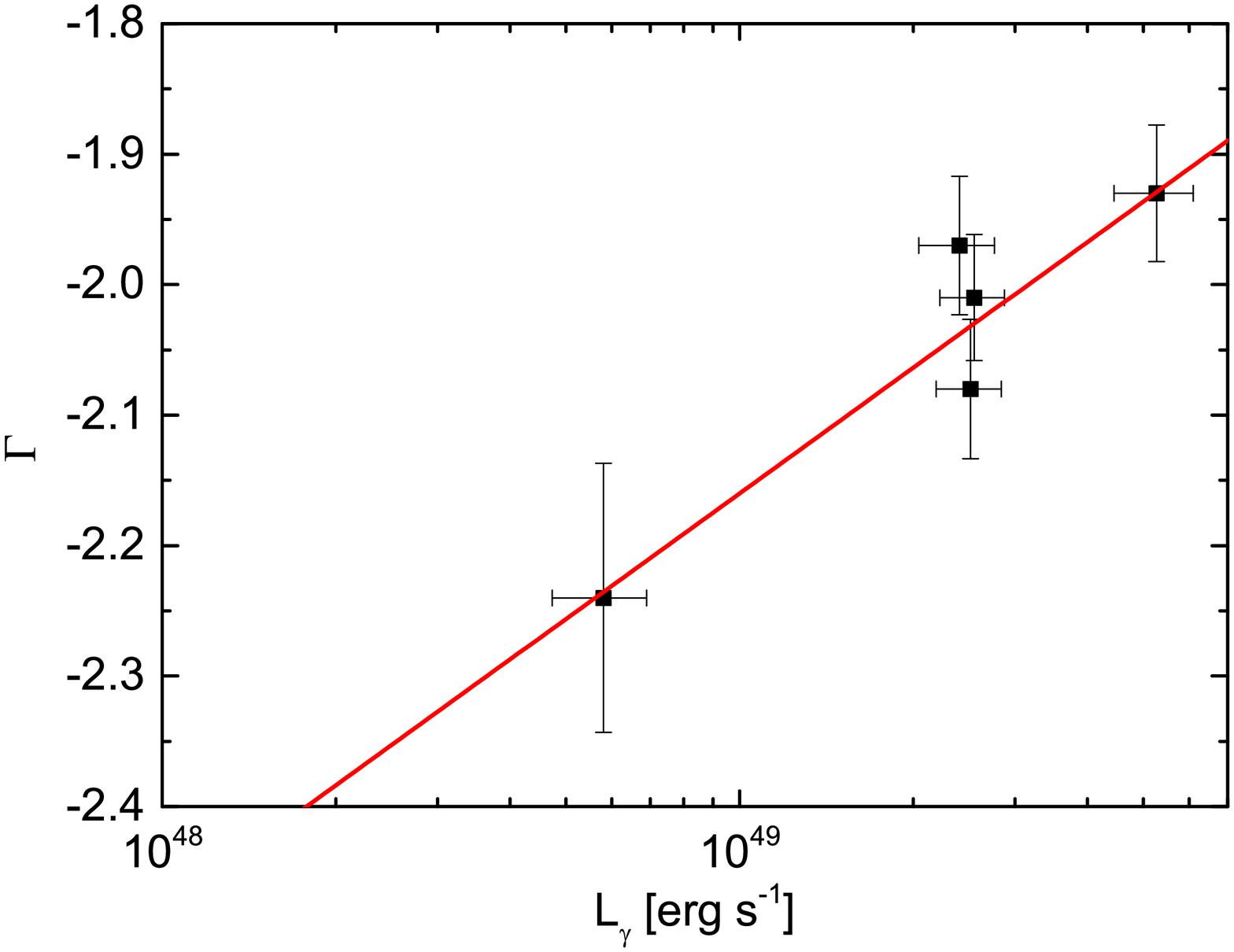}
\includegraphics[angle=0,scale=0.23]{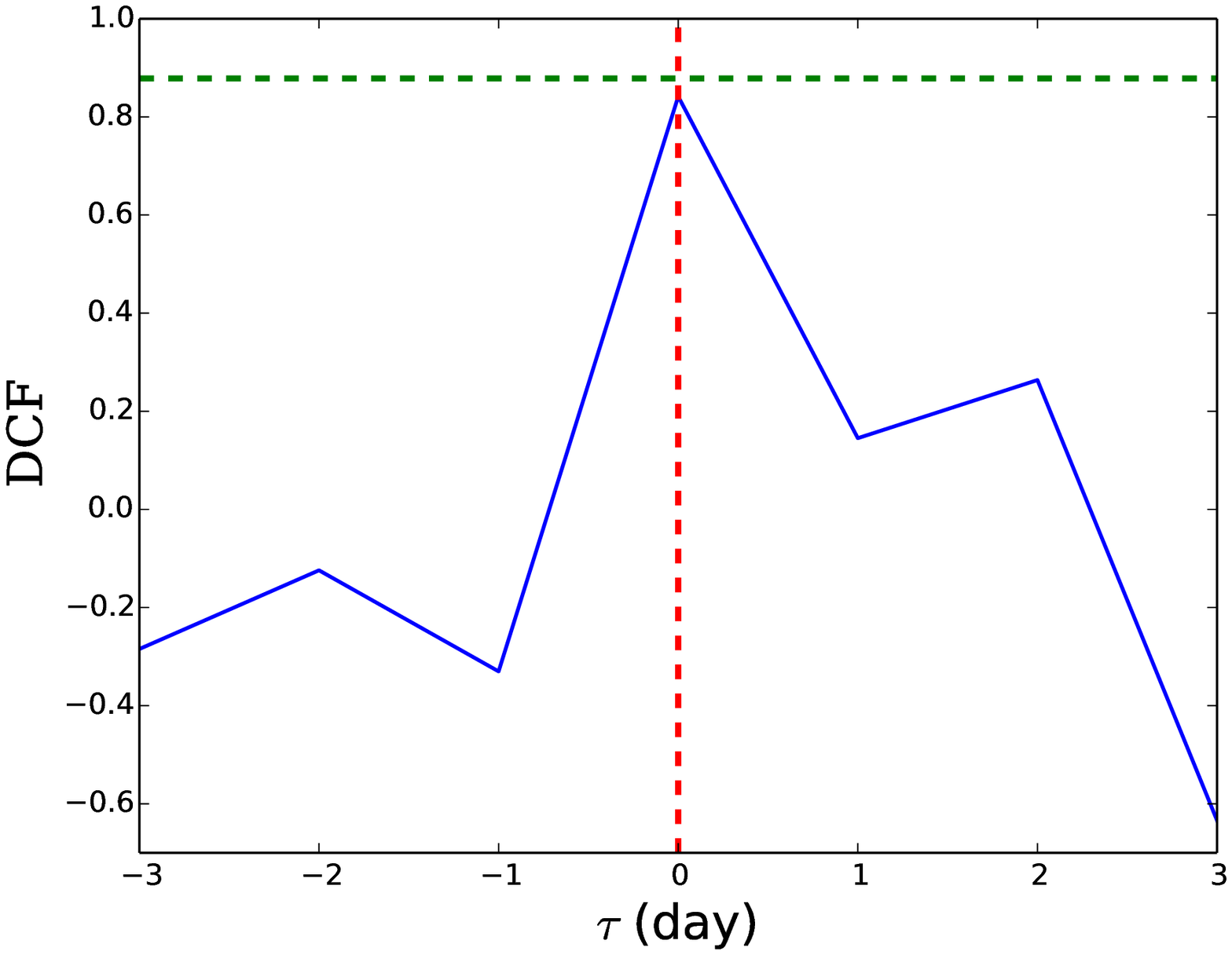}\\
\includegraphics[angle=0,scale=0.18]{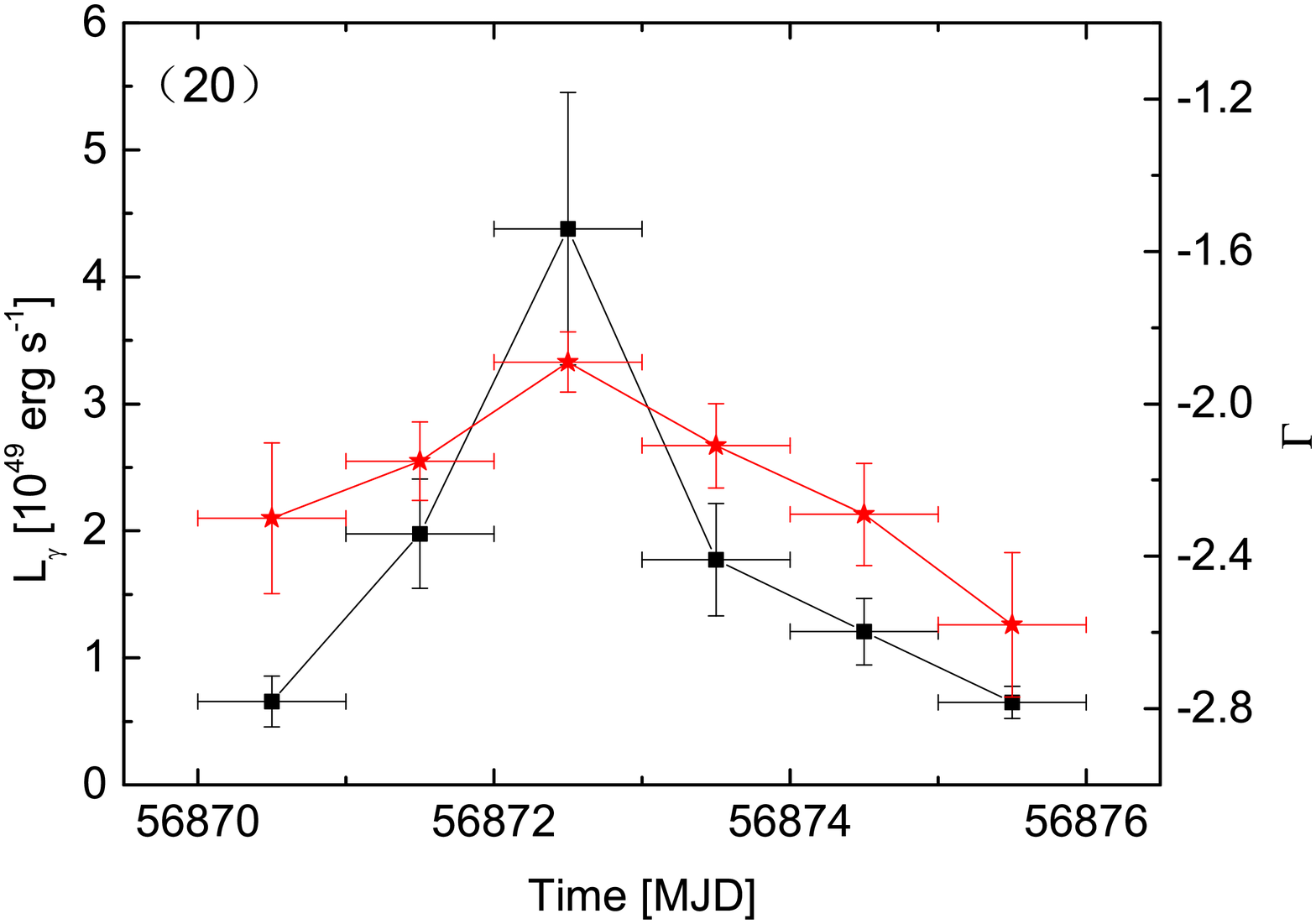}
\includegraphics[angle=0,scale=0.18]{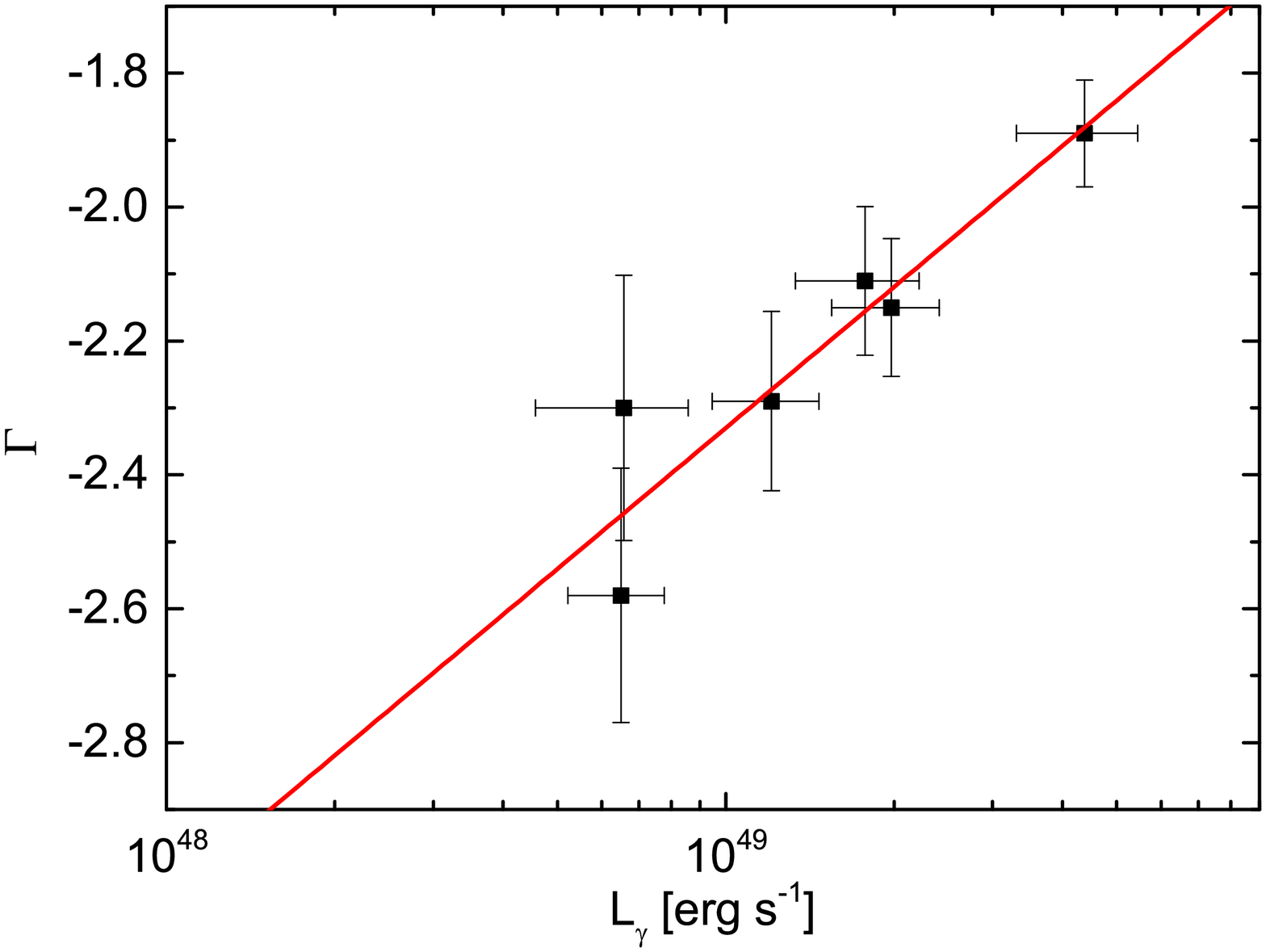}
\includegraphics[angle=0,scale=0.23]{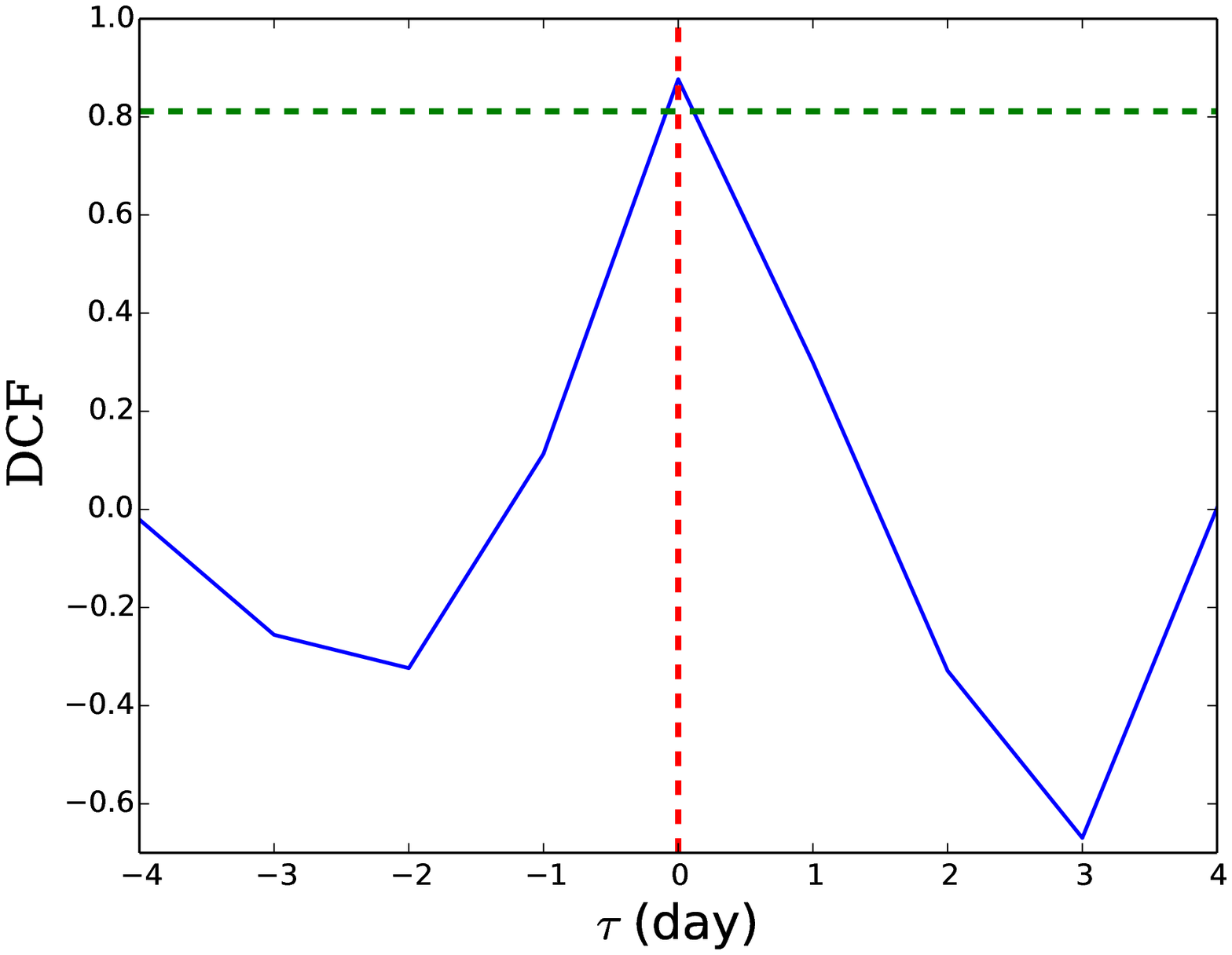}\\
\includegraphics[angle=0,scale=0.18]{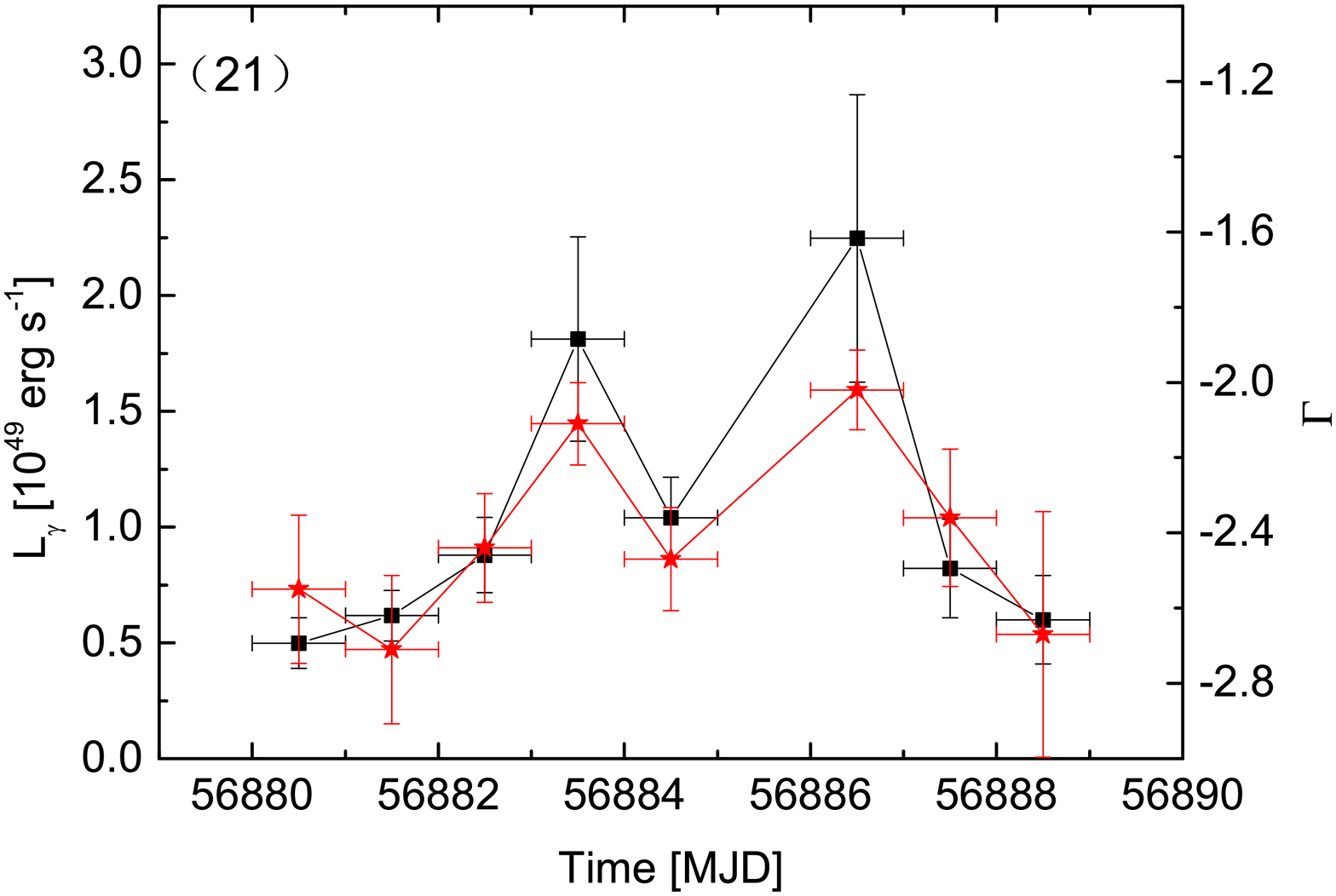}
\includegraphics[angle=0,scale=0.18]{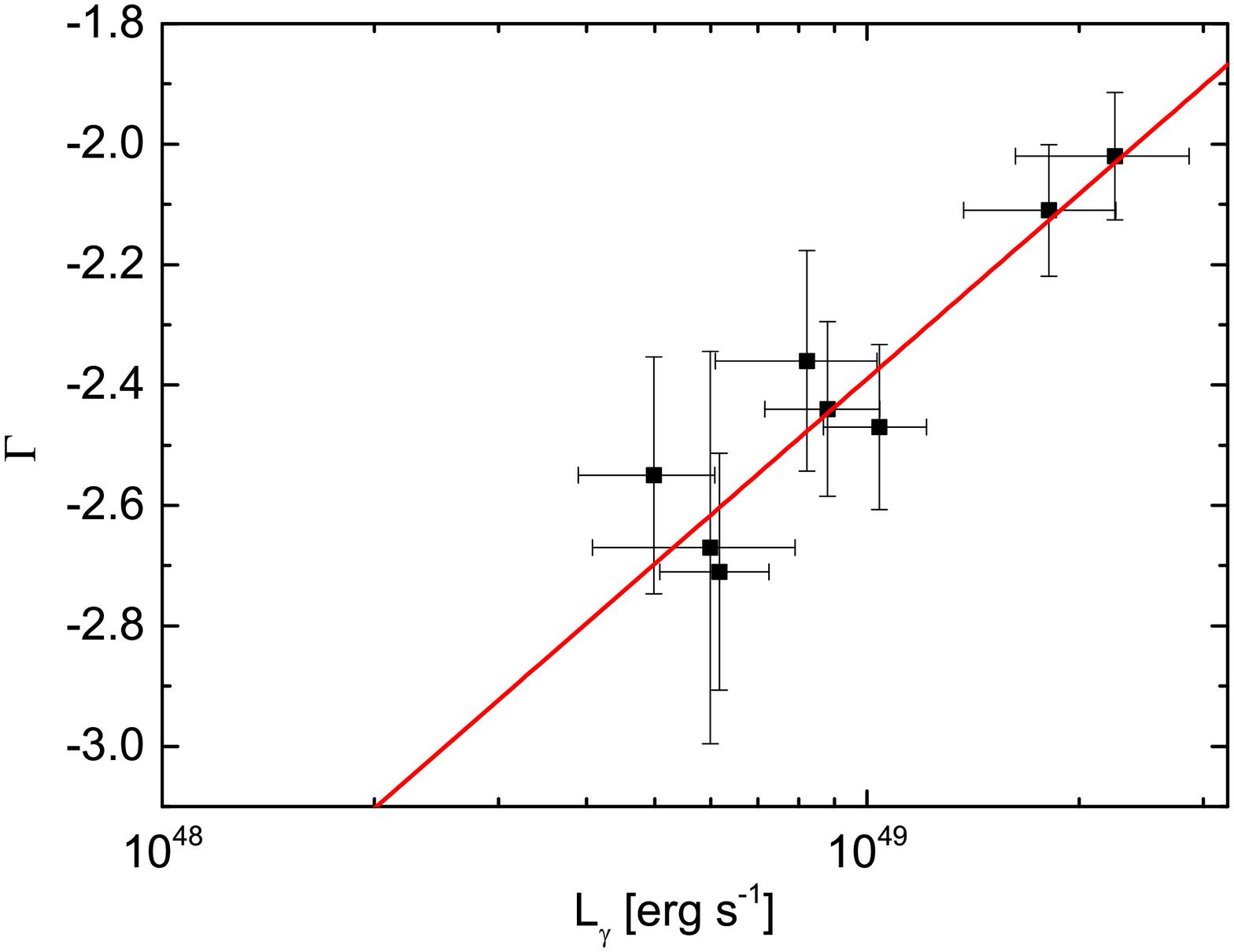}
\includegraphics[angle=0,scale=0.23]{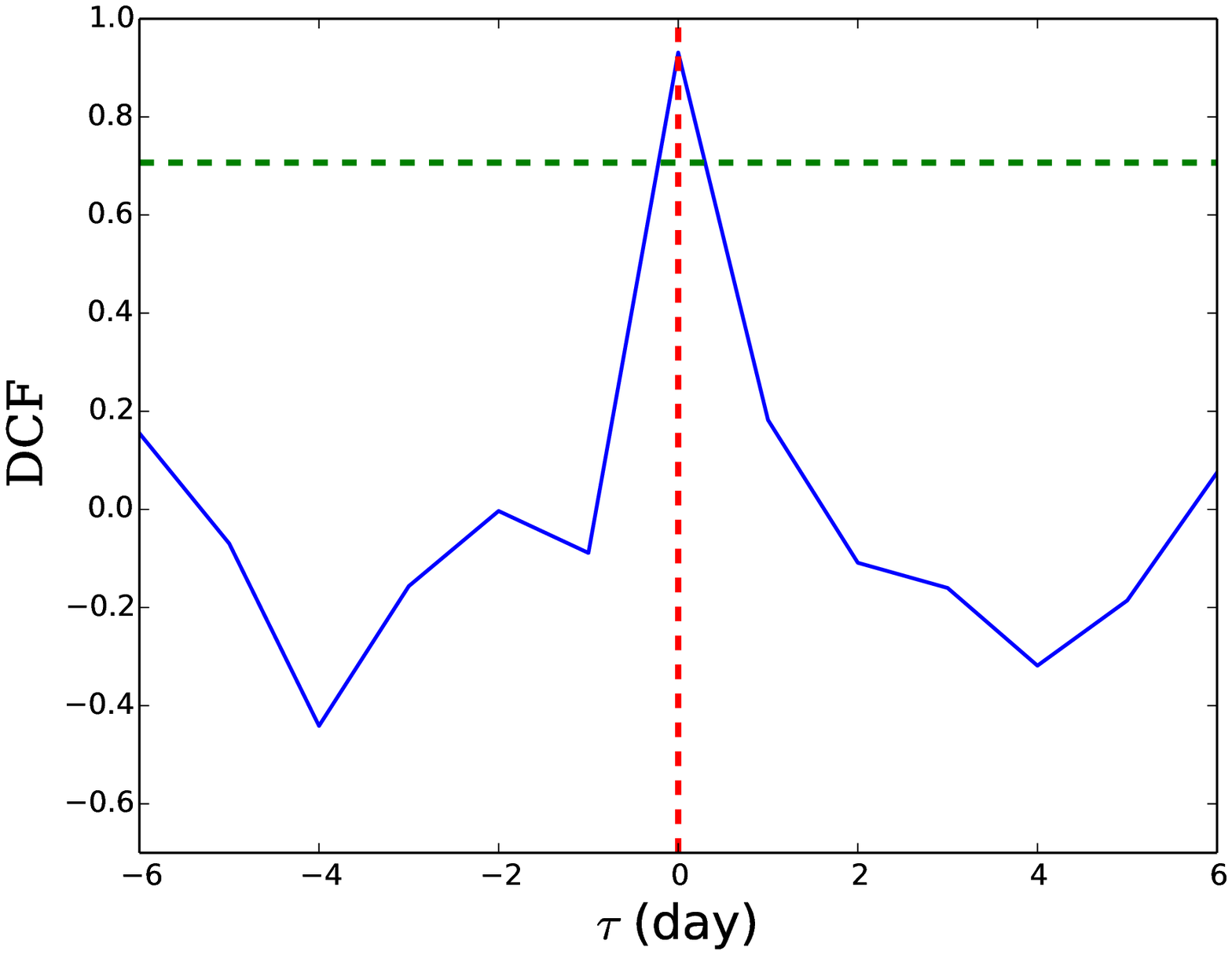}
\hfill\center{Fig.5---  continued}
\end{figure*}

\begin{figure*}
\includegraphics[angle=0,scale=0.18]{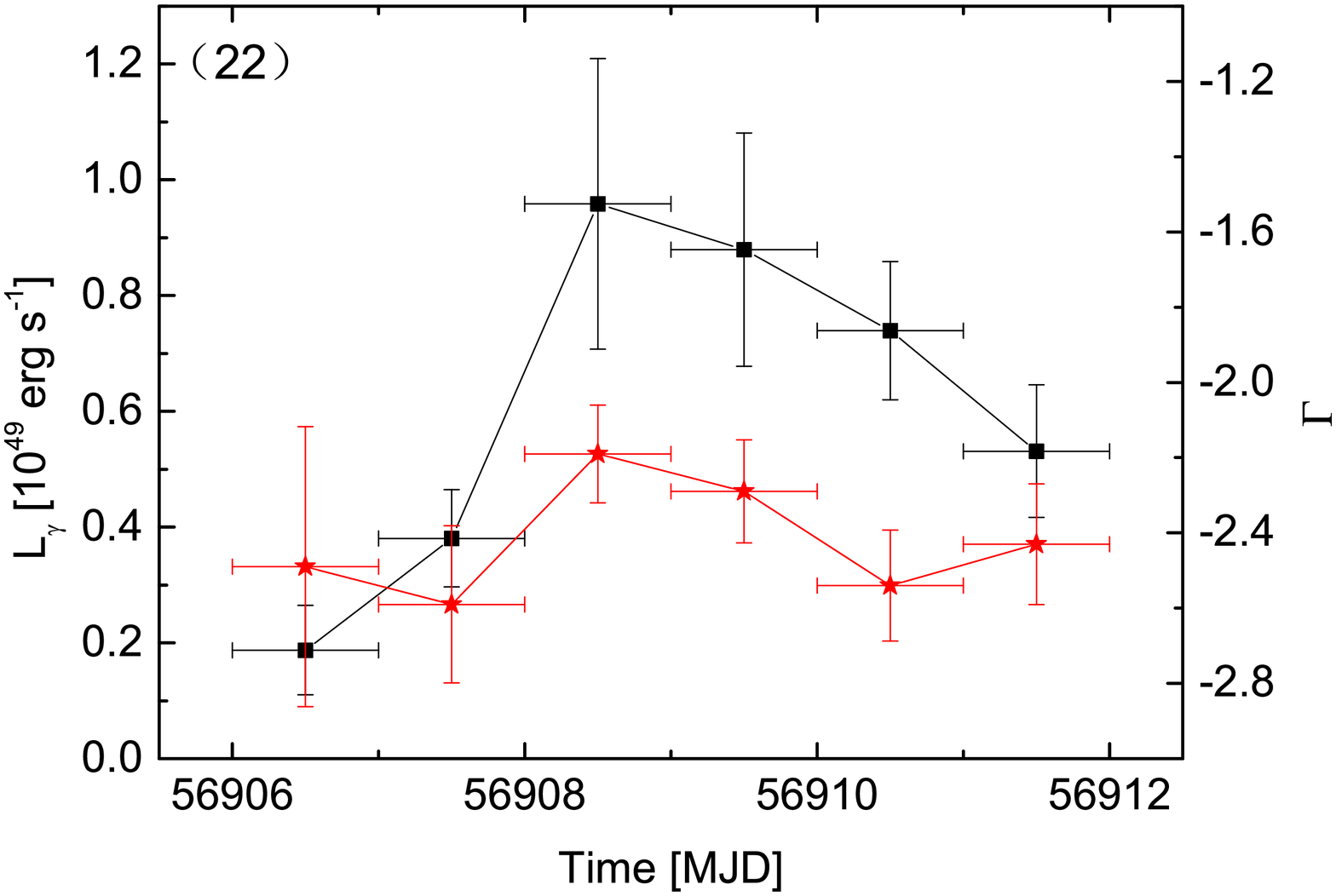}
\includegraphics[angle=0,scale=0.18]{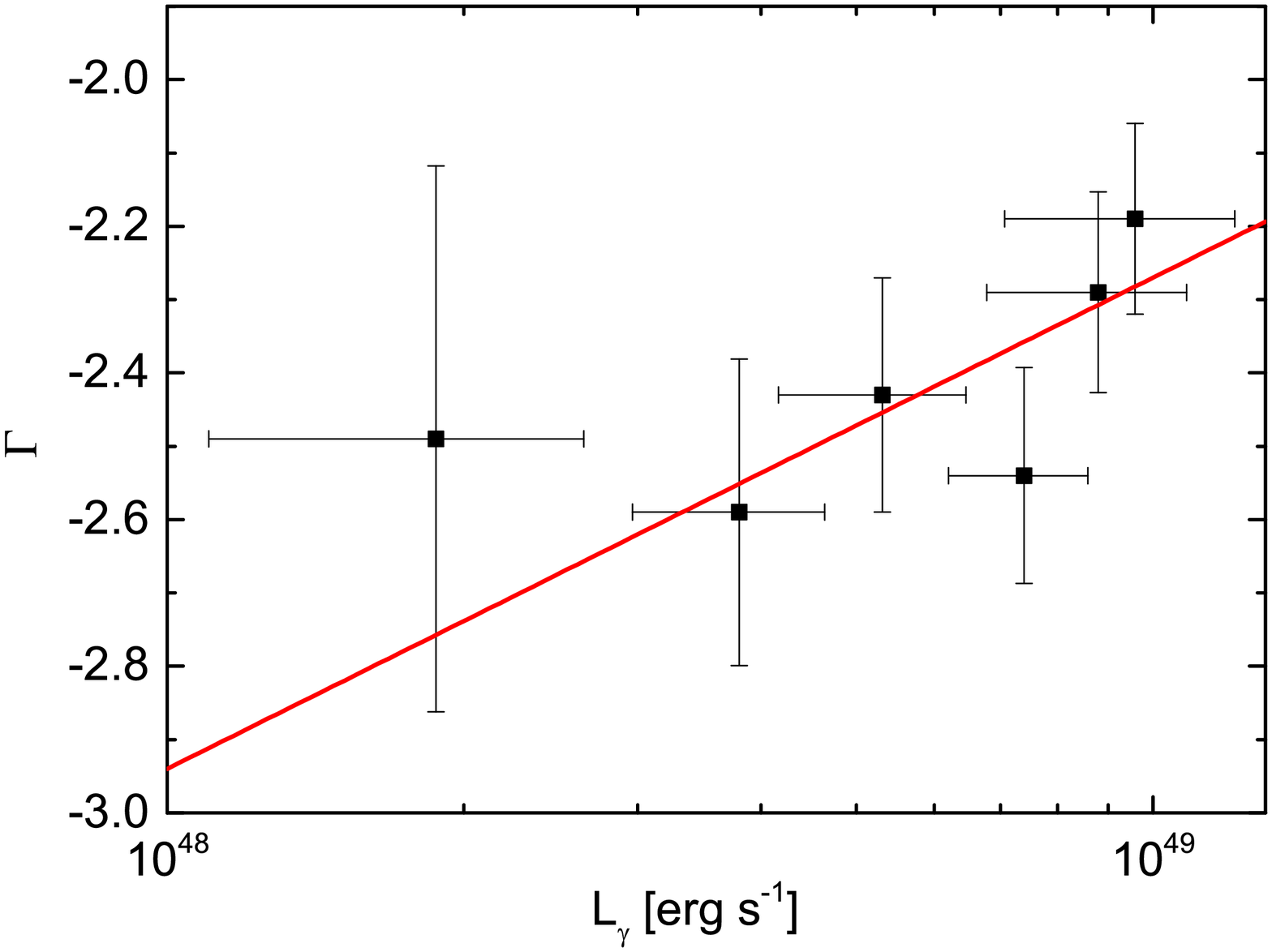}
\includegraphics[angle=0,scale=0.23]{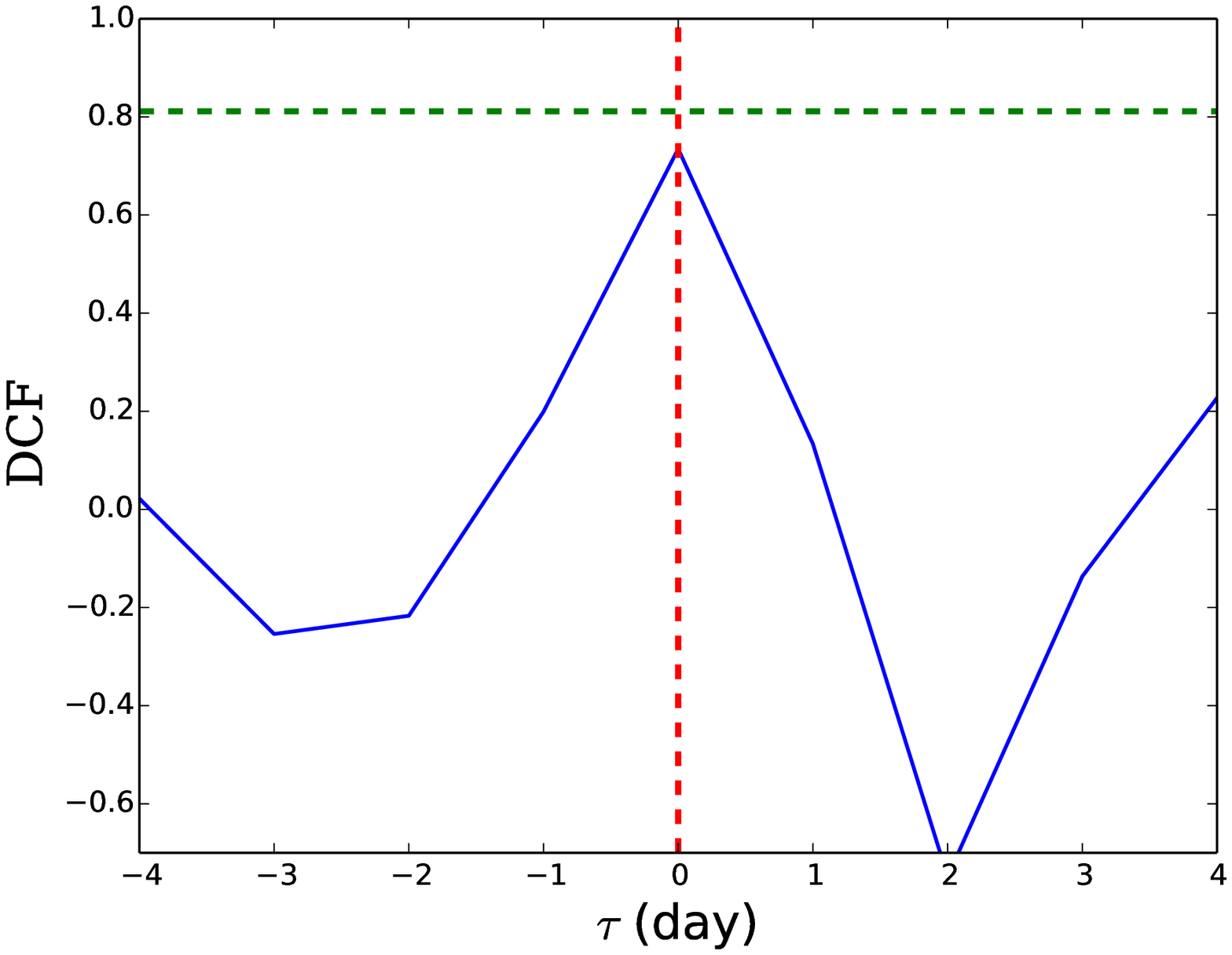}\\
\includegraphics[angle=0,scale=0.18]{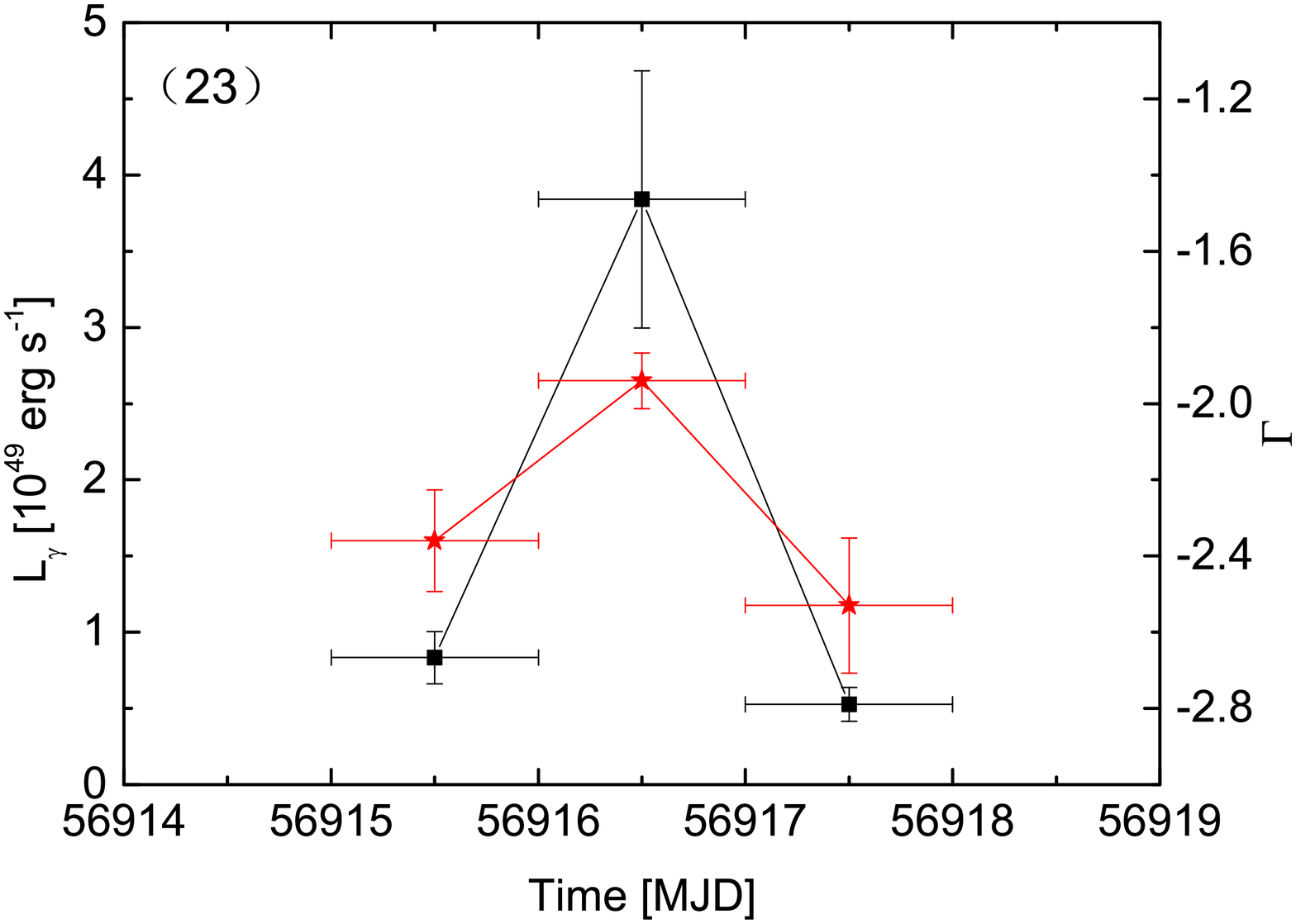}
\includegraphics[angle=0,scale=0.18]{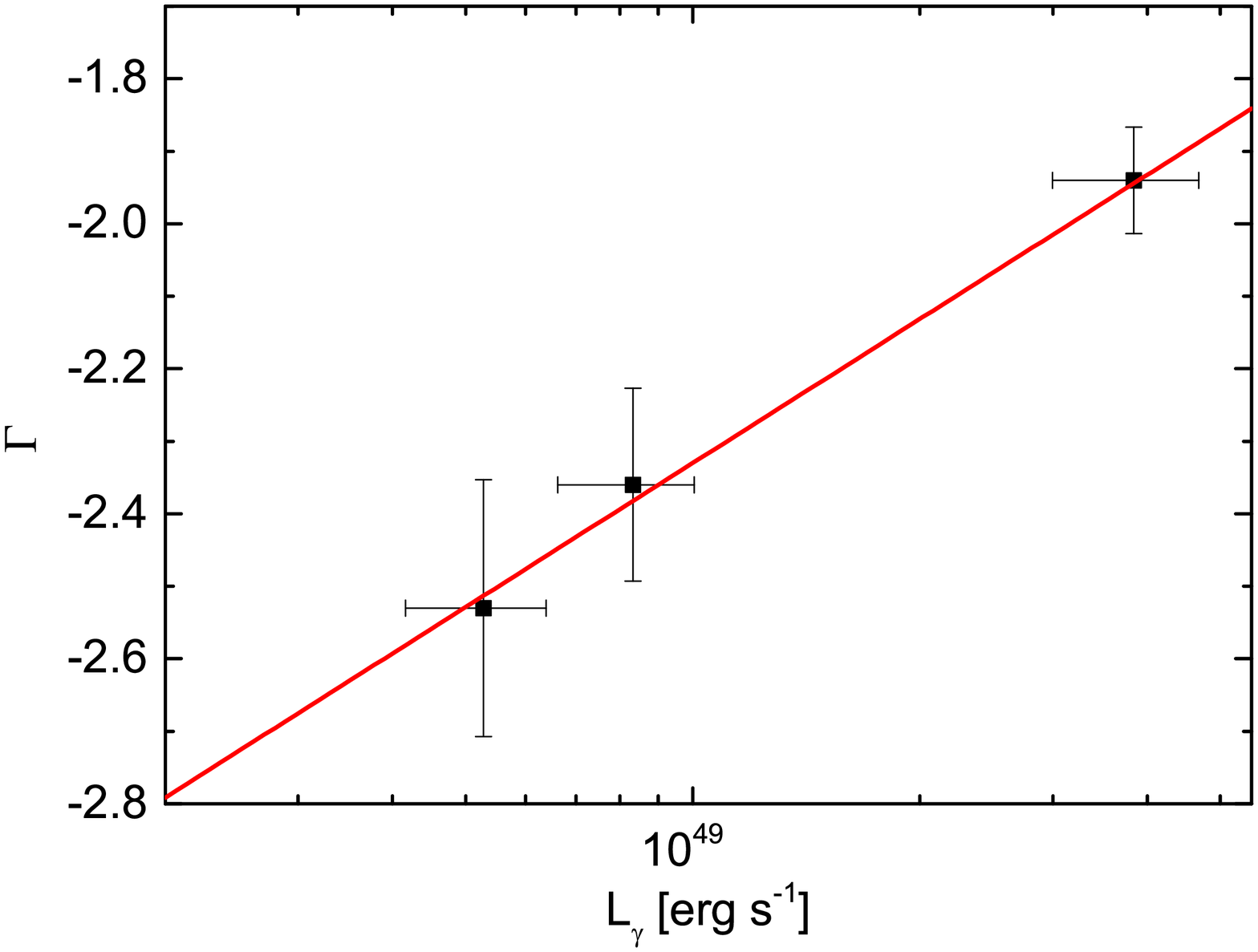}
\includegraphics[angle=0,scale=0.23]{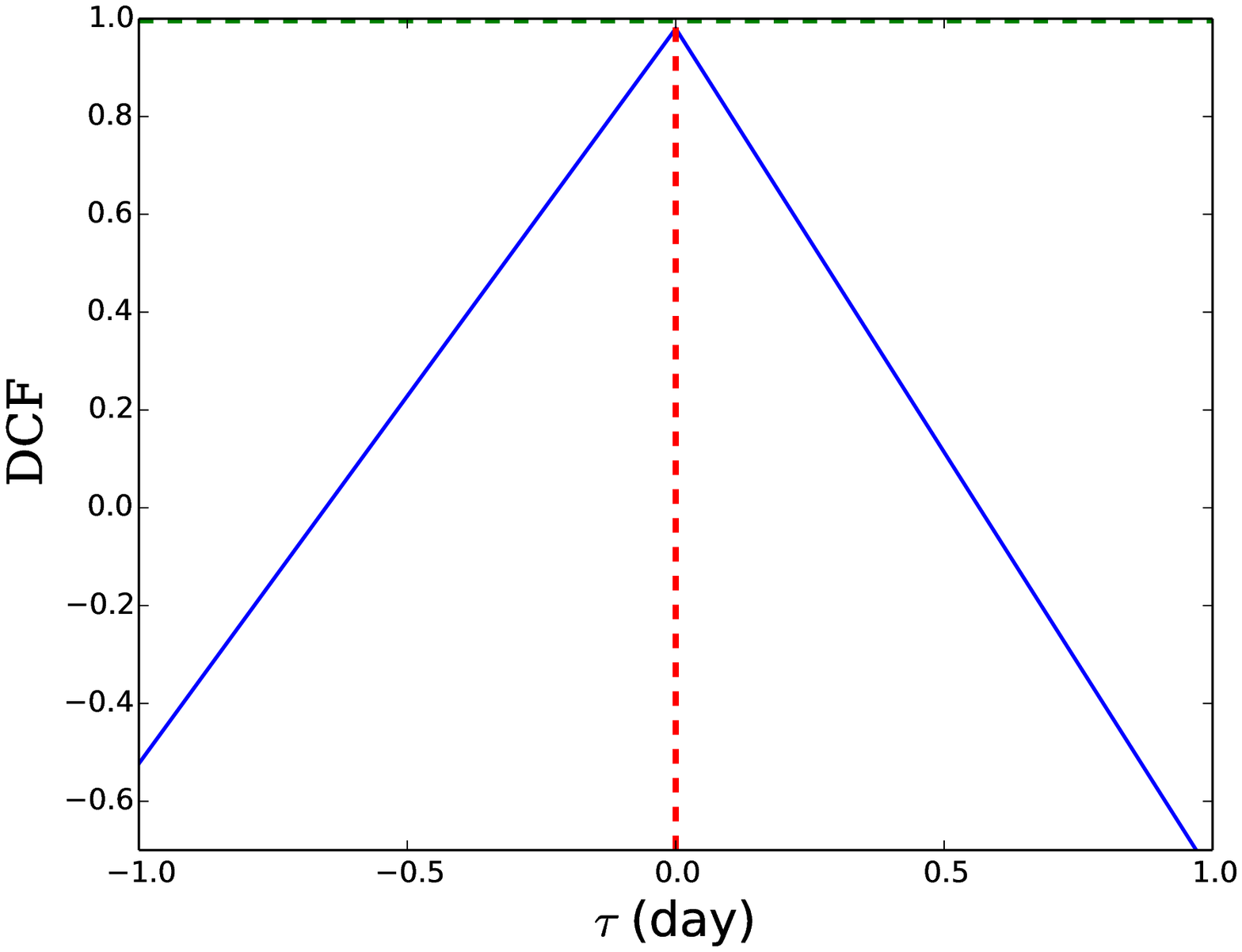}\\
\includegraphics[angle=0,scale=0.18]{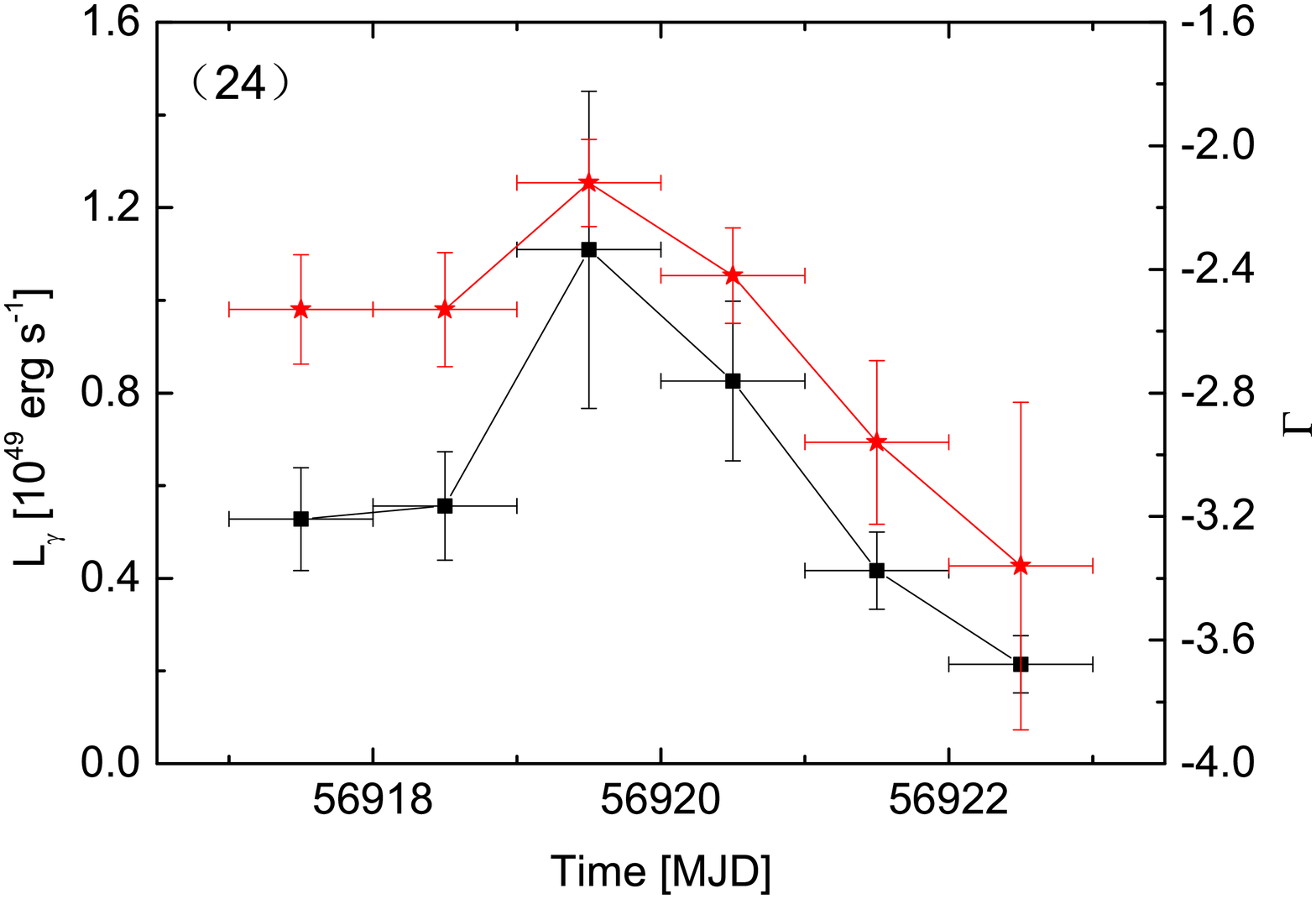}
\includegraphics[angle=0,scale=0.18]{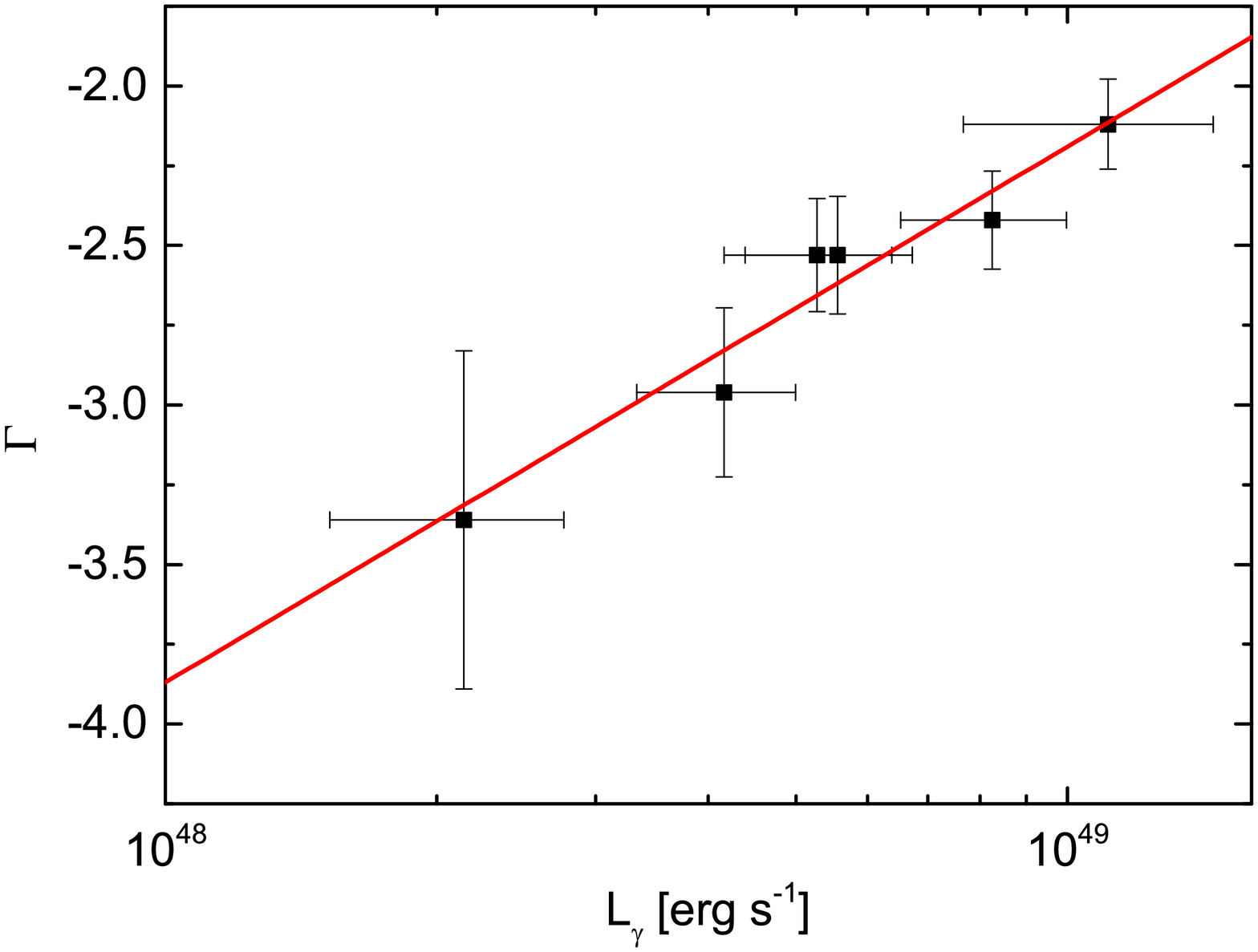}
\includegraphics[angle=0,scale=0.23]{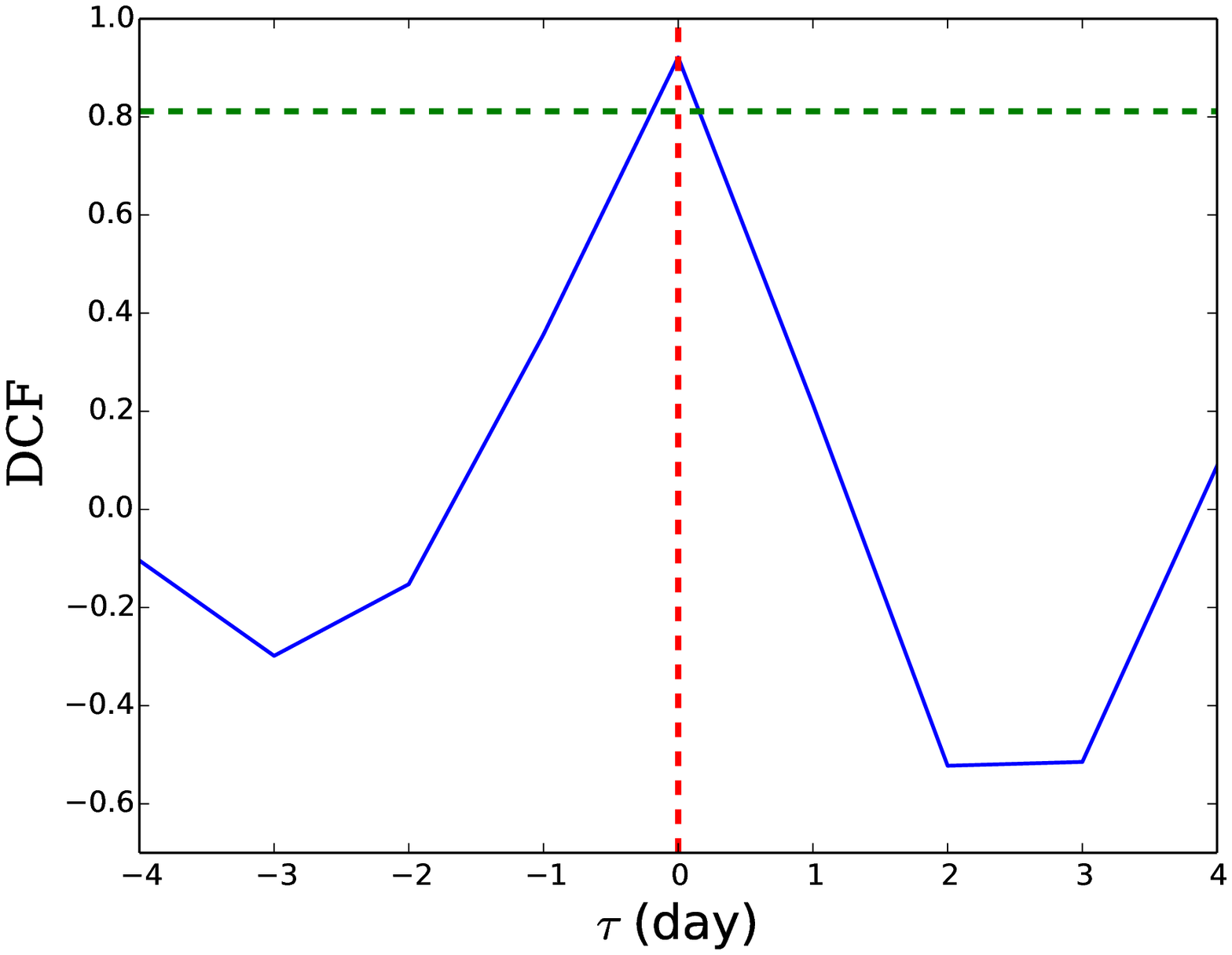}\\
\includegraphics[angle=0,scale=0.18]{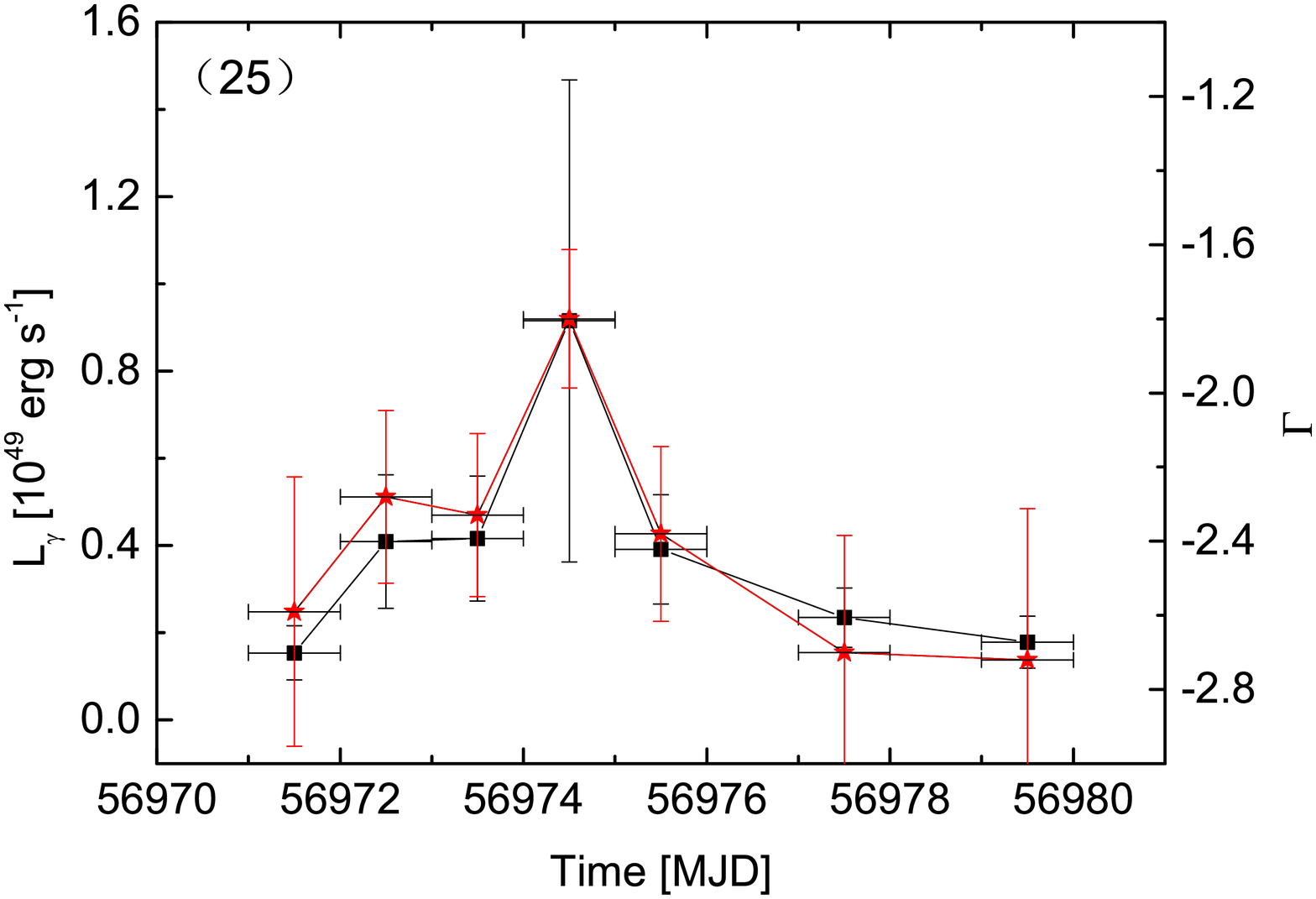}
\includegraphics[angle=0,scale=0.18]{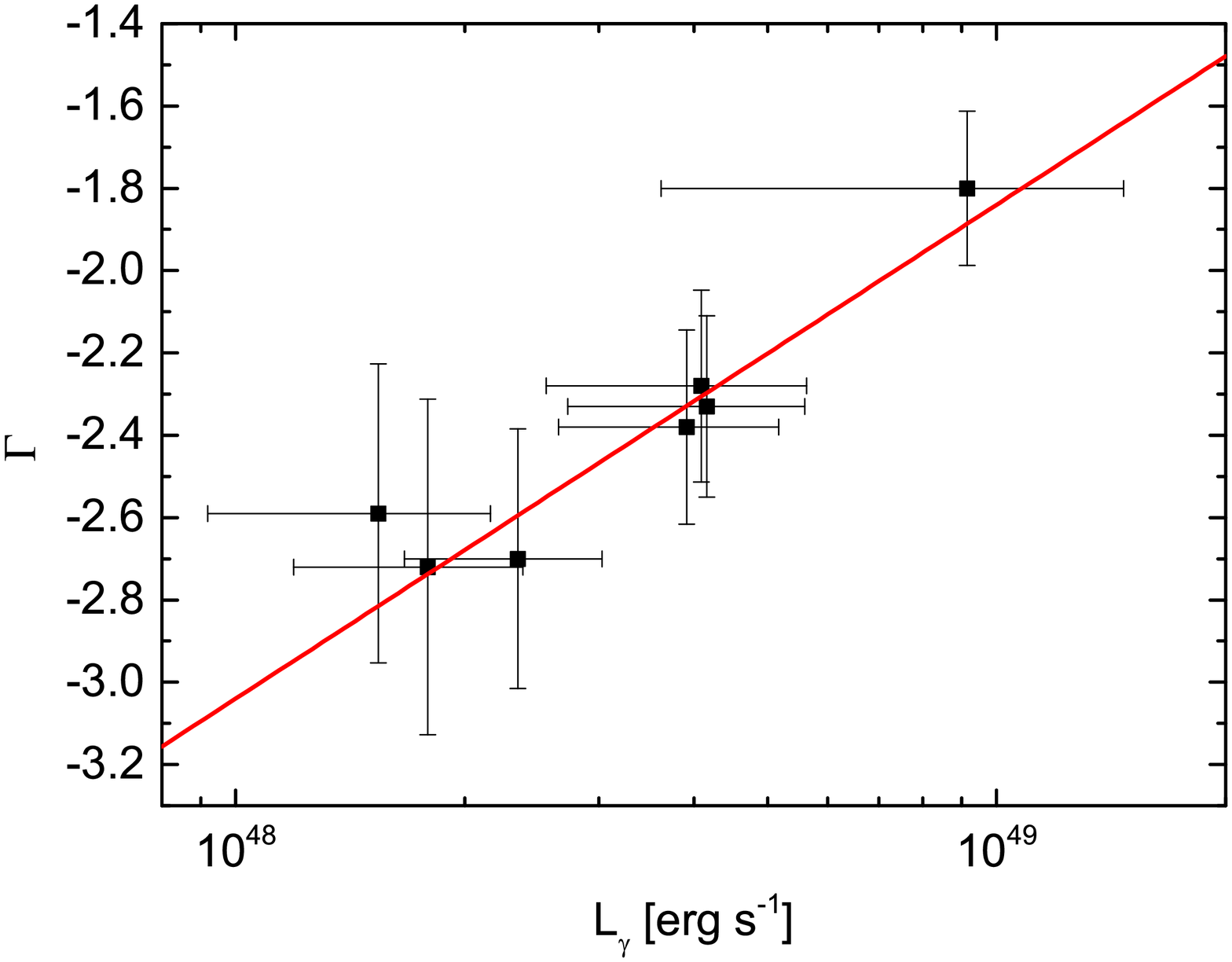}
\includegraphics[angle=0,scale=0.23]{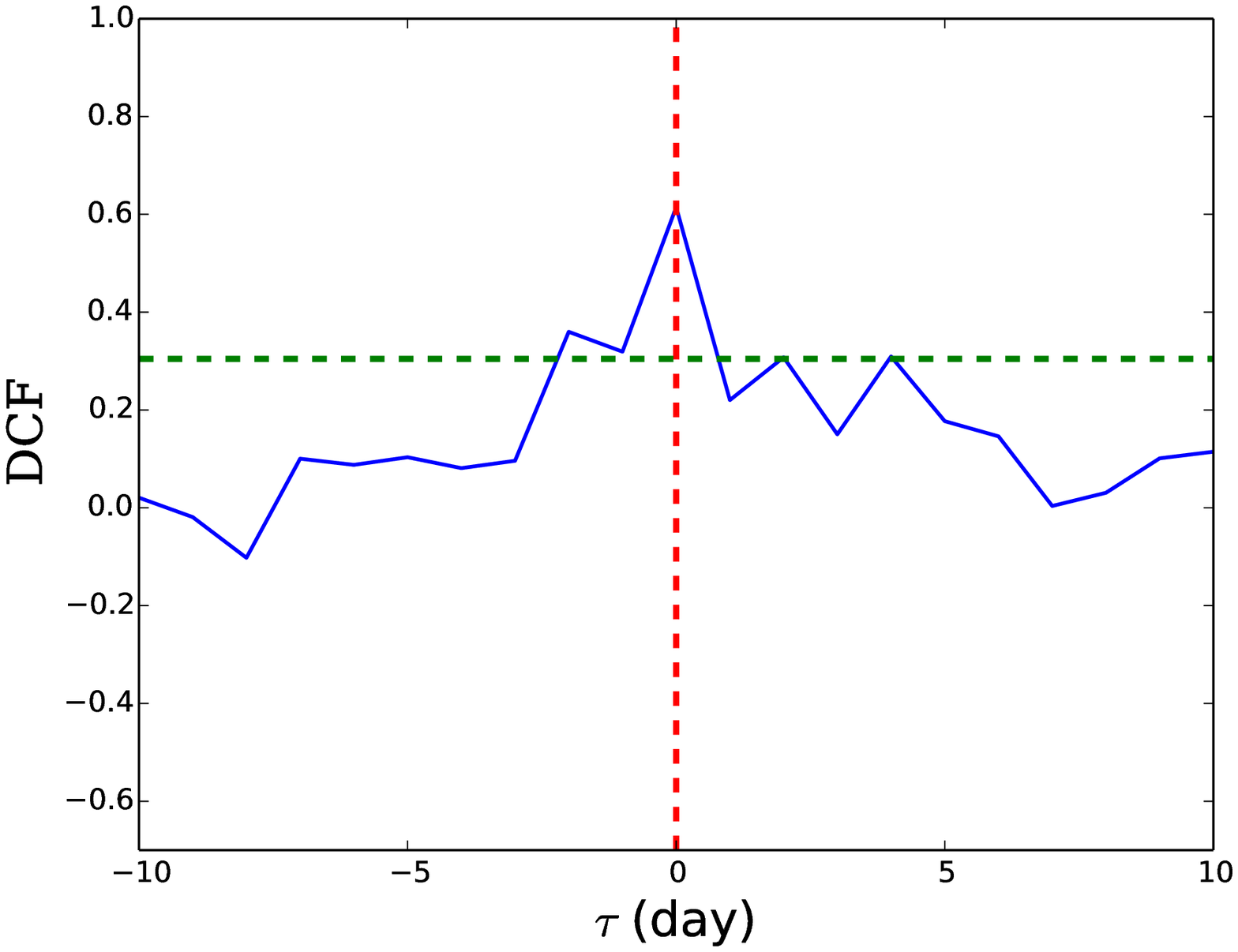}\\
\includegraphics[angle=0,scale=0.18]{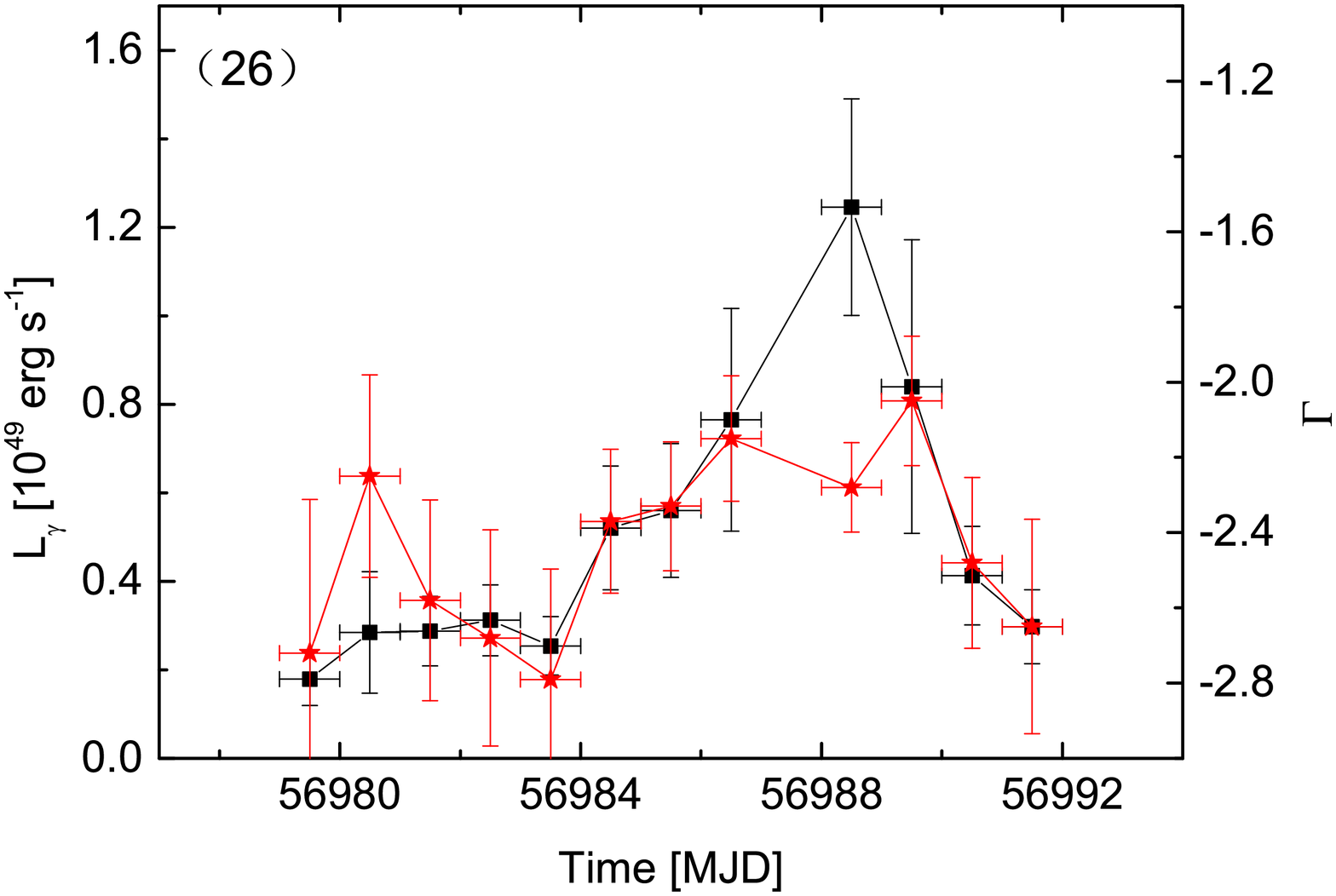}
\includegraphics[angle=0,scale=0.18]{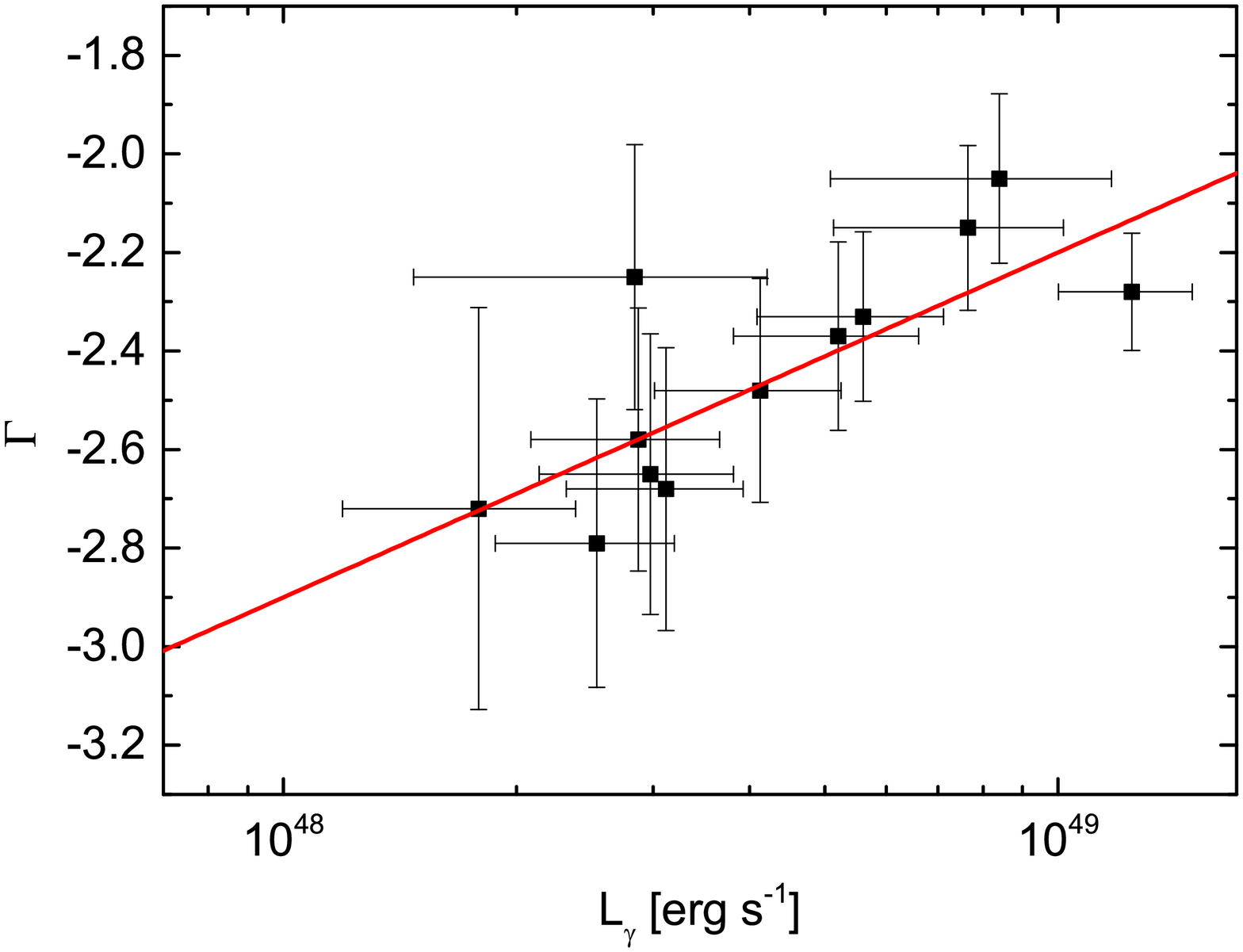}
\includegraphics[angle=0,scale=0.23]{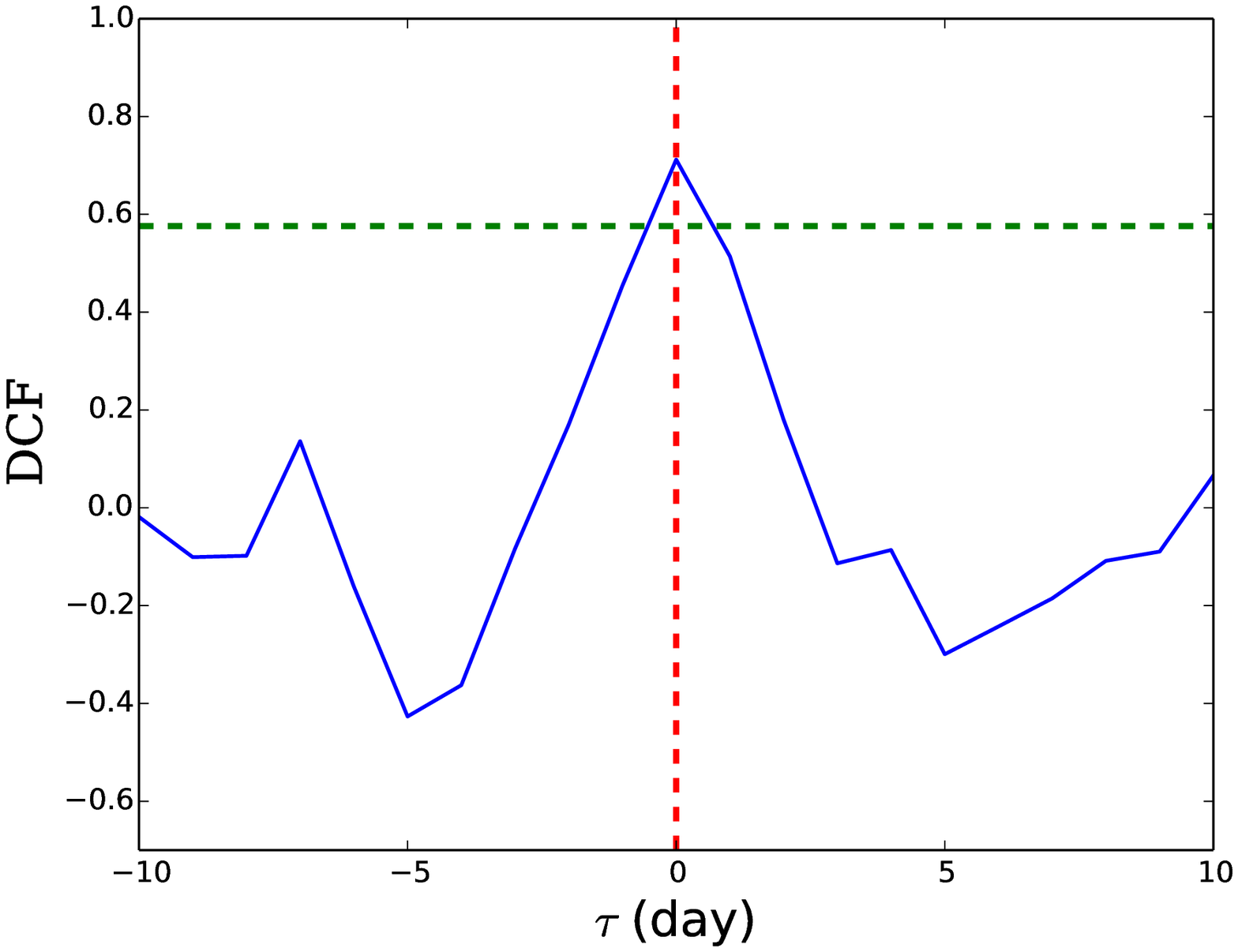}\\
\includegraphics[angle=0,scale=0.18]{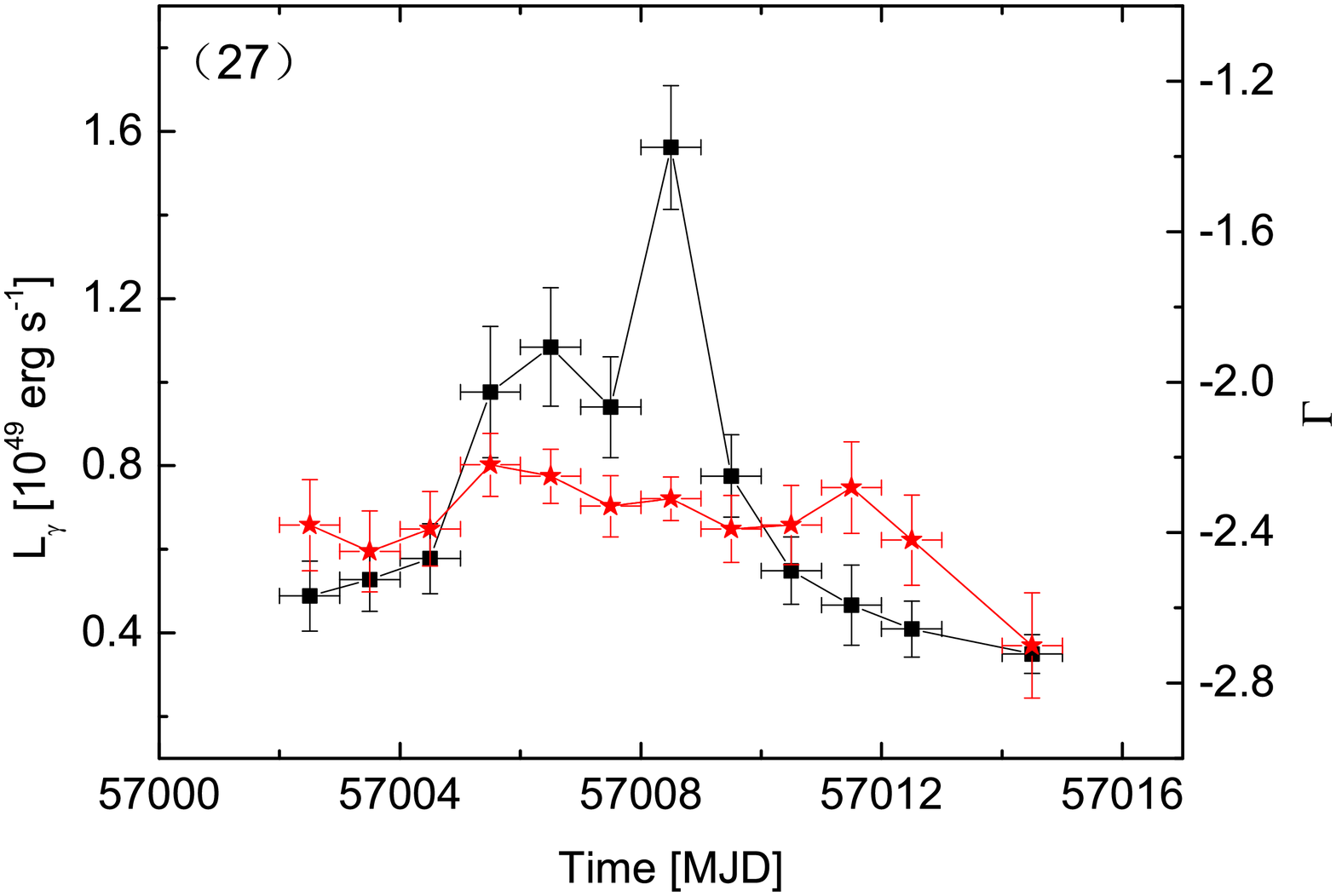}
\includegraphics[angle=0,scale=0.18]{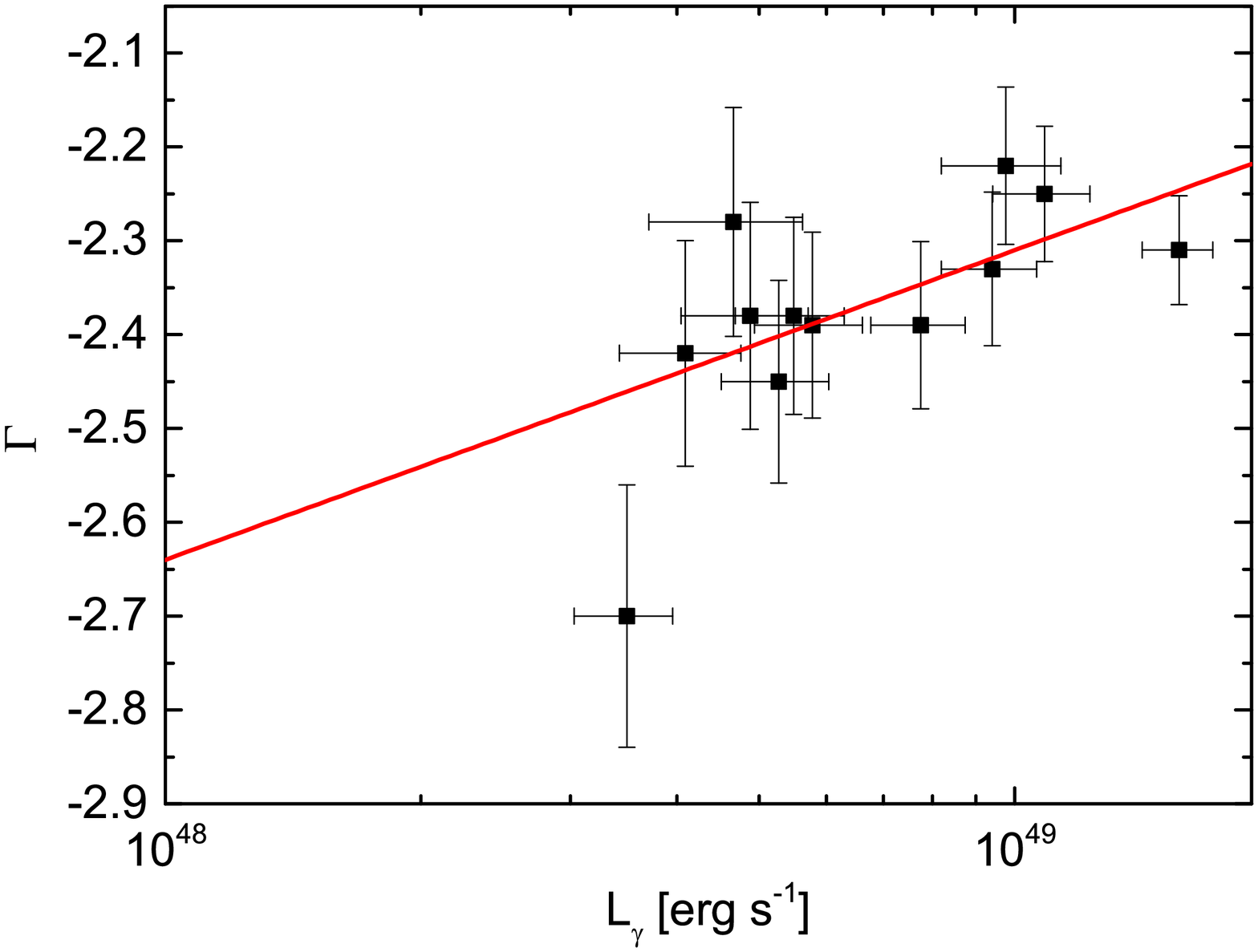}
\includegraphics[angle=0,scale=0.23]{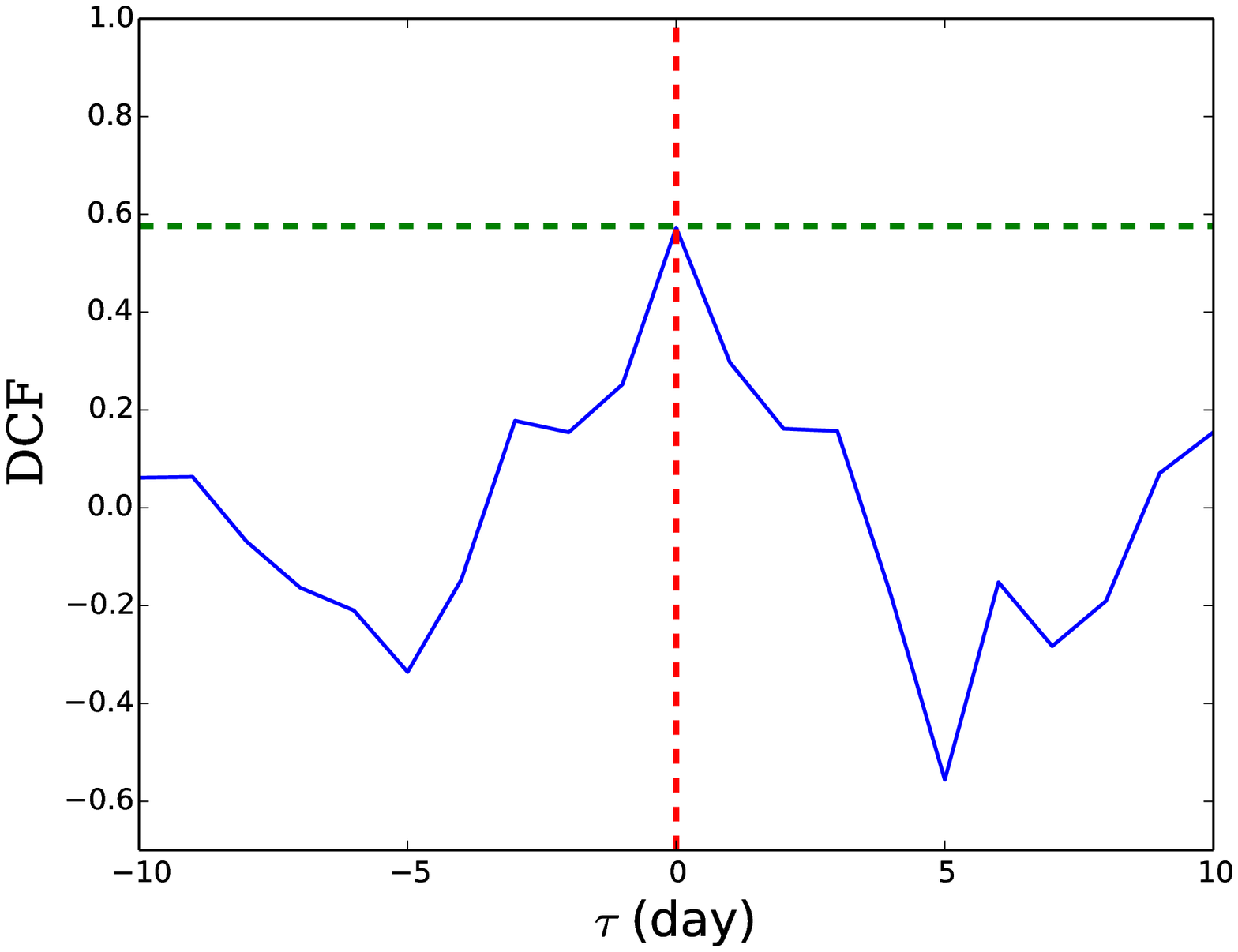}
\hfill\center{Fig.5---  continued}
\end{figure*}

\begin{figure*}
\includegraphics[angle=0,scale=0.18]{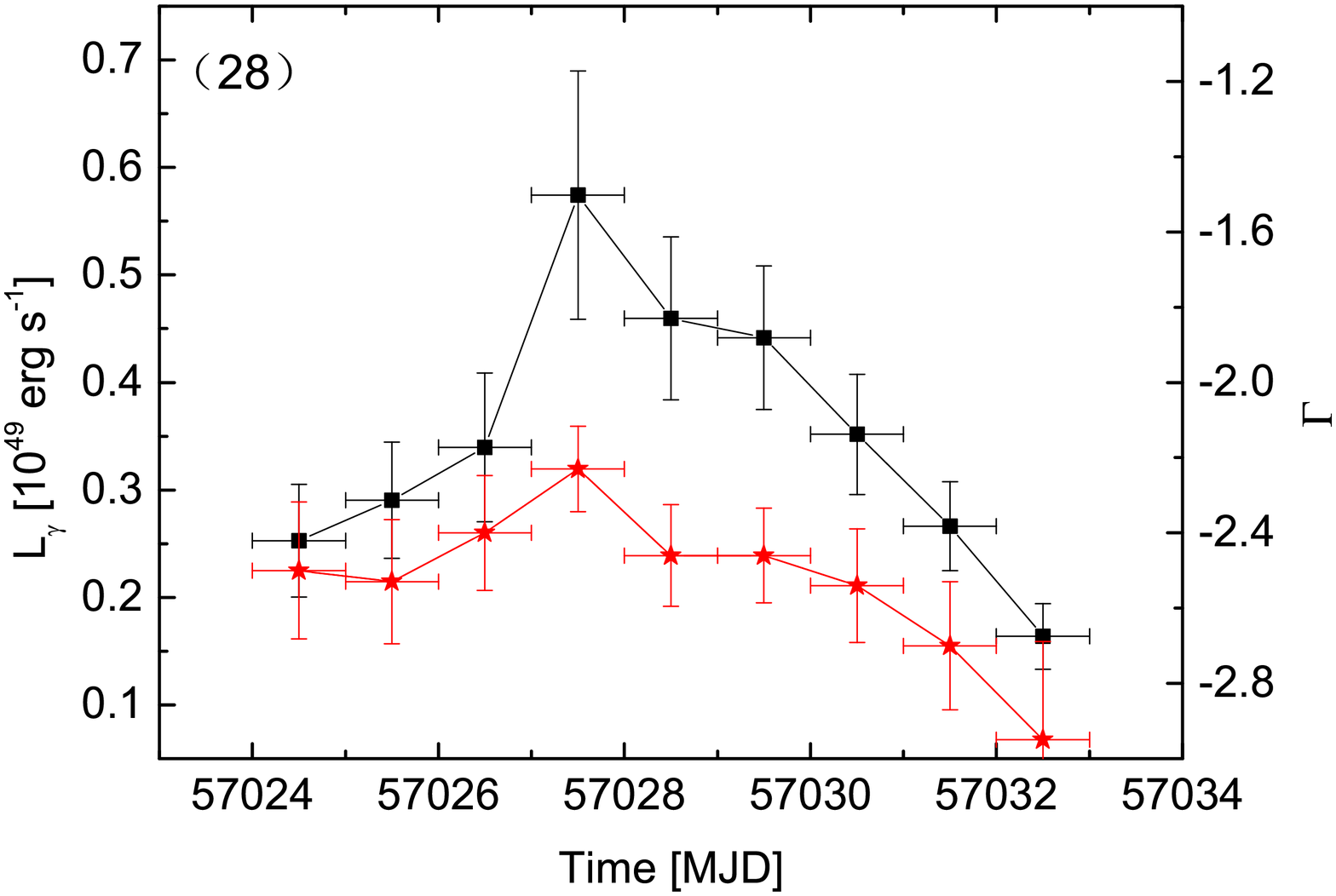}
\includegraphics[angle=0,scale=0.18]{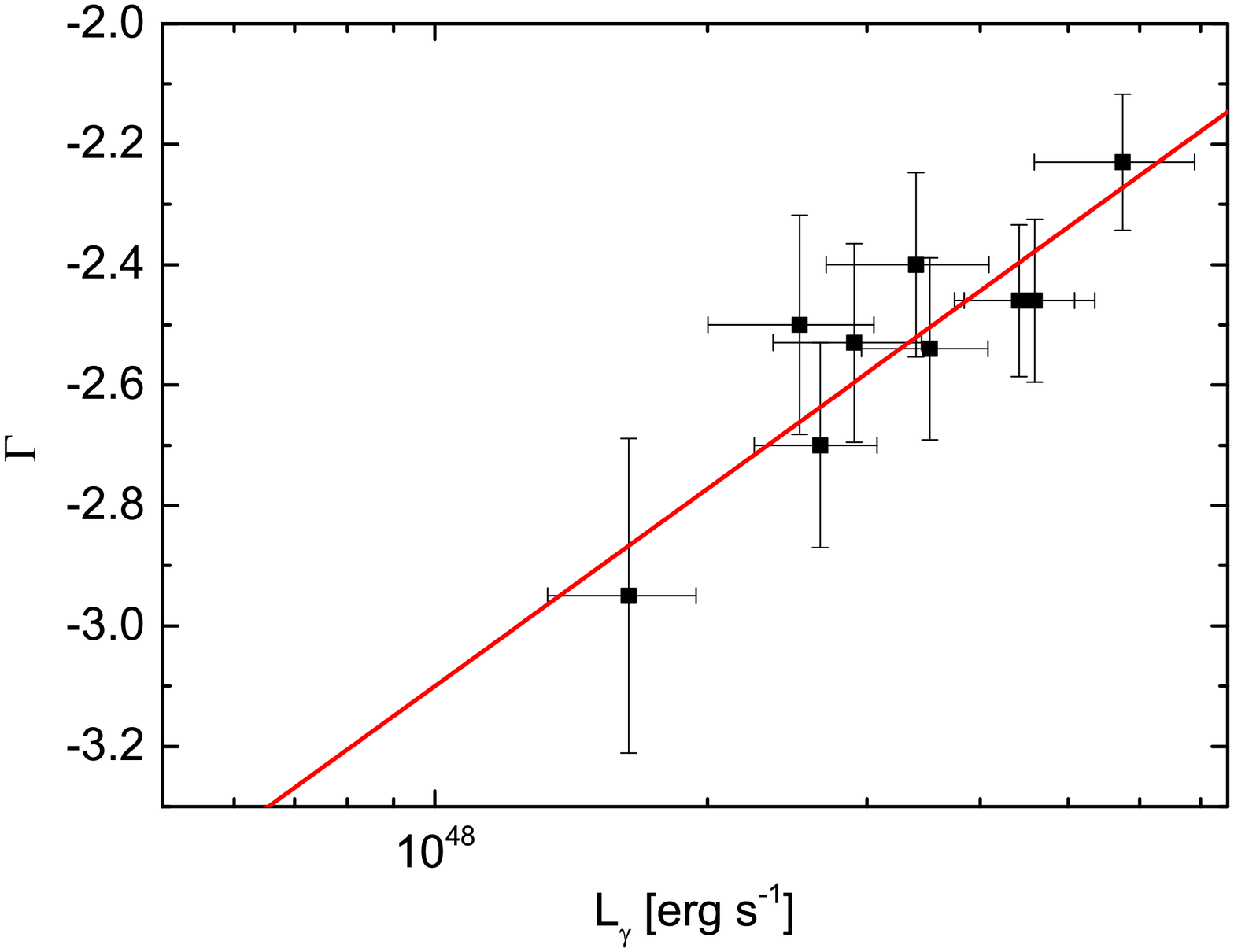}
\includegraphics[angle=0,scale=0.23]{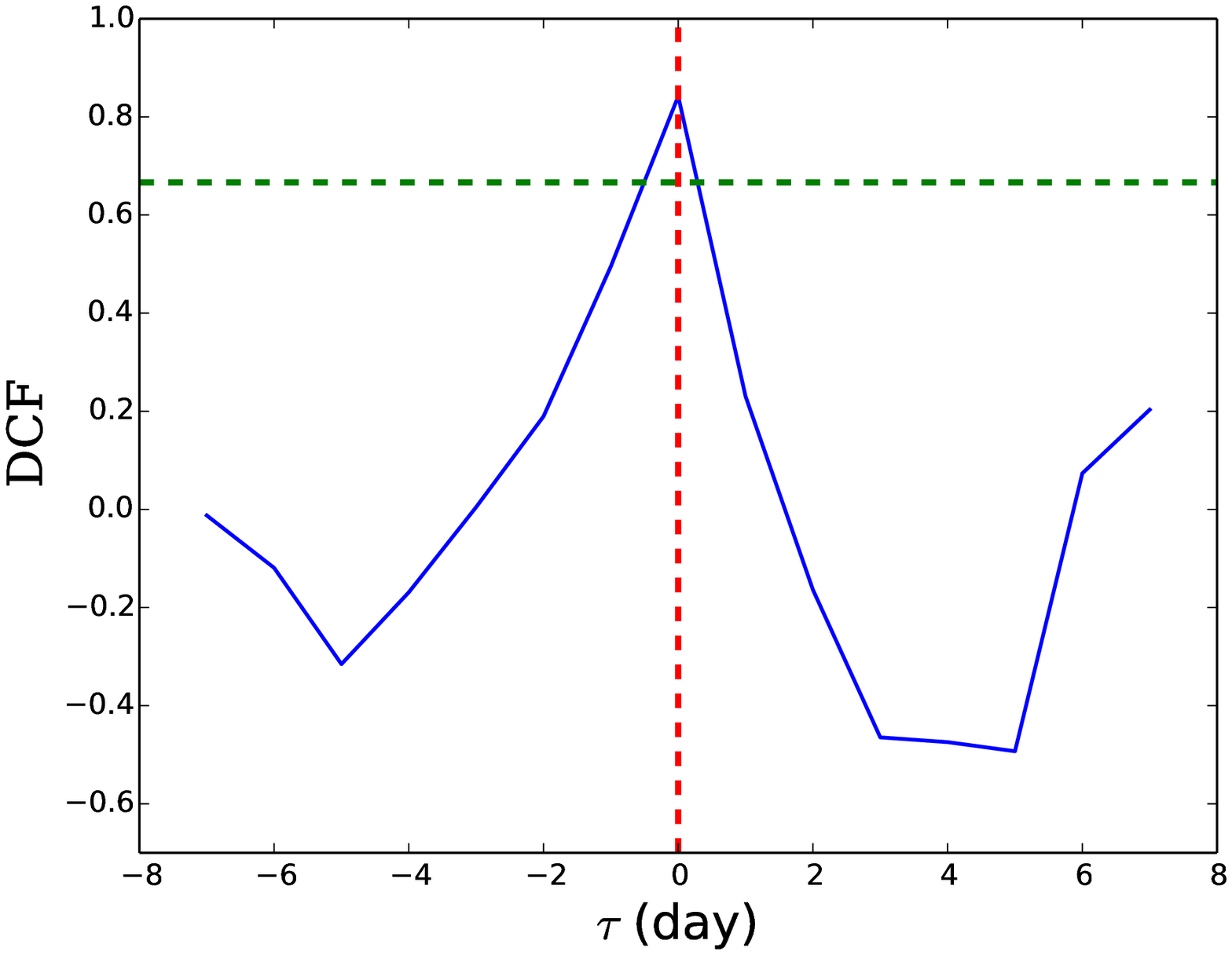}\\
\includegraphics[angle=0,scale=0.18]{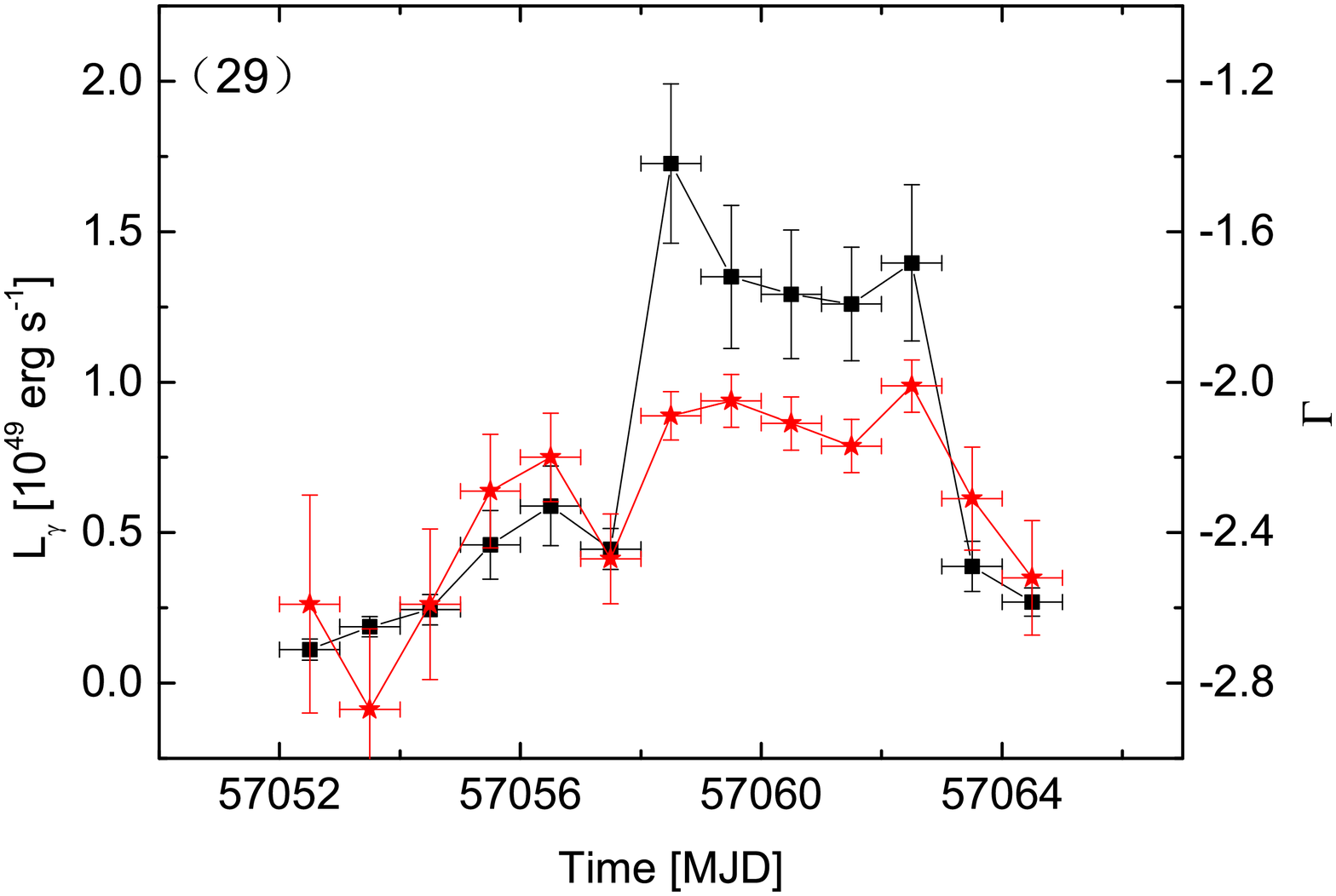}
\includegraphics[angle=0,scale=0.18]{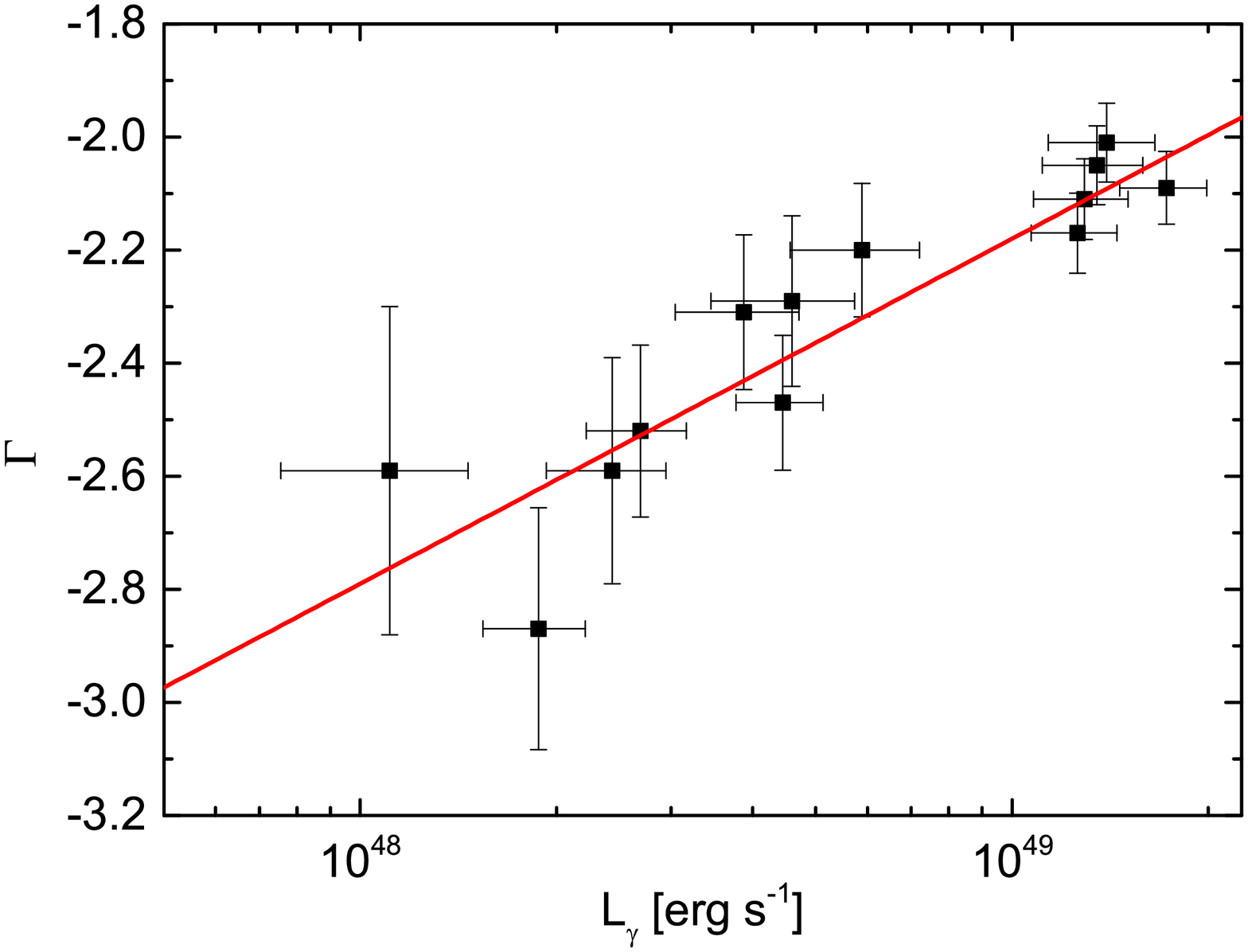}
\includegraphics[angle=0,scale=0.23]{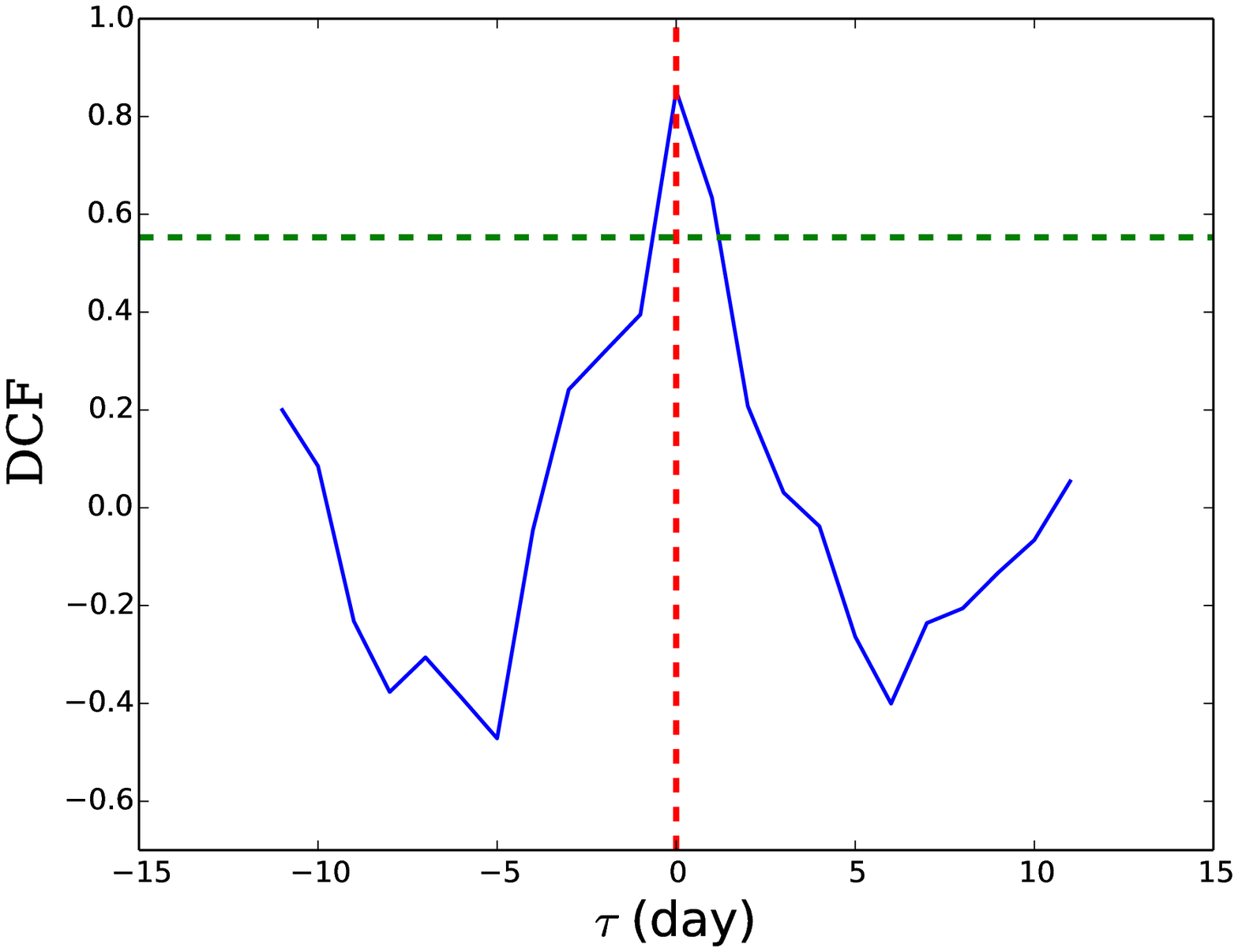}\\
\includegraphics[angle=0,scale=0.18]{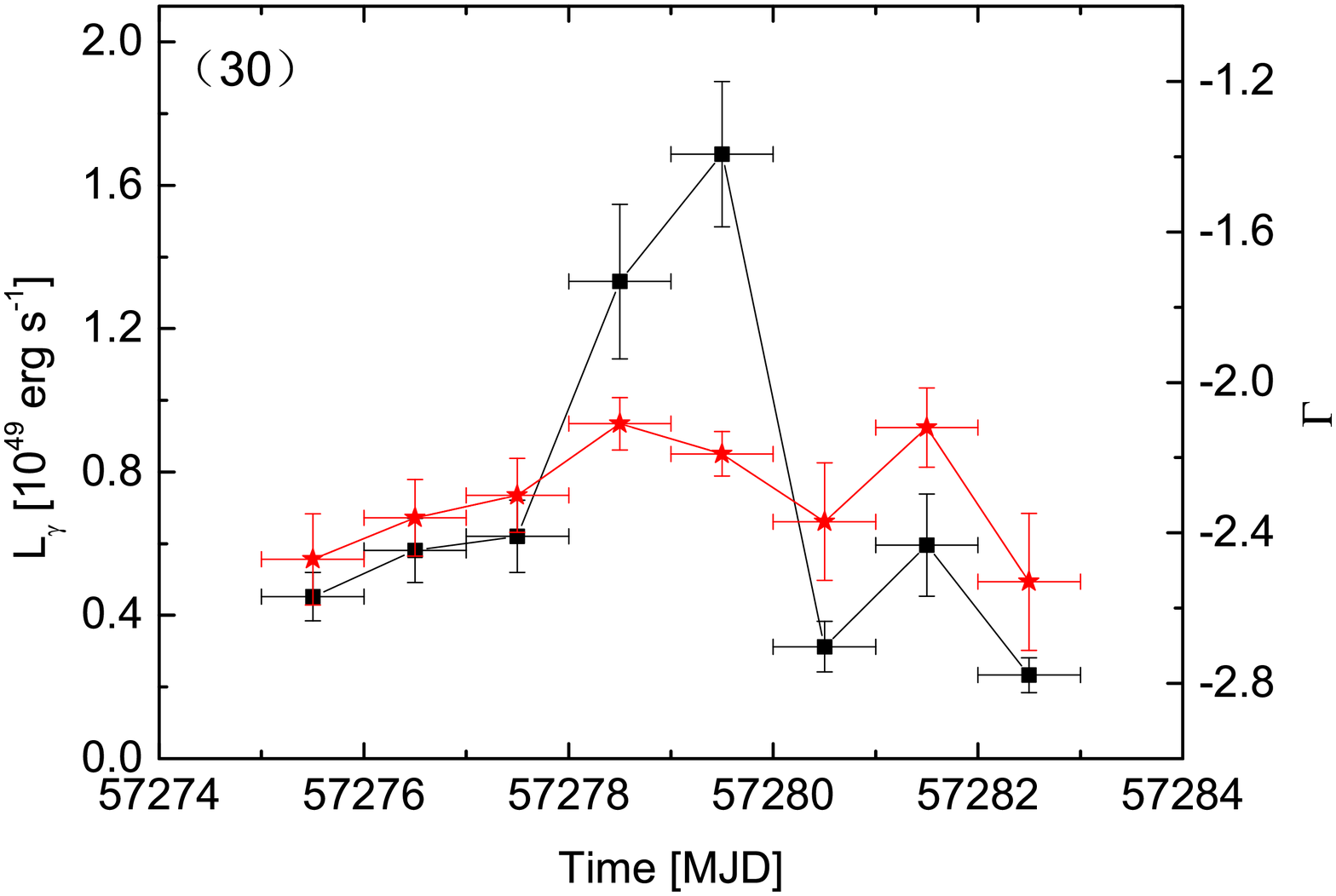}
\includegraphics[angle=0,scale=0.18]{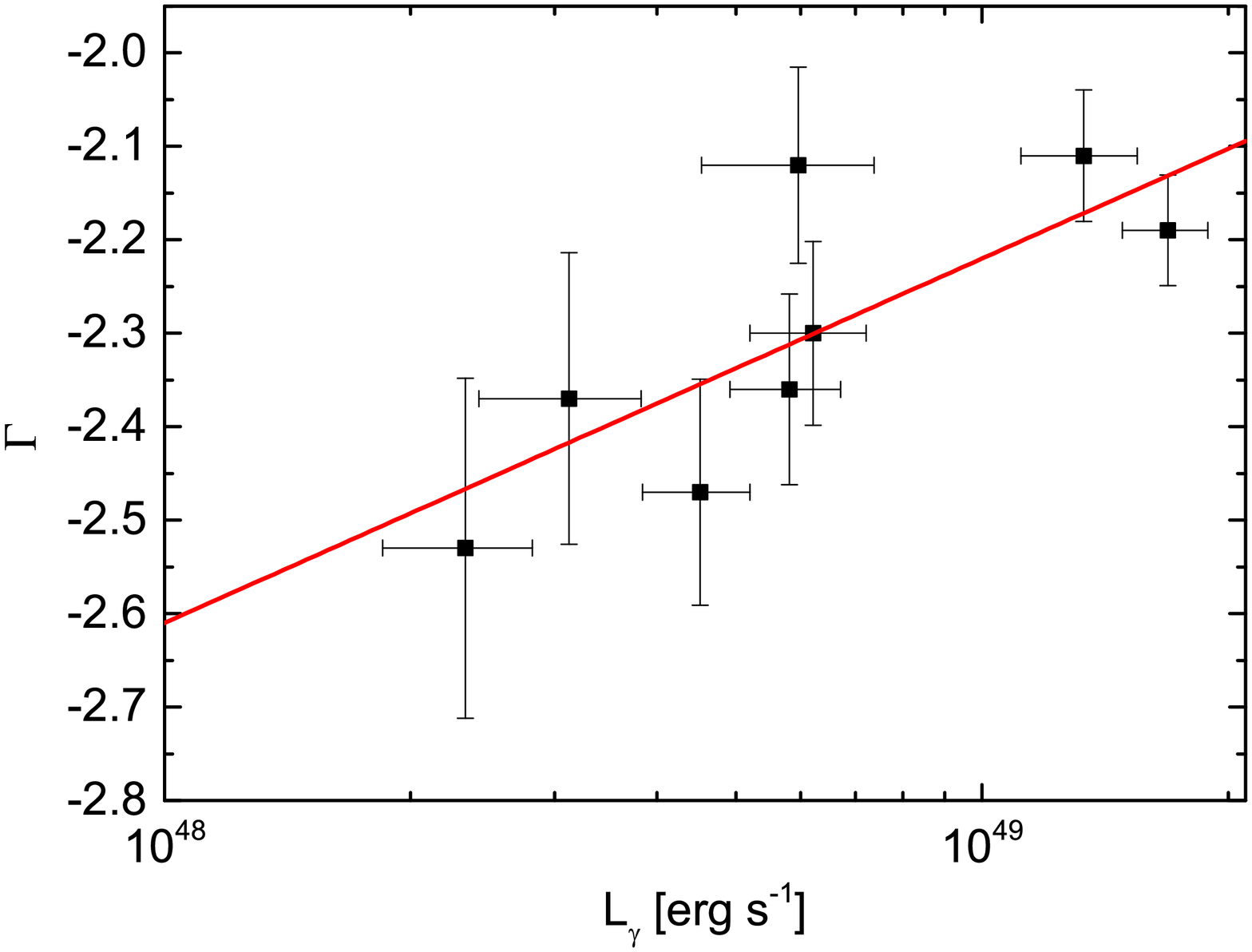}
\includegraphics[angle=0,scale=0.23]{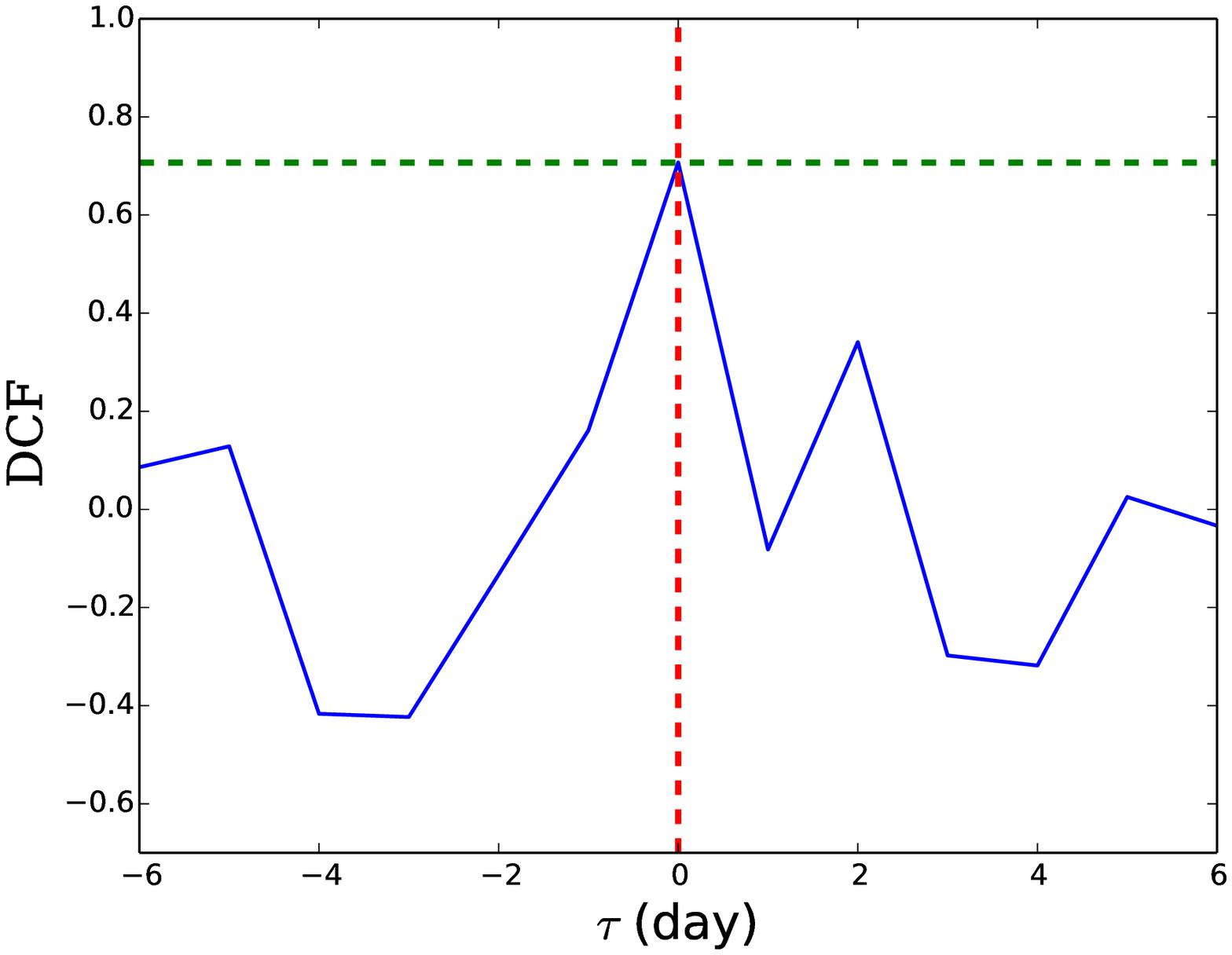}\\
\includegraphics[angle=0,scale=0.18]{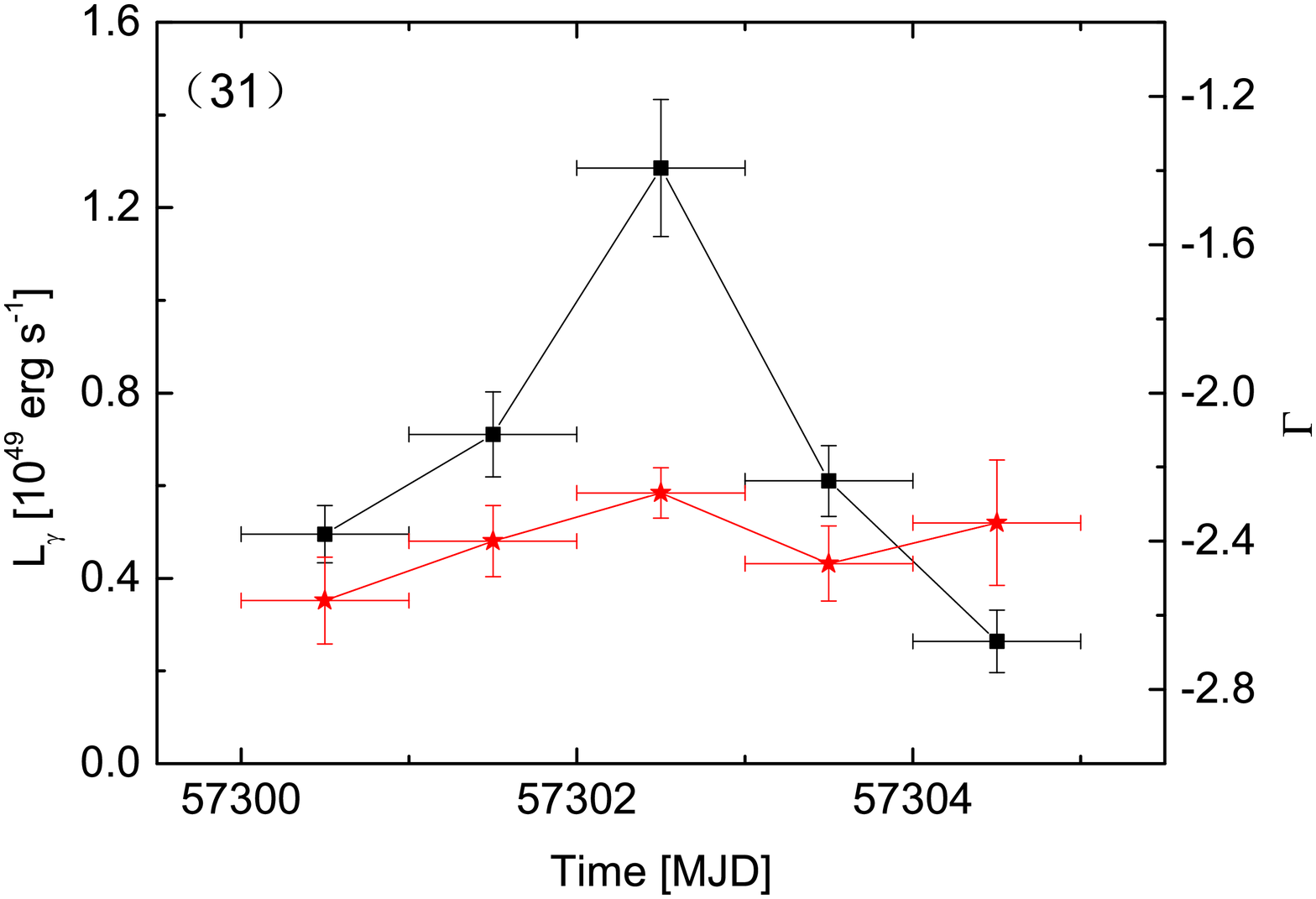}
\includegraphics[angle=0,scale=0.18]{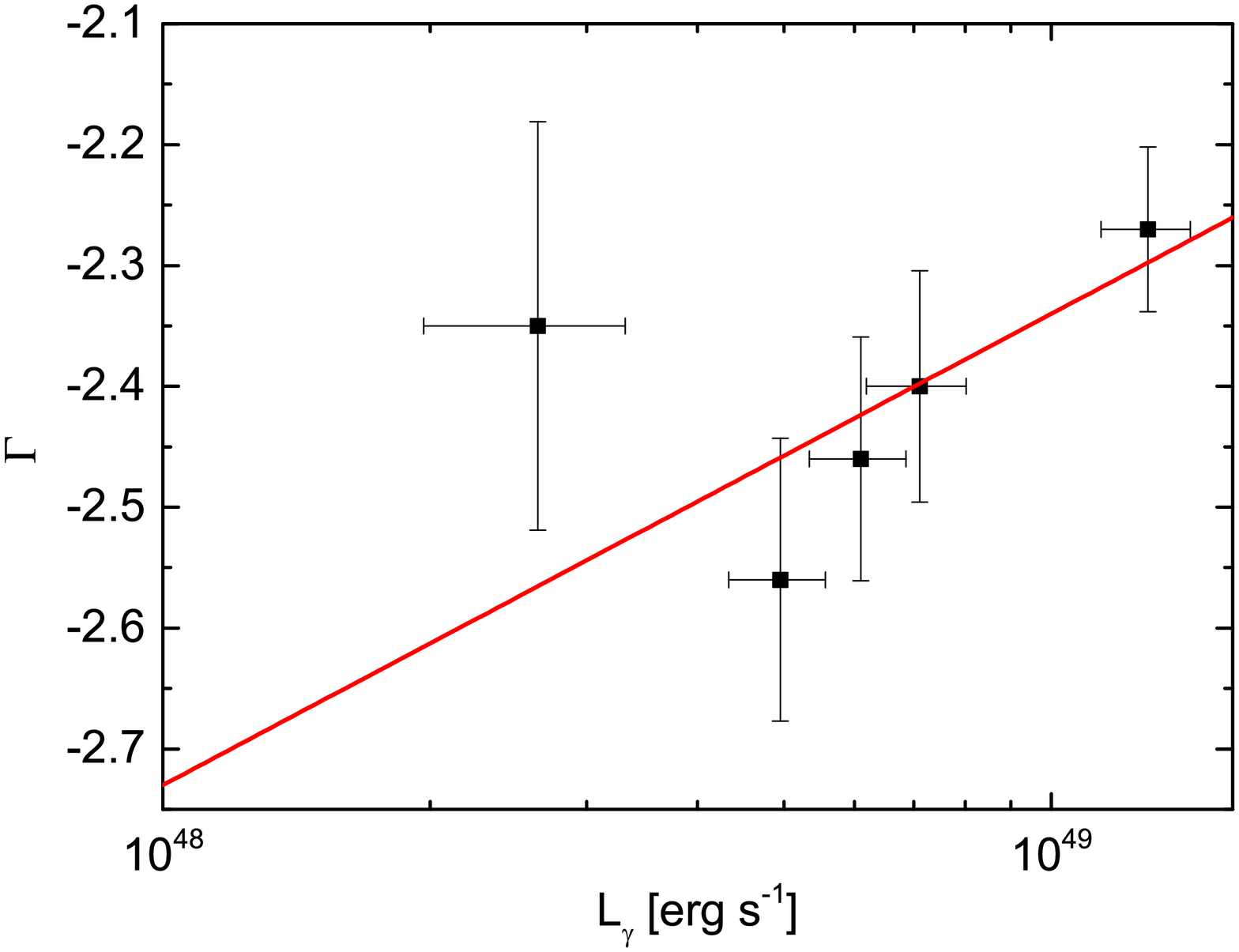}
\includegraphics[angle=0,scale=0.23]{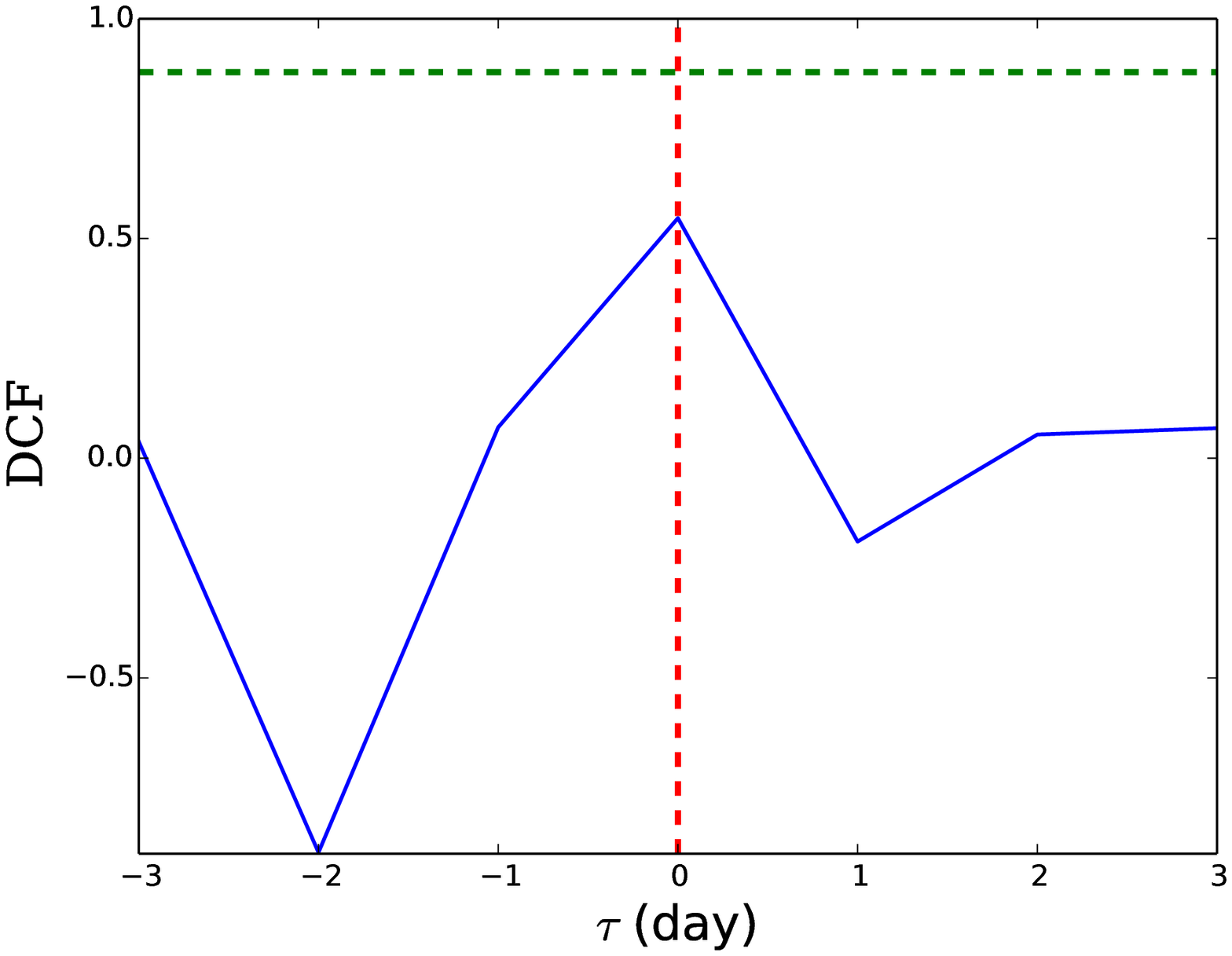}\\
\includegraphics[angle=0,scale=0.18]{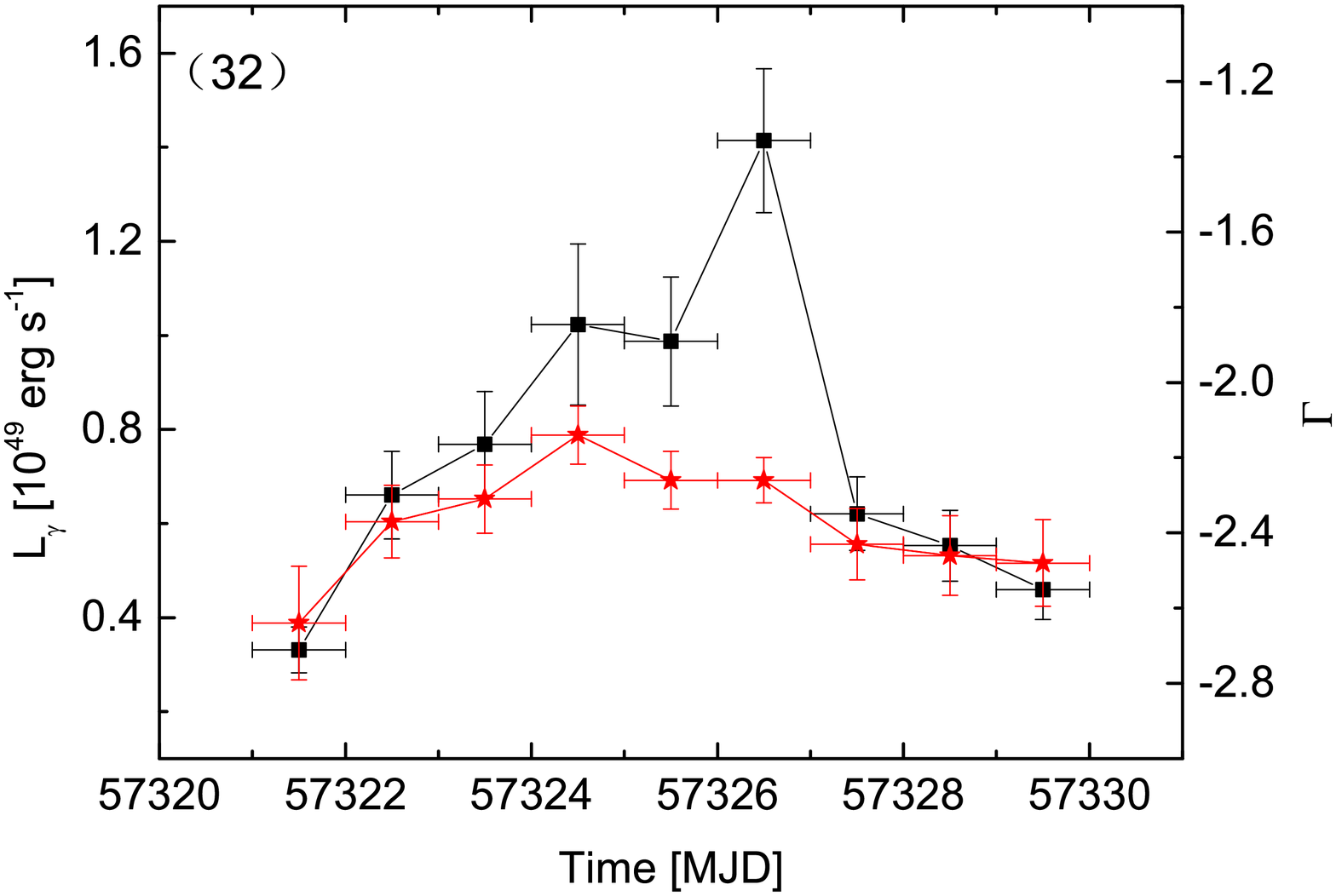}
\includegraphics[angle=0,scale=0.18]{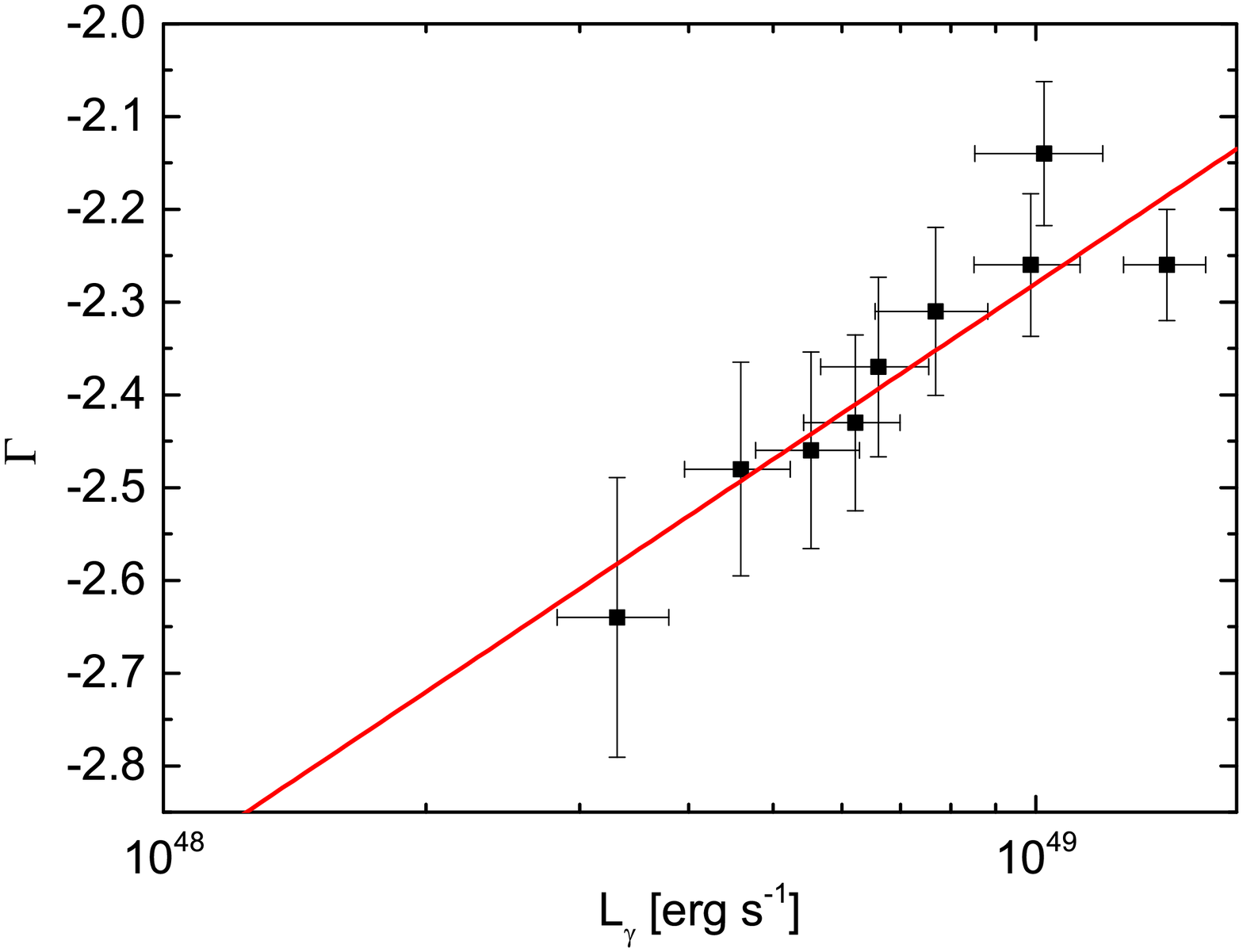}
\includegraphics[angle=0,scale=0.23]{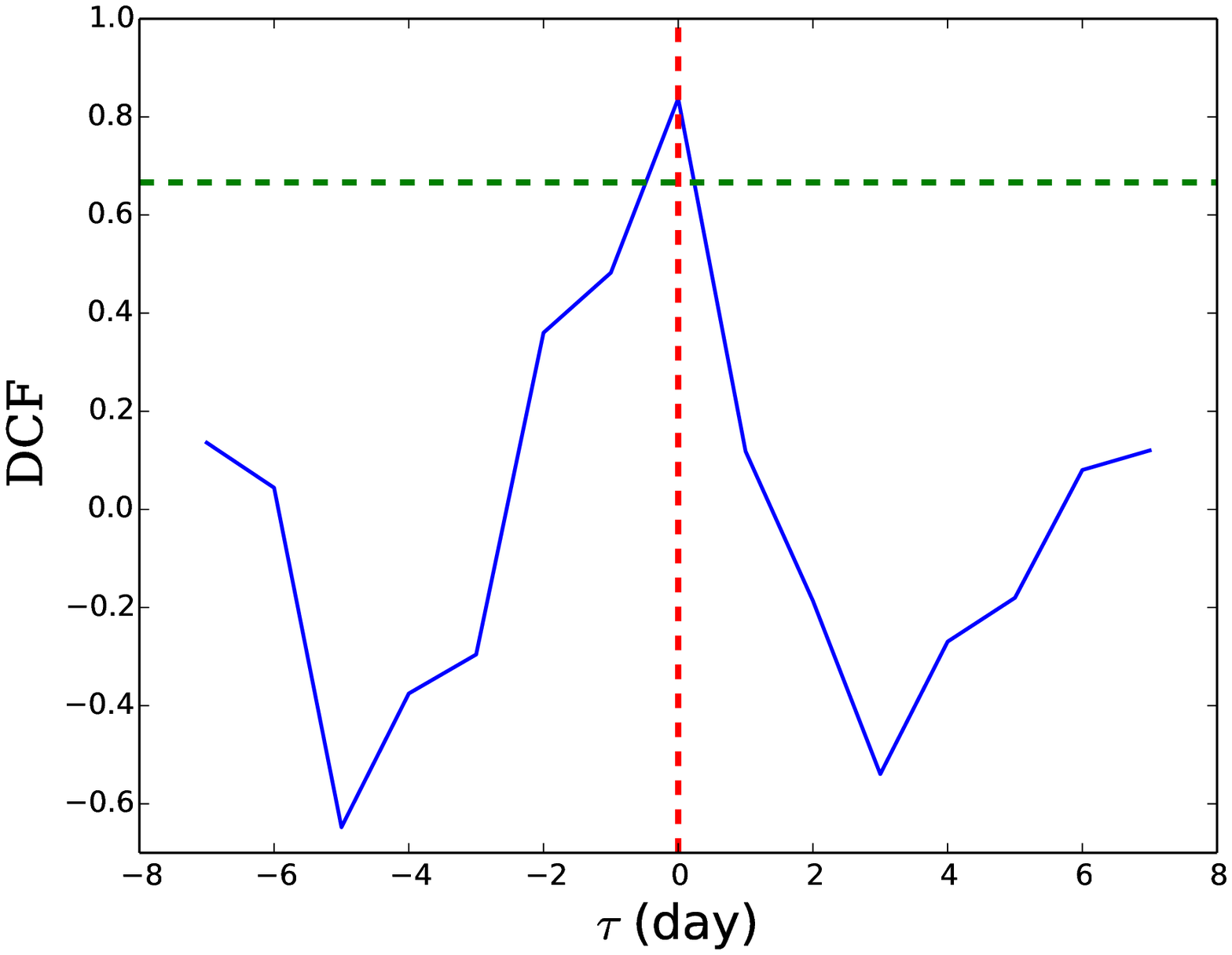}\\
\includegraphics[angle=0,scale=0.18]{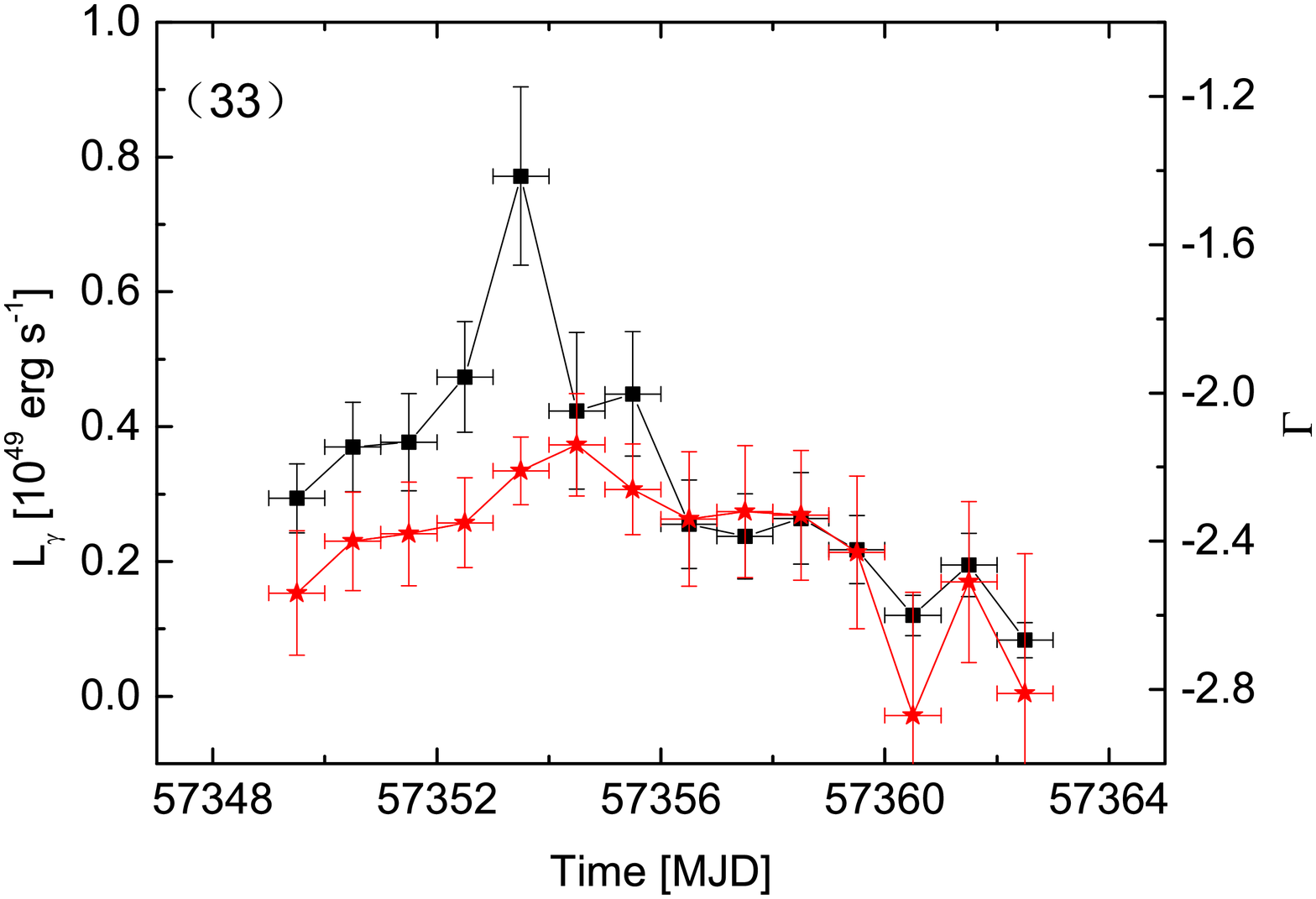}
\includegraphics[angle=0,scale=0.18]{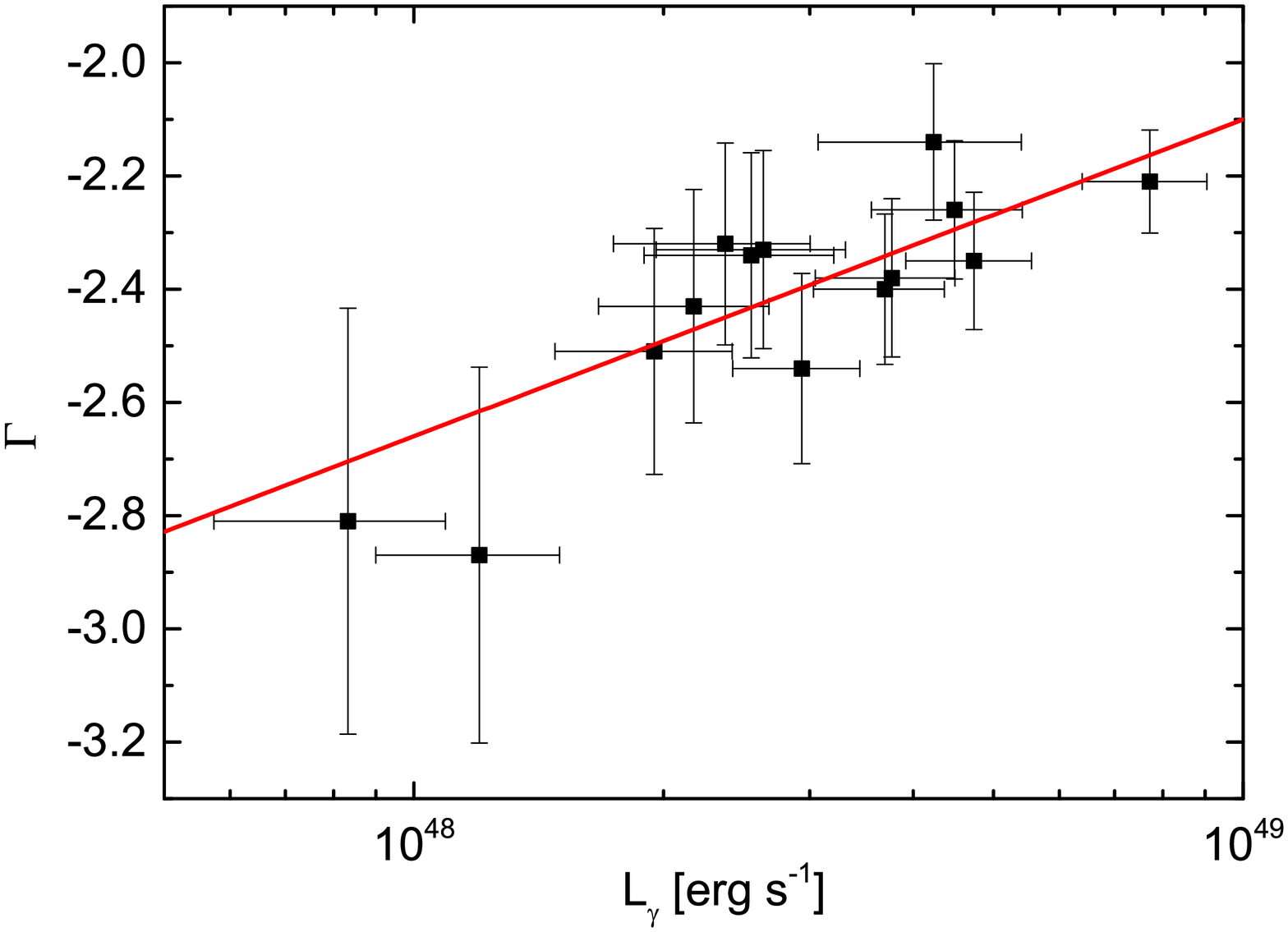}
\includegraphics[angle=0,scale=0.23]{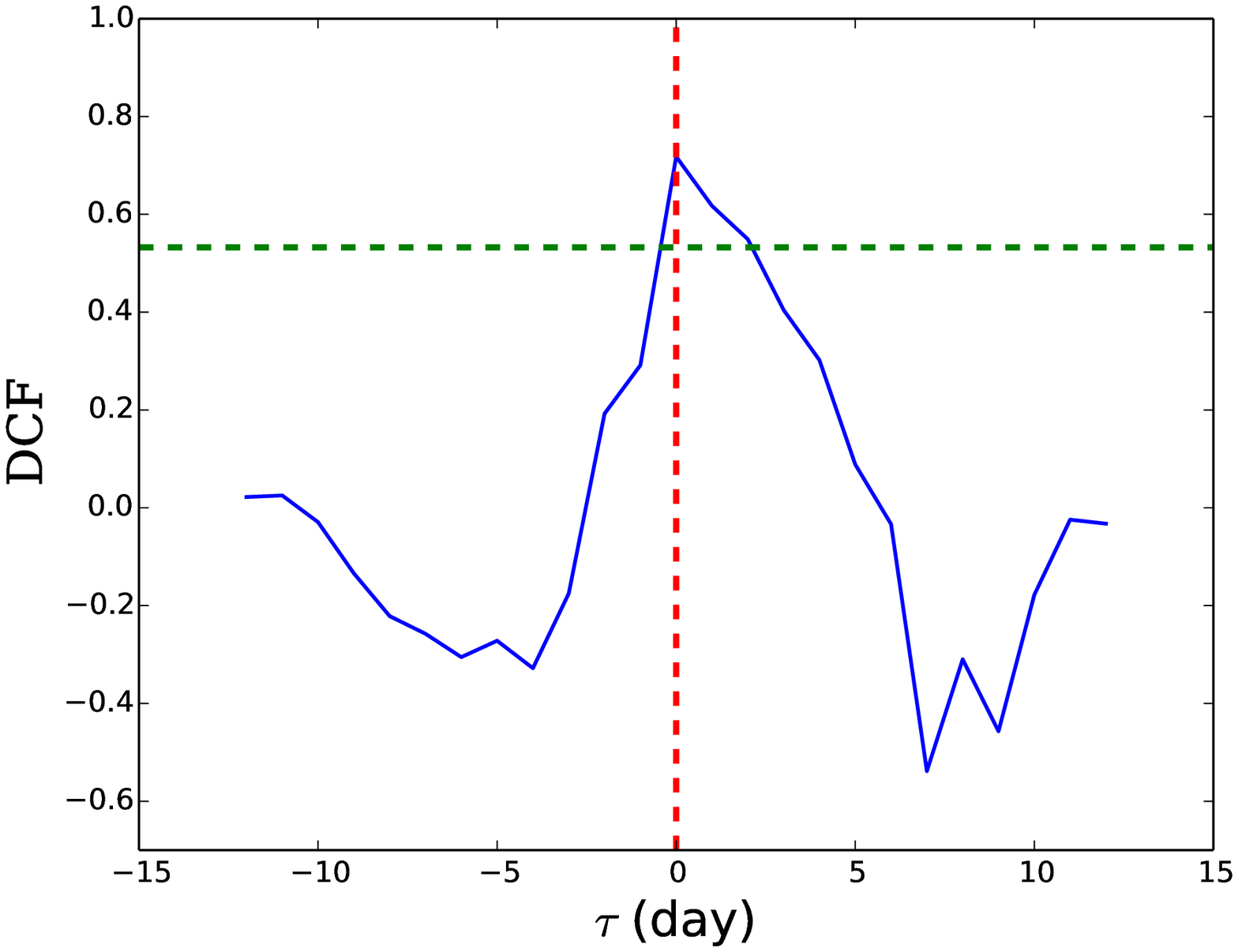}\\
\hfill\center{Fig.5---  continued}
\end{figure*}

\begin{figure*}
\includegraphics[angle=0,scale=0.18]{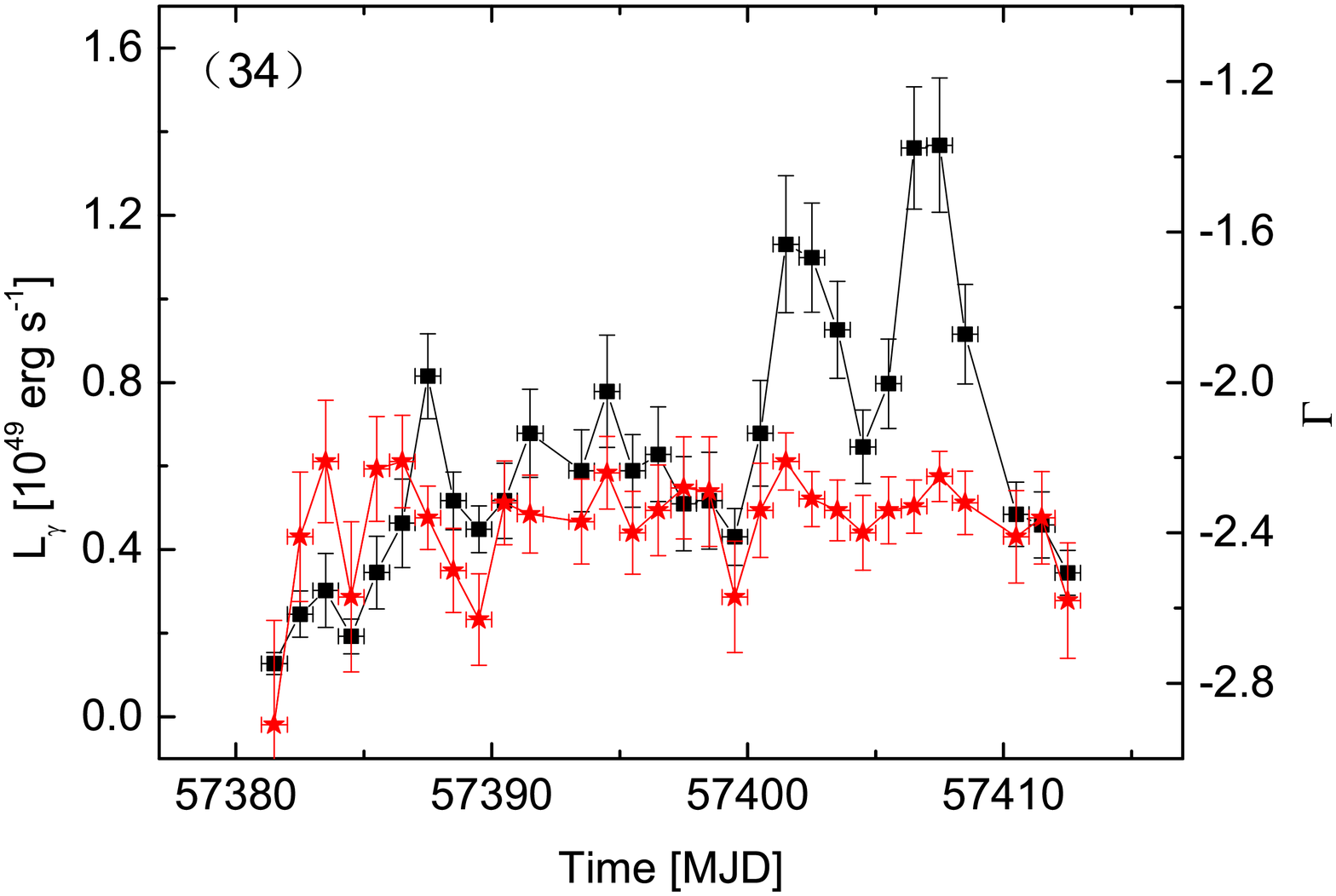}
\includegraphics[angle=0,scale=0.18]{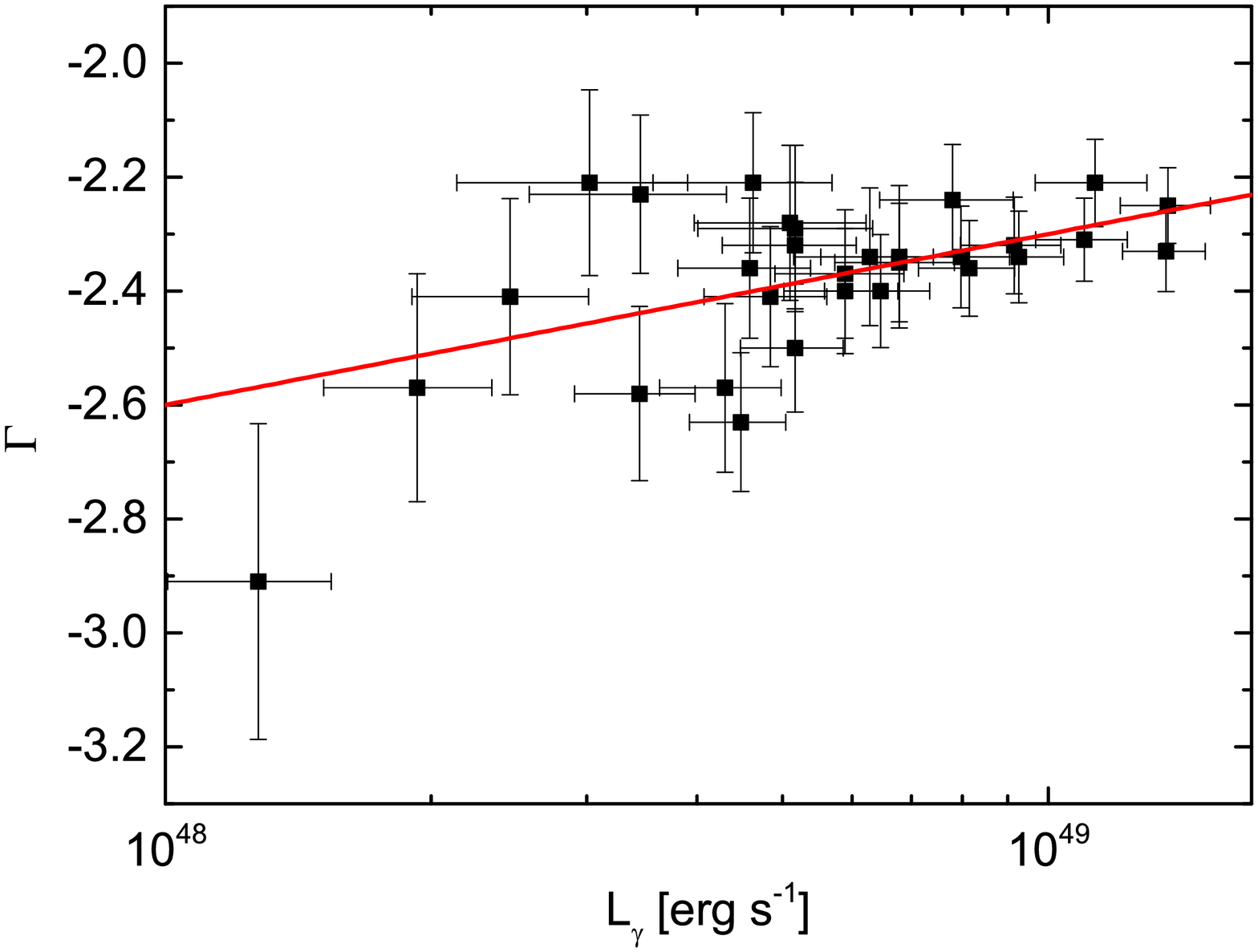}
\includegraphics[angle=0,scale=0.23]{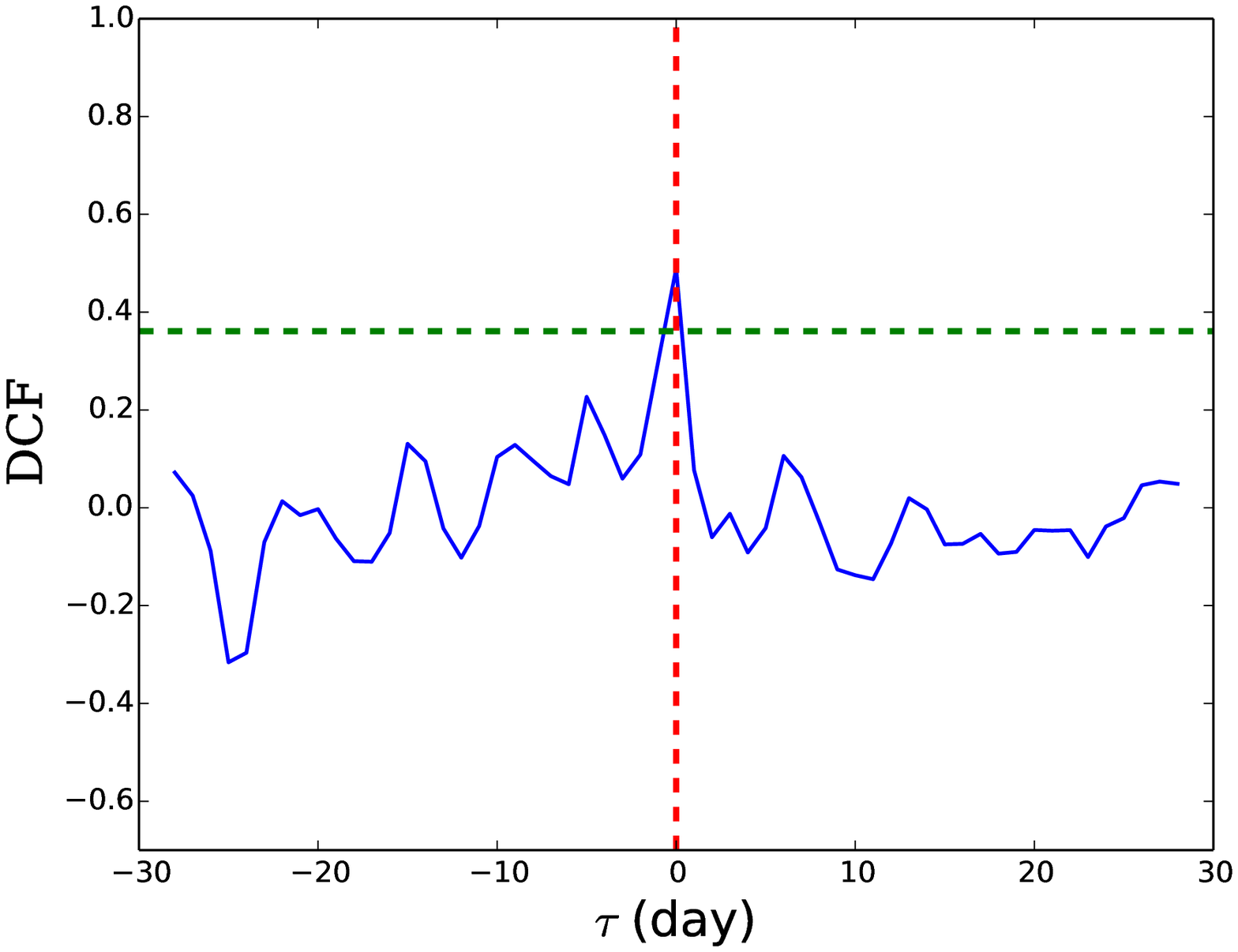}\\
\hfill\center{Fig.5---  continued}
\end{figure*}

\begin{figure*}
\includegraphics[angle=0,scale=0.43]{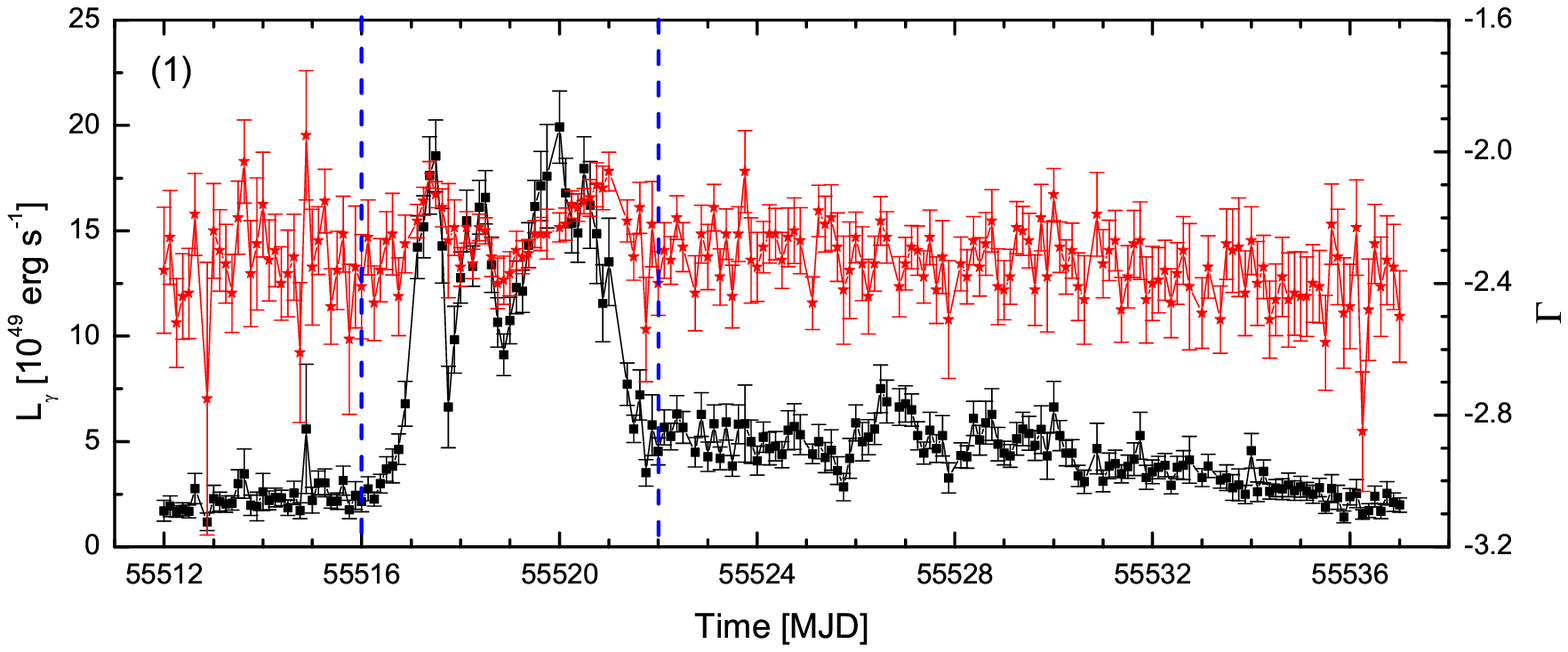}
\includegraphics[angle=0,scale=0.43]{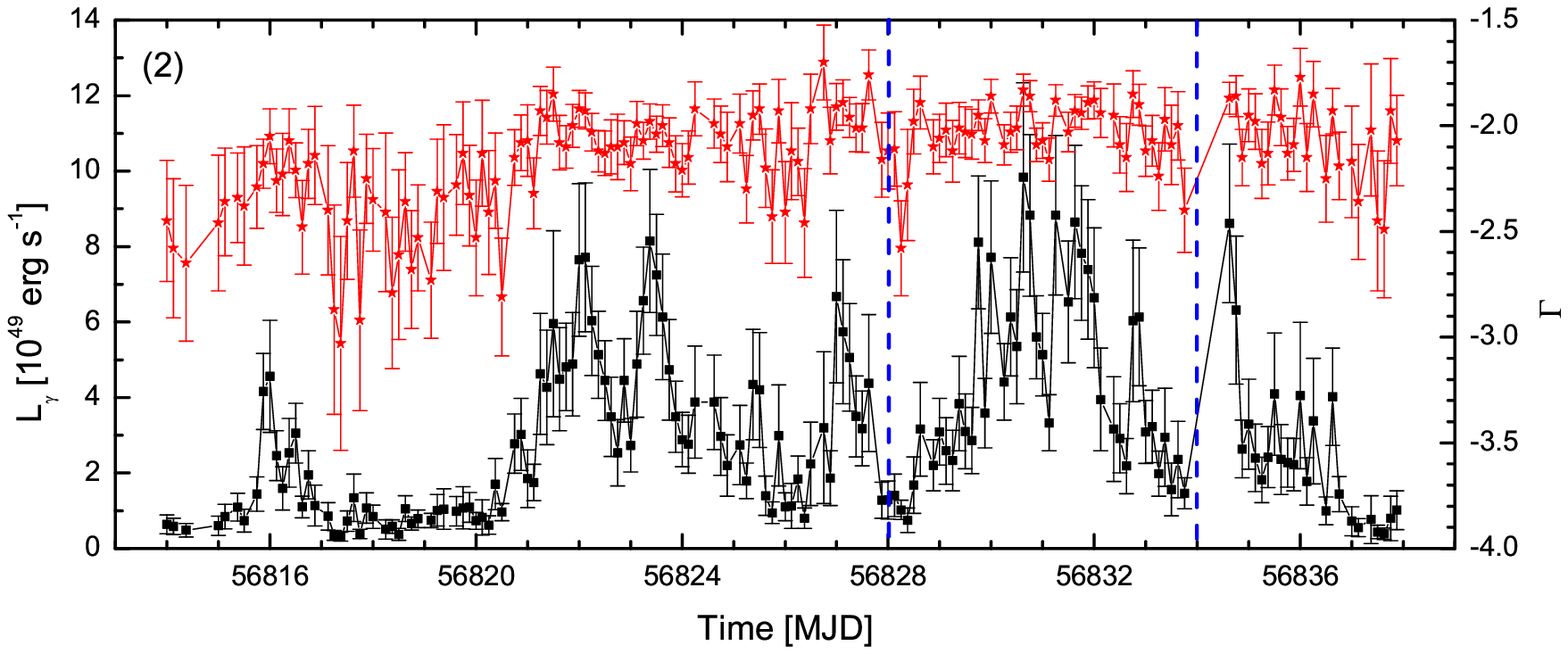}\\
\includegraphics[angle=0,scale=0.25]{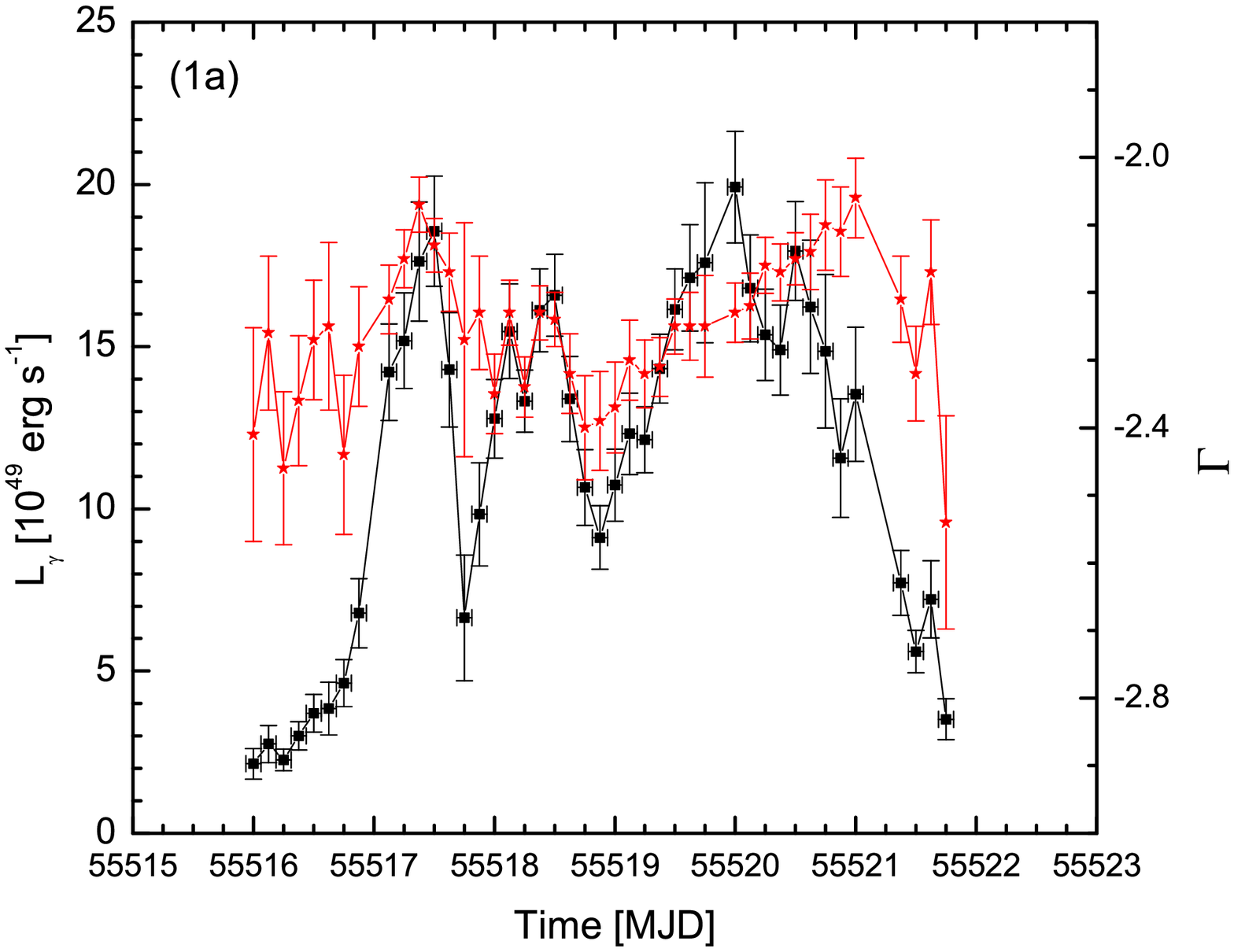}
\includegraphics[angle=0,scale=0.25]{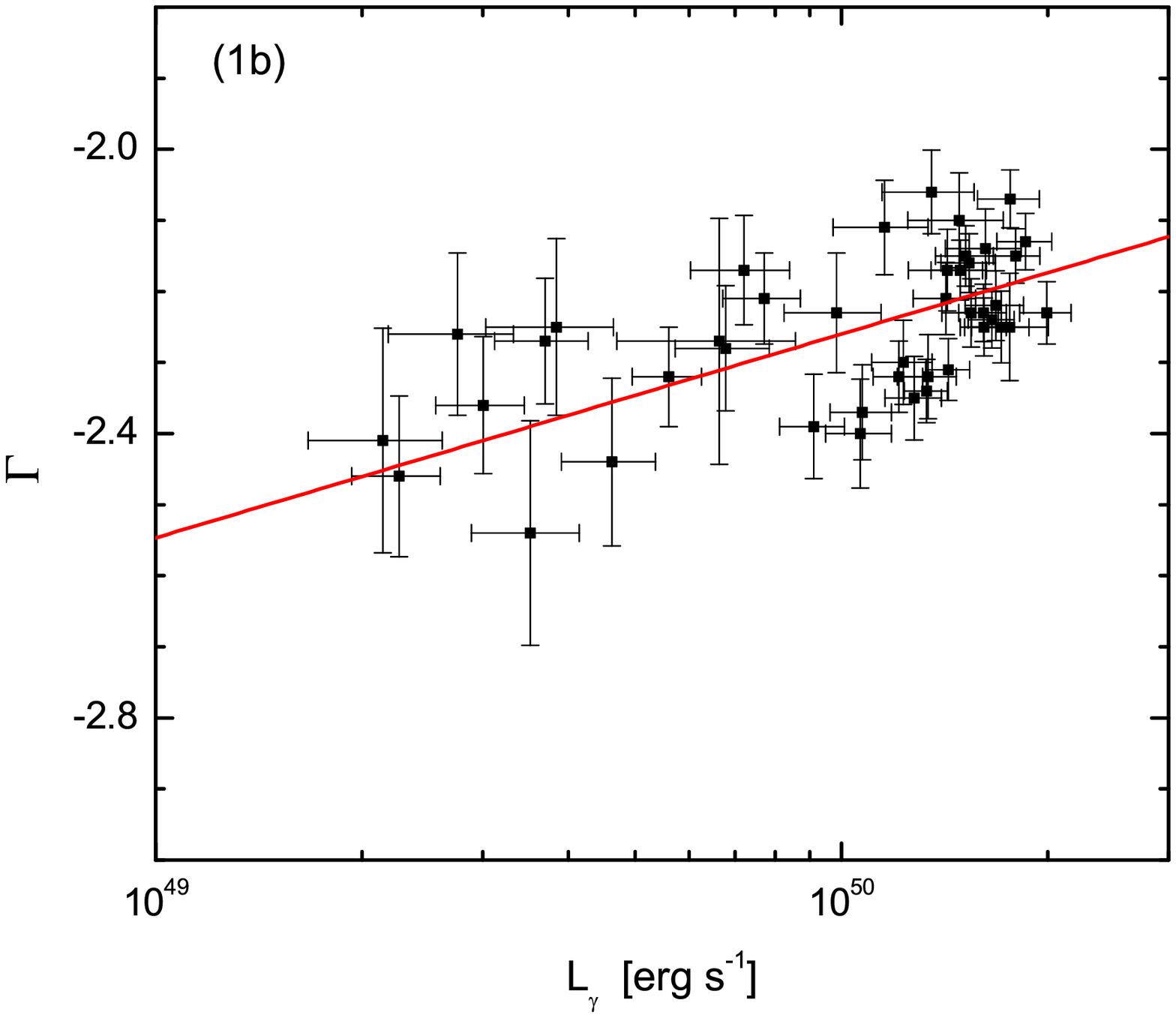}
\includegraphics[angle=0,scale=0.2]{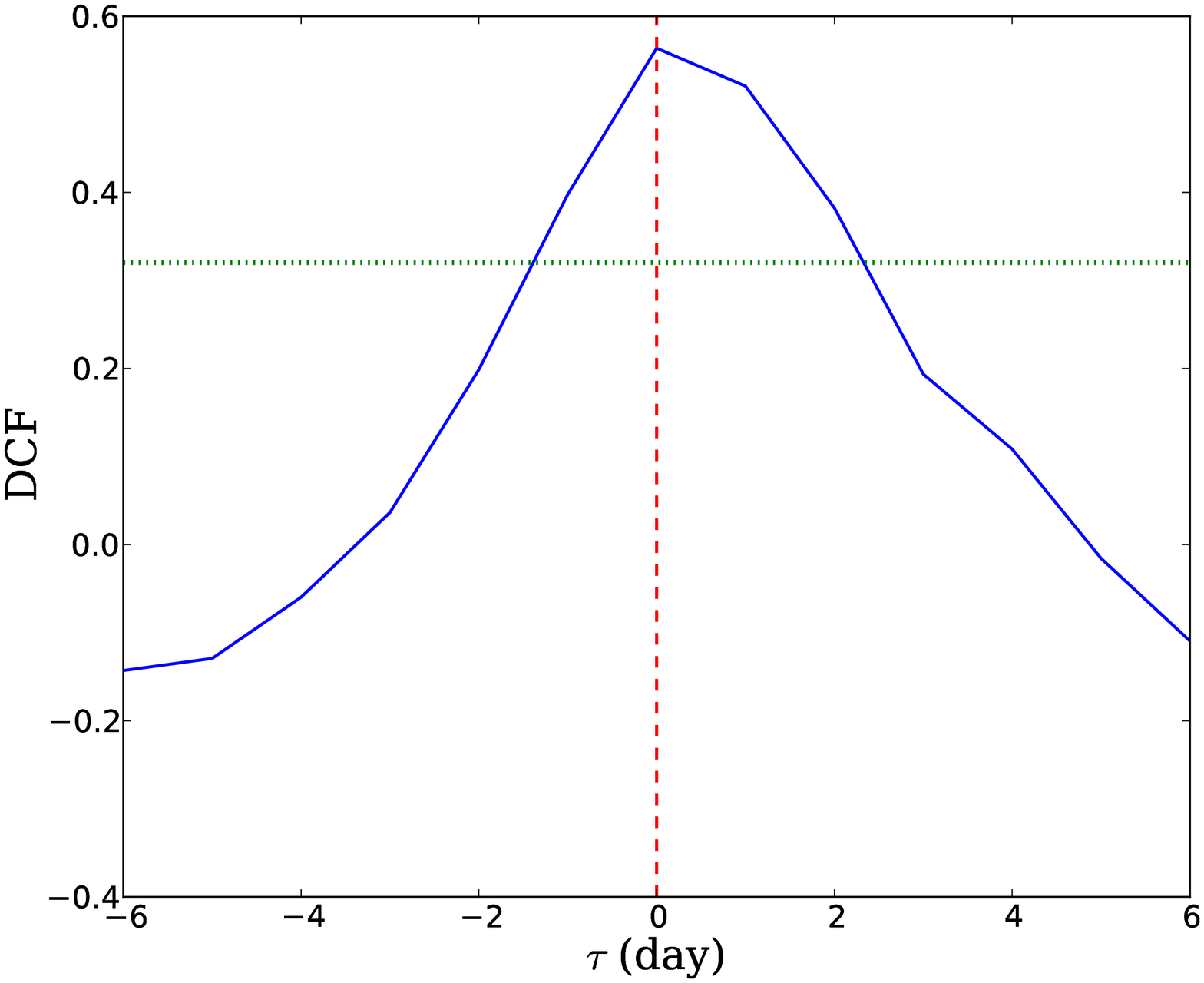}\\
\includegraphics[angle=0,scale=0.25]{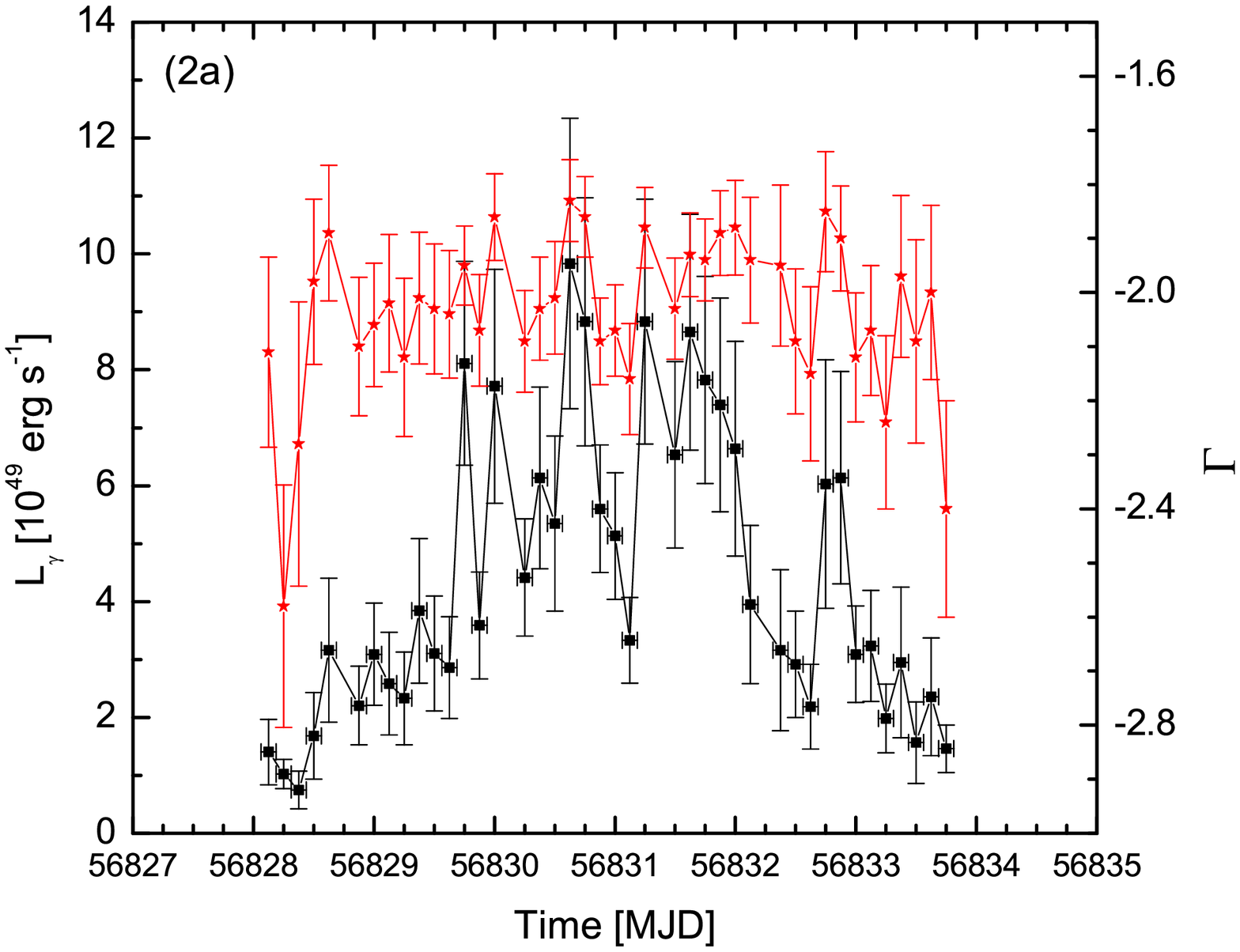}
\includegraphics[angle=0,scale=0.25]{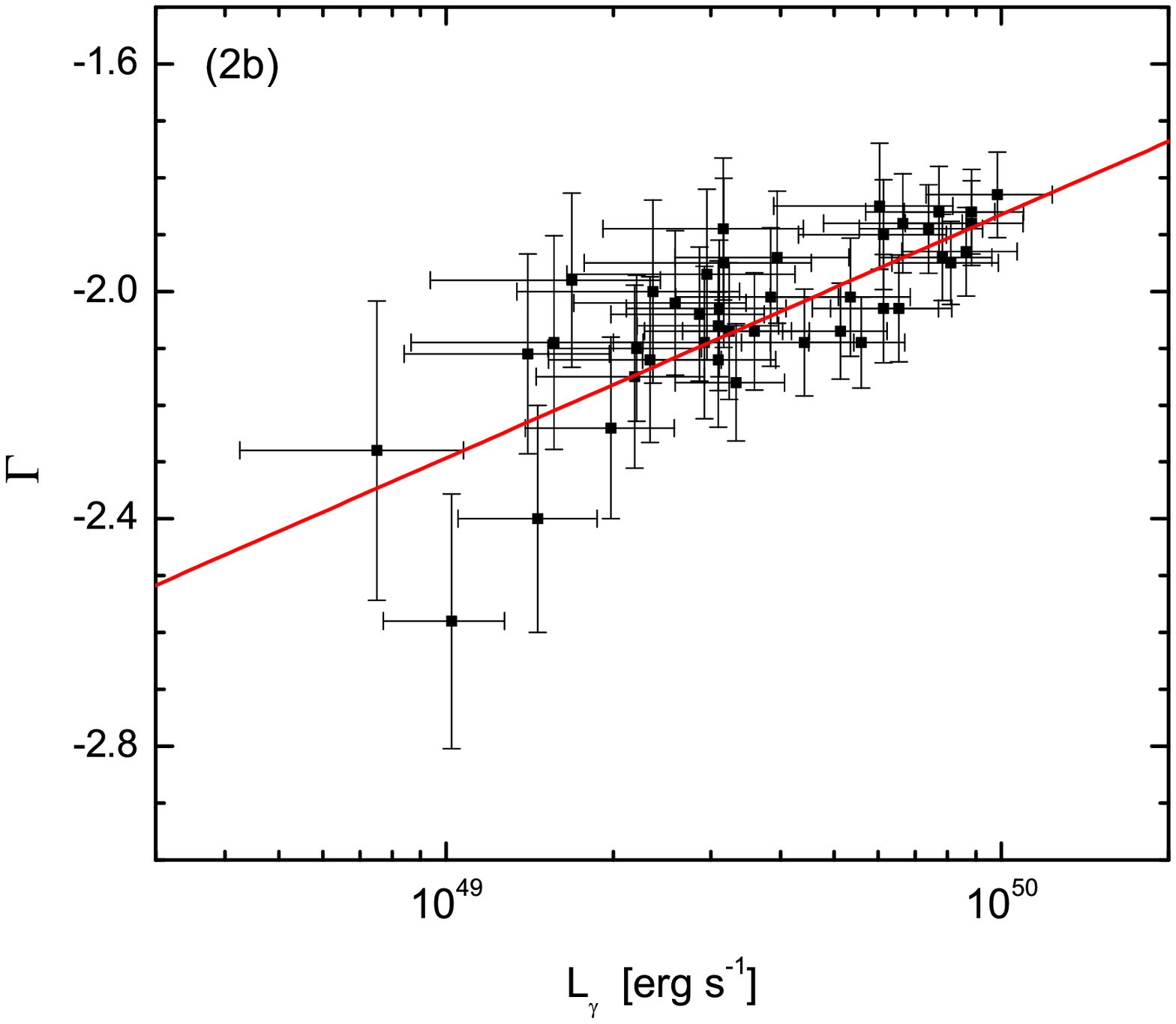}
\includegraphics[angle=0,scale=0.2]{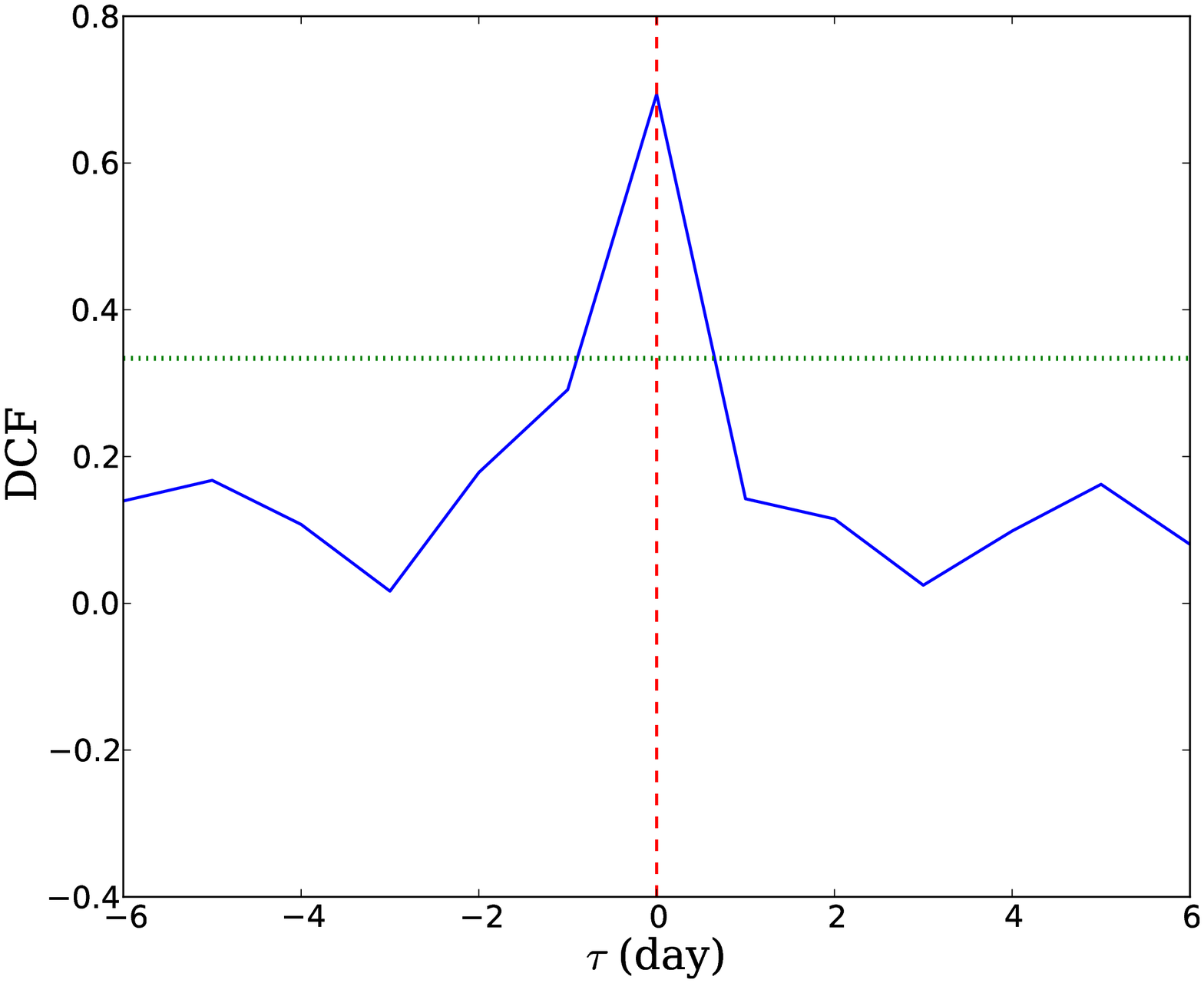}\\
\caption{\emph{Top Panels}---Temporal evolution of luminosity ($L_{\gamma}$, \emph{black squares}) and photon spectral index ($\Gamma$, \emph{red stars}) with 3-hour binning. The further analysis for the data between \emph{blue vertical dashed lines} is presented in below panels; \emph{left panels}---temporal evolution of $L_{\gamma}$ (\emph{black squares}) and $\Gamma$ (\emph{red stars}) with 3-hour binning. \emph{middle panels}---$\Gamma$ vs. $L_{\gamma}$ for two flares in the \emph{left panels}. The \emph{red lines} are the linear fitting lines by considering the errors of both $\Gamma$ and $L_{\gamma}$. \emph{right panels}---DCF results between $\Gamma$ and $L_{\gamma}$, where $\Gamma$ and $L_{\gamma}$ have been normalized with the Min-Max Normalization method before the DCF analysis. The \emph{green horizontal lines} are the 95\% confidence level lines.}\label{flare}
\end{figure*}

\begin{figure*}
\includegraphics[angle=0,scale=0.4]{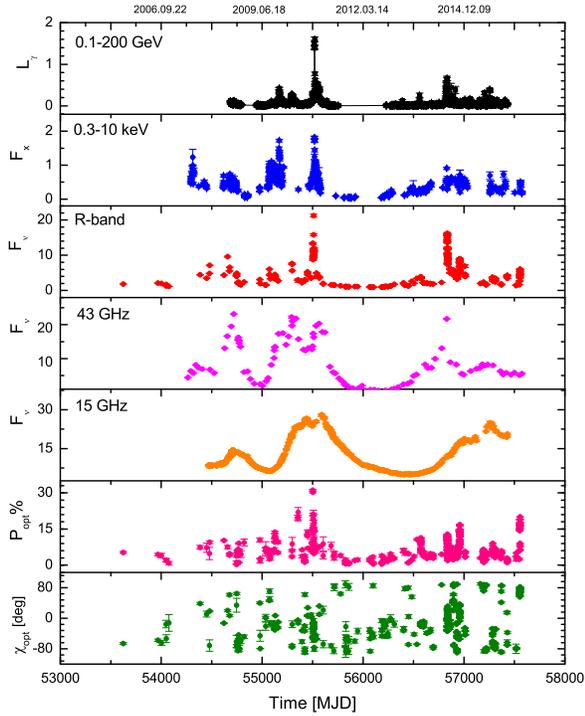}
\caption{Lightcurves of 3C 454.3 in multi-wavelength. The LAT lightcurve is the same as in Figure \ref{LAT}, in units of $10^{50}$ erg s$^{-1}$. The X-ray data are obtained with observations of the \emph{Swift} satellite (in units of $10^{-10}$ erg cm$^{-2}$ s$^{-1}$). The optical polarization and R-band data are taken from the 1.83 m Perkins telescope of Lowell Observatory. The 43 GHz data are from the VLBA observation. The 15 GHz data are from the OVRO 40 m radio telescope.}\label{lightcurve}
\end{figure*}

\begin{figure*}
\includegraphics[angle=0,scale=0.14]{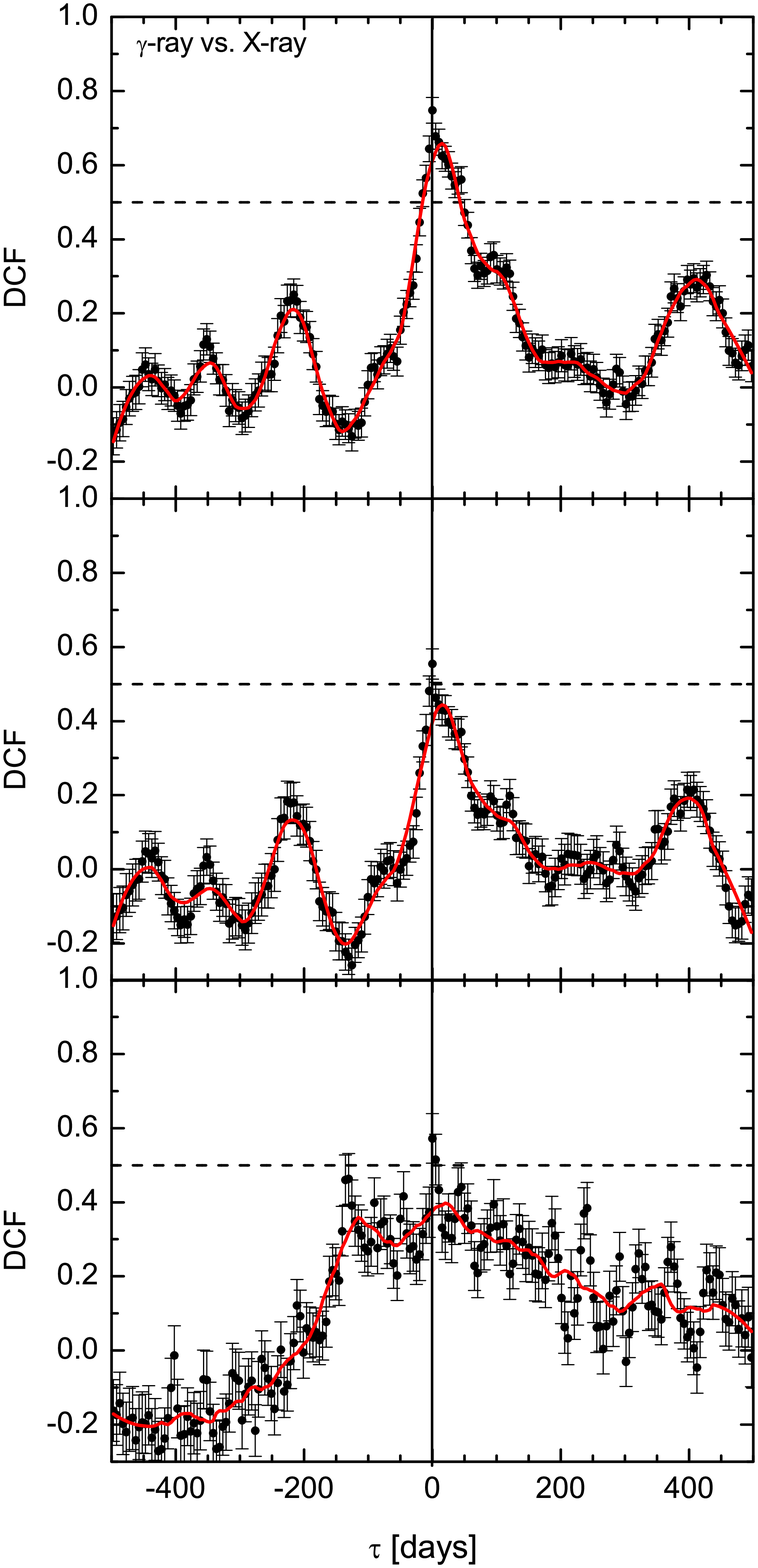}
\includegraphics[angle=0,scale=0.14]{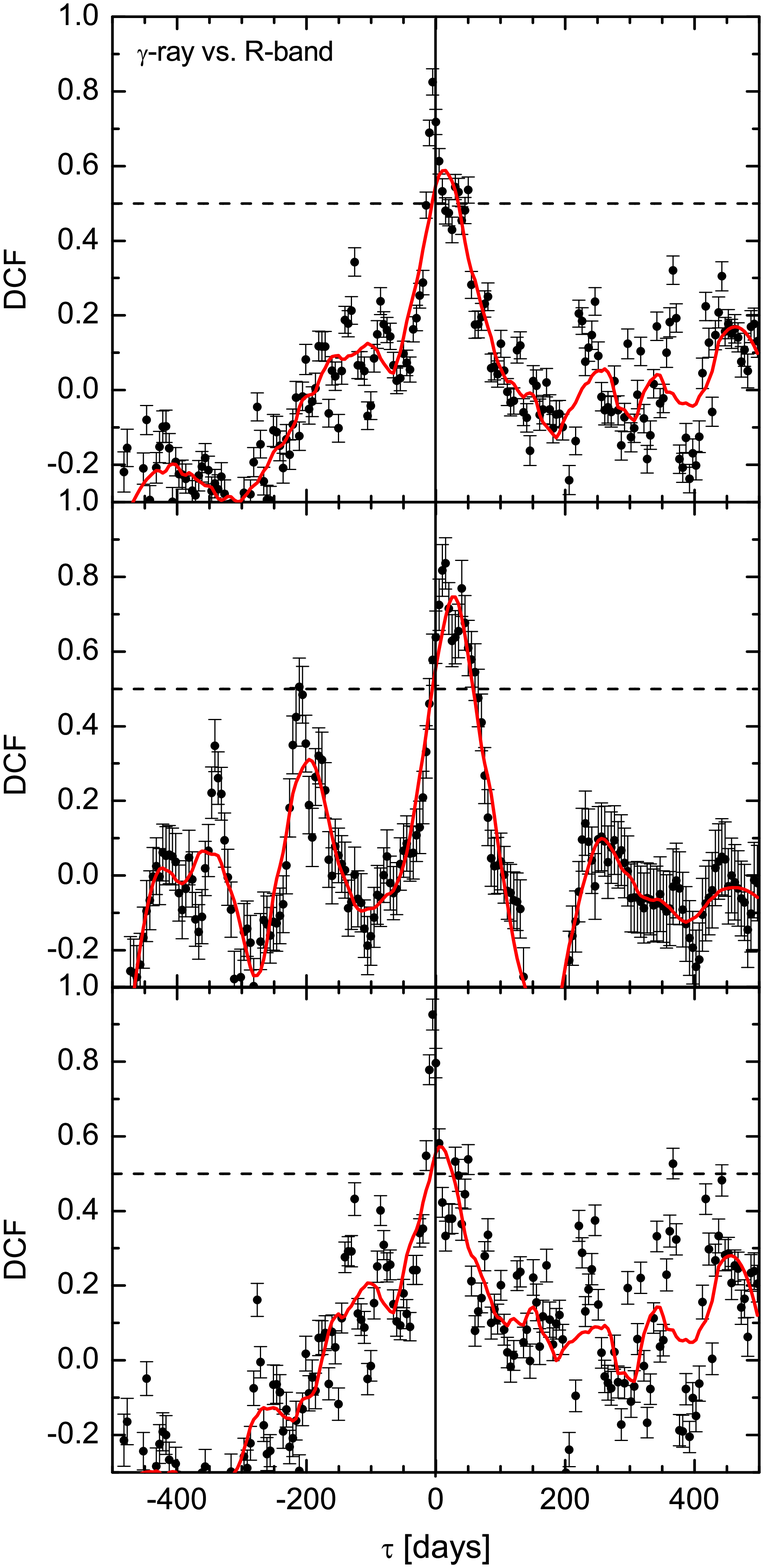}
\includegraphics[angle=0,scale=0.14]{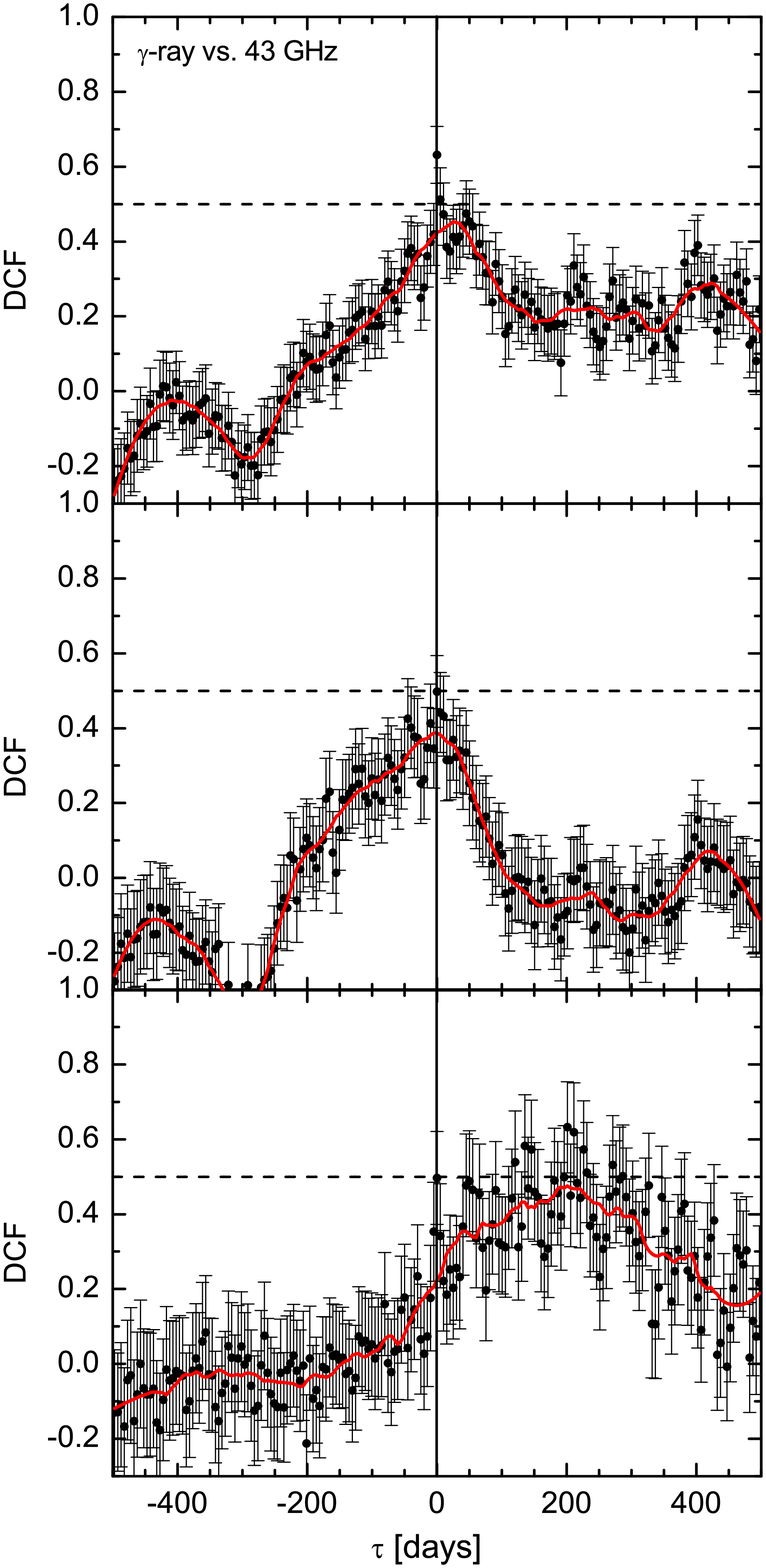}
\includegraphics[angle=0,scale=0.14]{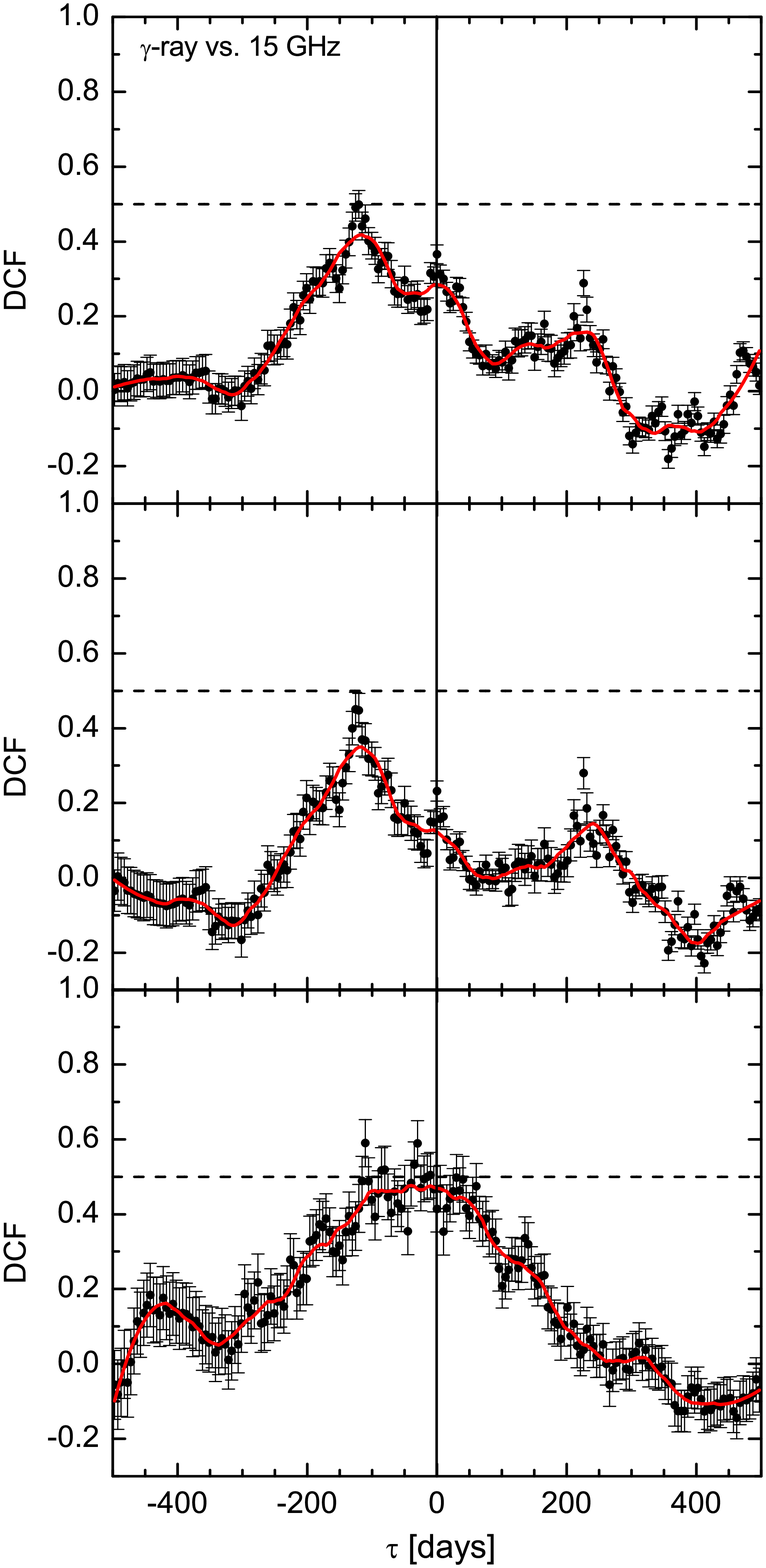}
\includegraphics[angle=0,scale=0.14]{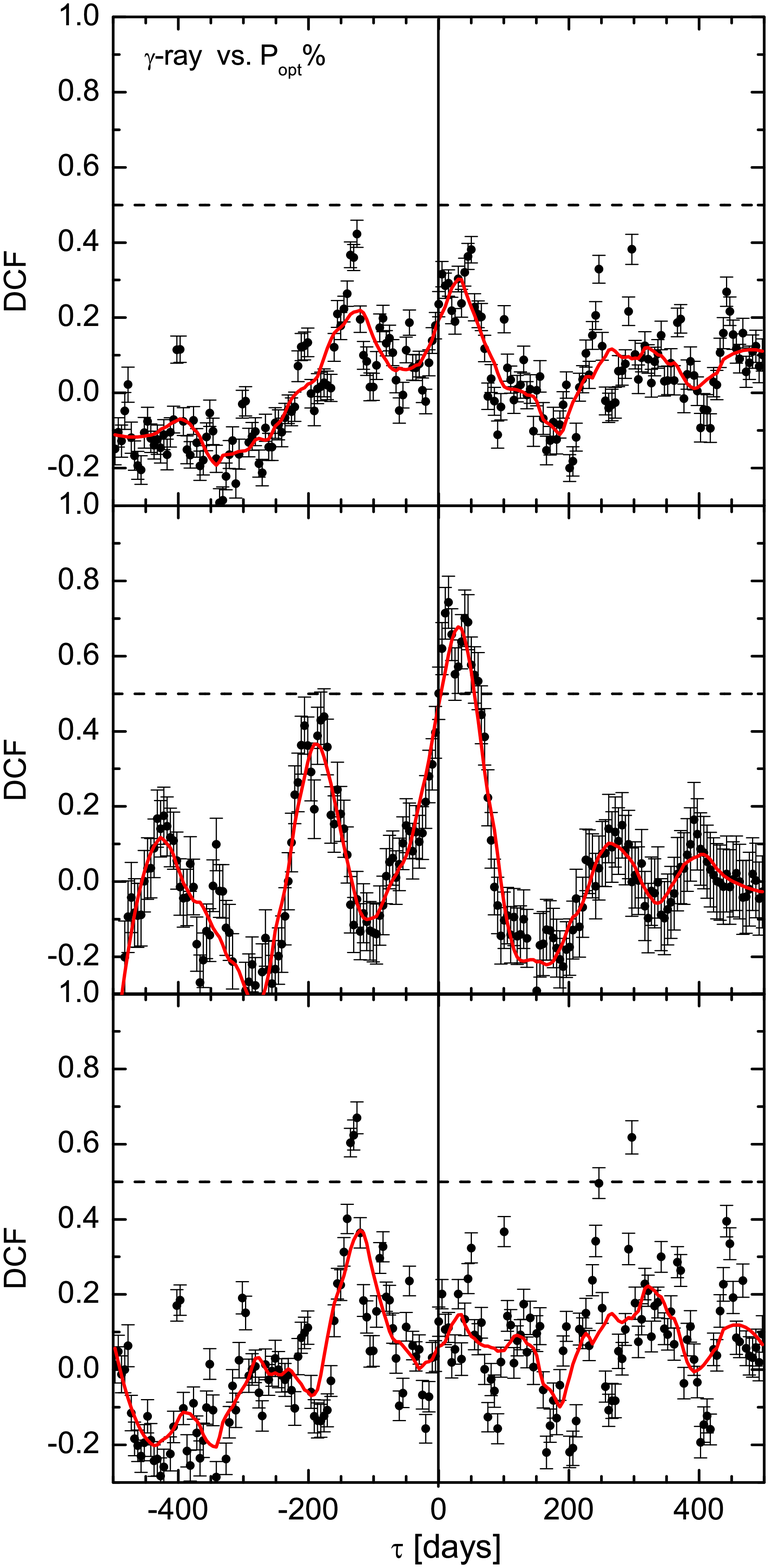}
\caption{The DCF results of variability between $\gamma$-ray and other bands. The global lightcurves are divided into two segments at MJD 56000. The DCF results of variability for the global lightcurves, the fist segment, and second segment are given in top panels, middle panels, and bottom panels, respectively. The vertical solid lines and horizontal dashed lines indicate the lag of zero and the correlation coefficient of 0.5. The red lines are the smoothed lines for the DCF curves.}\label{DCF}
\end{figure*}

\begin{figure*}
\includegraphics[angle=0,scale=0.2]{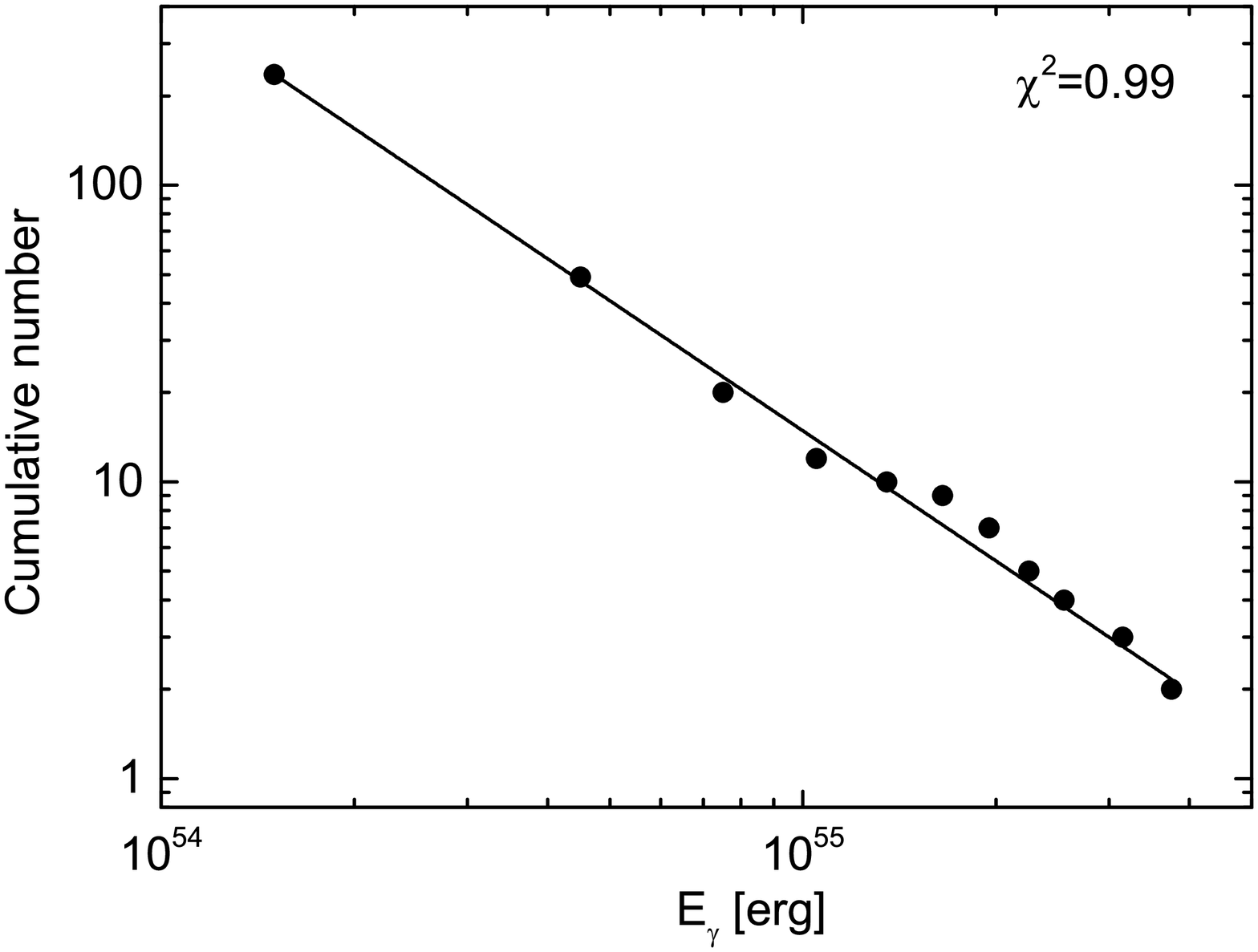}
\includegraphics[angle=0,scale=0.2]{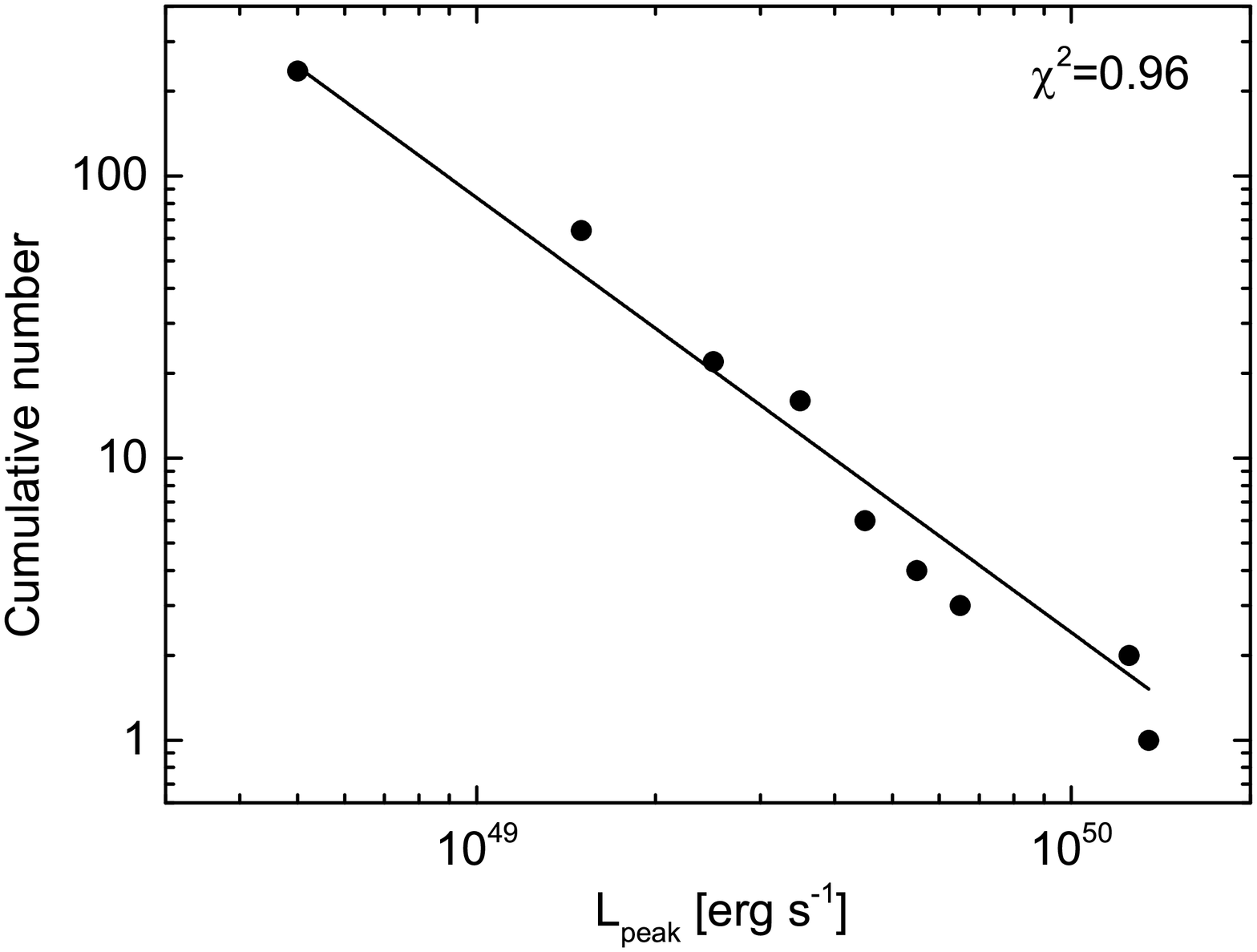}
\includegraphics[angle=0,scale=0.2]{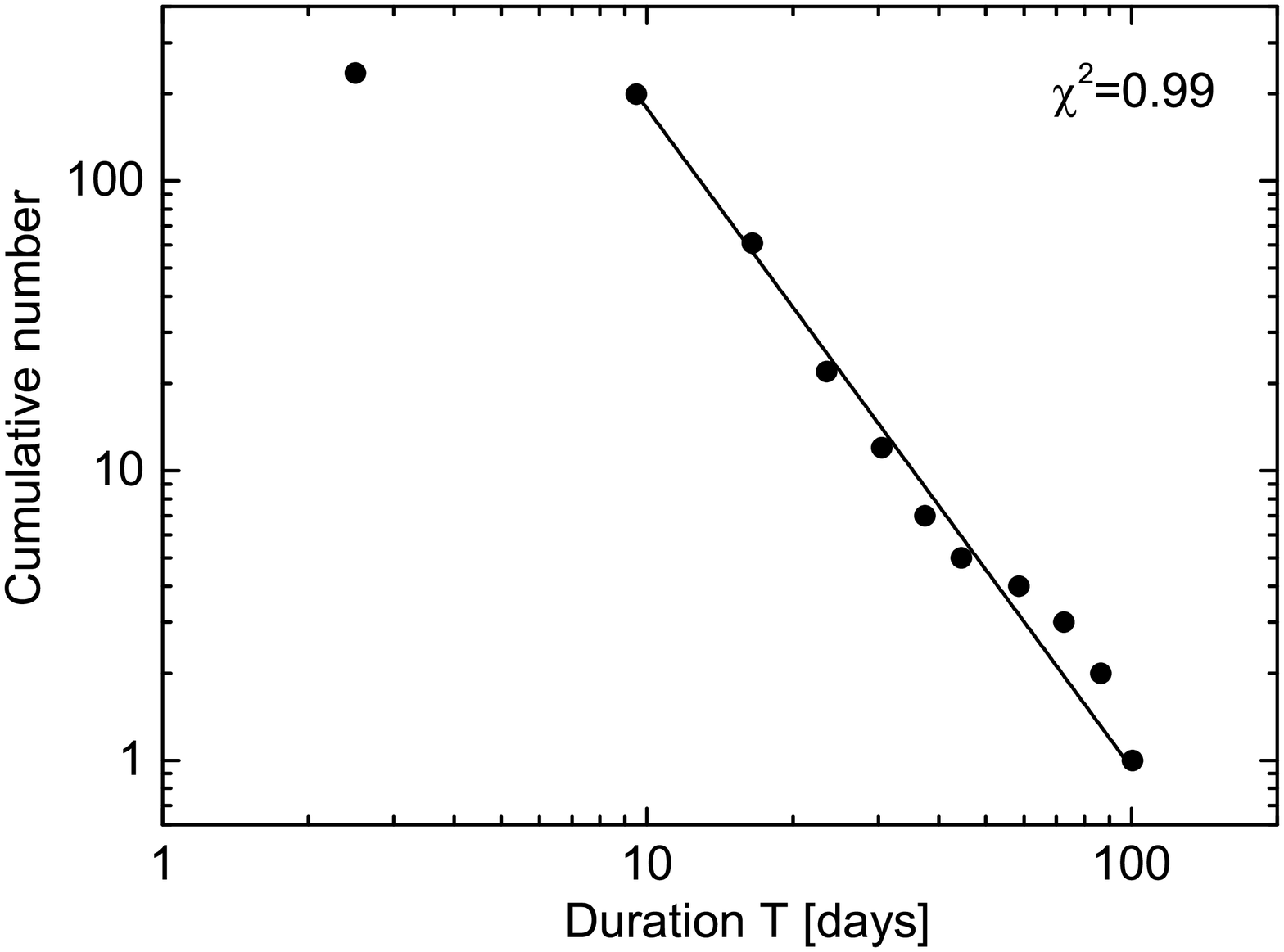}\\
\caption{Cumulative distributions of energy ($E_{\gamma}$), peak luminosity ($L_{\rm peak}$) and duration time ($T$) for the $\gamma$-ray flares. They are fitted with a power-law function, and the derived slopes are $\alpha_{\rm E}=1.46\pm0.02$, $\alpha_{\rm L}=1.54\pm0.08$, and $\alpha_{\rm T}=2.28\pm0.05$, respectively.}\label{SOC}
\end{figure*}
\clearpage

\begin{longtable}{lccc}
\caption{Fitting Results of the Global Lightcurve with the Multiple Gaussian Function}\\\hline
\centering
\label{tab:1}
Center& FWHM &$L_{\rm peak}$ & $E_{\gamma}$\\\hline
$\rm [MJD]$ & [d] & [$\times10^{48}$ erg s$^{-1}$] & [$\times10^{54}$ erg]\\\hline
\endhead
54717.49& 1.13& 3.00& 0.31\\
54718.50& 7.23& 6.83& 4.43\\
54719.47& 1.16& 4.19& 0.45\\
54721.65& 1.20& 5.09& 0.56\\
54723.51& 1.64& 8.36& 1.26\\
54726.16& 3.11& 4.74& 1.36\\
54729.58& 3.63& 8.28& 2.76\\
54734.59& 3.18& 3.24& 0.95\\
54738.17& 1.81& 5.94& 0.99\\
54742.46& 4.86& 5.14& 2.3\\
54748.54& 3.92& 3.41& 1.23\\
54753.46& 2.42& 1.43& 0.32\\
54756.68& 2.67& 0.47& 0.11\\
55066.56&	1.79&	2.64&	0.43\\
55069.64&	1.97&	3.33&	0.60\\
55072.63&	3.42&	5.17&	1.62\\
55075.83&	1.45&	2.57&	0.34\\
55078.47&	3.20&	3.88&	1.14\\
55081.79&	1.58&	2.46&	0.35\\
55084.92&	1.86&	2.66&	0.46\\
55089.15&	2.76&	4.88&	1.24\\
55091.63&	8.63&	3.33&	2.64\\
55091.64&	2.04&	4.84&	0.91\\
55096.24&	2.87&	1.75&	0.46\\
55102.26&	4.88&	5.52&	2.48\\
55104.70&	2.07&	3.38&	0.64\\
55107.26&	1.58&	5.13&	0.74\\
55110.10&	2.52&	5.64&	1.31\\
55114.32&	3.96&	3.03&	1.10\\
55117.21&	2.77&	2.20&	0.56\\
55121.61&	1.57&	3.09&	0.45\\
55123.47&	2.48&	5.21&	1.19\\
55126.50&	2.26&	7.30&	1.52\\
55129.83&	1.52&	7.48&	1.05\\
55132.55&	2.31&	4.34&	0.92\\
55137.20&	4.68&	1.97&	0.85\\
55145.16&	1.86&	4.10&	0.70\\
55149.31&	2.84&	4.41&	1.15\\
55155.09&	2.67&	12.96&	3.18\\
55162.30&	5.01&	12.56&	5.79\\
55167.75&	2.60&	31.80&	7.59\\
55170.55&	1.76&	30.28&	4.90\\
55171.18&	17.89&	8.40&	13.83\\
55172.78&	1.79&	21.99&	3.62\\
55174.17&	1.42&	15.30&	2.00\\
55175.41&	1.12&	9.34&	0.96\\
55178.55&	1.42&	12.48&	1.63\\
55180.48&	1.42&	14.43&	1.89\\
55182.90&	2.13&	14.27&	2.80\\
55185.59&	1.89&	12.87&	2.24\\
55188.39&	2.13&	9.84&	1.93\\
55192.34&	3.78&	6.89&	2.40\\
55195.64&	2.90&	12.52&	3.34\\
55198.84&	6.13&	4.78&	2.69\\
55205.67&	1.64&	4.60&	0.69\\
55208.28&	1.77&	3.80&	0.62\\
55210.07&	9.99&	2.88&	2.65\\
55212.51&	2.67&	5.55&	1.37\\
55214.84&	1.06&	6.99&	0.68\\
55216.77&	1.71&	9.04&   1.43\\
55220.42&	4.58&	8.89&	3.74\\
55225.84&	2.18&	3.66&	0.71\\
55282.52& 3.27& 9.21& 2.77\\
55285.19& 1.88& 3.58& 0.62\\
55288.53& 2.72& 10.81& 2.7\\
55293.94& 7.24& 17.12& 11.41\\
55298.99& 24.84& 8.87& 20.07\\
55299.39& 1.04& 6.5& 0.62\\
55301.55& 3.21& 13.47& 3.98\\
55306.15& 2.29& 8.15& 1.72\\
55313.48& 1.95& 5.38& 0.96\\
55320.58& 3.11& 8.66& 2.48\\
55323.77& 3.66& 11.43& 3.85\\
55327.4& 2.35& 10.44& 2.26\\
55329.51& 1.55& 7.28& 1.03\\
55332.17& 4.78& 5.85& 2.57\\
55335.01& 1.31& 3.09& 0.37\\
55339.24& 5.66& 4.29& 2.15\\
55502.67& 1.71& 16.48& 2.59\\
55505.42& 4.36& 19.99& 8.01\\
55510.47& 2.24& 8.01& 1.65\\
55517.29& 2.26& 121.44& 25.21\\
55520.48& 21.8& 19.33& 38.69\\
55520.48& 2.96& 135.86& 36.97\\
55523.94& 1.78& 18.17& 2.97\\
55526.5& 2.69& 34.52& 8.53\\
55528.41& 4.1& 9.9& 3.73\\
55529.88& 3.49& 22.79& 7.31\\
55532.98& 2.92& 12.84& 3.45\\
55534.56& 4.14& 7.61& 2.9\\
55537.43& 2.95& 4.91& 1.33\\
55541.38& 4.13& 16.42& 6.24\\
55544.67& 1.78& 17.99& 2.94\\
55549.41& 4.13& 24.79& 9.41\\
55549.51& 14.72& 22.54& 30.51\\
55552.7& 1.78& 15.05& 2.46\\
55560.4& 1.77& 9.79& 1.6\\
55563.12& 1.56& 10.78& 1.55\\
55566.96& 4.48& 36.73& 15.14\\
55575.18& 5.29& 11.32& 5.51\\
55583.71& 5.51& 12.17& 6.15\\
56551.59&	1.96&		3.01&	0.12\\
56554.48&	2.22&		2.74&	0.10\\
56556.42&	1.05&		1.45&	0.11\\
56559.29&	1.85&		37.30&	1.64\\	
56560.84&	5.01&		40.64&	0.66\\
56565.77&	1.92&		4.82&	0.20\\
56568.30&	2.01&		5.71&	0.23\\
56571.37&	2.58&		3.63&	0.11\\
56798.48& 2.95& 5.45& 1.47\\
56801.55& 2.6& 3.95& 0.94\\
56804.45& 2.54& 4.03& 0.94\\
56808.08& 2.84& 5.56& 1.45\\
56810.65& 1.25& 7.82& 0.9\\
56812.89& 1.99& 7.89& 1.44\\
56816.42& 1.89& 20.99& 3.64\\
56823.01& 4.4& 53.22& 21.52\\
56827.51& 1.1& 31.67& 3.2\\
56831.13& 3.37& 66.52& 20.59\\
56834.57& 1.27& 46.65& 5.43\\
56836.13& 1.73& 21.85& 3.48\\
56839.09& 2.2& 12.76& 2.58\\
56842.28& 3.03& 11.27& 3.14\\
56846.32& 2.91& 7.95& 2.13\\
56848.45& 1.07& 4.85& 0.48\\
56850.57& 2.19& 6.76& 1.15\\
56861.74& 1.6& 1.78& 0.25\\
56865.21& 1.48& 15& 2.04\\
56868.74& 1.91& 12.26& 2.15\\
56872.48& 1.69& 38.77& 6.04\\
56874.15& 1.55& 4.6& 0.65\\
56876.16& 0.96& 3.84& 0.34\\
56878.48& 2.12& 2.2& 0.43\\
56883.53& 1.61& 13.42& 1.99\\
56886.23& 1.64& 19.16& 2.89\\
56909.29& 2.83& 8.1& 2.08\\
56912.46& 1.43& 6.94& 0.91\\
56914.53& 1.44& 12.72& 1.69\\
56916.5& 1.04& 36.43& 3.48\\
56919.75& 2.26& 9.2& 1.92\\
56924.88& 1.65& 5.94& 0.9\\
56928.01& 2.49& 6.18& 1.42\\
56932.09& 2.84& 3.88& 1.01\\
56961.51& 1.68& 8.92& 1.38\\
56966.06& 3.6& 8.04& 2.67\\
56974.45& 1.67& 7.46& 1.15\\
56986.58& 8.04& 3.45& 2.55\\
56988.2& 2.7& 8.28& 2.05\\
57007.17& 6.93& 8.7& 5.54\\
57008.7& 0.8& 6.52& 0.48\\
57015.41& 1.6& 3.56& 0.52\\
57019.4& 4.24& 4.25& 1.66\\
57022.91& 1.75& 3.69& 0.59\\
57028.27& 4.54& 3.61& 1.49\\
57056.25& 2.52& 5.11& 1.18\\
57058.81& 1.35& 14.66& 1.82\\
57060.13& 1.57& 8.38& 1.21\\
57062.09& 2.58& 13.38& 3.17\\
57065.53& 1.26& 2.27& 0.26\\
57068.86& 2.31& 6.85& 1.46\\
57070.5& 7.08& 3.76& 2.45\\
57076.28& 3.49& 3.04& 0.98\\
57080.49& 2.26& 4.18& 0.87\\
57082.97& 1.84& 7.61& 1.29\\
57087.06& 4.5& 3.37& 1.4\\
57089.49& 1.42& 3.54& 0.46\\
57091.44& 1.78& 5.4& 0.88\\
57093.54& 1.39& 4.43& 0.57\\
57095.44& 2.26& 3.45& 0.72\\
57102.7& 5.99& 13.94& 7.69\\
57113.01& 4.39& 3.31& 1.34\\
57119.83& 4.69& 1.42& 0.59\\
57144.44& 2.02& 1.88& 0.34\\
57147.86& 2.1& 3.47& 0.67\\
57153.39& 2.71& 14.36& 3.58\\
57171.39& 1.68& 2.51& 0.39\\
57174.55& 1.51& 9.52& 1.32\\
57176.12& 0.9& 10.11& 0.84\\
57178.48& 2.25& 7.33& 1.52\\
57181.44& 3.3& 2.41& 0.73\\
57185.57& 0.84& 8.1& 0.63\\
57188.45& 1.66& 3.34& 0.51\\
57190.48& 0.73& 2.33& 0.16\\
57192.69& 2.51& 5.17& 1.19\\
57198.5& 3.43& 11.38& 3.59\\
57203.23& 2.32& 6.4& 1.36\\
57204.42& 1.21& 30.52& 3.39\\
57206.6& 1.57& 18.2& 2.62\\
57209.35& 2.46& 2.45& 0.54\\
57246& 5.23& 4.24& 2\\
57249.8& 2.16& 5.44& 1.08\\
57255.3& 4.55& 38.58& 16.15\\
57264.33& 1.82& 4.37& 0.73\\
57268.46& 2.86& 6.47& 1.7\\
57271.51& 2.12& 6.69& 1.3\\
57273.91& 1.64& 4.88& 0.74\\
57276.58& 3.15& 5.01& 1.45\\
57279.09& 1.84& 15.39& 2.6\\
57281.61& 1.06& 5.29& 0.52\\
57285.28& 4.29& 3& 1.18\\
57288.68& 2.07& 4.8& 0.92\\
57290.85& 1.75& 4.8& 0.77\\
57294.55& 3.5& 6.4& 2.06\\
57298.01& 1.61& 4.6& 0.68\\
57299.57& 1.94& 4.28& 0.76\\
57302.47& 2.18& 12.29& 2.47\\
57305.95& 2.38& 3.17& 0.69\\
57311.13& 4.35& 4.72& 1.89\\
57315.66& 2.89& 6.79& 1.81\\
57319.09& 3.11& 1.97& 0.56\\
57324.81& 5.79& 9.07& 4.83\\
57326.61& 1.39& 7.29& 0.93\\
57332.14& 4.96& 6.72& 3.06\\
57337.6& 2.07& 5.98& 1.14\\
57341.64& 3.03& 9.02& 2.51\\
57345.45& 2.15& 7.65& 1.51\\
57351.55& 11.96& 2.77& 3.01\\
57353.69& 1.88& 4.36& 0.75\\
57376.5& 2.66& 1.09& 0.26\\
57380.04& 2.06& 2.13& 0.4\\
57383.5& 1.9& 2.12& 0.37\\
57387.5& 2.68& 7.07& 1.74\\
57391.65& 3.63& 5.71& 1.9\\
57394.42& 1.09& 3.07& 0.31\\
57396.47& 4.09& 5.19& 1.95\\
57401.8& 3.23& 10.56& 3.13\\
57407.25& 4.37& 12.81& 5.15\\
57411.41& 1.76& 1.84& 0.3\\
57413.5& 0.97& 2.45& 0.22\\
57417& 6.27& 5.64& 3.25\\
57419.46& 0.81& 1.96& 0.15\\
57422.49& 2.71& 3.5& 0.87\\
57424.5& 0.77& 2.73& 0.19\\
57426.47& 2.65& 6.22& 1.52\\
57428.5& 1.22& 3.1& 0.35\\
57430.93& 4.71& 3.46& 1.35\\\hline
%\enddata

%\end{tabular}
\end{longtable}

\clearpage
\begin{table}
\begin{center}
\caption{Data of 34 Outburst Episodes}
\label{tab:2}
\setlength{\tabcolsep}{4pt}
\tiny
	\begin{tabular}{lcccccccccc}
	\hline\noalign{\smallskip}
Outbursts  & Date  & Duration &  $L_{\rm max}$ &  $L_{\rm min}$ & $\Gamma_{\rm max}$   & $\Gamma_{\rm min}$ & $L_{\rm max}/L_{\rm min}$ &$r$ & $p$ & $k$  \\
 & [MJD]  & [d]  &  [$\times10^{49}$ erg s$^{-1}$] &  [$\times10^{48}$ erg s$^{-1}$] &  &  &  &  & &  \\\hline
%\startdata
%
1&54714.5-54758.5&45&$1.22\pm0.19$&$1.50\pm0.34$&$-2.18\pm0.08$&$-2.69\pm0.17$&8.13&0.615&1.48e-5&$0.34\pm0.08$\\
2&55083.5-55097.5&15&$1.07\pm0.12$&$2.98\pm0.56$&$-2.32\pm0.08$&$-2.60\pm0.12$&3.59&0.404&0.152&$0.21\pm0.18$\\
3&55119.5-55142.5&24&$0.93\pm0.09$&$2.11\pm0.45$&$-2.29\pm0.09$&$-2.83\pm0.20$&4.41&0.663&5.68e-4&$0.52\pm0.16$\\
4&55151.5-55157.5&7&$1.48\pm0.23$&$3.41\pm0.65$&$-2.20\pm0.08$&$-2.43\pm0.10$&4.34&0.539&0.212&$0.29\pm0.21$\\
5&55157.5-55189.5&33&$4.20\pm0.30$&$6.25\pm0.85$&$-2.19\pm0.04$&$-2.56\pm0.06$&6.72&0.258&0.147&$0.10\pm0.05$\\
6&55189.5-55204.5&16&$1.76\pm0.18$&$4.63\pm0.72$&$-2.20\pm0.06$&$-2.47\pm0.09$&3.80&0.687&0.003&$0.32\pm0.12$\\
7&55274.5-55315.5&42&$2.77\pm0.17$&$2.69\pm0.42$&$-2.30\pm0.03$&$-2.70\pm0.17$&10.30&0.475&0.003&$0.14\pm0.05$\\
8&55315.5-55344.5&30&$1.53\pm0.12$&$3.63\pm0.60$&$-2.23\pm0.08$&$-2.62\pm0.15$&4.21&0.342&0.069&$0.14\pm0.09$\\
9&55492.5-55512.5&21&$3.13\pm0.28$&$6.57\pm0.67$&$-2.15\pm0.04$&$-2.58\pm0.10$&4.76&0.844&1.51e-6&$0.47\pm0.07$\\
10&55512.5-55525.5&14&$16.00\pm0.06$&$18.41\pm1.54$&$-2.15\pm0.02$&$-2.36\pm0.06$&8.69&0.706&6.52e-4&$0.14\pm0.03$\\
11&55525.5-55536.5&12&$6.28\pm0.34$&$20.35\pm1.41$&$-2.29\pm0.03$&$-2.41\pm0.04$&3.09&0.862&3.90e-4&$0.31\pm0.09$\\
12&55559.5-55593.5&35&$4.24\pm0.37$&$4.42\pm0.56$&$-2.21\pm0.04$&$-2.61\pm0.13$&9.59&0.574&0.227&$0.23\pm0.05$\\
13&55639.5-55649.5&11&$1.19\pm0.16$&$0.91\pm0.28$&$-2.20\pm0.10$&$-2.67\pm0.32$&13.08&0.668&0.028&$0.25\pm0.14$\\
14&56386.5-56394.5&9&$1.20\pm0.30$&$0.69\pm0.21$&$-1.84\pm0.21$&$-2.74\pm0.36$&17.39&0.524&0.227&$0.42\pm0.25$\\
15&56555.5-56569.5&15&$2.56\pm0.51$&$2.01\pm0.65$&$-1.81\pm0.06$&$-2.73\pm0.21$&12.74&0.750&0.001&$0.61\pm0.10$\\
16&56814.5-56818.5&5&$2.17\pm0.25$&$4.38\pm0.72$&$-2.18\pm0.06$&$-2.51\pm0.14$&4.95&0.992&8.65e-4&$0.53\pm0.17$\\
17&56818.5-56826.5&9&$5.46\pm0.49$&$6.35\pm0.81$&$-2.03\pm0.04$&$-2.49\pm0.10$&8.60&0.756&0.018&$0.34\pm0.08$\\
18&56828.5-56833.5&6&$6.60\pm0.56$&$16.30\pm2.14$&$-1.95\pm0.04$&$-2.11\pm0.06$&4.05&0.870&0.024&$0.24\pm0.09$\\
19&56833.5-56837.5&5&$5.28\pm0.83$&$5.81\pm0.11$&$-1.93\pm0.05$&$-2.24\pm0.10$&9.09&0.847&0.074&$0.32\pm0.13$\\
20&56870.5-56875.5&6&$1.98\pm0.43$&$6.50\pm0.13$&$-1.89\pm0.08$&$-2.58\pm0.19$&3.04&0.877&0.022&$0.70\pm0.22$\\
21&56880.5-56888.5&9&$2.25\pm0.62$&$4.99\pm1.95$&$-2.02\pm0.11$&$-2.71\pm0.20$&4.51&0.931&7.74e-4&$1.02\pm0.31$\\
22&56906.5-56911.5&6&$0.96\pm0.25$&$1.87\pm0.77$&$-2.19\pm0.13$&$-2.59\pm0.21$&5.13&0.735&0.096&$0.67\pm0.48$\\
23&56915.5-56917.5&3&$3.84\pm0.84$&$5.28\pm0.11$&$-1.94\pm0.07$&$-2.53\pm0.18$&7.27&0.980&0.127&$0.66\pm0.20$\\
24&56917.5-56922.5&6&$1.11\pm0.34$&$2.14\pm0.62$&$-2.12\pm0.14$&$-3.36\pm0.53$&5.19&0.921&0.009&$1.68\pm0.70$\\
25&56971.5-56979.5&9&$0.92\pm0.55$&$1.54\pm0.62$&$-1.80\pm0.19$&$-2.72\pm0.41$&5.97&0.968&3.38e-4&$1.20\pm0.56$\\
26&56979.5-56991.5&13&$1.25\pm0.24$&$1.79\pm0.60$&$-2.05\pm0.17$&$-2.79\pm0.29$&6.98&0.712&0.009&$0.70\pm0.29$\\
27&57002.5-57014.5&13&$1.56\pm0.15$&$3.50\pm0.47$&$-2.22\pm0.08$&$-2.70\pm0.14$&4.46&0.573&0.052&$0.33\pm0.14$\\
28&57024.5-57032.5&9&$0.57\pm0.12$&$1.64\pm0.31$&$-2.23\pm0.11$&$-2.95\pm0.26$&3.48&0.842&0.004&$1.09\pm0.44$\\
29&57052.5-57064.5&13&$1.73\pm0.26$&$1.11\pm0.35$&$-2.01\pm0.07$&$-2.87\pm0.21$&15.59&0.853&2.12e-4&$0.61\pm0.11$\\
30&57275.5-57282.5&8&$1.69\pm0.20$&$2.33\pm0.48$&$-2.11\pm0.07$&$-2.53\pm0.18$&7.25&0.707&0.050&$0.39\pm0.14$\\
31&57300.5-57304.5&5&$1.29\pm0.15$&$2.64\pm0.67$&$-2.27\pm0.07$&$-2.56\pm0.12$&4.89&0.546&0.341&$0.39\pm0.24$\\
32&57321.5-57329.5&9&$1.41\pm0.15$&$3.31\pm0.48$&$-2.14\pm0.08$&$-2.64\pm0.15$&4.26&0.837&0.005&$0.63\pm0.19$\\
33&57349.5-57362.5&14&$0.77\pm0.13$&$0.83\pm0.26$&$-2.14\pm0.14$&$-2.87\pm0.33$&9.28&0.718&0.004&$0.56\pm0.23$\\
34&57381.5-57412.5&32&$1.37\pm0.16$&$1.27\pm0.27$&$-2.21\pm0.08$&$-2.91\pm0.28$&10.79&0.490&0.006&$0.30\pm0.11$\\
\noalign{\smallskip}\hline
\end{tabular}
\end{center}
\end{table}

\end{document}